UNIVERSITA' DEGLI STUDI DI ROMA

" LA SAPIENZA"

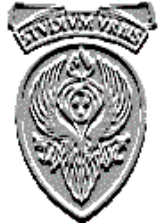

FACOLTA' DI INGEGNERIA ELETTRONICA

DIPARTIMENTO DI ENERGETICA

XVIII CICLO DI DOTTORATO DI RICERCA
IN ELETTROMAGNETISMO

TRIENNIO ACCADEMICO 2003-2005

# CLASSICAL AND QUANTUM APPROACH OF QUASI NORMAL MODES IN LINEAR OPTICAL REGIME: AN APPLICATION TO ONE DIMENSIONAL PHOTONIC CRYSTALS

Coordinatore  
Chiar$^{mo}$ Prof. Gerosa Giorgio

Relatore  
Chiar$^{mo}$ Prof. Bertolotti Mario

Dottorando  
Dott. Ing. Settimi Alessandro

Sigla di catalogazione:  
ING/INF-02

*In memory of my father*

*Franco,*

*nicknamed "Schizza" and "Pico della Mirandola" too.*

# INDEX

# INTRODUCTION

The definition of natural modes for confined structures is one of the central problems in physics, as in nuclear physics, astrophysics, etc. The main problem is due to the boundary conditions, when they are such to push out the problem from the class of Sturm-Liouville. This occurs when boundary conditions imply the presence of eigen-values, as for example when a scatterer excited from the outside gives rise to a transmitted and reflected field. An open cavity with an external or internal excitation represents a "non-canonical " problem, in the sense of a Sturm–Liouville's problem, due to the fact that cavity modes couple themselves with external modes. This problem is crucial when one intends to study light-matter interaction effects as absorption, spontaneous emission, stimulated emission, as they occur in micro-cavities.

## Quasi Normal Mode's (QNM's) approach.

The problem of the field description inside an optical cavity which radiates outside has been analyzed following several different points of view. The description of the e.m. field in an one side open and homogeneous cavity has been discussed in terms of Quasi Normal Modes (QNMs) in ref. [1]; the QNMs are defined inside a one dimensional leaky cavity, provided the open cavity is defined by a discontinuity in the refractive index which must approach its constant asymptotic value sufficiently rapidly.

In general, an optical cavity is not a conservative system, so the natural evolution of the e.m. field can not be described by an hermitian operator and the treatment of the e.m. field in terms of Normal Modes (NMs) has to be dropped. The QNM's approach considers the realistic situation in which the optical cavity is enclosed in an infinite external space. The lack of energy conservation for the open cavity gives complex, instead of real, eigen frequencies. The evolution operator for the system is not hermitian and the modes of the e.m. field are not Normal but Quasi Normal. In fact, under some conditions over the refractive index: (a) the evolution of the QNMs is similar to the hermitian one for the NMs, and (b) the e.m. field can be described as a superposition of QNMs only inside the open cavity and the QNM's functions form an orthogonal basis only according to a non-canonical metrics. The QNM's functions do not represent the e.m. field outside the cavity, while they represent the "not stationary" modes into the cavity. For the reason of not being complete in the whole space, they are called "Quasi-Normal" Modes.

Ho *et al.* in ref. [2] already made an essential first step towards the application of the QNMs to quantum electrodynamics phenomema in one side open and homogeneous cavities: the second quantization of the e.m. field in an open cavity is formulated, from first principles, in terms of the QNMs, which are eigen-solutions of the evolution equation, decaying exponentially in time as energy leaks to outside.

## One dimensional (1D) Photonic Band Gap (PBG) structures.

One dimensional (1D) Photonic Band Gap (PBG) structures [3]-[5] of finite length manifest all the aspects related to a class of problems which do not belong to the Sturm-Liouville's class: in fact, they behave as scattering objects when they are excited from outside and as open cavities when excited from inside. These structures can be realized by a stack of dielectric layers arranged according to some periodic or quasi-periodic sequence. Usually they are made alternating layers of high and low index materials of suitable thickness. If the 1D-PBG structure is periodic it may be thought as done by the superposition of a finite number of equal cells, each cell being usually constructed by two materials. The essential property of these 1D-PBGs is the existence of allowed and forbidden frequency bands and gaps, in analogy with energy bands and gaps of semiconductors. At the same time, the distribution of the e.m. field intensity along the structure strongly changes by changing the frequency or, in general if the e.m. field is coming from outside, the boundary conditions.

In the wake of theoretical promulgation for 1D-PBGs, concern has turned to the question of the behaviour of the atomic decay rate in such structures The density of states (DOS) is the fundamental feature that determines the behaviour of the system 'atom-field' and characterizes the various types of environments. The form and analytical properties of the DOS dictate the type of the approximations permissible in formulating the equations governing the time evolution of the system. When the DOS is a smooth function of frequency over the spectral range of atomic transition, the rate of spontaneous and stimulated emission is described by the Fermi's golden rule. Abrupt changes in the DOS and photon localization effects may drastically modify the emission dynamics. This modification takes the form of long time memory effects and non-Markovian behaviour in the atom-reservoir interaction.

A substantial modification in the DOS can be effected by means of PBG structures. It has been suggested that this would be accompanied by classical light localization, the inhibition of single-photon emission, fractionalized single-atom inversion, a photon-atom bound state and anomalously large vacuum Rabi's splitting.

# QNM's approach for 1D-PBGs.

In the period of three years, required for the Ph-D course in Electromagnetism, some papers have been published and recently submitted in order to extend the QNM's treatment for the description of the e.m. field inside 1D-PBGs. In ref. [6], a discussion is presented about the completeness of the QNM's representation and, moreover, a discussion on the complex frequencies, as well as the corresponding field distributions. In refs. [7], the role of the QNM's frequencies in the transmission coefficient is clarified; an application is performed for a symmetric quarter wave structure.

The second quantization based on the QNM's treatment has been extended to 1D-PBG structures. In ref. [8], starting from this representation, the Feynman's propagator is introduced to calculate the decay rate of a dipole inside a 1D-PBG, in presence of vacuum fluctuations outside the structure.

In ref. [9], applying the second quantization for QNMs, the problem of counter-propagation for two field pumps inside a 1D-PBG structure is discussed. The links between the usual annihilation and creation operators, describing the two pumping fields, and the not canonical QNM's operators of a 1D-PBG are investigated; an application is performed for a symmetric structure.

In ref. [10], the QNM's second quantization is applied to discuss the quantum problem of an atom embedded inside a 1D-PBG, pumped by two counter-propagating laser-beams. The e.m. field is quantized in terms of the QNMs in the structure and the atom, modelled as a two-level system, is assumed to be weakly coupled to just one of the QNMs.

In refs. [11], the stimulated emission is discussed, in strong coupling regime, for an atom embedded inside a 1D-PBG structure which is pumped by two counter-propagating laser beams. Quantum electro-dynamics is applied to model the atom-e.m. field coupling, by considering the atom as a two level system, the e.m. field as a superposition of normal modes, the coupling in dipole approximation, and the equations of motion in Wigner-Weisskopf's and rotating wave approximations. Besides, the QNM's quantization is adopted for the 1D-PBG, interpreting the local density of states (LDOS) as the local density of probability to excite one QNM of the structure; therefore, this LDOS depends on the phase difference of the two laser beams.

The present Ph-D thesis re-organizes the contents of the papers [6]-[11] in a systematic form, examining the QNM's approach for 1D-PBG structures still more thoroughly. This Ph-D thesis is subdivided in three parts and consists of seven chapters. The first part, concerning the QNM's approach for open cavities in classical electrodynamics, consists of three chapters. Chapter 1 is entitled "Quasi Normal Mode's (QNM's) approach for open cavities". Chapter 2 is "QNM's approach for open cavities in absence of external pumping: an application to one dimensional (1D)

Photonic Crystals (PCs)". Chapter 3, "QNM's approach for open cavities excited by an external pumping: Transmission properties of a 1D-PBG structure". The second part, concerning the QNM's approach for open cavities in quantum electrodynamics, consists of two chapters. Chapter 4 is entitled "Non canonical QNM' Quantization for open cavities". Chapter 5 is "Density of probability for QNMs: link with boundary conditions of open cavities". The third part, concerning the QNM's approach for emission processes in open cavities, consists of two chapters. Chapter 6 is entitled "Spontaneous Emission and coherent control of Stimulated Emission inside one dimensional Photonic Crystals (Weak Coupling regime)". Chapter 7, "Coherent Control of Stimulated Emission inside 1D-PBG structures (Strong Coupling regime)".

[1] P. T. Leung, S. Y. Liu, and K. Young, Phys. Rev. A **49**, 3057 (1994); P. T. Leung, S. S. Tong, and K. Young, J. Phys. A **30** 2139 (1997); P. T. Leung, S. S. Tong, and K. Young, J. Phys. A **30**, 2153 (1997); E. S. C. Ching, P. T. Leung, A. Maassen van der Brink, W. M. Suen, S. S. Tong, and K. Young, Rev. Mod. Phys. **70**, 1545 (1998); P. T. Leung, W. M. Suen, C. P. Sun and K. Young, Phys. Rev. E **57**, 6101 (1998).

[2] K. C. Ho, P. T. Leung, Alec Maassen van den Brink and K. Young, Phys. Rev. E **58**, 2965 (1998).

[3] S. John, Phys. Rev. Lett. **53**, 2169 (1984); S. John, Phys. Rev. Lett. **58**, 2486 (1987); E. Yablonovitch, Phys. Rev. Lett. **58**, 2059 (1987); E. Yablonovitch and T. J. Gmitter, Phys. Rev. Lett.. **63**, 1950 (1989).

[4] J. Maddox, Nature (London) **348**, 481 (1990); E. Yablonovitch and K.M. Lenny, Nature (London) **351**, 278, 1991; J. D. Joannopoulos, P. R. Villeneuve, and S. H. Fan, Nature (London) **386**, 143 (1997).

[5] J. D. Joannopoulos, *Photonic Crystals: Molding the Flow of Light* (Princeton University Press, Princeton, New York, 1995); K. Sakoda, *Optical properties of photonic crystals* (Springer Verlag, Berlin, 2001); K. Inoue and K. Ohtaka, *Photonic Crystals: Physics, Fabrication, and Applications* (Springer-Verlag, Berlin, 2004).

[6] A. Settimi, S. Severini, N. Mattiucci, C. Sibilia, M. Centini, G. D'Aguanno, M. Bertolotti, M. Scalora, M. Bloemer, C. M. Bowden, Phys. Rev. E **68**, 026614 (2003).

[7] S. Severini, A. Settimi, C. Sibilia, M. Bertolotti, A. Napoli, A. Messina, *Quasi Normal Frequencies in open cavities: an application to Photonic Crystals*, Acta Phys. Hung. B **23/3-4**, 135-142 (2005); A. Settimi, S. Severini, B. Hoenders, *Quasi Normal Modes description of transmission properties for Photonic Band Gap structures*, J. Opt. Soc. Am. B, **26**, 876-891 (2009).

[8] S. Severini, A. Settimi, C. Sibilia, M. Bertolotti, A. Napoli, A. Messina, Phys. Rev. E **70**, 056614 (2004).

[9] S. Severini, A. Settimi, C. Sibilia, M. Bertolotti, A. Napoli, A. Messina, *Quantum counter-propagation in open cavities via Quasi Normal Modes approach*, Laser Physics, **16**, 911-920 (2006).

[10] A. Settimi, S. Severini, C. Sibilia, M. Bertolotti, M. Centini, A. Napoli, N. Messina, Phys. Rev. E **71,** 066606 (2005).

[11] A. Settimi, S. Severini, C. Sibilia, M. Bertolotti, A. Napoli, A. Messina, *Coherent Control of stimulated emission inside one dimensional Photonic Crystals: strong coupling regime*, Eur. Phys. J. B. **50**, 379-391 (2006); A. Settimi, S. Severini, C. Sibilia, M. Bertolotti, A. Napoli, A. Messina, *ERRATUM Coherent control of stimulated emission inside one dimensional photonic crystals: strong coupling regime,* Eur. Phys. J. B. **69**, 613-614 (2009).

# First Part:

# QNM's approach for open cavities in classical electrodynamics.

# Chapter 1

# Quasi Normal Mode's approach for open cavities

## 1. Introductive outlines.

Resonators have the property to enhance the electromagnetic field in their interior, a property which has many applications and makes them essential elements of most lasers [1].
In the radio-frequency and microwave range, resonators are usually closed cavities with dimensions comparable with the wavelength. In the optical case at the beginning, it was neither possible nor convenient to make cavities of dimensions comparable with the light wavelength. This precluded the possibility of using closed resonators which are currently widely employed in the microwave range and which actually represent a cavity with reflecting walls, because the number of low-loss modes at optical frequencies would here be prohibitively large. To obtain a selection of the oscillating modes, it was therefore necessary to make recourse to open side cavities. These open resonators are without side walls and consist, in their simplest configuration, of two oppositely mounted mirrors between which the active medium is placed. It is the geometry of the mirror arrangement that accounts for radiation propagating in the preferred direction which, in its turn, reduces dramatically the number of low-loss modes compared with closed resonators.
The first resonator configuration employed in optics was the Fabry-Perot interferometer consisting of two parallel plane mirrors. One of them is partially transmitting to take out the generated radiation. It is still one of the most widely used types of laser resonator. Its popularity stems not only from its extreme simplicity, but also from the inherent possibility of obtaining high-energy outputs.

Many years were spent in establishing the very existence of the modes, their specific features in different resonators and the role played in their formation by the active medium, to learn whether the modes close in losses are excited separately or simultaneously, and how all this affects the characteristics of the laser radiation.
Underlying the theory of open resonators, like any other resonator devices, is the concept of fundamental oscillations, or modes. By the resonator mode one understands a field distribution whose dependence on time in the absence of external sources is described throughout the whole volume by the same factor $\exp(-i\omega t)$ where $\omega$ is the fundamental circular frequency which, in the general case, is complex: $\omega = \omega' + i\omega''$, being $\omega'$ and $\omega''$ real quantities. For empty resonators with sources of loss $\omega'' < 0$, i.e. the oscillations die out in time while the shape of the spatial field distribution remains unchanged.

The modes of an optical resonator can practically always be represented as a combination of several light beams transforming into one another on reflection from the mirrors or interfaces, thus ensuring reproducibility of the process in time. Thus the modes of the simplest linear resonators like the plane resonator, which is still in use at present, consist of two spatially matched beams propagating in opposite directions.

In this Ph-D thesis, only one-dimensional (1D) structures are considered. Several methods are available to derive the electromagnetic field inside the structures together with transmission and reflection coefficient. Two of these methods are the transfer matrix [1] and the ray method [2], which can be applied to any kind of 1D structure with piecewise constant refractive index.

## 1. 1. QNM's literature.

The definition of natural modes for confined structures is one of the central problems in physics [4], as in nuclear physics, astrophysics, etc. The main problem is due to the boundary conditions, when they are such to push out the problem from the class of Sturm-Liouville. This occurs when boundary conditions imply the presence of eigen-values, as for example when a scatterer excited from the outside [5] gives rise to a transmitted and reflected field. An open cavity with an external or internal excitation represents a "non-canonical " problem, in the sense of a Sturm–Liouville's problem, due to the fact that the cavity modes couple themselves with the external modes. This problem is crucial when one intends to study light-matter interaction effects as absorption, spontaneous emission, stimulated emission, as they occur in micro-cavities.

The problem of the field description inside an open cavity which radiates outside has been discussed following several different points of view [6].

The description of the e.m. field in an one side open and homogeneous cavity has been discussed in terms of Quasi Normal Modes (QNMs) in refs.[1]; QNMs are discussed in a one dimensional leaky cavity, provided the cavity is defined by a discontinuity in the refractive index which must approach its constant asymptotic value sufficiently rapidly.

In general, an optical cavity is not a conservative system, so the natural evolution of the e.m. field can not be described by an hermitian operator and the treatment of the e.m. field in terms of Normal Modes (NMs) [8] has to be dropped. The QNM's approach considers the realistic situation in which the open cavity is enclosed in an infinite external space. The lack of the energy conservation for the open cavity gives complex, instead of real, eigen frequencies. The evolution operator for the system is not hermitian and the modes of the e.m. field are not Normal but Quasi Normal. In fact, under some conditions over the refractive index: (a) the evolution of the QNMs is similar to the hermitian one for the NMs, and (b) the e.m. field can be described as a superposition of QNMs only inside the cavity and the QNM's functions form an orthogonal basis only according to a non-canonical metrics. The QNM's functions do not represent the e.m. field outside of the cavity while they represent the "not stationary" modes into the cavity. For the reason of not being complete in the whole space, they are called "Quasi-Normal" Modes.

In this chapter, some necessary results of ref. [6] are stated; in ref. [6], the QNM's approach has been discussed for double side open and inhomogeneous cavities. The description of the scalar field in 1D open cavities is represented in terms of QNMs; the QNM's description is extended from an one side open and homogeneous cavity to an open cavity from both ends, with an inhomogeneous distribution of refractive index. In section 2, the NMs of a closed system are compared with the QNMs of an open system, which QNMs are defined in absence of an external pumping. In section 3, the QNM's linear space is defined in terms of the QNM's inner product, eigen frequencies and functions. In section 4, the QNMs of a linear Fabry-Perot (FP) cavity are calculated as an useful example. Conclusions are given in section 5.

## 2. Closed and open systems.

To set the stage, start with the usual conservative case. For simplicity, consider an infinite interval $U = (-\infty, \infty)$ in one dimension, and scalar functions $\phi(x,t)$ defined on $U$ and vanishing at both ends $x = \pm\infty$ of the interval. It is straightforward to generalize to other boundary conditions, e.g. $d\phi/dx$ vanishing at both ends $x = \pm\infty$ of the interval. Then, if $H$ is an Hermitian operator bounded from below but unbounded from above, the family of eigen-functions $\{f_n(x)\}$ defined by $Hf_n(x) = \omega_n f_n(x)$ is complete and orthogonal, being the eigen-values $\omega_n$ real. Mathematically, completeness means that any function $\phi(x,t)$ of this class can be expanded as

$$\phi(x,t) = \sum_n a_n(t) f_n(x), \qquad (1.2.1)$$

whereas orthogonality ensures that the representation is unique, and also allows the coefficients $a_n(t)$ to be found by projecting in the standard way.

These elementary ideas are easily generalized to the wave equation

$$\left[\frac{\partial^2}{\partial x^2} - \rho(x)\frac{\partial^2}{\partial t^2}\right]\phi(x,t) = 0, \qquad (1.2.2)$$

where $\rho(x) = [n(x)/c]^2$, being $n(x)$ the refractive index and $c$ the velocity of light in vacuum. The eigen-functions and the eigen-values are defined by

$$\left[\frac{\partial^2}{\partial x^2} + \omega_n^2 \rho(x)\right] f_n(x) = 0. \tag{1.2.3}$$

First, consider these equations defined on a finite interval $C = [0, d]$, with $\phi(x,t)$ vanishing at both ends $x = 0$ and $x = d$ so that the system is closed and conservative. The operator $-\partial^2/\partial x^2$ is then Hermitian, positive definite and unbounded from above. By the same arguments, the family of eigen functions $\{f_n(x)\}$ is complete and eq. (1.2.1) holds. The eigen-values $\omega_n^2$ are real and positive, so one only needs the positive frequencies $0 \leq \omega_1 \leq \omega_2 \leq \ldots$ but not the corresponding set $\omega_{-n} = -\omega_n$, and this case can be emphasized writing eq. (1.2.1) as

$$\phi(x,t) = \sum_{n>0} a_n(t) f_n(x). \tag{1.2.4}$$

Inner products are defined by

$$\langle \phi | \psi \rangle = \int_0^d dx \, \phi^*(x) \rho(x) \psi(x), \tag{1.2.5}$$

under which the eigen-states $\{f_n(x)\}$ are mutually orthogonal.

Now, these notions may be generalized to open systems defined by the wave equation under suitable restrictions on $\rho(x)$ and on the class of functions to be represented. First, let $\rho(x)$, defined on $U = (-\infty, \infty)$, satisfy two conditions: (a) $\rho(x)$ has a step discontinuity or stronger discontinuity at $x = 0$ and $x = d$; (b) $\rho(x) = \rho_0$ for $x < 0$ and $x > d$. These condition are referred as the discontinuity condition and the "no tail" condition respectively [1]. The discontinuity marks the boundaries of the finite interval $C = [0, d]$, within which an eigen-function expansion is sought. There are advantages in considering a finite interval. Physically, the interval may describe a laser cavity, and it is desirable to describe its electrodynamics without reference to the outside. Secondly, attention is restricted to differentiable functions $\phi(x,t)$ not satisfying the nodal conditions, $\phi(0,t) \neq 0$ and $\phi(d,t) \neq 0$, but the outgoing wave conditions

$$\begin{cases} \partial_x \phi(x,t) = \sqrt{\rho_0} \partial_t \phi(x,t) & , \quad x = 0^- \\ \partial_x \phi(x,t) = -\sqrt{\rho_0} \partial_t \phi(x,t) & , \quad x = d^+ \end{cases}. \tag{1.2.6}$$

The escape of the waves to infinity characterizes an open system. The outgoing wave conditions (rather than the nodal conditions) at $x = 0^-$ and $x = d^+$ renders the operator $-\partial^2/\partial x^2$ non-Hermitian on the interval $C = [0, d]$, and the familiar proofs of completeness and orthogonality break down. For such open systems, completeness refers to an expansion in terms of the eigen-functions, but also with the outgoing wave conditions at $x = 0^-$ and $x = d^+$. Thus, the eigen-values $\omega_n$ are complex (with $\operatorname{Im}\omega_n < 0$, because the amplitude decays). The eigen-functions $\{f_n(x)\}$ are therefore not Normal Modes (NMs), but Quasi Normal Modes (QNMs) [1]. Quite generally the QNM's eigen frequencies exist in pairs, related by $\omega_{-n} = -\omega_n^*$, where by convention $0 \leq \operatorname{Re}\omega_1 \leq \operatorname{Re}\omega_2 \leq \ldots$. The case where one (or more) QNM's frequency falls on the imaginary axis is readily dealt with. But $\omega_{-n}^2 \neq \omega_n^2$, so the QNM's eigen-functions $f_{-n}(x)$ and $f_n(x)$ are linearly independent. Thus, the eigen-function expansion to be sought is (1.2.1) rather than (1.2.4), i.e. the full set of eigen-functions is necessary, not just those with $\operatorname{Re}\omega > 0$. Many models contain a parameter $\varepsilon \approx \operatorname{Im}\omega_n$, characterizing the amount of leakage. The above discussion shows that there is a fundamental difference between the NMs for $\varepsilon = 0$ and the QNMs for $\operatorname{Im}\omega_n \neq 0$—the latter are double in number. Under the conditions stated, the set of all QNMs $\{f_n(x)\}$ of such an open system is complete only in the interval $C = [0, d]$, in the sense of eq. (1.2.1).

Outgoing waves conditions (1.2.6), referred to the QNM's functions $\{f_n(x)\}$, assume the expressions:

$$\begin{cases} \partial_x f_n(x)\big|_{x=0} = -i\omega_n \sqrt{\rho_0} f_n(0) \\ \partial_x f_n(x)\big|_{x=d} = i\omega_n \sqrt{\rho_0} f_n(d) \end{cases}. \quad (1.2.7)$$

Applying the outgoing wave conditions (1.2.7) for the QNMs, it is easy to verify that, in a symmetric cavity, the QNM's function $f_n(x)$ satisfy the following relations:

$$\begin{cases} f_n(d) = (-1)^n f_n(0) \\ \partial_x f_n(x)\big|_{x=d} = -(-1)^n \partial_x f_n(x)\big|_{x=0} \end{cases}. \quad (1.2.8)$$

The outgoing wave conditions (1.2.8) for a symmetric cavity are incompatible with the cyclic boundary conditions, defining the space of NMs, which are used in canonical quantum electrodynamics [8]:

$$\begin{cases} f_n(d) = f_n(0) \\ \partial_x f_n(x)\big|_{x=d} = \partial_x f_n(x)\big|_{x=0} \end{cases}. \quad (1.2.9)$$

It is easy to prove that $Q \cap N = \varnothing$, being $Q$ the space of QNMs, defined by eq. (1.2.3) in addition to eq. (1.2.8), and being $N$ the space of NMs, defined by eq. (1.2.3) in addition to eq. (1.2.9). In fact, under the conditions stated, the set of all QNM's functions $\{f_n(x)\}$ of such an open system is complete in the interval $C = [0, d]$, in the sense of eq. (1.2.1); and, a scalar function $\phi(x, t)$ of the QNM's space $Q$ can not be represented in the NM's space $N$, because of:

$$\phi(d,t) = \sum_n a_n(t) f_n(d) = \sum_n a_n(t)(-1)^n f_n(0) \neq \sum_n a_n(t) f_n(0) = \phi(0,t). \qquad (1.2.10)$$

## 2. 1. Comparison of QNM's with other expansions.

There are many different ways of expanding a wave function $\phi(x,t)$, so it is necessary to spell out the unique features and advantages of the method developed here. For this purpose, three different expansion schemes [1] can be compared.

(A) The first is the scheme developed here in terms of the QNM's functions $f_n(x)$, using both the wave function $\phi(x,t)$ and the Lagrange conjugate momentum $\bar{\phi}(x,t) = \rho(x)\partial_t \phi(x,t)$, and with the coefficients $a_n(t)$ given by

$$a_n(t) = \frac{i}{2\omega_n} \int_{0^-}^{d^+} \left[ f_n(x)\bar{\phi}(x,t) + \bar{f}_n(x)\phi(x,t) \right] dx + \\ + \frac{i}{2\omega_n} \sqrt{\rho_0} \left[ f_n(d)\phi(d,t) + f_n(0)\phi(0,t) \right] \qquad (1.2.11)$$

being:

$$a_n(t) = a_n(t=0)\exp(-i\omega_n t) \\ a_{-n}(t) = a_n^*(t) \qquad (1.2.12)$$

(B) If one were to abandon the second component $\bar{\phi}(x,t)$, the expansion will involve a set of coefficients, $b_n(t=0)$, calculated from eq. (1.2.11), without the $\bar{\phi}(x,t)$ contribution. In other words, using $\bar{f}_n(x) = -i\omega_n \rho(x) f_n(x)$, it results

$$b_n(t=0) = \frac{1}{2} \int_{0^-}^{d^+} f_n(x)\rho(x)\phi(x,t=0) dx + \\ + \frac{i}{2\omega_n} \sqrt{\rho_0} \left[ f_n(d)\phi(d,t=0) + f_n(0)\phi(0,t=0) \right] \qquad (1.2.13)$$

Equation (1.2.13) can be regarded as the natural expansion (i.e. method A) applied, at the initial time $t=0$, to the pair $[\phi(x), \bar{\phi}(x)] = [\phi(x), 0]$, so the sum will certainly give $\phi(x)$ correctly.

(C) The third method makes use of the resolution of the identity [1]

$$\frac{1}{2}\rho(x)\sum_n f_n(x) f_n(x') = \delta(x - x'), \qquad (1.2.14)$$

which leads directly to an expansion with the coefficients

$$c_n(t=0) = \frac{1}{2} \int_{0^-}^{d^+} f_n(x)\rho(x)\phi(x,t=0) dx. \qquad (1.2.15)$$

In other words the surface terms in eq. (1.2.13) are ignored. It may seem surprising that, at the initial time $t=0$, neither the $\bar{\phi}(x)$ contribution nor the surface terms are necessary for representing $\phi(x)$; however, it is shown below that there are definite advantages when these are retained, as in method A.

Now, these methods of expansion can be compared, so explaining the advantages of the method developed here.

First and foremost, methods B and C will not solve the dynamical evolution for $t>0$ simply attaching phase factors $\exp(-i\omega_n t)$. This is hardly surprising since there is no knowledge of the initial $\partial_t \phi(x,t)$.

Methods A, B and C provide explicit projection formulae for the coefficients, respectively (1.2.11), (1.2.13) and (1.2.15). By a straightforward WKB analysis [1], it is easy to demonstrate several properties of these coefficients for functions $\phi(x)$ with bounded derivatives up to $\partial_x^2 \phi(x)$ and $\bar{\phi}(x)/\rho(x)$ being differentiable. (i) For a system with step discontinuities, these coefficients behave asymptotically as $|a_n f_n(x)| \sim n^{-3}$, $|b_n f_n(x)| \sim n^{-2}$, $|c_n f_n(x)| \sim n^{-1}$, showing that method A is the most rapidly convergent and thus the most effective in practice. (ii) The analogous sums for $\partial_x \phi(x)$ (or in the case of method A, also $\bar{\phi}(x)$) would involve $\partial_x f_n(x)$ (or $\bar{f}_n(x)$) rather than $f_n(x)$, differing by a factor of $-i\omega_n$ (or $-i\omega_n \rho(x)$) which is asymptotically $\propto n$. Thus, the terms in the sum go as $|a_n f_n'(x)| \sim n^{-2}$, $|b_n f_n'(x)| \sim n^{-1}$, $|c_n f_n'(x)| \sim n^0$. Method C fails completely for $\partial_x \phi(x)$. (iii) Whether method C converges for $\phi(x)$ and whether method B converges for $\partial_x \phi(x)$ depends on the phase of the summands, since their magnitudes go as $n^{-1}$. For $0<x<d$, the phases are oscillatory in $n$ (in fact linearly increasing with $n$, by an amount not equal to an integral multiple of $2\pi$), so the sum is conditionally convergent. But for $x=0$ and $x=d$, the phase is asymptotically constant, so these sums are logarithmically divergent.

In other words, method C (the naive projection using (1.2.15)) is the least convergent. Incorporation of the surface term (method B) improves it by one power; including the second component (method A) improves it by one more power. Thus, quite apart from dynamical evolution, method A is the best.

## 3. QNM's linear space.

The QNM's norm is defined by [1]

$$\langle f_n | f_n \rangle = 2\omega_n \int_0^d \rho(x) f_n^2(x) dx + i\sqrt{\rho_0}\left[ f_n^2(0) + f_n^2(d) \right]. \tag{1.3.1}$$

This generalized norm refers explicitly to $\omega_n$, and is therefore not immediately applicable to wave functions which are not eigen-functions, nor immediately generalizable to an inner product.

Ref. [10] develops further the linear space structure that supports these concepts. Central to these developments is the introduction of a generalized inner product. This has all the usual properties, except that it is linear, rather than conjugate linear, in the bra vector. This inner product has the following desirable properties: (a) it agrees with the generalized norm (1.3.1) for the inner product of an eigen-function with itself; (b) the projection of the e.m. field $\phi(x,t)$ is proportional to the inner product with the QNM's functions $f_n(x)$ [see eq. (1.2.11)]; (c) the time evolution equation can be written as a first order differential equation involving an Hamiltonian operator, $H$, formally analogous to the Schroedinger's equation. Most importantly, $H$ is self-adjoint under this inner product, even though the system is not conservative; (d) consequently, the eigen-functions are mutually orthogonal, and as an important corollary, the eigen-function expansion is unique.

The QNMs of such open systems are both complete and orthogonal. Compared with conservative systems described by Hermitian operators in the usual sense, the only missing element is the lack of positivity, e.g. in the generalized norm. So, most of the framework of mathematical physics based on eigenfunction expansion for *conservative* systems has been successfully generalized to a broad class of *not conservative* systems in which the loss is due to leakage.

### 3. 1. Linear space.

Consider the set, $\Gamma$, of function pairs $[\phi(x,t), \bar{\phi}(x,t)]$, where each of $\phi(x,t)$ and $\bar{\phi}(x,t)$ is defined on $[0,d]$, $\phi(x,t)$ and $\bar{\phi}(x,t)/\rho(x)$ are differentiable, and the two functions satisfy conditions (1.2.6). The set $\Gamma$ forms a linear space under addition and multiplication by complex scalars [10]. A ket vector is used to denote the column vector

$$|\Phi(t)\rangle = \begin{pmatrix} \phi(x,t) \\ \bar{\phi}(x,t) \end{pmatrix}. \tag{1.3.2}$$

A QNM's vector is a function pair $[f_n(x), \bar{f}_n(x)] \in \Gamma$ whose first component satisfies eq. (1.2.3), and whose second component is given by $\bar{f}_n(x) = -i\omega_n \rho(x) f_n(x)$, such that:

$$|f_n\rangle = \begin{pmatrix} f_n(x) \\ \hat{f}_n(x) \end{pmatrix}. \qquad (1.3.3)$$

The complex value $\omega_n$ in eq. (1.2.3) is the eigen-value, with $\text{Im}\,\omega_n < 0$ describing the rate of decay of the amplitude.

### 3. 2. Inner product.

Given two elements $[\psi(x,t), \bar{\psi}(x,t)]$ and $[\phi(x,t), \bar{\phi}(x,t)]$ of $\Gamma$, the generalized inner product is defined by [10]

$$\langle \Psi(t)|\Phi(t)\rangle = i\int_0^{d^+} \left[\bar{\psi}(x,t)\phi(x,t) + \psi(x,t)\bar{\phi}(x,t)\right]dx + i\sqrt{\rho_0}[\psi(d,t)\phi(d,t) + \psi(0,t)\phi(0,t)], \qquad (1.3.4)$$

which is symmetric and linear in both the bra and ket vectors (rather than conjugate linear in the bra vector). Using eq. (1.3.3), it is readily seen that the inner product of a QNM's function with itself agrees exactly with the generalized norm (1.3.1); the advantage of (1.3.4) is that it makes no reference to any eigen-values; this is possible only because the second component $\bar{\psi}(x,t)\,(\bar{\phi}(x,t))$ appears [10]. Moreover, the projection formula (1.2.11) for the eigen-function expansion can now be written as

$$a_n(t) = \frac{\langle f_n|\Phi(t)\rangle}{\langle f_n|f_n\rangle}. \qquad (1.3.5)$$

### 3. 3. Operators on linear space.

A linear operator is valid only if it maps $\Gamma$ into $\Gamma$, i.e. it maps into function pairs $[\phi(x,t), \bar{\phi}(x,t)]$ that satisfy conditions (1.2.6). It is readily seen that the following are valid operators: the identity operator $I$; the operator $\rho(x)I$, using the condition that $\rho(x) = \rho_0$ for $x < 0$ and $x > d$; and any 'potential' operator $V$ which vanishes outside $C = [0,d]$. Of particular importance is the time-dependent evolution, which can be written as [10]

$$\frac{\partial}{\partial t}|\Phi(t)\rangle = -iH|\Phi(t)\rangle, \qquad (1.3.6)$$

where

$$H = i\begin{pmatrix} 0 & \rho^{-1}(x) \\ \partial_x^2 & 0 \end{pmatrix}. \qquad (1.3.7)$$

The first component of eq. (1.3.7) reproduces the identification of $\bar{\phi}(x,t)$ as $\rho(x)\partial_t\phi(x,t)$. The set of valid operators forms an algebra under addition, multiplication by complex scalars, and composition.

### 3. 4. Self-adjoint operators.

Given the generalized inner product, the adjoint $A^\dagger$ of any operator $A$ is defined as follows

$$\langle\psi|\{A^\dagger|\phi\rangle\} = \langle\phi|\{A|\psi\rangle\} \tag{1.3.8}$$

and for a self-adjoint operator ($A^\dagger = A$), the following notation is adopted, which is suggestive of left–right symmetry

$$\langle\psi|\{A|\phi\rangle\} = \langle\phi|\{A|\psi\rangle\} \equiv \langle\psi|A|\phi\rangle. \tag{1.3.9}$$

The adjoint operator is defined without complex conjugation, so if $A$ is self-adjoint, then so is $\alpha A$ for any complex number $\alpha$. The time-evolution operator $H$ is self-adjoint [10]. Usually (say in quantum mechanics), the hermiticity of the Hamiltonian operator is intimately related to the conservation of probability; the QNM's approach succeeds in casting the dynamics of a not conservative system in terms of a self-adjoint evolution operator.

### 3. 5. Eigen frequencies and functions.

The QNM's frequencies and functions, can now be defined simply by [10]

$$H|f_n\rangle = \omega_n|f_n\rangle, \tag{1.3.10}$$

which incorporates both the differential equation for $f_n(x)$ as well as the definition $\bar{f}_n(x) = -i\omega_n\rho(x)f_n(x)$. Since $H$ is self-adjoint under the definition introduced, it is easily shown, by an usual procedure [10], that if $\omega_m \neq \omega_n$, then $\langle f_m|f_n\rangle = 0$. This leads immediately to the uniqueness of the completeness sum (1.2.1). It is seen that the mathematical structure is in place to carry over essentially all the familiar tools based on eigen-function expansions. The only exception is the lack of a positive-definite norm, and with it a simple probability or energy interpretation. This is hardly surprising since on the interval $C = [0,d]$ (as in any finite parts of space), probability or energy is not conserved.

### 3. 6. Damped harmonic oscillator.

It is interesting to observe that for each single QNM's function

$$\phi_n(x,t) = f_n(x)e^{-i\omega_n t}, \qquad (1.3.11)$$

three appropriate real constants $m, \gamma$ and $k$ exist in such a way that $\phi_n(x,t)$ satisfies the following damped harmonic oscillator equation [11]

$$m\partial_{tt}\phi_n + \gamma\partial_t\phi_n + k\phi_n = 0. \qquad (1.3.12)$$

This point can be easily understood substituting eq. (1.3.11) into eq. (1.3.12), obtaining

$$\begin{cases} m(\omega_{nR}^2 - \omega_{nI}^2) - \gamma\omega_{nI} - k = 0 \\ 2m\omega_{nR}\omega_{nI} + \gamma\omega_{nR} = 0 \end{cases} \qquad (1.3.13)$$

which is an always solvable system, where the complex frequency $\omega_n$ has been decomposed in its real and imaginary parts $\omega_n = \omega_{nR} + i\omega_{nI}$ (with $\omega_{nR} = \operatorname{Re}\omega_n$ and $\omega_{nI} = \operatorname{Im}\omega_n$). In conclusion, the temporal behaviour expressed by eq. (1.3.11) is strictly similar to the one coming from the damped harmonic oscillator theory [11]. For the QNM's approach, however, the origin of damping is not the presence of a first order time derivative for the field, but the escape of energy from the two sides of the structure, due to the outgoing wave conditions (1.2.6).

On the analogy of the quantum harmonic oscillator, it is suggestive that the Hamiltonian (1.3.7) can be expressed in the factorized form

$$H = A^\dagger A, \qquad (1.3.14)$$

being $A$ and $A^\dagger$ respectively the annihilation and creation operators

$$\begin{aligned} A &= \sum_n \sqrt{\omega_n} \, |f_{n-1}\rangle\langle f_n| \\ A^\dagger &= \sum_n \sqrt{\omega_{n+1}} \, |f_{n+1}\rangle\langle f_n| \end{aligned} \qquad (1.3.15)$$

defined in the QNM's linear space:

$$\begin{aligned} A|f_n\rangle &= \sqrt{\omega_n} \, |f_{n-1}\rangle \\ A^\dagger|f_n\rangle &= \sqrt{\omega_{n+1}} \, |f_{n+1}\rangle \end{aligned}. \qquad (1.3.16)$$

## 4. QNMs of a linear Fabry-Perot cavity.

In this section, the Quasi Normal Mode's (QNM's) approach is applied to a linear Fabry-Perot (FP) cavity, in order to discuss some properties characterizing the QNM's eigen frequencies and functions of the FP cavity.

As schematically reported in fig. 1.1, consider a particular class of refractive index:

$$n(x) = \begin{cases} n & , \quad x \in [0,d] \\ n_0 & , \quad \text{outside} \end{cases}. \quad (1.4.1)$$

If the e.m. field equation (1.2.3) is solved, adding the outgoing waves conditions (1.2.7), the QNM's frequencies of the linear and homogeneous cavity are calculated, as shown in the following relations:

$$\omega_k = k\alpha - i\beta, \quad (1.4.2)$$

where

$$\alpha = \frac{\pi}{(n/c)d} \quad , \quad \beta = \frac{1}{(n/c)d}\ln\left(\frac{n+n_0}{n-n_0}\right) \quad , \quad k \in \mathbb{Z} = \{0, \pm 1, \pm 2, ...\}, \quad (1.4.3)$$

being $n$ just the refractive index of linear FP cavity. The QNM's functions of the linear and homogeneous cavity are calculated as:

$$f_k(x) = K_0\left[\cos\left(n\frac{\omega_k}{c}x\right) - i\frac{n_0}{n}\sin\left(n\frac{\omega_k}{c}x\right)\right], \quad (1.4.4)$$

where $K_0$ is a suitable constant of normalization. The QNM's functions (1.4.4) can be re-expressed as in the following expression:

$$f_k(x) = Ae^{in\frac{\omega_k}{c}x} + Be^{-in\frac{\omega_k}{c}x}, \quad (1.4.5)$$

being the coefficients $A$ and $B$ equal to: $A = \frac{K_0}{2}\frac{n-n_0}{n}$ and $B = \frac{K_0}{2}\frac{n+n_0}{n}$.

Some useful results are reported, involving the QNM's frequencies (1.4.2)-(1.4.3):

$$\cos\left(\frac{\omega_k nd}{c}\right) = (-1)^k \frac{n^2+n_0^2}{n^2-n_0^2} \quad , \quad \sin\left(\frac{\omega_k nd}{c}\right) = -i(-1)^k \frac{2n}{n^2-n_0^2}$$

$$e^{i\frac{\omega_k nd}{c}} = e^{-i\frac{\omega_k^* nd}{c}} = \frac{n+n_0}{n-n_0}(-1)^k \quad , \quad e^{-i\frac{\omega_k nd}{c}} = e^{i\frac{\omega_k^* nd}{c}} = \frac{n-n_0}{n+n_0}(-1)^k \quad . \quad (1.4.6)$$

$$e^{i\frac{2\omega_k nd}{c}} = \left(\frac{n+n_0}{n-n_0}\right)^2 \quad , \quad e^{-i\frac{2\omega_k nd}{c}} = \left(\frac{n-n_0}{n+n_0}\right)^2$$

Applying eqs. (1.4.6), in the case of a linear FP cavity, it is immediately demonstrable that: the QNM's frequencies (1.4.2) and the QNM's functions (1.4.4) satisfy the general properties [1]

$$\omega_{-k} = -\omega_k^*$$
$$f_{-k}(x) = f_k^*(x) \quad ; \tag{1.4.7}$$

besides, the outgoing wave conditions (1.2.7), referred to the QNMs, can be specified as shown in the following relations:

$$\begin{cases} f_k(d) = (-1)^k f_k(0) \\ \partial_x f_k(x)\big|_{x=d} = -(-1)^k \partial_x f_k(x)\big|_{x=0} \end{cases} \tag{1.4.8}$$

An useful integral is $I_{n,m} = \int_0^d \rho(x) f_n(x) f_m(x) dx$ which can be calculated as

$$I_{n,m} = K_0^2 \frac{n_0}{c} \frac{1+(-1)^{n+m}}{i(\omega_n + \omega_m)} + \delta_{n,m} \frac{K_0^2}{2} \frac{n^2 - n_0^2}{c^2} d, \tag{1.4.9}$$

where $\delta_{n,m}$ is the Kronecker's delta. Eq. (1.4.9) is derived after some algebra, first inserting eqs. (1.4.5) and then reminding eqs. (1.4.6).

The norm of the QNM's function (1.4.4) is a complex number, given by the following expression [1]:

$$\langle f_k | f_k \rangle = 2\omega_k \int_0^d \rho(x) f_k^2(x) dx + i \frac{n_0}{c} [f_k^2(0) + f_k^2(d)], \tag{1.4.10}$$

being the main difference with the ordinary definition of norm the presence of $f_k^2(x)$ rather than $|f_k(x)|^2$ and the two additive "surfaces terms" $i(n_0/c) f_k^2(0)$ and $i(n_0/c) f_k^2(d)$.

Applying eq. (1.4.9), in the case of a linear and homogeneous cavity, it is immediately demonstrable that: the QNM's norm (1.4.10) for the linear and homogeneous cavity with the QNM's frequencies (1.4.2)-(1.4.3) and the QNM's functions (1.4.5) assumes the form:

$$\langle f_k | f_k \rangle = K_0^2 \cdot \frac{n^2 - n_0^2}{c^2} \omega_k d. \tag{1.4.11}$$

If a normalized version of the QNM's function is adopted, corresponding to $\langle f_k | f_k \rangle = 2\omega_k$, then the constant $K_0$ is fixed, $K_0^2 = \frac{2c^2}{(n^2 - n_0^2)d}$, so that:

$$f_k(0) = K_0 = c\sqrt{\frac{2}{(n^2 - n_0^2)d}} \quad , \quad f_k(d) = (-1)^k K_0. \tag{1.4.12}$$

Besides, applying eq. (1.4.9), the orthogonality condition can be immediately derived for the QNM's functions [1]:

$$\langle f_n | f_m \rangle = (\omega_n + \omega_m) \int_0^d f_n(x)\rho(x)f_m(x)dx + i\frac{n_0}{c}[f_n(0)f_m(0) + f_n(d)f_m(d)] =$$
$$= (\omega_n + \omega_m)K_0^2 \frac{n_0}{c} \frac{1+(-1)^{n+m}}{i(\omega_n + \omega_m)} + i\frac{n_0}{c}[K_0^2 + K_0^2(-1)^{n+m}] = 0 \quad , \quad n \neq m$$

(1.4.13)

## 5. Discussion and conclusions.

The e.m. field inside an open cavity can be obtained by suitable methods as the transfer matrix [1] or the ray method [2]. The representation of the e.m. field inside an open cavity can be given also as a superposition of Quasi Normal Modes (QNMs) [1] which describe the coupling between the cavity and the environment. The importance of the QNM's approach lies in the fact that it is possible to recover the orthogonal representation of the e.m. field, as it is necessary to consider quantum processes.

In general, an optical cavity is not a conservative system, so the natural evolution of the e.m. field can not be described by an hermitian operator and the treatment of the e.m. field in terms of Normal Modes (NMs) has to be dropped. The QNM's approach considers the realistic situation in which the open cavity is enclosed in an infinite external space. The lack of energy conservation for the open cavity gives complex, instead of real, eigen frequencies. The evolution operator for the system is not hermitian and the modes of the e.m. field are not normal but quasi-normal. In fact, under some conditions over the refractive index: (a) the evolution of the QNMs is similar to the hermitian one for the NMs, and (b) the e.m. field can be described as a superposition of QNMs only inside the cavity and the QNM's functions form an orthogonal basis only according to a non-canonical metrics. The QNM's functions do not represent the e.m. field outside of the cavity while they represent the "not stationary" modes into the cavity. For the reason of not being complete in the whole space, they are called "quasi-normal" modes. The QNM's approach is very different from the pseudo-mode one introduced by Dalton ed al. in ref. [3]. In fact, the pseudo-modes are obtained by a Fano's transformation of the NMs, they are defined as pseudo-modes because they depend on the external pumping, and they use an ordinary metric; the QNMs are actual modes because they are defined in absence of an external pumping, and they use a specific definition of complex metric.

In this chapter, some necessary results of ref. [6] have been stated; in ref. [6], the QNM's approach is discussed for double side open and inhomogeneous cavities. The description of the scalar field in 1D open cavities has been represented in terms of QNMs; the QNM's description has been extended from an one side open and homogeneous cavity [1] to an open cavity from both ends, with an inhomogeneous distribution of refractive index.

# References

[1] D.R. Hall and P.E. Jackson, *The Physics and Technology of Laser Resonators* (Adam Hilger, Bristol, 1989); Y.A. Anan'ev, *Laser Resonators and the Beam Divergence Problem* (Adam Hilger, Bristol, 1992).

[2] M. Born and E. Wolf, *Principles of Optics* (Macmillan, New York, 1964).

[3] A.W. Crook, J. Opt. Soc. Am. A **38**, 954 (1948).

[4] D.N. Paltanayou and E. Wolf, Phys. Rev. D **13**, 913 (1976); B.J. Hoenders, J. Math. Phys. **20**, 329 (1979).

[5] Donald G. Dudley, *Mathematical Foundations for Electromagnetic Theory* (IEEE Press, New York, 1994).

[6] F. DeMartini, G. Innocenti, G.R. Jacobovitz, and P. Mataloni, Phys. Rev. Lett. **59**, 2955 (1987); D.J. Heinzen, J.J. Childs, J.E. Thomas, and M.S. Feld, Phys. Rev. Lett. **58**, 1320 (1987); A. E. Siegmann, Phys. Rev. A **39**, 1253 (1989).

[7] P. T. Leung, S. Y. Liu, and K. Young, Phys. Rev. A, **49**, 3057 (1994); P. T. Leung, S. S. Tong, and K. Young, J. Phys. A **30**, 2139 (1997); P. T. Leung, S. S. Tong, and K. Young, J. Phys. A **30**, 2153 (1997); E. S. C. Ching, P. T. Leung, A. Maassen van der Brink, W. M. Suen, S. S. Tong, and K. Young, Rev. Mod. Phys., **70**, 1545 (1998).

[8] C. Cohen-Tannoudji, B. Diu, F. Laloe, *Quantum Mechanics* (John Wiley, New York, 1977); W. H. Louisell, *Quantum Statistical Properties of Radiation* (John Wiley, New York, 1973).

[9] A. Settimi, *Studio del campo elettromagnetico alle frequenze ottiche nei cristalli fotonici uni-dimensionali tramite la teoria dei quasi normal modes*, Master Thesis (University of Rome "La Sapienza", Rome, 2002).

[10] P. T. Leung, W. M. Suen, C. P. Sun and K. Young, Phys. Rev. E, **57**, 6101 (1998).

[11] R. Banerjee, P. Mukherjee, J. Phys. A **35**, 5591 (2002).

[12] B. J. Dalton, Stephen M. Barnett, and B. M. Garraway, Phys. Rev. A. **64**, 053813 (2001).

# Figure and caption

Figure 1.1.

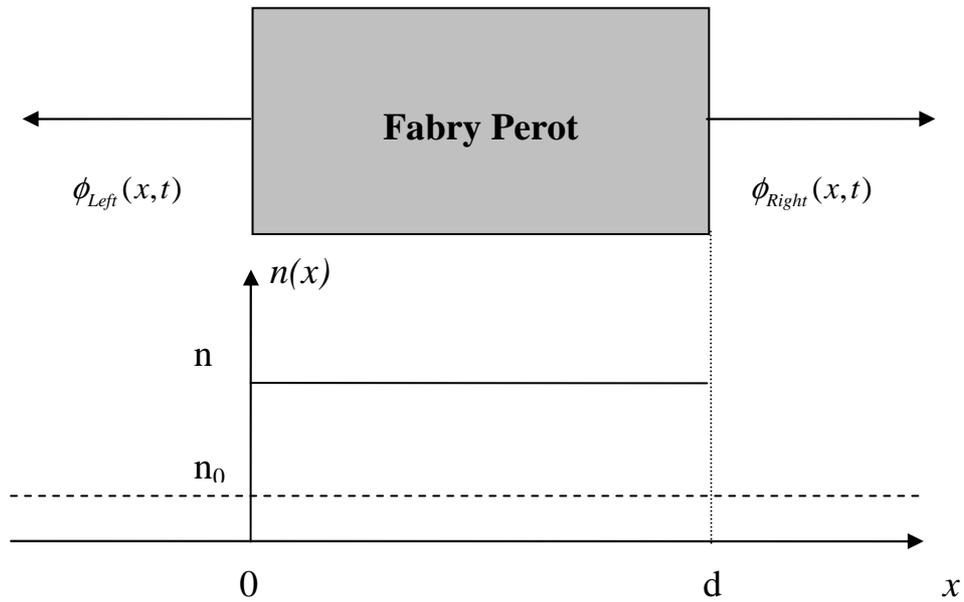

Figure 1.1. A linear Fabry-Perot (FP) cavity of length $d$, with refractive index $n$, in absence of external pumping so the e.m. field $\phi(x,t)$ satisfies the outgoing waves conditions (1.2.6).

# Chapter 2

# QNM's approach for open cavities
# in absence of external pumping:
# an application to one dimensional Photonic Crystals.

## 1. Introduction.

The past two decades have witnessed an intense investigation on electromagnetic propagation phenomena at optical frequencies in periodic structures, usually referred to as one dimensional (1D) Photonic Band Gap (PBG) structures [3]-[5]. These structures can be realized by a stack of dielectric layers, arranged according to some periodic or quasi-periodic sequence. Usually they are made alternating layers of high and low index materials of suitable thickness. If the 1D-PBG structure is periodic it may be thought as done by the superposition of a finite number of equal cells, each cell being usually constructed by two materials. The essential property of these 1D-PBG structures is the existence of allowed and forbidden frequency bands and gaps, in analogy with the energy bands and gaps in the semiconductors. At the same time, the distribution of the e.m. field intensity along the structure strongly changes by changing the frequency or, in general if the e.m. field is coming from outside, the boundary conditions. The 1D-PBG structures can be considered a special case of an open cavity and have great importance in nonlinear processes [4]-[6].

The dispersive properties are usually evaluated assuming an infinite periodic structure [7]. The finite dimensions of a 1D-PBG structure conceptually modify the calculation and the nature of the dispersive properties: this is mainly due to the existence of an energy flow into and out of the crystal. The application of the effective-medium approach to a 1D-PBG structure is discussed, and the analogy of a 1D-PBG to a simple Fabry-Perot is developed by Sipe et al. in ref. [8]. A phenomenological approach to the 1D-PBG's dispersive properties has been presented in ref. [9].

In this chapter, the Quasi Normal Mode's (QNM's) approach [1], even if limited to open cavities in absence of an external pumping, is applied to 1D-PBG structures as open cavities from both sides [6]. The chapter is organized as follows. In section 2, the QNM's approach is deeper examined [7] so that the definition of the QNM's eigen frequencies and functions is better clarified respect to ref.

[6]. In section 3, the completeness of the QNM's representation is proved inside 1D-PBG structures. In section 4, the equation for the QNM's eigen frequencies is derived for symmetric Quarter Wave (QW) 1D-PBGs. Conclusions are given in section 5.

## 2. Examining deeper QNM's approach.

In this section, the role of the complex Green's function formalism is put into evidence for the definition of the Quasi Normal Modes (QNMs); these considerations are here reported in more details respect to ones already presented in ref [7]. With reference to figure 2.1, consider now an open cavity of length $d$, filled with a refractive index $n(x)$, which is enclosed in an infinite external space. The cavity includes also the terminal surfaces, so it is represented as $C = [0,d]$ and the rest of universe as $U = (-\infty, 0) \cup (d, \infty)$.

The e.m. field $E(x,t)$ in the cavity satisfies the equation [1]

$$\left[\frac{\partial^2}{\partial x^2} - \rho(x)\frac{\partial^2}{\partial t^2}\right] E(x,t) = 0, \qquad (2.2.1)$$

where $\rho(x) = [n(x)/c]^2$, and $c$ is the speed of light in vacuum. If there is no external pumping, the e.m. field satisfies suitable "outgoing waves conditions" [1]

$$\begin{cases} \partial_x E(x,t) = \sqrt{\rho_0}\,\partial_t E(x,t) & \text{for } x < 0 \\ \partial_x E(x,t) = -\sqrt{\rho_0}\,\partial_t E(x,t) & \text{for } x > d \end{cases}, \qquad (2.2.2)$$

where $\rho_0 = (n_0/c)^2$, and $n_0$ is the outside refractive index. To take the cavity leakages into account, the Laplace's transform of the e.m. field is considered

$$\tilde{E}(x,\omega) = \int_0^\infty E(x,t)\exp(i\omega t)\,dt, \qquad (2.2.3)$$

where $\omega$ is a complex frequency. The e.m. field has to satisfy the Sommerfeld's radiative condition:

$$\lim_{x \to \pm\infty} \tilde{E}(x,\omega) = 0. \qquad (2.2.4)$$

The Laplace's transform of the e. m. field converges to an analytic function $\tilde{E}(x,\omega)$ only over the half-plane of convergence $\operatorname{Im}\omega > 0$. In fact, if the Laplace transform (2.2.3) is applied to the "outgoing waves conditions" (2.2.2), it follows

$$\begin{cases} \partial_x \tilde{E}(x,\omega) = -i\omega\sqrt{\rho_0}\,\tilde{E}(x,\omega) & \text{for } x < 0 \\ \partial_x \tilde{E}(x,\omega) = i\omega\sqrt{\rho_0}\,\tilde{E}(x,\omega) & \text{for } x > d \end{cases}, \qquad (2.2.5)$$

and solving the last equation of (2.2.5),

$$\tilde{E}(x,\omega) \propto \exp(i\omega\sqrt{\rho_0}\,x) = \exp(i\,\mathrm{Re}\,\omega\sqrt{\rho_0}\,x)\exp(-\mathrm{Im}\,\omega\sqrt{\rho_0}\,x) \quad \text{for} \quad x > d. \tag{2.2.6}$$

The Sommerfeld radiative condition (2.2.4) can be satisfied only if $\mathrm{Im}\,\omega > 0$.

The transformed Green's function $\tilde{G}(x,x',\omega)$ can be defined by [1]

$$\left[\frac{\partial^2}{\partial x^2} + \omega^2 \rho(x)\right]\tilde{G}(x,x',\omega) = -\delta(x-x'); \tag{2.2.7}$$

it is an e.m. field so, over the half-plane of convergence $\mathrm{Im}\,\omega > 0$, it satisfies the Sommerfeld's radiative conditions

$$\tilde{G}(x,x',\omega) = \begin{cases} \exp(i\omega\sqrt{\rho_0}\,x) \to 0 & \text{for } x \to \infty \\ \exp(-i\omega\sqrt{\rho_0}\,x) \to 0 & \text{for } x \to -\infty \end{cases}. \tag{2.2.8}$$

Two "auxiliary functions" $g_{\pm}(x,\omega)$ can be defined by

$$\left[\frac{\partial^2}{\partial x^2} + \omega^2 \rho(x)\right]g_{\pm}(x,\omega) = 0; \tag{2.2.9}$$

they are not defined as e.m. fields, because, over the half-plane of convergence $\mathrm{Im}\,\omega > 0$, they satisfy only the "asymptotic conditions" [1]:

$$\begin{cases} g_+(x,\omega) = \exp(i\omega\sqrt{\rho_0}\,x) \to 0 & \text{for } x \to \infty \\ g_-(x,\omega) = \exp(-i\omega\sqrt{\rho_0}\,x) \to 0 & \text{for } x \to -\infty \end{cases}. \tag{2.2.10}$$

However, the transformed Green's function $\tilde{G}(x,x',\omega)$ can be calculated in terms of the "auxiliary functions" $g_{\pm}(x,\omega)$. In fact, it can be shown that [1] the Wronskian associated to the two "auxiliary functions" $g_{\pm}(x,\omega)$ is $x$-independent

$$W(\omega) = g_+(x,\omega)g'_-(x,\omega) - g_-(x,\omega)g'_+(x,\omega), \tag{2.2.11}$$

and for the transformed Green's function:

$$\tilde{G}(x,x',\omega) = \begin{cases} -\dfrac{g_-(x,\omega)g_+(x',\omega)}{W(\omega)} & \text{for } x < x' \\ -\dfrac{g_+(x,\omega)g_-(x',\omega)}{W(\omega)} & \text{for } x' < x \end{cases}. \tag{2.2.12}$$

In what follows it is proved that, just for the "asymptotic conditions" (2.2.10), the "auxiliary functions" $g_{\pm}(x,\omega)$ are linearly independent over the half-plane of convergence $\mathrm{Im}\,\omega > 0$, and so the transformed Green function $\tilde{G}(x,x',\omega)$ is analytic over $\mathrm{Im}\,\omega > 0$.

The "asymptotic conditions" establish that, only over the half-plane of convergence $\mathrm{Im}\,\omega > 0$, the auxiliary function $g_+(x,\omega)$ acts as an e.m. field for large $x$, because it is exponentially decaying; in fact, from eq. (2.2.10): $g_+(x,\omega) = \exp(i\omega\sqrt{\rho_0}\,x) = \exp(i\,\mathrm{Re}\,\omega\sqrt{\rho_0}\,x)\exp(-\mathrm{Im}\,\omega\sqrt{\rho_0}\,x) \to 0$ for $x \to \infty$. Then, still over the half-plane of convergence $\operatorname{Im}\omega > 0$, the other auxiliary function $g_-(x,\omega)$ in general does not act as an e.m. field for large $x$, so it is exponentially increasing; in fact, according to eq. (2.2.10): $g_-(x,\omega) = A(\omega)\exp(i\omega\sqrt{\rho_0}x) + B(\omega)\exp(-i\omega\sqrt{\rho_0}x)$ for $x \to \infty$, with $B(\omega) \neq 0$, and so $g_-(x,\omega) \approx B(\omega)\exp(-i\omega\sqrt{\rho_0}x) = B(\omega)\exp(-i\operatorname{Re}\omega\sqrt{\rho_0}x)\exp(\operatorname{Im}\omega\sqrt{\rho_0}x) \to \infty$, for $x \to \infty$. It follows that the "auxiliary functions" $g_\pm(x,\omega)$ are linearly independent over the half-plane of convergence $\operatorname{Im}\omega > 0$, because the Wronskian $W(\omega)$ is not null; in fact, from eq. (2.2.11): $W(\omega) = \lim_{x \to \infty}[g_+(x,\omega)g'_-(x,\omega) - g_-(x,\omega)g'_+(x,\omega)] = 2i\sqrt{\rho_0}\omega B(\omega) \neq 0$ over $\operatorname{Im}\omega > 0$. So, the transformed Green's function $\tilde{G}(x,x',\omega)$ is analytic over the half-plane of convergence $\operatorname{Im}\omega > 0$, where $\tilde{G}(x,x',\omega)$ does not diverge; in fact, from eq. (2.2.12): $\tilde{G}(x,x',\omega) \propto 1/W(\omega)$, with $W(\omega) \neq 0$ over $\operatorname{Im}\omega > 0$.

The transformed Green's function $\tilde{G}(x,x',\omega)$ can be extended also over the lower complex half-plane $\operatorname{Im}\omega < 0$, for analytic continuation [14]. According to [15], over the lower complex half-plane $\operatorname{Im}\omega < 0$, it is always possible to define an infinite set of frequencies which are the poles of the transformed Green's function $\tilde{G}(x,x',\omega)$; in other words, there exists an infinite set of complex frequencies $\omega_n \in \mathbb{Z} = \{0, \pm 1, \pm 2, \ldots\}$, with negative imaginary parts $\operatorname{Im}\omega_n < 0$, for which the Wronskian (2.2.11) is null:

$$W(\omega_n) = 0. \qquad (2.2.13)$$

The poles of the transformed Green's function are referred as Quasi-Normal-Mode's eigen frequencies [15]. The definition of the QNM's frequencies implies that the "auxiliary functions" $g_\pm(x,\omega)$ become linearly dependent when they are calculated in the QNM's frequencies $\omega_n \in \mathbb{Z} = \{0, \pm 1, \pm 2, \ldots\}$; so, the auxiliary functions in the QNM's frequencies are such that

$$g_+(x,\omega_n) = c(\omega_n)g_-(x,\omega_n) = f_n(x), \qquad (2.2.14)$$

where $c(\omega_n)$ is an suitable complex constant. The above functions $f_n(x)$ are referred Quasi-Normal-Mode eigen functions [15].
Applying the QNM's condition (2.2.14) to the equation for the "auxiliary functions" (2.2.9), it follows that the QNMs $[\omega_n, f_n(x)]$ satisfy the equation:

$$\left[\frac{\partial^2}{\partial x^2} + \omega_n^2 \rho(x)\right] f_n(x) = 0. \qquad (2.2.15)$$

Moreover, applying the QNM's condition (2.2.14) to the "asymptotic conditions" for the "auxiliary functions" (2.2.10), it follows that, outside the cavity, the QNMs do not represent e.m. fields because they satisfy the QNM's "asymptotic conditions"

$$f_n(x) = \exp(\pm i\omega_n \sqrt{\rho_0} x) \to \infty \quad \text{for} \quad x \to \pm\infty, \quad (2.2.16)$$

while, into the cavity, the QNMs represent not stationary modes, so the QNM's "outgoing waves conditions" can be imposed at the terminal surfaces:

$$\begin{cases} \partial_x f_n(x)\big|_{x=0} = -i\omega\sqrt{\rho_0} f_n(0) \\ \partial_x f_n(x)\big|_{x=d} = i\omega\sqrt{\rho_0} f_n(d) \end{cases}. \quad (2.2.17)$$

Under the condition of QNM's completeness [1], the Green's function is calculated as superposition of QNMs, only inside the cavity

$$G(x, x'; t) = \frac{i}{2} \sum_n \frac{F_n(x) F_n(x')}{\omega_n} e^{-i\omega_n t}, \quad \text{for} \quad x, x' \in [0, d], \quad (2.2.18)$$

where are introduced the normalized QNM's functions $F_n(x) = f_n(x)\sqrt{2\omega_n / \langle f_n | f_n \rangle}$, denoting the QNM's norm as $\langle f_n | f_n \rangle$.

## 3. Completeness of QNM's representation inside 1D-PC cavities.

In this section, the completeness of Quasi Normal Modes's (QNM's) representation is proved inside one dimensional (1D) Photonic Crystal (PC) cavities (see ref. [6]). As depicted in figure 2.2.a, consider now a 1D-PC as a cavity open at both ends, with refractive index which is continuous in some intervals:

$$n(x) = \begin{cases} n_0(x), & \text{for} \quad x < x_0 \\ n_k(x), & \text{for} \quad x_{k-1} < x < x_k, \quad \text{where} \quad k \in [1, N]. \\ n_{N+1}(x), & \text{for} \quad x > x_N \end{cases} \quad (2.3.1)$$

Outside the 1D-PC cavity there is a medium with an asymptotic refractive index:

$$\lim_{x \to -\infty} n_0(x) = \lim_{x \to \infty} n_{N+1}(x) = n_0. \quad (2.3.2)$$

The QNM's representation is complete only inside the 1D-PC cavity. In what follows it is proved that, inside a 1D-PC cavity, the condition of QNM's completeness is valid, i.e. the behaviour of $\tilde{G}(x, x', \omega)$ for large $|\omega|$ is (see ref. [1] for details):

$$\lim_{|\omega| \to \infty} \tilde{G}(x, x', \omega) = 0, \quad \forall \omega \big| \text{Im}(\omega) < 0. \quad (2.3.3)$$

The proof of QNM's completeness is based on the application of the WKB method extended to optical regime. The WKB method was proposed to solve the Schrödinger's equation (so applied to the Planck's constant $\hbar$, considering $\hbar \to 0$) and here is proposed in Optics (but applied to the e.m. field wavelength λ, considering λ→0) [6]. With reference to the 1D-PBG of refractive index (2.3.1), as depicted in figure 2.1.a, the equation (2.2.9) can not be solved exactly adding the asymptotic conditions (2.2.10); however, a WKB-like method can be used in every period of the 1D-PC cavity, if the wavelength $\lambda$ is supposed so small to verify:

$$\left|\frac{dn_k(x)}{dx}\right| << \frac{4\pi}{\lambda}, \quad \text{for} \quad x_{k-1} < x < x_k, \quad \text{where} \quad k \in [1, N]. \tag{2.3.4}$$

For a 1D-PC cavity, whose refractive index is given by eq. (2.3.1), the following expressions are obtained for the auxiliary function $g_-(x,\omega)$ (see ref. [6] for details)

$$\begin{cases} g_-(x,\omega) = A_k(\omega)e^{i\frac{\omega}{c}\int_x^{x_k} n(\xi)d\xi} + B_k(\omega)e^{-i\frac{\omega}{c}\int_x^{x_k} n(\xi)d\xi} & , \quad x_{k-1} < x < x_k \\ g_-(x,\omega) = e^{-i\frac{\omega}{c}\int_x^{x_0} n(\xi)d\xi} & , \quad x < x_0 \end{cases} \tag{2.3.5}$$

where $k \in [1, N+1]$ and $x_{N+1} = +\infty$. For the auxiliary function $g_+(x,\omega)$ the following expressions are obtained

$$\begin{cases} g_+(x,\omega) = C_k(\omega)e^{i\frac{\omega}{c}\int_{x_{k-1}}^x n(\xi)d\xi} + D_k(\omega)e^{-i\frac{\omega}{c}\int_{x_{k-1}}^x n(\xi)d\xi} & , \quad x_{k-1} < x < x_k \\ g_+(x,\omega) = e^{i\frac{\omega}{c}\int_{x_N}^x n(\xi)d\xi} & , \quad x > x_N \end{cases} \tag{2.3.6}$$

where $k \in [0, N]$ and $x_{-1} = -\infty$. Under the conditions of continuity for the auxiliary functions $g_\pm(x,\omega)$ and their derivatives in the points $x = x_k$, it results

$$\begin{pmatrix} A_{k+1} \\ B_{k+1} \end{pmatrix} = S_k \begin{pmatrix} A_k e^{-i\vartheta_k} + R_k B_k e^{-i\vartheta_k} \\ R_k A_k e^{i\vartheta_k} + B_k e^{i\vartheta_k} \end{pmatrix}, \quad k \in [0, N]. \tag{2.3.7}$$

$$\begin{pmatrix} C_k \\ D_k \end{pmatrix} = S'_k \begin{pmatrix} C_{k+1} e^{-i\vartheta_{k-1}} - R_k D_{k+1} e^{-i\vartheta_{k-1}} \\ -R_k C_{k+1} e^{i\vartheta_{k-1}} + D_{k+1} e^{i\vartheta_{k-1}} \end{pmatrix}, \quad k \in [0, N], \tag{2.3.8}$$

where

$$\begin{cases} \vartheta_k = \frac{\omega}{c} \int_{x_k}^{x_{k-1}} n(x)dx \\ R_k = \frac{[n(x_k^+) - n(x_k^-)]}{[n(x_k^+) + n(x_k^-)]} \\ S_k = \frac{[n(x_k^+) + n(x_k^-)]}{2n(x_k^+)} \\ S'_k = \frac{[n(x_k^+) + n(x_k^-)]}{2n(x_k^-)} \end{cases}, \quad k \in [0, N]. \tag{2.3.9}$$

Now, only inside the 1D-PC cavity, i.e. $\forall (x, x') \mid x_0 < x' \leq x < x_N$, there exists a suitable value of $k$, such that $x_0 < x' \leq x_k$, $x_{k-1} \leq x < x_N$, so it can be found a couple $(n, m)$, such that $1 \leq n \leq k$ and $k \leq m \leq N$, and it results:

$$g_-(x', \omega) = A_n(\omega) e^{i\frac{\omega}{c}\int_{x'}^{x_n} n(\xi)d\xi} + B_n(\omega) e^{-i\frac{\omega}{c}\int_{x'}^{x_n} n(\xi)d\xi}, \qquad (2.3.10)$$

$$g_+(x, \omega) = C_m(\omega) e^{i\frac{\omega}{c}\int_{x_{m-1}}^{x} n(\xi)d\xi} + D_m(\omega) e^{-i\frac{\omega}{c}\int_{x_{m-1}}^{x} n(\xi)d\xi}. \qquad (2.3.11)$$

Then, the Fourier's transform of the Green's function $\tilde{G}(x, x', \omega)$ for $x' \leq x$ has the following expression [see eq. (2.2.12)]:

$$\tilde{G}(x, x', \omega) = -\frac{\left[ A_n(\omega) e^{i\frac{\omega}{c}\int_{x'}^{x_n} n(\xi)d\xi} + B_n(\omega) e^{-i\frac{\omega}{c}\int_{x'}^{x_n} n(\xi)d\xi} \right]\left[ C_m(\omega) e^{i\frac{\omega}{c}\int_{x_{m-1}}^{x} n(\xi)d\xi} + D_m(\omega) e^{-i\frac{\omega}{c}\int_{x_{m-1}}^{x} n(\xi)d\xi} \right]}{2i\frac{\omega}{c}n(x)\left[ D_m(\omega)B_n(\omega) e^{-i\frac{\omega}{c}\int_{x_{m-1}}^{x_n} n(\xi)d\xi} - C_m(\omega)A_n(\omega) e^{i\frac{\omega}{c}\int_{x_{m-1}}^{x_n} n(\xi)d\xi} \right]}. \qquad (2.3.12)$$

The coefficients $A_n$, $B_n$, $C_m$, $D_m$ are obtained from eqs. (2.3.7) and (2.3.8) after some algebra:

$$\begin{pmatrix} A_n \\ B_n \end{pmatrix} = \prod_{k=0}^{n-1} \begin{pmatrix} A_k \\ B_k \end{pmatrix} \cong \prod_{k=0}^{n-1} S_k \begin{pmatrix} R_n e^{i\sum_{k=0}^{n-2} \vartheta_k - \vartheta_{n-1}} \\ e^{i\sum_{k=0}^{n-1} \vartheta_k} \end{pmatrix}, \qquad (2.3.13)$$

$$\begin{pmatrix} C_m \\ D_m \end{pmatrix} = \prod_{k=m+1}^{N} \begin{pmatrix} C_k \\ D_k \end{pmatrix} \cong \prod_{k=m+1}^{N} S'_k \begin{pmatrix} R_N R_m e^{i\sum_{k=m}^{N-1} \vartheta_k - \vartheta_{m-1}} \\ -R_N e^{i\sum_{k=m-1}^{N-1} \vartheta_k} \end{pmatrix}. \qquad (2.3.14)$$

If the refractive index is supposed such that $\left| n(x_k^+) - n(x_k^-) \right| \leq \Delta n \leq 1$, $\forall k \in [0, N]$, then $R_k \leq \Delta n / [n(x_k^+) + n(x_k^-)] \leq 1 / [n(x_k^+) + n(x_k^-)] < 1/2$, $\forall k \in [0, N]$. If $R_k$, $\forall k \in [0, N]$, is close to 0 than to 1, then, from eq. (2.3.13) and eq. (2.3.14), it follows that $B_n$ is dominant with respect to $A_n$, and $D_m$ is dominant with respect to $C_m$, so eq. (2.3.12) becomes:

$$\tilde{G}(x, x', \omega) \cong -\frac{e^{-i\frac{\omega}{c}\left[\int_{x'}^{x_n} n(\xi)d\xi + \int_{x_{m-1}}^{x} n(\xi)d\xi\right]}}{2i\frac{\omega}{c}n(x)e^{-i\frac{\omega}{c}\int_{x_{m-1}}^{x_n} n(\xi)d\xi}}. \qquad (2.3.15)$$

Since for $x' \leq x$, it follows

$$\int_{x'}^{x_n} n(\xi)d\xi + \int_{x_{m-1}}^{x} n(\xi)d\xi = \int_{x_{m-1}}^{x_n} n(\xi)d\xi + \int_{x'}^{x} n(\xi)d\xi \leq \int_{x_{m-1}}^{x_n} n(\xi)d\xi. \qquad (2.3.16)$$

and the transformed Green's function in a 1D-PC cavity has the following behaviour:

$$\tilde{G}(x,x',\omega) \to 0 \quad \text{for} \quad |\omega| \to \infty. \tag{2.3.17}$$

Therefore, the QNM's completeness in 1D-PC cavities is proved.

## 4. QNM's frequencies for 1D-PBG structures.

The previous considerations can be specified for a symmetric 1D-PBG structure, consisting on $N$ periods plus one layer; every period is composed of two layers respectively with lengths $h$ and $l$ and with refractive indices $n_h$ and $n_l$, while the added layer is with parameters $h$ and $n_h$. With reference to fig. 2.2.b, consider now a symmetric 1D-PBG structure, consisting of $2N+1$ layers with a total length $d = N(h+l)+h$; if the two layers external to the symmetric 1D-PBG structure are counted, the 1D-space $x$ can be divided into $2N+3$ layers: they are $L_k = [x_{k-1}, x_k]$, $k = 0,1,\ldots,2N+1,2N+2$, with $x_{-1} = -\infty$, $x_0 = 0$, $x_{2N+1} = d$, $x_{2N+2} = +\infty$. The refractive index $n(x)$ takes a constant value $n_k$ in every layer $L_k$, $k = 0,1,\ldots,2N+1,2N+2$, i.e.

$$n(x) = \begin{cases} 1 & \text{for } x \in L_0, L_{2N+2} \\ n_h & \text{for } x \in L_k, k = 1,3,\ldots,2N-1,2N+1 \\ n_l & \text{for } x \in L_k, k = 2,4,\ldots,2N \end{cases} \tag{2.4.1}$$

In Appendix A, it is proved that, introducing the two phase-terms

$$\begin{cases} \delta_l = q_l l = n_l l \dfrac{\omega}{c} \\ \delta_h = q_h h = n_h h \dfrac{\omega}{c} \end{cases}, \tag{2.4.2}$$

the QNM's frequencies can be found by solving the following transcendental equation

$$\alpha \sum_{k=0}^{\left[\frac{N-1}{2}\right]} \frac{(-1)^k}{k!} \frac{(N-1-k)!}{(N-1-2k)!}(\gamma)^{N-1-2k} + \beta \sum_{k=0}^{\left[\frac{N-2}{2}\right]} \frac{(-1)^k}{k!} \frac{(N-2-k)!}{(N-2-2k)!}(\gamma)^{N-2-2k} = 0, \tag{2.4.3}$$

where the coefficients $\alpha, \beta, \gamma$ are parameters related to the refractive indices of each layer:

$$\begin{cases}
\alpha = \frac{1}{4}\Bigg\{\left[\left(n_l+\frac{1}{n_l}\right)+2\left(n_h+\frac{1}{n_h}\right)-4-2\left(\frac{n_h}{n_l}+\frac{n_l}{n_h}\right)+\left(\frac{n_h^2}{n_l}+\frac{n_l}{n_h^2}\right)\right]e^{2i\delta_h+i\delta_l} \\
\quad +\left[\left(n_l+\frac{1}{n_l}\right)-2\left(n_h+\frac{1}{n_h}\right)-4+2\left(\frac{n_h}{n_l}+\frac{n_l}{n_h}\right)+\left(\frac{n_h^2}{n_l}+\frac{n_l}{n_h^2}\right)\right]e^{-2i\delta_h+i\delta_l} \\
\quad +\left[2\left(n_l+\frac{1}{n_l}\right)-2\left(\frac{n_h^2}{n_l}+\frac{n_l}{n_h^2}\right)\right]e^{i\delta_l} \\
\quad +\left[-\left(n_l+\frac{1}{n_l}\right)+2\left(n_h+\frac{1}{n_h}\right)-4+2\left(\frac{n_h}{n_l}+\frac{n_l}{n_h}\right)-\left(\frac{n_h^2}{n_l}+\frac{n_l}{n_h^2}\right)\right]e^{2i\delta_h-i\delta_l} \\
\quad +\left[-\left(n_l+\frac{1}{n_l}\right)-2\left(n_h+\frac{1}{n_h}\right)-4-2\left(\frac{n_h}{n_l}+\frac{n_l}{n_h}\right)-\left(\frac{n_h^2}{n_l}+\frac{n_l}{n_h^2}\right)\right]e^{-2i\delta_h-i\delta_l} \\
\quad +\left[-2\left(n_l+\frac{1}{n_l}\right)+2\left(\frac{n_h^2}{n_l}+\frac{n_l}{n_h^2}\right)\right]e^{-i\delta_l}\Bigg\} \\
\beta = \left[2-\left(n_h+\frac{1}{n_h}\right)\right]e^{i\delta_h}+\left[2+\left(n_h+\frac{1}{n_h}\right)\right]e^{-i\delta_h} \\
\gamma = \frac{1}{4n_h n_l}\left\{(n_h+n_l)^2 e^{i(\delta_h+\delta_l)}-(n_h-n_l)^2 e^{i(\delta_h-\delta_l)}-(n_h-n_l)^2 e^{i(\delta_l-\delta_h)}+(n_h+n_l)^2 e^{-i(\delta_h+\delta_l)}\right\}
\end{cases} \quad (2.4.4)$$

More details are given in Appendix A. For a symmetric quarter wave (QW) 1D-PBG with $N$ periods and $\omega_{ref}$ as reference frequency, there are $2N+1$ families of QNMs, i.e. $G_n^{QNM}$, $n \in [0, 2N]$; the $G_n^{QNM}$ family of QNMs consists on infinite QNM's frequencies, i.e. $\omega_{n,m}$, $m \in \mathbb{Z} = \{0, \pm 1, \pm 2, \ldots\}$, which have the same imaginary part, i.e. $\operatorname{Im}\omega_{n,m} = \operatorname{Im}\omega_{n,0}$, $m \in \mathbb{Z}$, and are aligned by a step $\Delta = 2\omega_{ref}$, i.e. $\operatorname{Re}\omega_{n,m} = \operatorname{Re}\omega_{n,0} + m\Delta$, $m \in \mathbb{Z}$. It follows that, if the complex plane is divided into some ranges, i.e. $R_m = \{m\Delta \leq \operatorname{Re}\omega < (m+1)\Delta\}$, $m \in \mathbb{Z}$, each of the QNM's family $G_n^{QNM}$ drops only one QNM's frequency over the range $R_m$, i.e. $\omega_{n,m} = (\operatorname{Re}\omega_{n,0} + m\Delta,\ \operatorname{Im}\omega_{n,0})$; so, there are $2N+1$ QNM's frequencies over the range $R_m$ and they can be referred as $\omega_{n,m} = \omega_{n,0} + m\Delta$, $n \in [0, 2N]$. It results that there are $2N+1$ QNM's frequencies over the basic range, i.e. $S_0 = \{0 \leq \operatorname{Re}\omega < \Delta\}$, and they correspond to $\omega_{n,0}$, $n \in [0, 2N]$.

The QNM's frequencies are not uniformly distributed in the complex plane, but they arrange themselves in order to form permitted and forbidden bands, in agreement with the known characteristics of the QW 1D-PBG structures [16]. In figure 2.3.a, the QNM's frequencies are plotted for a symmetric QW 1D-PBG, where the reference wavelength is $\lambda_{ref} = 1\mu m$, the number of periods is *N=4* and the two used refractive indices are *$n_h$=1.5, $n_l$=1*. Simple inspection of figure 2.3.a. shows that next to the gap, the QNM's frequencies have the smallest imaginary part in modulus, and hence have the narrowest resonance lines. In figure 2.3.b, the QNM's frequencies are shown for the same cavity of fig.2.3.a, but with $n_h=2$, $n_l=1$. Contrasting figs. 2.3.a. and 2.3.b, it can be seen that as the difference between the refractive indices of adjacent layers is increased, the width of the gap increases. This also entails that the magnitude of the imaginary part of the QNM's frequency decreases in modulus, and the resonance peaks become tighter. In figure 2.3.c, the QNM's frequencies are shown for the same cavity as above, with an increased number of periods $N=8$ and $n_h=2$, $n_l=1$. Contrasting figs. 2.3.b. and 2.3.c, it can be seen that as the number of periods is increased, the position of the gap remains the same: as in the previous case, the imaginary parts of the QNM's frequencies decrease in modulus, and the resonance peaks become narrower.

## 5. Conclusions.

One-dimensional (1D) Photonic Band Gap (PBG) structures are particular optical cavities, with both sides open to the external environment and a stratified material inside. A 1D-PBG structure is finite in space and, working with e.m. pulses of a spatial extension longer than the length of the 1D-PBG, the open cavity cannot be studied as infinite: rather, the boundary conditions must be considered at the two ends of the cavity.

In this chapter, the Quasi Normal Mode's (QNM's) approach, even if limited to open cavities in absence of an external pumping, has been applied to 1D-PBG structures as open cavities from both sides. The QNM's approach has been deeper examined, so that the definition of the QNM's frequencies and functions has been better clarified. The completeness of the QNM's representation has been proved inside 1D-PBG structures. The equation for the QNM's frequencies has been derived for symmetric Quarter Wave (QW) 1D-PBGs. The QNM's frequencies are not uniformly distributed in the complex plane, but they arrange themselves in order to form permitted and forbidden bands, in agreement with the known characteristics of the QW 1D-PBG; it has been proved that, for a symmetric QW1D-PBG, with $N$ periods and $\omega_{ref}$ as reference frequency, there are exactly $2N+1$ QNM's frequencies in the $[0, 2\omega_{ref})$ range.

# Appendix A.

This appendix describes how to obtain the equation of the Quasi Normal Mode's (QNM's) frequencies (2.4.3)-(2.4.4) for the symmetric one dimensional (1D) Photonic Band Gap (PBG) structure of fig. 2.2.b. with refractive index (2.4.1). Then, this equation is solved for a quarter-wave (QW) 1D-PBG.

The matrix method [17] is used for describing the 1D-PBG structures. The transmission matrix for a single period of the 1D-PBG cavity has the form

$$M = \begin{pmatrix} \mu_{11} & \mu_{12} \\ \mu_{21} & \mu_{22} \end{pmatrix}, \qquad (2.A.1)$$

where

$$\begin{cases} \mu_{11} = \cos\delta_h \cos\delta_l - \dfrac{q_h}{q_l}\sin\delta_h \sin\delta_l \\ \mu_{12} = \dfrac{1}{q_h}\sin\delta_h \cos\delta_l + \dfrac{1}{q_l}\sin\delta_l \cos\delta_h \\ \mu_{21} = -q_h \sin\delta_h \cos\delta_l - q_l \sin\delta_l \cos\delta_h \\ \mu_{22} = \cos\delta_h \cos\delta_l - \dfrac{q_l}{q_h}\sin\delta_h \sin\delta_l \end{cases} \qquad (2.A.2)$$

In the expression (2.A.2), the propagation constants in the two layers of a period $q_h = n_h(\omega/c)$, $q_l = n_l(\omega/c)$ and the respective phases $\delta_h = q_h h$, $\delta_l = q_l l$ appear.

The transmission matrix of a symmetric 1D-PBG structure has the form

$$M_{PBG} = \begin{pmatrix} m_{11} & m_{12} \\ m_{21} & m_{22} \end{pmatrix}, \qquad (2.A.3)$$

where

$$\begin{cases} m_{11} = \cos\delta_h[\mu_{11}U_{N-1}(\vartheta) - U_{N-2}(\vartheta)] + \dfrac{\sin\delta_h}{q_h}\mu_{21}U_{N-1}(\vartheta) \\ m_{12} = \cos\delta_h\mu_{12}U_{N-1}(\vartheta) + \dfrac{\sin\delta_h}{q_h}[\mu_{22}U_{N-1}(\vartheta) - U_{N-2}(\vartheta)] \\ m_{21} = -q_h \sin\delta_h[\mu_{11}U_{N-1}(\vartheta) - U_{N-2}(\vartheta)] + \cos\delta_h \mu_{21}U_{N-1}(\vartheta) \\ m_{22} = -q_h \sin\delta_h \mu_{12}U_{N-1}(\vartheta) + \cos\delta_h[\mu_{22}U_{N-1} - U_{N-2}(\vartheta)] \end{cases} \qquad (2.A.4)$$

In expression (A.4.) the Chebyshev's polynomials $U_N(\vartheta) = \dfrac{\sin[(N+1)\vartheta]}{\sin\vartheta}$ with argument $\cos\vartheta = \dfrac{\mu_{11} + \mu_{22}}{2}$ appear.

As depicted in fig. 2.2.b, if $f(x)$ is the wave function of a generic QNM and $f'(x)$ is its spatial derivative, then $(f_0, f'_0)$ and $(f_{2N+1}, f'_{2N+1})$ are their values at the input and output of a 1D-PBG structure, such that:

$$\begin{pmatrix} f_{2N+1} \\ f'_{2N+1} \end{pmatrix} = M_{PBG} \begin{pmatrix} f_0 \\ f'_0 \end{pmatrix}. \tag{2.A.5}$$

Referring to fig. 2.2.b, if the equation $\left[\partial_x^2 + \rho(x)\omega^2\right] f(x) = 0$ of the e.m. field is solved in every layer, the expression for the wave function of a generic QNM is obtained for a symmetric 1D-PBG structure:

$$\begin{aligned}
f(x) &= \left( A_0 e^{i\frac{\omega}{c}x} + B_0 e^{-i\frac{\omega}{c}x} \right) \vartheta(-x) + \\
&+ \sum_{k=0}^{N} \left( A_{2k+1} e^{i\frac{\omega}{c} n_h x} + B_{2k+1} e^{-i\frac{\omega}{c} n_h x} \right) \vartheta[x - k(h+l)] \vartheta[k(h+l) + h - x] + \\
&+ \sum_{k=0}^{N-1} \left( A_{2k+2} e^{i\frac{\omega}{c} n_l x} + B_{2k+2} e^{-i\frac{\omega}{c} n_l x} \right) \vartheta[x - k(h+l) - h] \vartheta[(k+1)(h+l) - x] + \\
&+ \left( A_{2N+2} e^{i\frac{\omega}{c}x} + B_{2N+2} e^{-i\frac{\omega}{c}x} \right) \vartheta[x - N(h+l) - h]
\end{aligned} \tag{2.A.6}$$

where $\vartheta(x)$ is the unitary step function. By imposing the conditions for QNM's frequencies, $f(x) = \exp(\pm i \omega x/c)$ to $x \to \pm\infty$, it results $A_0(\omega) = 0$ and

$$B_{2N+2}(\omega) = \frac{i(\omega/c) f_{2N+1}(\omega) - f'_{2N+1}(\omega)}{2i(\omega/c)} e^{i(\omega/c)[N(h+l)+h]} = 0. \tag{2.A.7}$$

Eqs. (2.4.3) and (2.4.4) are obtained after some algebra from eq. (2.A.7), by inserting eq. (2.A.5), (2.A.4) and (2.A.2).

Now, working with a QW 1D-PBG structure, it results

$$n_l l = n_h h = \frac{\lambda_{ref}}{4}, \tag{2.A.8}$$

where $\lambda_{ref}$ is a reference wavelength. With these assumptions, the phases $\delta_h, \delta_l$ become

$$\delta_l = \delta_h = \delta, \tag{2.A.9}$$

where the phase $\delta = (\lambda_{ref}/4)(\omega/c)$ is introduced. Equation (2.4.3) becomes a real coefficient polynomial equation (instead of a transcendental equation) of degree $2N+1$ in the variable $e^{i\delta}$. There are $2N+1$ families of QNMs and, if a generic one is picked, all the QNM's frequencies have the same imaginary part and are distributed with a step $\Delta = 2\omega_{ref}$, where $\omega_{ref} = 2\pi c/\lambda_{ref}$. Then, there are exactly $2N+1$ QNM's frequencies in the $[0, 2\omega_{ref})$ range. From eq. (2.A.6), the QNM's functions $f_n(x)$ corresponding to the QNM's frequencies $\omega = \omega_n$ can be constructed. By imposing the conditions of continuity for the wave function $f(x)$ and its spatial derivative $g(x) = f'(x)$ at interfaces of the symmetric QW 1D-PBG cavity, it results

$$\begin{cases}
\begin{pmatrix} A_1 \\ B_1 \end{pmatrix} = \frac{1}{2} \begin{pmatrix} e^{-in_h \frac{\omega}{c} x_0} & \frac{1}{n_h} e^{-in_h \frac{\omega}{c} x_0} \\ e^{in_h \frac{\omega}{c} x_0} & -\frac{1}{n_h} e^{in_h \frac{\omega}{c} x_0} \end{pmatrix} \begin{pmatrix} e^{i\frac{\omega}{c} x_0} & e^{-i\frac{\omega}{c} x_0} \\ e^{i\frac{\omega}{c} x_0} & -e^{-i\frac{\omega}{c} x_0} \end{pmatrix} \begin{pmatrix} A_0 \\ B_0 \end{pmatrix} \\
\begin{pmatrix} A_{2k+1} \\ B_{2k+1} \end{pmatrix} = \frac{1}{2} \begin{pmatrix} e^{-in_h \frac{\omega}{c}(x_0+k(h+l))} & \frac{1}{n_h} e^{-in_h \frac{\omega}{c}(x_0+k(h+l))} \\ e^{in_h \frac{\omega}{c}(x_0+k(h+l))} & -\frac{1}{n_h} e^{in_h \frac{\omega}{c}(x_0+k(h+l))} \end{pmatrix} \begin{pmatrix} e^{in_l \frac{\omega}{c}(x_0+k(h+l))} & e^{-in_l \frac{\omega}{c}(x_0+k(h+l))} \\ n_l e^{in_l \frac{\omega}{c}(x_0+k(h+l))} & -n_l e^{-in_l \frac{\omega}{c}(x_0+k(h+l))} \end{pmatrix} \begin{pmatrix} A_{2k} \\ B_{2k} \end{pmatrix} \\
\begin{pmatrix} A_{2k+2} \\ B_{2k+2} \end{pmatrix} = \frac{1}{2} \begin{pmatrix} e^{-in_l \frac{\omega}{c}(x_0+k(h+l)+h)} & \frac{1}{n_l} e^{-in_l \frac{\omega}{c}(x_0+k(h+l)+h)} \\ e^{in_l \frac{\omega}{c}(x_0+k(h+l)+h)} & -\frac{1}{n_l} e^{in_l \frac{\omega}{c}(x_0+k(h+l)+h)} \end{pmatrix} \begin{pmatrix} e^{in_h \frac{\omega}{c}(x_0+k(h+l)+h)} & e^{-in_h \frac{\omega}{c}(x_0+k(h+l)+h)} \\ n_h e^{in_h \frac{\omega}{c}(x_0+k(h+l)+h)} & -n_h e^{-in_h \frac{\omega}{c}(x_0+k(h+l)+h)} \end{pmatrix} \begin{pmatrix} A_{2k+1} \\ B_{2k+1} \end{pmatrix} \\
\begin{pmatrix} A_{2N+2} \\ B_{2N+2} \end{pmatrix} = \frac{1}{2} \begin{pmatrix} e^{-i\frac{\omega}{c}(x_0+N(h+l)+h)} & e^{-i\frac{\omega}{c}(x_0+N(h+l)+h)} \\ e^{i\frac{\omega}{c}(x_0+N(h+l)+h)} & -e^{i\frac{\omega}{c}(x_0+N(h+l)+h)} \end{pmatrix} \begin{pmatrix} e^{in_h \frac{\omega}{c}(x_0+N(h+l)+h)} & e^{-in_h \frac{\omega}{c}(x_0+N(h+l)+h)} \\ n_h e^{in_h \frac{\omega}{c}(x_0+N(h+l)+h)} & -n_h e^{-in_h \frac{\omega}{c}(x_0+N(h+l)+h)} \end{pmatrix} \begin{pmatrix} A_{2N+1} \\ B_{2N+1} \end{pmatrix}
\end{cases}$$

(2.A.10)

where $A_0(\omega_n) = 0$ and $B_{2N+2}(\omega_n) = 0$.

[1] S. John, Phys. Rev. Lett. **53**, 2169 (1984); S. John, Phys. Rev. Lett. **58**, 2486 (1987); E. Yablonovitch, Phys. Rev. Lett. **58**, 2059 (1987); E. Yablonovitch and T. J. Gmitter, Phys. Rev. Lett.. **63**, 1950 (1989).

[2] J. Maddox, Nature (London) **348**, 481 (1990); E. Yablonovitch and K.M. Lenny, Nature (London) **351**, 278, 1991; J. D. Joannopoulos, P. R. Villeneuve, and S. H. Fan, Nature (London) **386**, 143 (1997).

[3] J. D. Joannopoulos, *Photonic Crystals: Molding the Flow of Light* (Princeton University Press, Princeton, New York, 1995); K. Sakoda, *Optical properties of photonic crystals* (Springer Verlag, Berlin, 2001); K. Inoue and K. Ohtaka, *Photonic Crystals: Physics, Fabrication, and Applications* (Springer-Verlag, Berlin, 2004).

[4] J. P. Dowling and C. M. Bowden, Phys. Rev. A **46**, 612 (1992); M. J. Bloemer and M. Scalora, Appl. Phys. Lett. **72**, 1676 (1998); I. S. Fogel, J. M. Bendickson, M. D. Tocci, M. J. Bloemer, M. Scalora, C. M. Bowden, and J. P. Dowling, Pure Appl. Opt. **7**, 393 (1998); C. M. Bowden and A. Zheltikov, J. Opt. Soc. Am. B **19**, 2046 (2002).

[5] M. Centini, C. Sibilia, M. Scalora, G. D'Aguanno, M. Bertolotti, M. J. Bloemer, C. M. Bowden, and I. Nefedov, Phys.Rev E **60**, 4891 (1999); Y. Dumeige, P. Vidakovic, S. Sauvage, I. Sagnes, J. A. Levenson, C. Sibilia, M. Centini, G. D'Aguanno, and M. Scalora, Appl. Phys. Lett. **78**, 3021 ( 2001); M. Scalora, M. J. Bloemer, C. M. Bowden, G. D'Aguanno, M. Centini, C. Sibilia, M. Bertolotti, Y. Dumeige, I. Sagnes, P. Vidakovic, and A. Levenson, Opt. Photonics News **12**, 36 (2001).

[6] *Nanoscale Linear and Nonlinear Optics*, edited by M. Bertolotti, C. M. Bowden, and C. Sibilia, AIP Conf. Proc. No. 560 (AIP, Melville, New York, 2001).

[7] P. Yeh, *Optical Waves in Layered Media* (Wiley, New York, 1988).

[8] J. E. Sipe, L. Poladian, C. Martijn de Sterke, J. Opt. Soc. Am. A **11**, 1307 (1994).

[9] G. D'Aguanno, M. Centini, M. Scalora, C. Sibilia, M.J. Bloemer, C.M. Bowden, J.W. Haus, M. Bertolotti, Phys. Rev. *E* **63**, 036610 (2001).

[10] P. T. Leung, S. Y. Liu, and K. Young, Phys. Rev. A, **49**, 3057 (1994); P. T. Leung, S. S. Tong, and K. Young, J. Phys. A **30**, 2139 (1997); P. T. Leung, S. S. Tong, and K. Young, J. Phys. A **30**, 2153 (1997); E. S. C. Ching, P. T. Leung, A. Maassen van der Brink, W. M. Suen, S. S. Tong, and K. Young, Rev. Mod. Phys. **70**, 1545 (1998); P. T. Leung, W. M. Suen, C. P. Sun and K. Young, Phys. Rev. E, **57**, 6101 (1998).

[11] A. Settimi, S. Severini, N. Mattiucci, C. Sibilia, M. Centini, G. D'Aguanno, M. Bertolotti, M. Scalora, M. Bloemer, C. M. Bowden, Phys. Rev. E, **68**, 026614 (2003).

[12] S. Severini, A. Settimi, C. Sibilia, M. Bertolotti, A. Napoli, A. Messina, *Quasi Normal Frequencies in open cavities: an application to Photonic Crystals*, Acta Phys. Hung. B **23/3-4**, 135-142 (2005); A. Settimi, S. Severini, B. Hoenders, *Quasi Normal Modes description of transmission properties for Photonic Band Gap structures*, J. Opt. Soc. Am. B, **26**, 876-891 (2009).

[13] M. Born and E. Wolf, *Principles of Optics* (Macmillan, New York, 1964).

[14] G. F. Carrier, M. Krook, C. E. Pearson, *Functions of a complex variable – theory and technique* (McGraw-Hill Book Company, New York, 1983).

[15] A. Bachelot, A. Motet-Bachelot, Ann. Inst. Henry Poincare **59**, 3, (1993).

[16] D. W. L. Sprung, H. Wu, and J. Martorell, Am. J. Phys. **61**, 1118 (1993); M. G. Rozman, P. Reineken, and R. Tehver, Phys. Lett. A **187**, 127 (1994).

[17] J. Lekner, *Theory of Reflection* (Martinus Nijhoff, Dordrecht, 1987).

# Figures and captions

Figure 2.1.

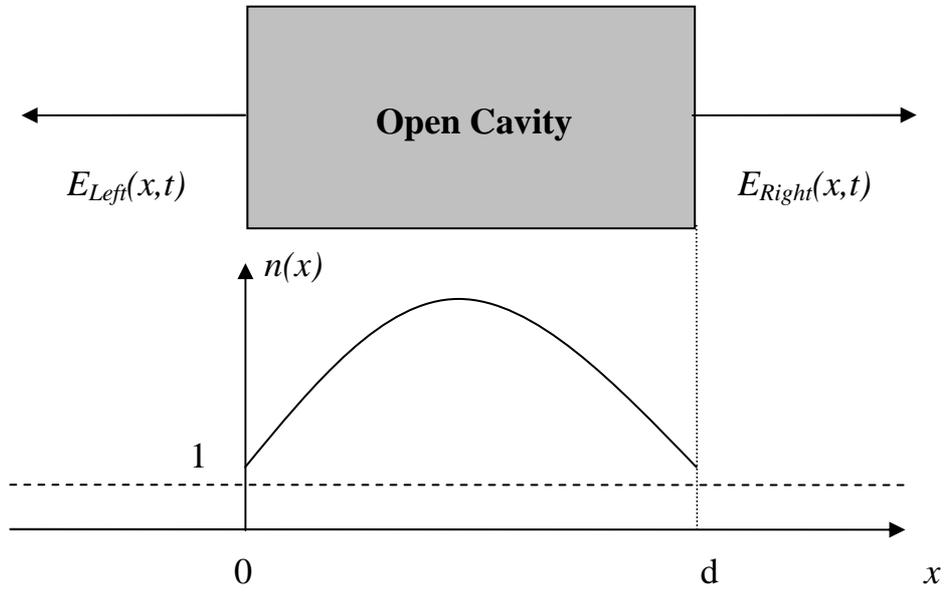

Figure 2.2.a.

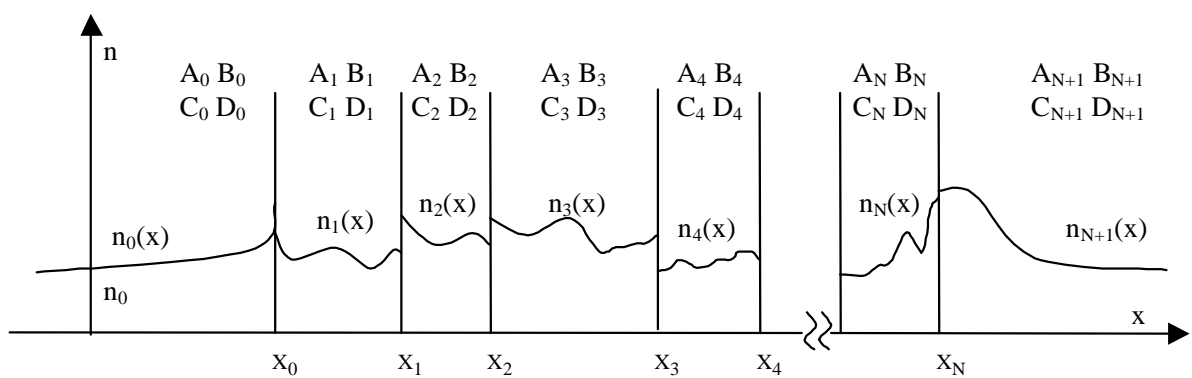

Figure 2.2.b.

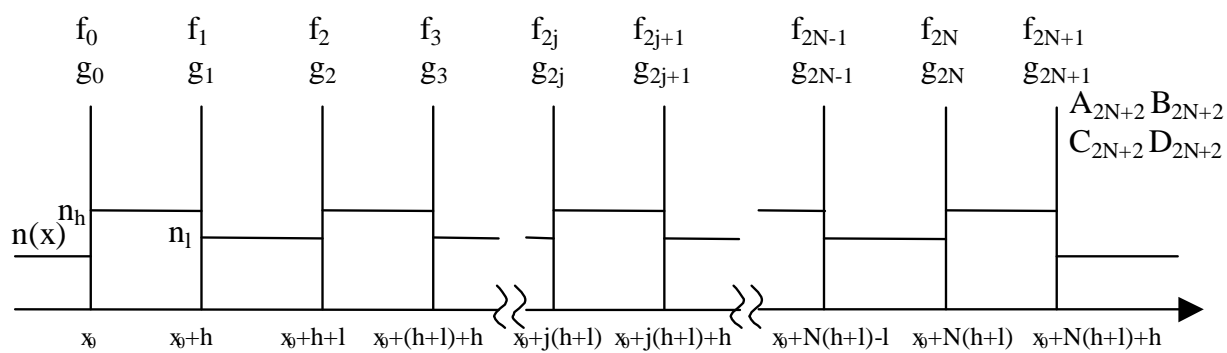

Figure 2.3.a.

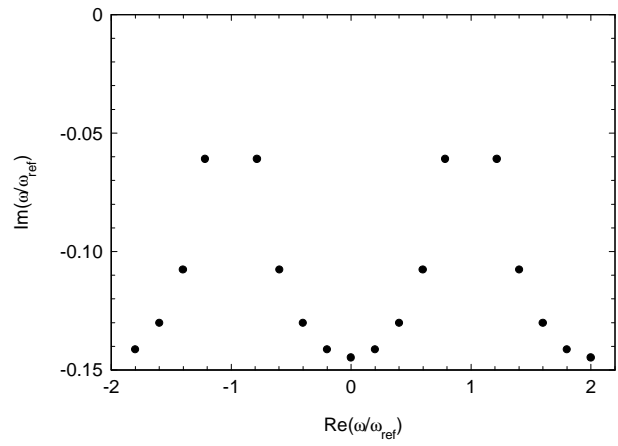

Figure 2.3.b.

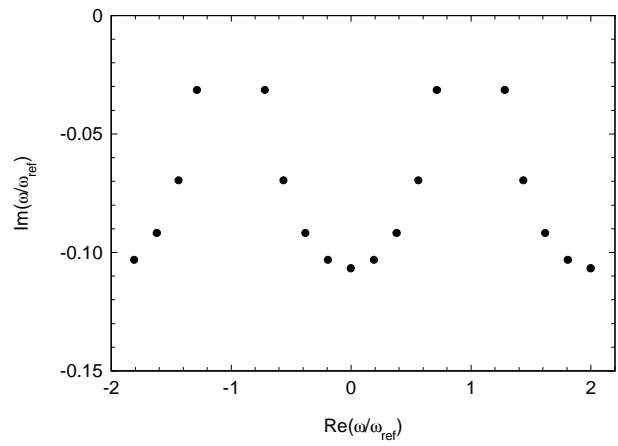

Figure 2.3.c.

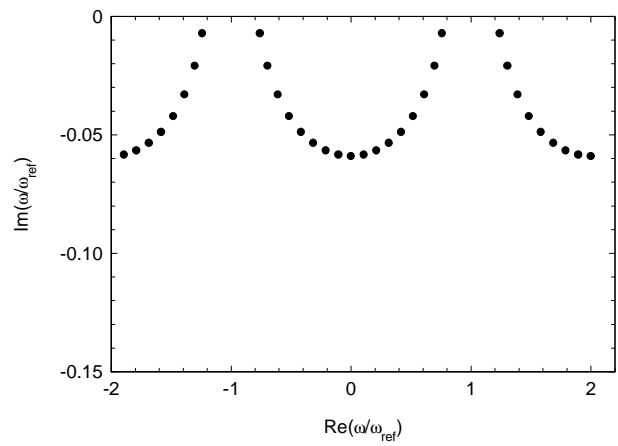

Figure 2.1. A cavity opened at both ends of length *d* with a refractive index $n(x)$, in absence of external pumping so the e.m. field satisfies the "outgoing waves" conditions (2.2.2).

Figure 2.2: (a) Typical behaviour of the refractive index $n(x)$ in a one dimensional (1D) Photonic Crystal (PC) cavity. The index $n(x)$ is generally continuous with *N* discontinuities and it is asymptotic for large distance; (b) refractive index $n(x)$ for a symmetric 1D Photonic Band Gap (PBG) structure.

Figure 2.3: (a) Quasi Normal Mode's (QNM's) frequencies for a symmetric quarter-wave (QW) one dimensional (1D) Photonic Band Gap (PBG) structure with reference wavelength $\lambda_{ref} = 1\mu m$, number of periods *N=4* and refractive indices $n_h$=1.5, $n_l$=1; (b) QNM's frequencies for a symmetric QW 1D-PBG structure with reference wavelength $\lambda_{ref} = 1\mu m$, number of periods *N=4* and refractive indices $n_h$=2, $n_l$=1; (c) QNM's frequencies for a symmetric QW 1D-PBG structure with reference wavelength $\lambda_{ref} = 1\mu m$, number of periods *N=8* and refractive indices $n_h$=2, $n_l$=1.

# Chapter 3

# QNM's approach for open cavities excited by an external pumping: Transmission properties of a 1D-PBG structure.

## 1. Introduction.

The definition of natural modes for confined structures is one of the central problems in physics [4], as in nuclear physics, astrophysics, etc. The main problem is due to the boundary conditions, when they are such to push out the problem from the class of Sturm-Liouville. This occurs when boundary conditions imply the presence of eigen-values, as for example when a scatterer excited from the outside [5] gives rise to a transmitted and reflected field. An open cavity with an external or internal excitation represents a "non-canonical " problem, in the sense of a Sturm–Liouville's problem, due to the fact that the cavity modes couple themselves with the external modes.

One dimensional (1D) Photonic Band Gap (PBG) structures [3]-[5] of finite length manifest all the aspects related to a class of problems which do not belong to the Sturm-Liouville's class: in fact, they behave as scattering objects when they are excited from outside and as open cavities when excited from the inside.

In this chapter, the Quasi Normal Mode's (QNM's) approach, defined for an open cavity in absence of an external pumping [1], is extended to the e.m. problem of a cavity which is excited by an external pump [7]; then, the QNM's approach is applied to a symmetric quarter wave (QW) 1D-PBG structure to discuss its transmission properties [6]. The chapter is organized as follows. In section 2, the transmission coefficient for an open cavity is calculated as a QNM's super-position. In section 3, a transmission coefficient formula is obtained for a symmetric QW 1D-PBG structure. Moreover, the transmission properties are discussed in terms of the QNM's eigen frequencies. Then, the "e.m. modes", i.e. the field distributions at the transmission resonances, are obtained as super-positions of the QNM's eigen functions. Finally, the external pumping of the 1D-PBG structure is generalized as two counter-propagating field pumps, showing the e.m. interference inside the cavity of the two field pumps. Conclusions are given in section 4.

## 2. Extending QNM's approach:
## e.m. problem of an open cavity excited by an external pump.

In this section, the Quasi Normal Mode's (QNM's) approach is extended to the e.m. problem of an open cavity which is excited by an external pump; these considerations are here reported in more details respect to ones already presented in ref [7]. With reference to fig. 3.1., consider now an open cavity of length $d$, with a refractive index $n(x)$ and excited by a pump $E^{(P)}(x,t)$ incoming from the left side.

Under these conditions, the QNM's functions form a complete basis only inside the open cavity, and the e.m. field can be calculated as a superposition of QNMs [2]

$$E(x,t) = \sum_n a_n(t) F_n(x), \quad \text{for} \quad 0 \leq x \leq d, \tag{3.2.1}$$

where are introduced the normalized QNM's functions $F_n(x) = f_n(x)\sqrt{2\omega_n/\langle f_n | f_n \rangle}$, denoting the QNM's norm as $\langle f_n | f_n \rangle$. The superposition coefficients satisfy the dynamic equation [2]

$$\dot{a}_n(t) + i\omega_n a_n(t) = \frac{i}{2\omega_n \sqrt{\rho_0}} F_n(0) J(t), \tag{3.2.2}$$

where is defined a driving force: $J(t) = -2\sqrt{\rho_0}\, \partial_x E^{(P)}(x,t)\big|_{x=0}$. The left-pump satisfies the "incoming wave condition" [2]:

$$J(t) = -2\sqrt{\rho_0}\, \partial_x E^{(P)}(x,t)\big|_{x=0} = 2\rho_0 \partial_t E^{(P)}(0,t). \tag{3.2.3}$$

Each QNM is driven by the driving force $J(t)$ and decays at the same time because of $\text{Im}\,\omega_n < 0$. The coupling to the force is determined by the surface value $F_n(0)$ of the QNM's function. The open cavity is in a steady state, so the Fourier's transform with a real frequency $\tilde{E}(x,\omega) = \int_{-\infty}^{\infty} E(x,t)\exp(i\omega t) dt$ can be applied to equations (3.2.1)-(3.2.3), and it follows

$$\begin{cases} \tilde{E}(x,\omega) = \sum_n \tilde{a}_n(\omega) F_n(x) \\ \tilde{a}_n(\omega) = \frac{F_n(0)}{2\omega_n \sqrt{\rho_0}} \frac{\tilde{J}(\omega)}{\omega_n - \omega}, \\ \tilde{J}(\omega) = -2i\rho_0 \omega \tilde{E}^{(P)}(0,\omega) \end{cases} \tag{3.2.4}$$

so

$$\tilde{E}(x,\omega) = \tilde{E}_P(0,\omega) i\omega \sqrt{\rho_0} \sum_n \frac{F_n(0)F_n(x)}{\omega_n(\omega-\omega_n)}. \tag{3.2.5}$$

With reference to figure 3.1, the e.m. field is continuous on the cavity surfaces $x=0$ and $x=d$, so $\tilde{E}(0^-,\omega)=\tilde{E}(0^+,\omega)$ and $\tilde{E}(d^-,\omega)=\tilde{E}(d^+,\omega)$, and the e.m. field $\tilde{E}(0,\omega)$ on the surface $x=0$ is the superposition of the incoming pump $\tilde{E}^{(P)}(0,\omega)$ and the reflected field $\tilde{E}^{(R)}(0,\omega)$, so $\tilde{E}(0,\omega)=\tilde{E}^{(P)}(0,\omega)+\tilde{E}^{(R)}(0,\omega)$, while the e.m. field $\tilde{E}(d,\omega)$ on the surface $x=d$ is only the transmitted field $\tilde{E}_T(d,\omega)$, so $\tilde{E}(d,\omega)=\tilde{E}^{(T)}(0,\omega)$. It follows that the transmission coefficient $t(\omega)$ for an open cavity of length $d$ can be defined as the ratio between the transmitted field $\tilde{E}(d,\omega)$ on the surface $x=d$ and the incoming pump $\tilde{E}^{(P)}(0,\omega)$ on the surface $x=0$:

$$t(\omega) = \frac{\tilde{E}(d,\omega)}{\tilde{E}^{(P)}(0,\omega)}. \tag{3.2.6}$$

The transmission coefficient is obtained as superposition of QNMs inserting eq. (3.2.5) in eq. (3.2.6):

$$t(\omega) = i\omega\sqrt{\rho_0} \sum_n \frac{F_n(0)F_n(d)}{\omega_n(\omega-\omega_n)}. \tag{3.2.7}$$

Applying the QNM's completeness condition [1] $\sum_n \frac{F_n(x)F_n(x')}{\omega_n}=0$ for $0 \le x, x' \le d$, the transmission coefficient simplifies as:

$$t(\omega) = i\sqrt{\rho_0} \sum_n \frac{F_n(0)F_n(d)}{\omega-\omega_n}. \tag{3.2.8}$$

Inserting the normalized QNM's functions $F_n(x) = f_n(x)\sqrt{2\omega_n / \langle f_n | f_n \rangle}$, with $f_n(0)=1$, equation (3.2.8) becomes:

$$t(\omega) = i2\sqrt{\rho_0} \sum_n \frac{f_n(d)}{\langle f_n | f_n \rangle} \frac{\omega_n}{\omega-\omega_n}. \tag{3.2.9}$$

For a symmetric cavity, such that $f_n(d)=(-1)^n f_n(0)=(-1)^n$, finally

$$t(\omega) = \sum_n \frac{(-1)^n}{\gamma_n} \frac{\omega_n}{\omega-\omega_n}, \tag{3.2.10}$$

where $\gamma_n = \langle f_n | f_n \rangle / 2i\sqrt{\rho_0}$. Therefore, the transmission coefficient $t(\omega)$ of an open cavity can be calculated as a superposition of suitable functions, with the QNM's norms $\langle f_n | f_n \rangle$ as weighting coefficients and the QNM's frequencies $\omega_n$ as parameters.

# 3. An application of QNM's approach:
## Transmission properties of a symmetric QW 1D-PBG structure.

In this section, the Quasi Normal Mode's (QNM's) approach is applied to discuss the transmission properties of one dimensional (1D) Photonic Crystal (PC) structures (see refs. [7][6]). For the symmetric 1D-PBG of fig. 2.2.b, with refractive index (2.4.1), the QNM's norm $\langle f_n | f_n \rangle$ can be obtained in terms of the $\omega_n$ frequency and of the $A_{2N+2}(\omega)$, $B_{2N+2}(\omega)$ coefficients for the $g_-(x,\omega)$ auxiliary function in the $L_{2N+2}$ layer on the right-hand side of the 1D-PBG (see Appendix B):

$$\langle f_n | f_n \rangle = 2i \frac{\omega_n}{c} A_{2N+2}(\omega_n) \left( \frac{dB_{2N+2}}{d\omega} \right)_{\omega=\omega_n}. \tag{3.3.1}$$

A symmetric quarter wave (QW) 1D-PBG, with $N$ periods and $\omega_{ref}$ as reference frequency, is characterized by the QNM's frequencies $\omega_{n,m} = \omega_{n,0} + m\Delta$, where $\Delta = 2\omega_{ref}$, being $n \in [0, N]$ and $m \in \mathbb{Z} = \{0, \pm 1, \pm 2, \ldots\}$; the QNM's norm (3.3.1) becomes:

$$\langle f_{n,m} | f_{n,m} \rangle = 2i \frac{\omega_{n,m}}{c} A_{2N+2}(\omega_{n,m}) \left( \frac{dB_{2N+2}}{d\omega} \right)_{\omega=\omega_{n,m}}. \tag{3.3.2}$$

The coefficients $A_{2N+2}(\omega)$, $B_{2N+2}(\omega)$ are of type $\exp(i\delta)$, where $\delta = 2(\omega/\omega_{ref})$ (see ref. [6]). If the coefficients $A_{2N+2}(\omega)$, $B_{2N+2}(\omega)$ are calculated at the QNM's frequencies, it follows $A_{2N+2}(\omega_{n,m}) = A_{2N+2}(\omega_{n,0})$ and $(dB_{2N+2}/d\omega)_{\omega=\omega_{n,m}} = (dB_{2N+2}/d\omega)_{\omega=\omega_{n,0}}$, so from eq. (3.3.2):

$$\gamma_{n,m} = \frac{\langle f_{n,m} | f_{n,m} \rangle}{2i(1/c)} = \frac{\gamma_{n,0}}{\omega_{n,0}} (\omega_{n,0} + m\Delta). \tag{3.3.3}$$

It results that, for the $G_n^{QNM}$ QNM's family, the $\gamma_{n,m}$ norm is periodic with a step $\Gamma = \gamma_{n,0}(\Delta/\omega_{n,0})$. The transmission coefficient (3.2.10) becomes:

$$t(\omega) = \sum_{n=0}^{2N} \sum_{m=-\infty}^{\infty} \frac{(-1)^{n+m}}{\gamma_{n,m}} \frac{\omega_{n,0} + m\Delta}{\omega - \omega_{n,0} - m\Delta}, \tag{3.3.4}$$

and, reminding eq. (3.3.3),

$$t(\omega) = \sum_{n=0}^{2N} (-1)^n \frac{\omega_{n,0}}{\gamma_{n,0}} \sum_{m=-\infty}^{\infty} (-1)^m \frac{1}{\omega - \omega_{n,0} - m\Delta}, \tag{3.3.5}$$

it converges to:

$$t(\omega) = \frac{\pi}{\Delta} \sum_{n=0}^{2N} (-1)^n \frac{\omega_{n,0}}{\gamma_{n,0}} \csc[\frac{\pi}{\Delta}(\omega - \omega_{n,0})]. \tag{3.3.6}$$

## 3. 1. QNM's frequencies and transmission resonances.

As from eq. (3.3.6), the transmission coefficient of a symmetric QW 1D-PBG structure, with *N* periods and $\omega_{ref}$ as reference frequency, is calculated as a superposition of *2N+1* QNM's cosecants, which correspond to the number of QNMs in the $[0, 2\omega_{ref})$ range.

In figure 3.2.a, the transmission coefficient predicted by QNM's approach and then by numeric methods existing in literature [10] are plotted for a symmetric QW 1D-PBG, where the reference wavelength is $\lambda_{ref}=1\mu m$, the number of periods is *N=6* and the values of refractive indices are $n_h=2$, $n_l=1.5$. In figure 3.2.b., the two transmission coefficients are plotted for the same structure of fig. 3.2.a. but with $n_h=3$, $n_l=2$. In figure 3.2.c, the two transmission coefficients are plotted for the same structure as in figure 3.2.b., with an increased number of periods *N=7*. An excellent agreement can be noted between the transmission coefficient predicted by the QNM's theory and that obtained by the numeric methods of ref. [10].

In the tables 3.3, the same examples of symmetric QW 1D-PBG structures are considered: (a) $\lambda_{ref}=1\mu m$, *N=6*, $n_h=2$, $n_l=1.5$, (b) $\lambda_{ref}=1\mu m$, *N=6*, $n_h=3$, $n_l=2$, (c) $\lambda_{ref}=1\mu m$, *N=7*, $n_h=3$, $n_l=2$. For each of the three examples, the low and high frequency band-edges are described in terms of their resonances ($\omega_{Band\text{-}Edge}/\omega_{ref}$) and phases $\angle t(\omega_{B.E.}/\omega_{ref})$; besides, there is one QNM next to every single band-edge: the real part of the QNM's frequency $Re(\omega_{QNM}/\omega_{ref})$ is reported together with the relative shift from the band-edge resonance $(Re\omega_{QNM}-\omega_{B.E.})/\omega_{B.E.}$. From tables 3.3.a. and 3.3.b. (*N=6*), the low and high frequency band-edges are characterized by negative phases, which sum is (-π), while, from table 3.3.c. (*N=7*), their respective phases are positive with a sum π. In fact, a symmetric QW 1D-PBG, with *N* periods and $\omega_{ref}$ as reference frequency, presents a transmission spectrum with *2N+1* transmission resonances in the *[0,2$\omega_{ref}$)* range, i.e. $\exists \omega_n^{peak} \Rightarrow t(\omega_n^{peak})=\exp(i\varphi_n)$, $n=0,1,\ldots,2N$; some numerical simulations on (3.3.6) prove that:

$$\begin{cases} \varphi_0 = 0 \\ \varphi_n + \varphi_{(2N+1)-n} = (-1)^{n+1}\pi, \ n=1,\ldots,N \end{cases} \quad (3.3.7)$$

So, the lowest resonance $\omega_0^{peak}=0$ in the range *[0,2$\omega_{ref}$)* has always a phase $\varphi_0=0$. The phases $\varphi_N$ and $\varphi_{N+1}$ of the low and high frequency band-edges $\omega_N^{peak}$ and $\omega_{N+1}^{peak}$ are such that $\varphi_N + \varphi_{N+1}=(-1)^{N+1}\pi$; so, for a symmetric QW 1D-PBG with *N* periods, the phases of the two band-edges are such that their sum is π when *N* is an odd number or (–π) if *N* is an even number. Again from tables 3.3, there is one QNM next to every single band-edge, with a relative shift from the band-edge. In fact, a symmetric QW 1D-PBG, with *N* periods and $\omega_{ref}$ as reference frequency, presents a transmission coefficient which is a superposition of *2N+1* QNM's cosecants (3.3.6), centred to the $2N+1$ QNM's frequencies in $[0,2\omega_{ref})$ range, i.e. $\omega_n^{QNM}$, $n=0,1,\ldots,2N$. The QNM's cosecants are not sharp functions, so, when they are superposed, an aliasing occurs; the peak of the $n^{th}$ QNM-cosecant can see the tails of the $(n-1)^{th}$ and $(n+1)^{th}$ QNM's cosecants, so the $n^{th}$ resonance of the transmission spectrum results effectively shifted from the $n^{th}$ QNM's frequency i.e. $\Delta\omega_n = \text{Re}\,\omega_{n,0} - \omega_n^{peak} \neq 0$, $\forall n = 0,1,\ldots,2N$. Comparing values in the tables 2.5.a. ($\Delta n=n_h-n_l=0.5$) and table 2.5.b. ($\Delta n=1$), the relative shift $(Re\,\omega_{QNM}-\omega_{B.E.})/\omega_{B.E.}$ decreases with the refractive index step $\Delta n$, while, comparing values in tables 3.3.b. ($N=6$) and 3.3.c. ($N=7$), the same shift increases with the number of periods $N$. Some numerical simulations on (3.3.6) prove that, the more the 1D-PBG presents a large number of periods ($d>>\lambda_{ref}$) with an high refractive index step ($\Delta n>>0$), the more the $n^{th}$ QNM describes the $n^{th}$ transmission peak in the sense that $Re\,\omega_{n,0}$ comes near to $\omega_n^{peak}$.

### 3. 2. QNM's functions and transmission peaks.

Consider now one monochromatic field pump, incoming from the left side, with unit amplitude and zero initial phase:

$$\tilde{E}_\omega^{(\rightarrow)}(x) = \exp(i\frac{\omega}{c}x). \qquad (3.3.8)$$

The e.m. field inside the symmetric QW 1D-PBG structure, with $N$ periods and $\omega_{ref}$ as reference frequency, can be calculated as a superposition of the QNMs $[\omega_{n,m}, f_{n,m}(x)]$, being $\omega_{n,m}= \omega_{n,0}+m\Delta$, $\Delta=2\omega_{ref}$, $n\in[0,2N]$ and $m \in \mathbb{Z} = \{0,\pm 1,\pm 2,\ldots\}$ [see sections 2. and 3.]:

$$\tilde{E}'_\omega(x) = \sum_{n=0}^{2N} \frac{\omega_{n,0}}{\gamma_{n,0}} \sum_{m=-\infty}^{\infty} \frac{f_{n,m}(x)}{\omega - \omega_{n,0} - m\Delta}. \qquad (3.3.9)$$

If the field pump (3.3.8) incoming from the left side is tuned at the transmission peak of the 1D-PBG which is next to the $n^{th}$ QNM frequency of the $[0,2\omega_{ref})$ range, i.e. $\omega_n^{peak} \approx \text{Re}\,\omega_{n,0}$ where $n=0,1,\ldots,2N$, then the "e.m. mode" inside the structure can be calculated from (3.3.9) as a superposition only of the dominant terms

$$\tilde{E}'_n(x) = \sum_{m=-\infty}^{\infty} \tilde{a}_{n,m}(\omega_n^{peak}) f_{n,m}(x), \qquad (3.3.10)$$

where

$$\tilde{a}_{n,m}(\omega_n^{peak}) = \frac{\omega_{n,0}/\gamma_{n,0}}{\omega_n^{peak} - \omega_{n,0} - m\Delta}. \qquad (3.3.11)$$

So the field distribution $\tilde{E}'_\omega(x)$ inside the 1D-PBG structure, tuned to one of the transmission resonance $\omega_n^{peak} \approx \text{Re}\,\omega_{n,0}$, $n=0,1,\ldots,2N$ within the range $[0,2\omega_{ref})$, can be calculated as a superposition of the QNM's functions $f_{n,m}(x)$, $m \in \mathbb{Z} = \{0, \pm 1, \pm 2, ...\}$ which belong to the $n^{th}$ QNM's family; moreover, the weigh-coefficients $\tilde{a}_{n,m}(\omega_n^{peak})$ of the superposition are calculated in the transmission resonance $\omega_n^{peak}$ and depend from the $n^{th}$ QNM's family.

Some numeric simulations prove that, when the transmission resonance is close enough to the QNM's frequency, i.e. $\omega_n^{peak} \approx \text{Re}\,\omega_{n,0}$, the field distribution (3.3.10) reduces to its first order approximation:

$$\tilde{E}'_n(x) = \tilde{a}_{n,0}(\omega_n^{peak}) f_{n,0}(x). \qquad (3.3.12)$$

Moreover, for large number of periods ($d >> \lambda_{ref}$) and high refractive index step ($\Delta n >> 0$), the weigh-coefficient $\tilde{a}_{n,0}(\omega_n^{peak})$ converges to the value 1 and the first order approximation of (3.3.10) becomes the QNM's approximation:

$$\tilde{E}'_n(x) \approx f_{n,0}(x). \qquad (3.3.13)$$

So, the more the 1D-PBG structure presents a large number of periods with an high refractive index step, the more the QNMs describe the transmission peaks in the sense that the QNM's functions approximate the field distributions in the transmission resonances.

In fig. 3.4, with reference to a symmetric QW 1D-PBG structure *($\lambda_{ref}=1\mu m$, $N=6$, $n_h=3$, $n_l=2$)*, the "e.m. mode" intensity $I_{II\,band-edge}$ at the high frequency band edge *($\omega_{II\,band-edge}/\omega_{ref}=1.178$)*, in units of the intensity $I_{pump}$ for an incoming field pump, is plotted as a function of the dimensionless space $x/d$, where $d$ is the length of the 1D-PBG structure. It is clear the shifting between the QNM's approximation (3.3.12) (the field intensity distribution $I_{II\,band-edge}$ is approximated by the QNM's function corresponding the QNM's frequency next to the high frequency band-edge) and the first order approximation (3.3.13) (a superior order approximation, the field intensity distribution $I_{II\,band-edge}$ is the product of the same QNM's function for a weigh coefficient which takes into account the shift between the band-edge resonance and the QNM's frequency).

### 3. 3. 1D-PBG structures excited by two counter-propagating field pumps.

Consider now a second field pump $\tilde{E}_\omega^{(\leftarrow)}(x)$ at frequency $\omega$, incoming from the right side, with unit amplitude and constant phase $\Delta\varphi$:

$$\tilde{E}_\omega^{(\leftarrow)}(x) = \exp[-i\frac{\omega}{c}(x-d)]\exp[-i\Delta\varphi]. \qquad (3.3.14)$$

If both the two counter-propagating field pumps (3.3.8) and (3.3.14) outside the 1D-PBG structure are tuned at the $n^{th}$ transmission resonances of the *[0,$2\omega_{ref}$)* range, i.e. $\omega = \omega_n^{peak}$ such that $t(\omega_n^{peak}) = \exp(i\varphi_n)$, $n = 0,1,\ldots,2N$, the "e.m. mode" intensity inside the symmetric 1D-PBG can be calculated as

$$\tilde{I}''_n(x) = 4\tilde{I}'_n(x)\cos^2[\phi'_n(x) + \frac{\Delta\varphi - \varphi_n}{2}], \qquad (3.3.15)$$

if the field pump incoming from the left side excites the field distribution (3.3.10)-(3.3.11), inside the structure, represented by $\tilde{E}'_n(x) = \sqrt{\tilde{I}'_n(x)}\exp[i\phi'_n(x)]$. Eq (3.3.15) shows clearly the e.m. interference inside the 1D-PBG structure of the two counter-propagating field pumps.

In fig. 3.5, with reference to a symmetric QW 1D-PBG structure *($\lambda_{ref}=1\mu m$, N=6, $n_h=3$, $n_l=2$)*, the "e.m. mode" intensity excited inside the open cavity by one field pump (- - -) coming from the left side [see eq. (3.3.13)] and by two counter-propagating field pumps in phase (—) [see (3.3.15)] are compared when each of the two field pumps are tuned in (a) the low frequency band-edge *($\omega_{I\ band\text{-}edge}/\omega_{ref}=0.822$)* or (b) the high frequency band-edge *($\omega_{II\ band\text{-}edge}/\omega_{ref}=1.178$)*. The field intensity distributions $I_{I\ band\text{-}edge}$ and $I_{II\ band\text{-}edge}$, in units of the intensity for the two pumps $I_{pump}$, are plotted as functions of the dimensionless space *x/d*, where *d* is the length of the 1D-PBG structure. If the two counter-propagating field pumps in phase are tuned at the low frequency band-edge, there is a constructive interference and the field distribution in that band-edge is reminiscent of the distribution excited by one field pump in the same transmission peak; while, if the two pumps are tuned at the high frequency band-edge, there is a destructive interference and almost no e.m. field penetration occurs in the structure.

Some numerical simulations on (3.3.15) prove that, when a symmetric QW 1D-PBG structure with *N* periods is excited by two counter-propagating field pumps with a phase-difference *Δφ*, if *N* is an even number, the field distribution in the low (high) frequency band-edge increases in strength when the two field pumps are in phase *Δφ=0* (out of phase *Δφ=π*) and almost flags when the two pumps are out phase *Δφ=π* (in phase *Δφ=0*). Moreover, if *N* is an odd number, the two field distributions in the low and high frequency band-edges exchange their physical response with respect to the phase-difference of the two pumps. The duality between the field distributions in the two band-edges, for which the one increases in strength when the other almost flags, can be explained recalling that the transmission phases of the two band edges are such that their sum is π (-π) when *N* is an odd (even) number.

Therefore, if a 1D-PBG structure is excited by two counter-propagating field pumps which are tuned at one transmission resonance, no "e.m. mode" could be produced because it depends not only from the boundary conditions (the two field pumps) but also from the initial conditions (the phase-difference of the two pumps). The "density of modes" must be considered as a dynamic variable which has the flexibility to adjust with respect to the boundary conditions as well as the initial conditions.

## 4. Conclusions.

One-dimensional (1D) Photonic Band Gap (PBG) structures are particular optical cavities, with both sides open to the external environment and a stratified material inside. A 1D-PBG is finite in space and, working with e.m. pulses of a spatial extension longer than the length of the structure, the open cavity cannot be studied as infinite: rather, the boundary conditions must be considered at the two ends of the cavity.

In this chapter, the Quasi Normal Mode's (QNM's) approach, defined for an open cavity in absence of an external pumping, has been extended to the e.m. problem of a cavity which is excited by an external pump; then, the QNM's approach has been applied to a symmetric quarter wave (QW) 1D-PBG structure to discuss its transmission properties. The transmission coefficient $t(\omega)$ of an open cavity has been calculated as a superposition of suitable functions, with the QNM's norms $\langle f_n | f_n \rangle$ as weighting coefficients and the QNM's frequencies $\omega_n$ as parameters; the transmission coefficient of a symmetric QW 1D-PBG structure, with $N$ periods and $\omega_{ref}$ as reference frequency, has been calculated as a superposition of $2N+1$ QNM's cosecants, which correspond to the number of QNMs in the $[0, 2\omega_{ref})$ range. The more the 1D-PBG presents a large number of periods ($d > \lambda_{ref}$) with an high refractive index step ($n_h - n_l > n_0$), the more the $n^{th}$ QNM $[\omega_n, f_n(x)]$ describes the $n^{th}$ transmission peak in the sense that: $Re(\omega_n)$ comes near to the resonance frequency of the $n^{th}$ transmission peak; $|Im(\omega_n)|$ approximates the FWHM of the $n^{th}$ transmission peak; and $|f_n(x)|^2$ approximates the field intensity distribution inside the 1D-PBG at the $n^{th}$ transmission peak. Finally, the external pumping of the 1D-PBG has been generalized as two counter-propagating field pumps, anticipating that the "density of modes" must be considered as a dynamic variable which has the flexibility to adjust with respect to the boundary conditions (i.e. the two field pumps) as well as the initial conditions (i.e. the phase-difference between the two pumps).

# Appendix B.

This appendix describes how to obtain the Quasi Normal Mode's (QNM's) norm (3.3.1) for the symmetric one dimensional (1D) Photonic Band Gap (PBG) structure of fig. 2.2.b. with refractive index (2.4.1).

The homogeneous equation for the auxiliary functions (2.2.9) has been solved, adding the "asymptotic conditions" (2.2.10): the auxiliary function $f(x) = g_-(x,\omega)$ is given by eq. (2.A.6), while the auxiliary function $g_+(x,\omega)$ has a similar expression to eq. (2.A.6), where some coefficients $C_k(\omega)$ and $D_k(\omega)$ replace $A_k(\omega)$ and $B_k(\omega)$ with $k = 0,1,\ldots,2N+1, 2N+2$; the auxiliary function $g_-(x,\omega)$ is a "right to left wave" for $x < 0$, so $A_0(\omega) = 0$, while the auxiliary function $g_+(x,\omega)$ is a "left to right wave" for $x > d$, so $D_{2N+2}(\omega) = 0$. If the continuity conditions are imposed on the 1D-PBG surfaces for the auxiliary function $g_-(x,\omega)$ and its spatial derivative $\partial_x g_-(x,\omega)$, it results eq. (2.A.10), while the continuity conditions for $g_+(x,\omega)$ are similar to (2.A.10), but the coefficients $C_k(\omega)$ and $D_k(\omega)$ replace $A_k(\omega)$ and $B_k(\omega)$ with $k = 0,1,\ldots,2N+1, 2N+2$: the $B_0(\omega)$ coefficient is fixed choosing a normalization condition and all the $[A_k(\omega), B_k(\omega)]$ couples with $k = 1,\ldots,2N+2$ are determined applying the continuity conditions (2.A.10); similarly for the $C_{2N+2}(\omega)$ coefficient and the $[C_k(\omega), D_k(\omega)]$ couples with $k = 0,1,\ldots,2N+1$.

The QNM's frequencies can be calculated, if suitable conditions are imposed on the $g_\pm(x,\omega)$ coefficients. At the QNM's frequencies $\omega = \omega_n$, where $\mathrm{Im}\,\omega_n < 0$, the auxiliary functions $g_\pm(x,\omega)$ are linearly dependent $g_-(x,\omega) \propto g_+(x,\omega)$; they are "right to left waves" for $x < 0$, so $C_0(\omega_n) = 0$, and "left to right waves" for $x > d$, so:

$$\begin{cases} B_{2N+2}(\omega_n) = 0 \\ A_{2N+2}(\omega_n) = C_{2N+2}(\omega_n) \end{cases}. \qquad (3.B.1)$$

Finally, the QNM's norm can be calculated for the symmetric 1D-PBG structure with refractive index (2.4.1). The Wronskian $W(x,\omega)$ [see eq. (2.2.11)] of the auxiliary functions $g_\pm(x,\omega)$ [see eq. (2.A.6)] can be calculated as:

$$W(x,\omega) = \begin{cases} 2i\dfrac{\omega}{c}B_0(\omega)C_0(\omega) & ,\ x \in L_0 \\[2pt] 2in_h\dfrac{\omega}{c}[-A_{2k+1}(\omega)D_{2k+1}(\omega)+B_{2k+1}(\omega)C_{2k+1}(\omega)] & ,\ x \in L_{2k+1},\ k=0,\ldots,N \\[2pt] 2in_l\dfrac{\omega}{c}[-A_{2k+2}(\omega)D_{2k+2}(\omega)+B_{2k+2}(\omega)C_{2k+2}(\omega)] & ,\ x \in L_{2k+2},\ k=0,\ldots,N-1 \\[2pt] 2i\dfrac{\omega}{c}B_{2N+2}(\omega)C_{2N+2}(\omega) & ,\ x \in L_{2N+2} \end{cases} \quad (3.B.2)$$

The Wronskian $W(x,\omega)$ is $x$-independent, for the equivalence of all the expressions (3.B.2), as it can be proved using eq. (2.A.10), so:

$$W(\omega) = 2i\frac{\omega}{c}B_{2N+2}(\omega)C_{2N+2}(\omega), \quad \forall x \in \mathbb{R}. \tag{3.B.3}$$

The QNM's norm $\langle f_n | f_n \rangle = (dW/d\omega)_{\omega=\omega_n}$ (see ref. [1]) is obtained from eq.(3.B.3), using eq (3.B.1):

$$\langle f_n | f_n \rangle = 2i\frac{\omega_n}{c}A_{2N+2}(\omega_n)\left(\frac{dB_{2N+2}}{d\omega}\right)_{\omega=\omega_n}. \tag{3.B.4}$$

# References.

[1] D.N. Paltanayou and E. Wolf, Phys. Rev. D **13**, 913 (1976); B.J. Hoenders, J. Math. Phys. **20**, 329 (1979).

[2] Donald G. Dudley, *Mathematical Foundations for Electromagnetic Theory* (IEEE Press, New York, 1994).

[3] S. John, Phys. Rev. Lett. **53**, 2169 (1984); S. John, Phys. Rev. Lett. **58**, 2486 (1987); E. Yablonovitch, Phys. Rev. Lett. **58**, 2059 (1987); E. Yablonovitch and T. J. Gmitter, Phys. Rev. Lett. **63**, 1950 (1989).

[4] J. Maddox, Nature (London) **348**, 481 (1990); E. Yablonovitch and K.M. Lenny, Nature (London) **351**, 278, 1991; J. D. Joannopoulos, P. R. Villeneuve, and S. H. Fan, Nature (London) **386**, 143 (1997).

[5] J. D. Joannopoulos, *Photonic Crystals: Molding the Flow of Light* (Princeton University Press, Princeton, New York, 1995); K. Sakoda, *Optical properties of photonic crystals* (Springer Verlag, Berlin, 2001); K. Inoue and K. Ohtaka, *Photonic Crystals: Physics, Fabrication, and Applications* (Springer-Verlag, Berlin, 2004).

[6] P. T. Leung, S. Y. Liu, and K. Young, Phys. Rev. A **49**, 3057 (1994); P. T. Leung, S. S. Tong, and K. Young, J. Phys. A **30**, 2139 (1997); P. T. Leung, S. S. Tong, and K. Young, J. Phys. A **30**, 2153 (1997); E. S. C. Ching, P. T. Leung, A. Maassen van der Brink, W. M. Suen, S. S. Tong, and K. Young, Rev. Mod. Phys. **70**, 1545 (1998); P. T. Leung, W. M. Suen, C. P. Sun and K. Young, Phys. Rev. E **57**, 6101 (1998); M. Maksimovic, M. Hammer, E. Van Groesen, Optical Engineering 47(11), 114601 (2008); M. Maksimovic, M. Hammer, E. Van Groesen, Opt. Comm. **281**, 1401 (2008).

[7] S. Severini, A. Settimi, C. Sibilia, M. Bertolotti, A. Napoli, A. Messina, *Quasi Normal Frequencies in open cavities: an application to Photonic Crystals*, Acta Phys. Hung. B **23/3-4**, 135-142 (2005); A. Settimi, S. Severini, B. Hoenders, *Quasi Normal Modes description of transmission properties for Photonic Band Gap structures*, J. Opt. Soc. Am. B, **26**, 876-891 (2009).

[8] A. Settimi, S. Severini, N. Mattiucci, C. Sibilia, M. Centini, G. D'Aguanno, M. Bertolotti, M. Scalora, M. Bloemer, C. M. Bowden, Phys. Rev. E **68**, 026614 (2003).

[9] K. C. Ho, P. T. Leung, Alec Maassen van den Brink, and K. Young, Phys. Rev. E **58**, 2965 (1998).

[10] M. Bendickson, J.P. Dowling, and M. Scalora, Phys. Rev. E **53**, 4107 (1996).

# Figures and captions

Figure 3.1.

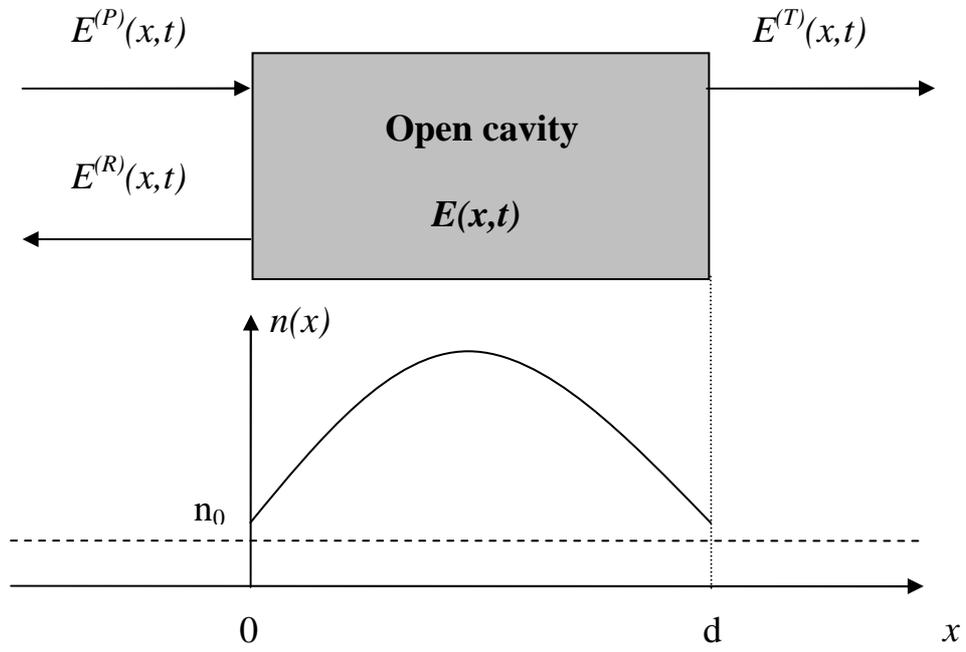

Figure 3.2.a.

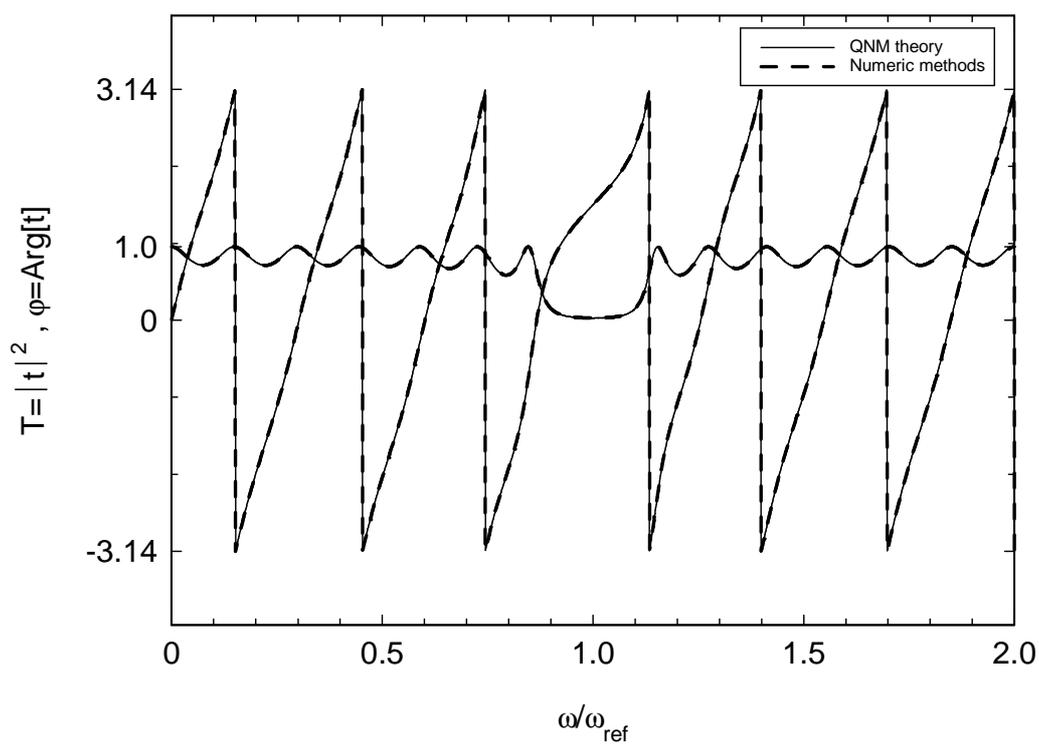

Figure 3.2.b.

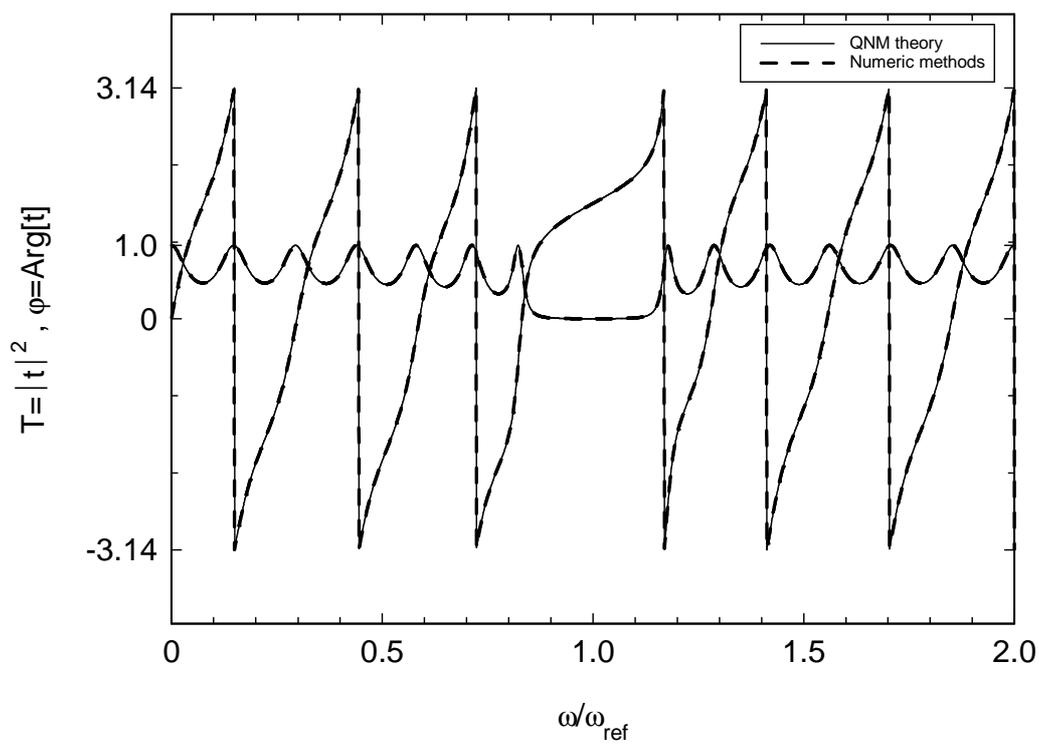

Figure 3.2.c.

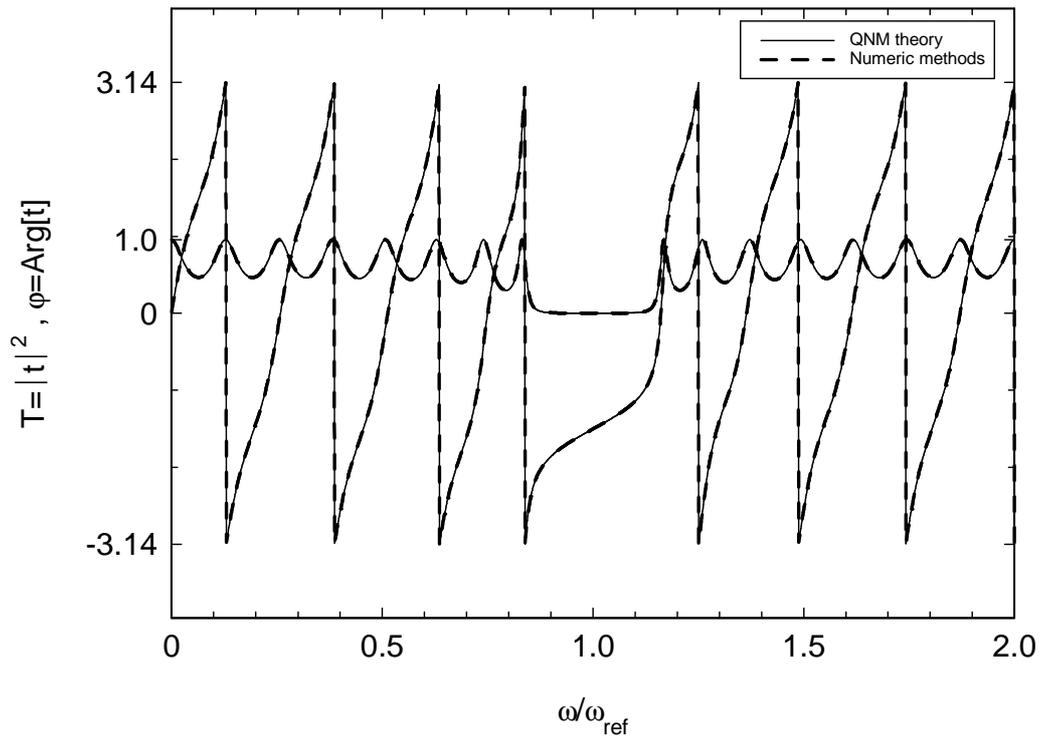

Table 3.3.a.

| $\lambda_{ref}=1\mu m$, N=6, $n_h$=2, $n_l$=1.5 | Re($\omega_{QNM}/\omega_{ref}$) | $\omega_{Band-Edge}/\omega_{ref}$ | (Re$\omega_{QNM}$-$\omega_{B.E.}$)/$\omega_{B.E.}$ | $\angle t(\omega_{B.E.}/\omega_{ref})$ |
|---|---|---|---|---|
| Low frequency band-edge | 0.834 | 0.846 | -0.014 | -0.8011 |
| High frequency band-edge | 1.166 | 1.154 | 0.010 | -2.341 |

Table 3.3.b.

| $\lambda_{ref}=1\mu m$, N=6, $n_h$=3, $n_l$=2 | Re($\omega_{QNM}/\omega_{ref}$) | $\omega_{Band-Edge}/\omega_{ref}$ | (Re$\omega_{QNM}$-$\omega_{B.E.}$)/$\omega_{B.E.}$ | $\angle t(\omega_{B.E.}/\omega_{ref})$ |
|---|---|---|---|---|
| Low frequency band-edge | 0.825 | 0.822 | 0.003 | -0.6342 |
| High frequency band-edge | 1.175 | 1.178 | -0.002 | -2.507 |

Table 3.3.c.

| $\lambda_{ref}=1\mu m$, N=7, $n_h$=3, $n_l$=2 | Re($\omega_{QNM}/\omega_{ref}$) | $\omega_{Band-Edge}/\omega_{ref}$ | (Re$\omega_{QNM}$-$\omega_{B.E.}$)/$\omega_{B.E.}$ | $\angle t(\omega_{B.E.}/\omega_{ref})$ |
|---|---|---|---|---|
| Low frequency band-edge | 0.853 | 0.832 | 0.025 | 2.528 |
| High frequency band-edge | 1.147 | 1.168 | -0.015 | 0.6135 |

Figure 3.4.

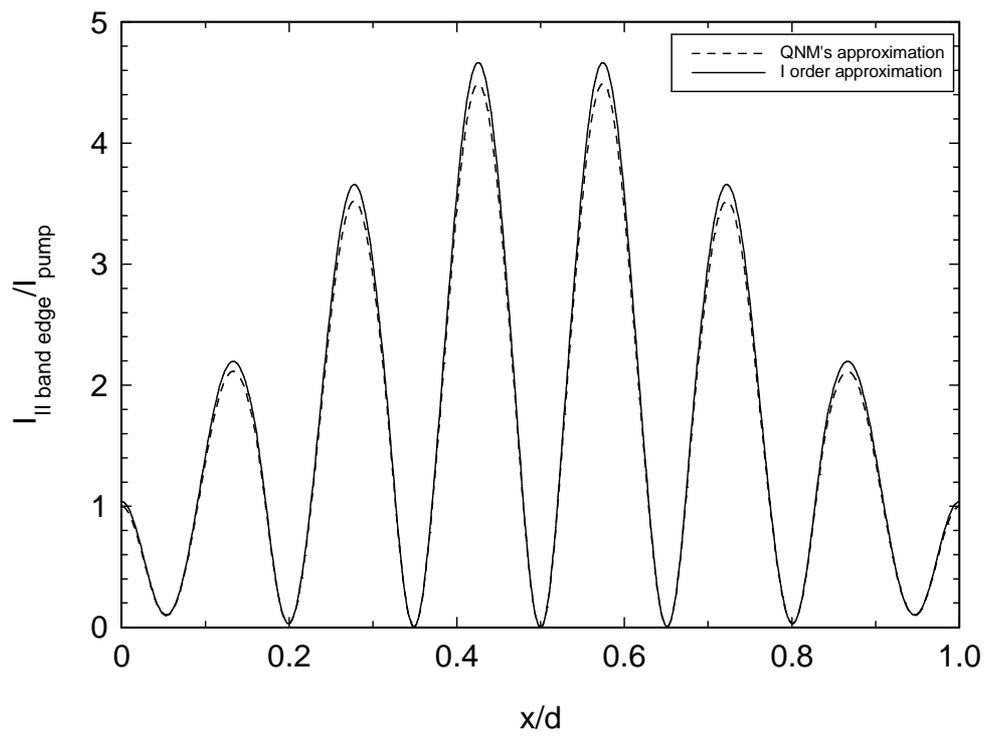

Figure 3.5.a.

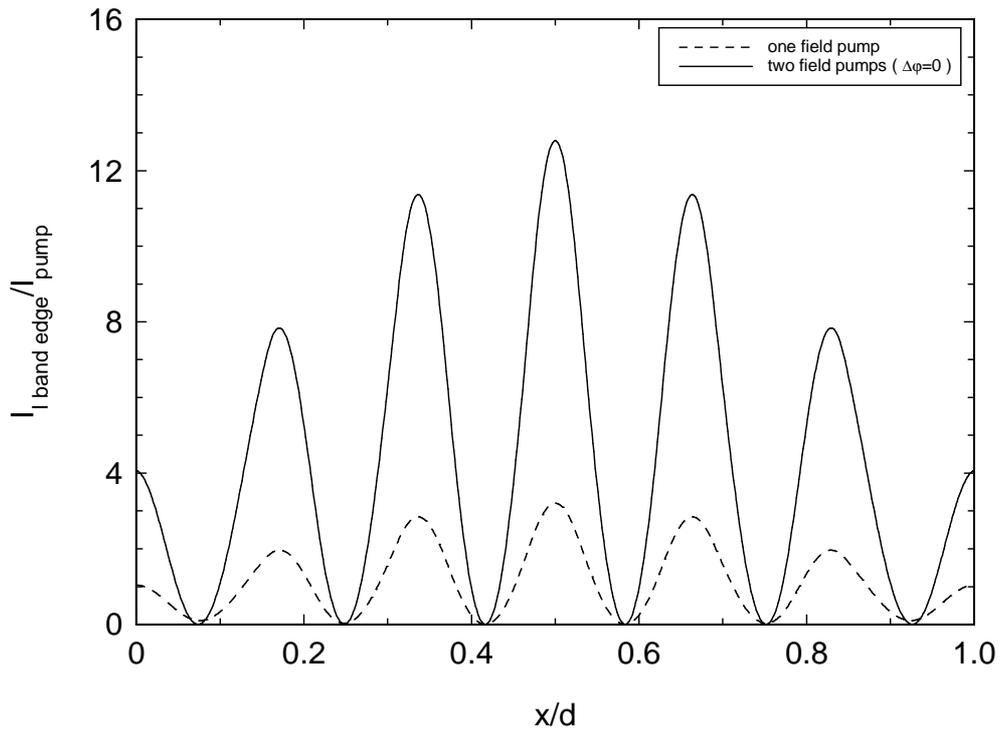

Figure 3.5.b.

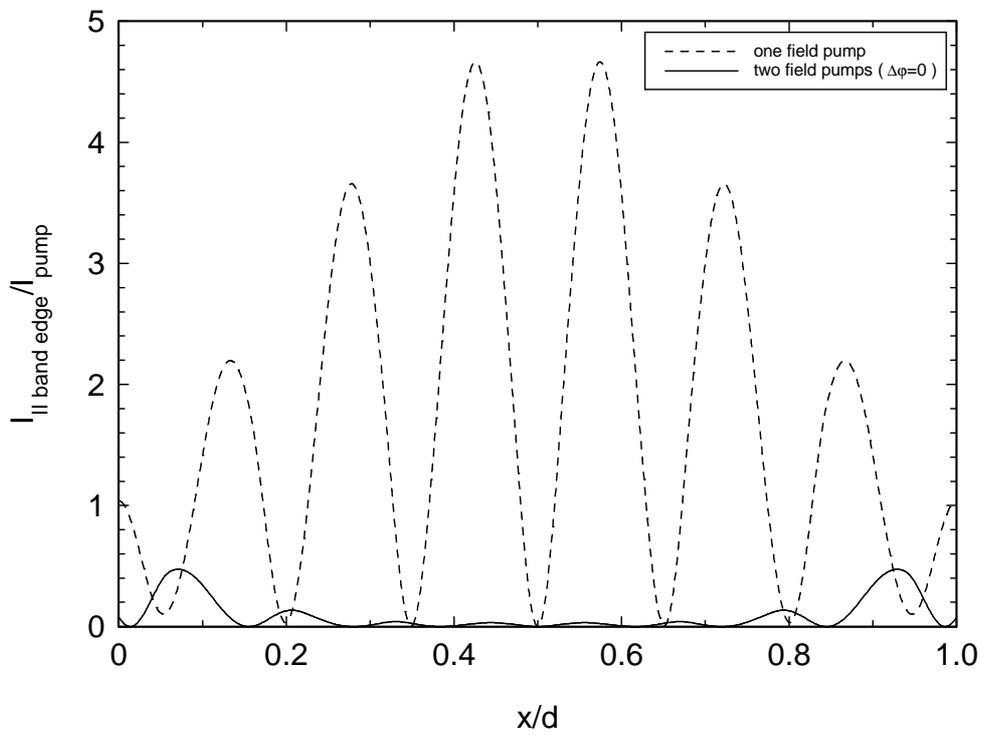

Figure 3.1. A cavity opened at both ends, of length *d* with a refractive index *n(x)*, excited by a left-pump $E^{(P)}(x,t)$, satisfying the "incoming wave" condition (3.2.3).

Figure 3.2. Transmission spectrum (squared modulus and phase) predicted by Quasi Normal Mode's (QNM's) approach (—) and by numeric methods existing in the literature [10] (- - -), for symmetric QW 1D-PBG structures, with reference wavelength $\lambda_{ref} = 1 \mu m$ and: (a) number of periods $N = 6$, refractive indices $n_h = 2$, $n_l = 1.5$; (b) $N = 6$, refractive indices $n_h = 3$, $n_l = 2$; (c) $N = 7$, $n_h = 3$, $n_l = 2$.

Table 3.3. Three examples of symmetric QW 1D-PBG structures are considered: (a) $\lambda_{ref}=1\mu m$, $N=6$, $n_h=2$, $n_l=1.5$, (b) $\lambda_{ref}=1\mu m$, $N=6$, $n_h=3$, $n_l=2$, (c) $\lambda_{ref}=1\mu m$, $N=7$, $n_h=3$, $n_l=2$. For each of the three examples, the low and high frequency band-edges are described in terms of their resonances ($\omega_{Band-Edge}/\omega_{ref}$) and phases $\angle t(\omega_{B.E.}/\omega_{ref})$; besides, there is one QNM's frequency next to every single band-edge: the real part of the QNM's frequency $Re(\omega_{QNM}/\omega_{ref})$ is reported together with the relative shift from the band-edge resonance $(Re\omega_{QNM}-\omega_{B.E.})/\omega_{B.E.}$.

Figure 3.4. With reference to a symmetric QW 1D-PBG structure *($\lambda_{ref}=1\mu m$, $N=6$, $n_h=3$, $n_l=2$)*, the "e.m. mode" intensity $I_{II\,band-edge}$ at the high frequency band edge *($\omega_{II\,band-edge}/\omega_{ref}=1.178$)*, in units of the intensity $I_{pump}$ for a field pump, is plotted as a function of the dimensionless space *x/d*, where *d* is the length of the 1D-PBG structure. It is clear the shifting between the QNM's approximation (3.3.12) (the field intensity distribution $I_{II\,band-edge}$ is approximated by the QNM's function corresponding the QNM's frequency next to the high frequency band-edge) and the first order approximation (3.3.13) (a superior order approximation, the field intensity distribution $I_{II\,band-edge}$ is the product of the same QNM's function for a weigh coefficient which takes into account the shift between the band-edge resonance and the QNM's frequency).

Figure 3.5. With reference to a symmetric QW 1D-PBG structure *($\lambda_{ref}=1\mu m$, $N=6$, $n_h=3$, $n_l=2$)*, the "e.m. mode" intensities excited inside the open cavity by one field pump (- - -) coming from the left side [see eq. (3.3.13)] and by two counter-propagating field pumps in phase (—) [see (3.3.15)] are compared when each of the two field pumps are tuned in (a) the low frequency band-edge *($\omega_{I\,band-edge}/\omega_{ref}=0.822$)* or (b) the high frequency band-edge *($\omega_{II\,band-edge}/\omega_{ref}=1.178$)*. The field intensity distributions $I_{I\,band-edge}$ and $I_{II\,band-edge}$, in units of the intensity for the two pumps $I_{pump}$, are plotted as functions of the dimensionless space *x/d*, where *d* is the length of the 1D-PBG structure. So, two counter-propagating field pumps tuned to the same transmission resonance do not necessarily excite the field distribution that one might expect at that frequency if the same field pumps have an appropriate phase-difference.

**Second Part:**

**QNM's approach for open cavities in quantum electrodynamics.**

# Chapter 4

# Non canonical Quasi Normal Mode's Quantization for open cavities.

## 1. Introductive discussion.

The calculations in quantum optics have traditionally used a form of quantum electrodynamics in which the e.m. field is assumed to be confined within an optical cavity [1][2]. The e.m. field is quantized in terms of a complete set of discrete eigen-modes of the cavity. When appropriate boundary conditions are applied at the cavity walls, the discrete modes can have the form of a standing or running wave. However, the vast majority of optical experiments have no identifiable cavity, but rather the optical energy flows from sources through some kind of interaction region to a set of detectors. For systems of this nature, it is preferable to quantize the electromagnetic field in free space with a set of e.m. modes characterized by a continuous wave vector [3][4].

Blow *et al.* have formulated in ref. [5] the quantum theory of optical wave propagation without recourse to the cavity quantization. This approach avoids the introduction of a box-related mode spacing and enables to use a continuum frequency space description. Ref. [5] introduces a complete orthonormal set of operators that can describe states of finite energy. The set is countable and the operators have all the usual properties of the single-mode frequency operators. With the use of these operators a generalization of the single-mode normal-ordering theorem is proved. The formalism presented in ref. [5] is suited to the treatment of quantum optical systems where the propagation effectively takes place in one dimension with no restriction on the optical length of the one dimensional axis. Finally it should be emphasized that the removal of the notional cavity is accomplished without any increase in the mathematical complexity of the formalism.

However, with the advent of modern optical materials, such as optical fibres, the problem of quantization of the e.m. field in dielectrics has become an important subject and much activity has been taking place in this field.

In the case of linear conservative dielectrics, the quantization is well-known for systems where the permittivity (electric permeability) $\varepsilon$ is a real constant [1][2]. For inhomogeneous media the situation is more complicated, because the permittivity is a real function of space, i.e. $\varepsilon = \varepsilon(x)$.

Recently two methods have been developed for the quantization of the e.m. field in inhomogeneous linear dielectrics.

The first [referred to as the Langevin noise (LN) method] [6] is based upon the introduction of Langevin noise current (and charge) densities, as dictated by the fluctuation-dissipation theorem [4], into the classical Maxwell equations which can then be transferred to quantum theory by conversion of the field quantities into operators. In this scheme, the dyadic Green's function associated with the classical (inhomogeneous) Helmholtz equation plays a prominent role. Its properties come into play by deriving the equal-time commutators for the e.m. fields, given those of the noise current operator.

The second [referred to as the auxiliary field (AF) method] [7] also starts off from the classical Maxwell equations. Here, the introduction of a set of auxiliary fields (instead of a noise current) allows the replacement of the Maxwell's equations, which features a time convolution term relating the polarization to the electric field, by a new set of equations for the combined set of electromagnetic and auxiliary fields but without time convolutions. The formalism can then be cast into a Hamiltonian form which, upon quantization, features a noise current with the same commutation properties as in the LN formalism [8]. For ease of comparison, a generalized temporal gauge has been adopted, but the actual choice does of course not affect the equal-time commutator.

A third method uses the discrete (dissipative) Quasi Normal Modes (QNMs) of a leaky cavity rather than the continuous (Hermitian) NMs of universe [1]-[4]; Ho *et al.* in ref. [2] already made an essential first step towards the application of QNMs to quantum electrodynamics phenomena in one side open and homogeneous cavities: the second quantization of the e.m. field in a leaky cavity is formulated, from first principles, in terms of the QNMs, which are eigen-solutions of the evolution equation, decaying exponentially in time as the energy leaks to outside [1] . The QNM's method is usually justified in the following way: let the central system $C = [0, d]$ (here the open cavity) be described by a Hamiltonian $H_C$ and the bath $B = (-\infty, 0) \cup (d, \infty)$ (the rest of universe) be described by a Hamiltonian $H_B$, with the two subsystems coupled by an interaction term $H_I$, $H = H_C + H_B + H_I$; the effect of $H_I$ is then eliminated and expressed as a dissipation and a fluctuation force acting on *C* [11]. There appears to be a novelty in applying such arguments to problems of the present type, which can be characterized in general as involving leaky cavity; in fact, the degrees of freedom of *C* are $\phi(x,t)$, $0 \leq x \leq d$, while those of *B* are $\phi(x,t)$, $x < 0$, $x > d$. The two are coupled not by a term in the Hamiltonian but by boundary conditions [12]: the Hilbert space of wave functions satisfying the boundary conditions, hence the subspaces for *C* and *B*, "know" about each other through the conditions at the interfaces.

In ref. [8], the second quantization scheme based on QNM's approach has been extended to double side open and inhomogeneous cavities, excited by vacuum fluctuations of the universe. In ref. [9], the QNM's second quantization has been applied to leaky cavities, but excited by two counter-propagating field pumps. The QNM's quantization succeeds in incorporating the external pumping fields tracing back their effect to that of two assigned electrical currents [15] existing only on the two limiting surfaces of the open cavity; as a consequence, ref. [9] enables to establish, for the first time, a direct link between the internal QNM's operators of ref. [2] and the external NM's ones of ref. [5] at a generic frequency. Besides, QNM's commutation rule is not canonical, and it depends on the geometry of the leaky cavity and the phase-difference of the two pumps; ref. [9] confirms that the third method of ref. [2] does not obey to canonical commutation rules adopted by the first and second methods of ref. [6]-[8], so QNM's method, respect to the other methods, paves naturally the way to interesting investigations like, for example, the modified density of modes inside the open cavity [16], as well as the consequent modification of the spontaneous and stimulated emission processes of an atom placed inside the leaky cavity.

## 2. Incoming Waves conditions.

The expansion of a classical field sketched in chapter 1 is restricted to outgoing waves: the linear space $\Gamma$ of function pairs $[\phi(x,t), \bar{\phi}(x,t)]$ is considered, where the scalar field $\phi(x,t)$ and its Lagrange conjugate momentum $\bar{\phi}(x,t)$ is defined on the interval $C=[0,d]$ of the open cavity; the two functions $\phi(x,t)$ and $\bar{\phi}(x,t)$ are differentiable and they satisfy the outgoing waves conditions (1.2.6). In preparing the ground for the expansion of a quantum field [1]-[4], it is necessary to remove this restriction, for the simple reason that zero-point quantum fluctuations will inevitably contain incoming waves as well. Moreover, one would wish that the ensuing theory should be applicable to situations where there are incoming pump fields.

Thus, the equation (1.2.2) of the scalar field $\phi(x,t)$ is studied for the open cavity $C=[0,d]$ together with the outside ''bath,'' i.e. on the half-lines $x<0$ and $x>d$, with relative dielectric constant $\rho(x) \propto n^2(x)$ satisfying the discontinuity and "no tail" conditions [1]: $\rho(x)=\rho_0$ for $x<0$ and $x>d$. The initial conditions are now arbitrary and accordingly the outgoing waves conditions (1.2.6) are abandoned, i.e. the restriction of the scalar field to the open cavity need not lie in the linear space $\Gamma$. As in ref. [2], the expansion formula [see eqs. (1.2.1) and (1.2.11)], and the inner product definition and notation [see eq. (1.3.4)] are *retained* even though $[\phi(x,t), \bar{\phi}(x,t)] \notin \Gamma$. The sum for the first component converges to $\phi(x,t)$ everywhere, while the sum for the second component converges to $\bar{\phi}(x,t)$ everywhere except at $x=0$ and $x=d$. This flaw on a set of measure zero does not lead to problems, however, for the projection formula (1.2.11) renders the coefficients $a_n(t)$ well defined irrespective of the convergence of the series for $\phi(x,t)$ and $\bar{\phi}(x,t)$.

## 2. 1. Extending QNM's approach:
### e.m. problem of an open cavity excited
### by two classical counter-propagating incoming waves.

In this section, the QNM's approach is extended to solve the e.m. problem of two classical counter-propagating waves exciting an open cavity; this theoretical extension, here reported, has been presented in ref [9]. The idea of counter-propagating field pumps is schematically reported in fig. 4.1; an open cavity of length $d$, filled with a medium having refractive index $n(x)$, is excited by two field pumps which obviously have to satisfy the field equation (1.2.2): the first one, denoted as $\phi^{(\rightarrow)}(x,t)$, travels in the positive $x$ direction and enters the cavity from the left side, whereas the second one $\phi^{(\leftarrow)}(x,t)$, in the negative $x$ direction, enters the cavity from the right side. The two pumps $\phi^{(\rightarrow)}(x,t)$ and $\phi^{(\leftarrow)}(x,t)$ satisfy some boundary conditions, defined incoming wave conditions

$$\begin{cases} \partial_x \phi^{(\rightarrow)}(x,t) = -\sqrt{\rho_0} \partial_t \phi^{(\rightarrow)}(x,t) & \text{for } x < 0 \\ \partial_x \phi^{(\leftarrow)}(x,t) = \sqrt{\rho_0} \partial_t \phi^{(\leftarrow)}(x,t) & \text{for } x > d \end{cases}, \qquad (4.2.1)$$

where $\rho_0 = (n_0/c)^2$ and $n_0$ is the outside refractive index. Owing to the theorem of equivalence for the e.m. sources [15], the two real pumps in the free space can be substituted by two fictitious electrical currents on the surfaces of the cavity:

$$\begin{cases} J_0(t) = -2\sqrt{\rho_0}\, \partial_x \phi^{(\rightarrow)}(x,t)\big|_{x=0} = 2\rho_0 \partial_t \phi^{(\rightarrow)}(0,t) \\ J_d(t) = 2\sqrt{\rho_0}\, \partial_x \phi^{(\leftarrow)}(x,t)\big|_{x=d} = 2\rho_0 \partial_t \phi^{(\leftarrow)}(d,t) \end{cases}. \qquad (4.2.2)$$

So, respect to ref. [2], the fictitious electrical currents on the surfaces of the open cavity have been now justified. The advantage of this replacement is the possibility to develop the QNM's approach only into the spatial domain $C=[0,d]$, including the driving sources now traced back to the two currents on the surfaces of the cavity. In such driven situation, the boundary conditions associated to the equation (1.2.2) of the scalar field, first expressed by the outgoing waves conditions (1.2.6), now can be cast in the following form:

$$\begin{cases} \partial_x \phi(x,t) - \sqrt{\rho_0} \partial_t \phi(x,t) = J_0(t) & \text{for } x=0 \\ \partial_x \phi(x,t) + \sqrt{\rho_0} \partial_t \phi(x,t) = J_d(t) & \text{for } x=d \end{cases}. \qquad (4.2.3)$$

The refractive index $n(x)$ of the open cavity satisfies the discontinuity and no-tail conditions [1], so, inside the cavity, the QNM's functions form an orthogonal basis [2] and the scalar field can be calculated as a superposition of QNMs

$$\phi(x,t) = \sum_{n=-\infty}^{\infty} a_n(t) F_n(x) = \sum_{n=0}^{\infty} [a_n(t) F_n(x) + a_n^*(t) F_n^*(x)], \quad (4.2.4)$$

where are introduced the normalized QNM's functions $F_n(x) = f_n(x)\sqrt{2\omega_n/\langle f_n | f_n \rangle}$, denoting the QNM's norm as $\langle f_n | f_n \rangle$. The coefficients $a_n(t)$ of the QNM's superposition satisfy the dynamic equation:

$$\dot{a}_n(t) + i\omega_n a_n(t) = \frac{i}{2\omega_n \sqrt{\rho_0}} [F_n(0) J_0(t) + F_n(d) J_d(t)]. \quad (4.2.5)$$

In this new scenario, it is easy to verify that the QNM's coefficients $a_n(t) = a_n(0)\exp(-i\omega_n t)$, in absence of an external pumping [see eq. (1.2.12)], can be deduced as a particular case of eq. (4.2.5), imposing $J_0(t) = J_d(t) = 0$. The symmetry property $a_{-n}(t) = a_n^*(t)$ [see eq. (1.2.12)] holds also in presence of counter-propagating pumps, as it can be easily observed from eq. (4.2.5), reminding that $J_0(t)$ and $J_d(t)$ are real functions. From a simple inspection of eq. (4.2.5), all the coefficients $a_n(t)$ are driven by the two currents $J_0(t)$ and $J_d(t)$; even if each coefficient has the property $\mathrm{Im}(\omega_n) < 0$, the incoming waves conditions (4.2.3) ensure the presence of quasi-stationary regime solutions. The coupling between the QNMs inside the cavity and the currents $J_0(t)$ and $J_d(t)$ is determined by the surface values of the QNM's functions $F_n(0)$ and $F_n(d)$. So, the Fourier's transform with a real frequency,

$$\Im[\ ] = \frac{1}{\sqrt{2\pi}} \int_{-\infty}^{\infty} [\ ] \exp(i\omega t) dt, \quad (4.2.6)$$

can be applied to eqs. (4.2.4)-(4.2.5) and (4.2.2). If the two pumps in the free space are tuned at the frequency $\omega$, the scalar field inside the cavity is

$$\tilde{\phi}_\omega(x) = \sum_{n=-\infty}^{\infty} \tilde{a}_n(\omega) F_n(x) = \sum_{n=0}^{\infty} [\tilde{a}_n(\omega) F_n(x) + \tilde{a}_n^*(\omega) F_n^*(x)], \quad (4.2.7)$$

where

$$\tilde{a}_n(\omega) = \frac{F_n(0) \tilde{J}_0(\omega) + F_n(d) \tilde{J}_d(\omega)}{2\sqrt{\rho_0}\omega_n(\omega_n - \omega)}, \quad (4.2.8)$$

being

$$\begin{cases} \tilde{J}_0(\omega) = -2\sqrt{\rho_0}\, \partial_x \tilde{\phi}_\omega^{(\rightarrow)}(x)\big|_{x=0} = 2\rho_0(-i\omega)\tilde{\phi}_\omega^{(\rightarrow)}(0) \\ \tilde{J}_d(\omega) = 2\sqrt{\rho_0}\, \partial_x \tilde{\phi}_\omega^{(\leftarrow)}(x)\big|_{x=d} = 2\rho_0(-i\omega)\tilde{\phi}_\omega^{(\leftarrow)}(d) \end{cases}. \quad (4.2.9)$$

## 3. Second quantization.

Having outlined, in the previous section, necessary aspects of the classical QNM's approach, let turn now to the quantum description of the e.m. field inside and outside an open cavity under incoming waves conditions of two counter-propagating pump fields. In this section, the second quantization of e.m. field in its simplest kind is applied, being described classically by one Lorentz invariant quantity [1] at each space-time point.

In what follows, let report schematically some necessary definitions about the so called scalar fields. The scalar field can be described by an action which is defined as the time integral of a Lagrangian. The Lagrangian density $L$ is a function of the space-time point $(x,t)$ through its dependence on the scalar field $\phi(x,t)$ and its derivatives $\partial_x \phi$ and $\partial_t \phi$. The Lagrangian for the field equation [see eq. (1.2.2)] is calculated as:

$$L(x,t) = \left[ \rho(x)(\partial_t \phi(x,t))^2 - (\partial_x \phi(x,t))^2 \right]/2. \tag{4.3.1}$$

A canonical momentum of the scalar field $\phi(x,t)$ can be defined, in analogy to the systems with a finite number of degrees of freedom, as

$$\bar{\phi}(x,t) = \frac{\delta L}{\delta(\partial_t \phi)}, \tag{4.3.2}$$

and (see ref. [8]) it is calculated as an immediate consequence of its definition (4.3.2)-(4.3.1),

$$\bar{\phi}(x,t) = \rho(x)\partial_t \phi(x,t), \tag{4.3.3}$$

i.e. the product of the refractive index $\rho(x)$ and the time derivative $\partial_t \phi$ of the scalar field.

An Hamiltonian density $h(x,t)$ can be introduced, so that the integral of $h(x,t)$ is the total Hamiltonian $H(t)$. The Hamiltonian density $h(x,t)$ can be defined in terms of the canonical momenta as

$$h(x,t) = \bar{\phi}(x,t)\partial_t \phi(x,t) - L(x,t), \tag{4.3.4}$$

so the total Hamiltonian $H(t)$ is

$$H(t) = \int_{-\infty}^{\infty} h(x,t)dx. \tag{4.3.5}$$

The Hamiltonian density (4.3.4) associated to the Lagrangian (4.3.1) and the canonical momentum (4.3.3) is calculated as

$$h(x,t) = \frac{[\partial_x \phi(x,t)]^2}{2} + \rho(x)\frac{[\partial_t \phi(x,t)]^2}{2}. \tag{4.3.6}$$

As discussed in [2], the total Hamiltonian is time-independent, whether it satisfies the Sommerfeld's radiative conditions [15] of the free space or the boundary conditions of an open cavity with narrow resonances, i.e. [8]

$$H(t) \approx H \qquad (4.3.7)$$

in conservative limit $|Im\omega_n| << Re\omega_n$.

For a quantum mechanics treatment, the equal-time canonical quantization rules are adopted

$$\left[\phi(x,t), \overline{\phi}(x',t)\right] = i\hbar\delta(x-x'), \qquad (4.3.8)$$

where $\delta(x)$ is the Dirac's delta distribution, being $\hbar$ the Planck's constant. Eq (4.3.8) implies that $\phi(x,t)$ and $\overline{\phi}(x,t)$ are no more considered as classical scalar fields, but they should be treated as field operators, denoted as $\hat{\phi}(x,t)$ and $\hat{\overline{\phi}}(x,t)$. The quantization procedure is a direct consequence of the equal-time canonical quantization rules.

## 3. 1. Canonical commutation rules in free space:
## quantum apparatus for counter-propagation of incoming waves.

In this section, the theory of continuum fields in quantum optics [5] is applied to discuss the e.m. problem of two quantum counter-propagating waves exciting an open cavity; this theoretical discussion, here reported, has been presented in ref [9].

The electric field operator $\hat{\phi}^{(\rightarrow)}(x,t)$ associated to a propagation in the positive $x$ direction can be obtained as

$$\hat{\phi}^{(\rightarrow)}(x,t) = \hat{\phi}_{+}^{(\rightarrow)}(x,t) + \hat{\phi}_{-}^{(\rightarrow)}(x,t), \qquad (4.3.9)$$

where

$$\hat{\phi}_{+}^{(\rightarrow)}(x,t) = i\sqrt{\frac{\hbar}{2\varepsilon_0 n_0^2}} \int_0^\infty d\omega \sqrt{\rho_0^{1/2}} \omega \hat{b}(\omega) g_\omega(x) \exp(-i\omega t) \qquad (4.3.10)$$

and, obviously,

$$\hat{\phi}_{-}^{(\rightarrow)}(x,t) = [\hat{\phi}_{+}^{(\rightarrow)}(x,t)]^{\dagger}, \qquad (4.3.11)$$

which are respectively called positive and negative part of the electric field operator. As it can be easily verified, the two quantities alone (4.3.10) and (4.3.11) are not observables. In eq. (4.3.10), $g_\omega(x) = 1/\sqrt{2\pi} \exp(i\omega\sqrt{\rho_0}x)$ is the Normal Mode's (NM's) function, $\varepsilon_0$ the dielectric constant in vacuum and $\hbar = h/2\pi$, being $h$ the Planck's constant. The Fourier's transform (4.2.6) of the field operator (4.3.9) can be expressed as

$$\hat{\phi}_{\omega}^{(\rightarrow)}(x) = i\sqrt{\frac{\hbar}{2\varepsilon_0 n_0^2}} \sqrt{\rho_0^{1/2}\omega} \hat{b}(\omega)\exp(i\omega\sqrt{\rho_0}x), \tag{4.3.12}$$

satisfying, as property,

$$\hat{\phi}_{-\omega}^{(\rightarrow)}(x) = [\hat{\phi}_{\omega}^{(\rightarrow)}(x)]^{\dagger}. \tag{4.3.13}$$

It is easy to verify that the operator (4.3.12) satisfies the incoming wave condition in quantum domain:

$$\partial_x \hat{\phi}_{\omega}^{(\rightarrow)}(x) = i\omega\sqrt{\rho_0}\hat{\phi}_{\omega}^{(\rightarrow)}(x). \tag{4.3.14}$$

In the context of the problem examined in this section, eq. (4.3.12) defines an incoming wave, left to right (LTR) propagating, in the zone $x<0$.

The electric field operator $\hat{\phi}^{(\leftarrow)}(x,t)$ associated to the propagation in the opposite direction, can be obtained applying the parity operator $\pi$ [2] to eqs. (4.3.9)-(4.3.11); the electric field operator propagating in the negative $x$ direction is

$$\hat{\phi}^{(\leftarrow)}(x,t) = \pi^{\dagger}\hat{\phi}^{(\rightarrow)}(x,t)\pi = \hat{\phi}^{(\rightarrow)}(-x,t). \tag{4.3.15}$$

Applying the parity operator $\pi$ to eq. (4.3.12), the Fourier's transform (4.2.6) of the field operator (4.3.12) can be expressed as

$$\hat{\phi}_{\omega}^{(\leftarrow)}(x) = i\sqrt{\frac{\hbar}{2\varepsilon_0 n_0^2}} \sqrt{\rho_0^{1/2}\omega} \hat{b}(\omega)\exp(-i\omega\sqrt{\rho_0}x), \tag{4.3.16}$$

satisfying, as property,

$$\hat{\phi}_{-\omega}^{(\leftarrow)}(x) = [\hat{\phi}_{\omega}^{(\leftarrow)}(x)]^{\dagger}. \tag{4.3.17}$$

By construction, the operator (4.3.16) satisfies the incoming wave condition in quantum domain:

$$\partial_x \hat{\phi}_{\omega}^{(\leftarrow)}(x) = -i\omega\sqrt{\rho_0}\hat{\phi}_{\omega}^{(\leftarrow)}(x). \tag{4.3.18}$$

Eq. (4.3.18) is valid for all $x$ values but, in the context of the problem examined in this section, it defines an incoming wave, right to left (RTL) propagating, in the zone $x<d$.

In all the observables introduced, the NM's annihilation and creation operators, respectively $\hat{b}(\omega)$ and $\hat{b}^{\dagger}(\omega)$, satisfy the canonical commutation rules:

$$[\hat{b}(\omega), \hat{b}^{\dagger}(\omega')] = \delta(\omega - \omega'). \tag{4.3.19}$$

One possible experimental apparatus realizing the incoming waves conditions (4.3.14) and (4.3.18) of two counter-propagating field pumps is shown in fig. 4.2. The fifty-fifty beam splitter *BS* splits an input field $\hat{\phi}^{(in)}(x,t)$ in two output fields, the LTR propagating field pump $\hat{\phi}^{(\rightarrow)}(x,t)$ and the auxiliary field $\hat{\phi}^{(aux)}(x,t)$; after a spatial translation and inversion, the auxiliary field $\hat{\phi}^{(aux)}(x,t)$ gives rise to the RTL propagating field pump $\hat{\phi}^{(\leftarrow)}(x,t)$. This is a possible unitary transformation to realize a quantum counter-propagation of incoming waves. In the quantum domain, the effect of noise incoming from the other input of beam splitter *BS* can not be neglected; so, another input field $\hat{\phi}^{(noise)}(x,t)$ must be considered.

As a quantum model of the beam splitter *BS*, a generic version of the 50/50 beam splitter is considered with a *SU(2)* transfer matrix [2], i.e.

$$B = \frac{1}{\sqrt{2}}\begin{pmatrix} 1 & -e^{i\Delta\varphi} \\ e^{-i\Delta\varphi} & 1 \end{pmatrix}, \qquad (4.3.20)$$

where $\Delta\varphi$ is regarded as the preponderant term, due to the beam splitter *BS*, of the phase difference between the two counter-propagating field pumps, on the surfaces of the open cavity. The transfer matrix (4.3.20), if $\Delta\varphi = -\pi/2$, becomes the usual one, i.e. $\frac{1}{\sqrt{2}}\begin{pmatrix} 1 & i \\ i & 1 \end{pmatrix}$. So, the expression for the Fourier transform of the LTR field pump $\hat{\phi}^{(\rightarrow)}_\omega(x)$, entering the cavity from the left side, is given by the following relation:

$$\hat{\phi}^{(\rightarrow)}_\omega(x) = i\sqrt{\frac{\hbar}{2\varepsilon_0 n_0^2}}\sqrt{\rho_0^{1/2}}\omega \frac{\hat{b}(\omega) - \hat{b}_{noise}(\omega)\exp(i\Delta\varphi)}{\sqrt{2}}\exp(i\omega\sqrt{\rho_0}x), \qquad (4.3.21)$$

where $\hat{b}_{noise}(\omega)$ is the noise field operator, satisfying the canonical commutation relation (4.3.19). The auxiliary field $\hat{\phi}^{(aux)}(x,t)$, outgoing from the beam splitter *BS*, is spatially translated by a system of mirrors (fig. 2.4.). From an operator point of view, these processes can be viewed as an application of two unitary transformations on the field $\hat{\phi}^{(aux)}(x,t)$: the first one transformation, indicated by $T_d$, induces a spatial translation of the cavity length *d*, whereas the second one, $\pi$, makes a spatial inversion; summarizing, the RTL field pump $\hat{\phi}^{(\leftarrow)}(x,t)$ can be viewed as $\pi^\dagger T_d^\dagger \hat{\phi}^{(aux)}(x,t) T_d \pi$ [2]. In order to point out the role of the beam splitter *BS* which introduces the degree of freedom $\Delta\varphi$, the phase term $\Delta\Phi$ due to the propagation of the auxiliary field $\hat{\phi}^{(aux)}(x,t)$ is regarded negligible respect to the phase term $\Delta\varphi$ due the beam splitter *BS*, i.e. $\Delta\Phi << \Delta\varphi$. So, the Fourier's transform of the RTL field pump $\hat{\phi}^{(\leftarrow)}_\omega(x)$, entering the cavity from the left side, is given by:

$$\hat{\phi}^{(\leftarrow)}_\omega(x) = i\sqrt{\frac{\hbar}{2\varepsilon_0 n_0^2}}\sqrt{\rho_0^{1/2}}\omega \frac{\hat{b}(\omega)\exp(-i\Delta\varphi) + \hat{b}_{noise}(\omega)}{\sqrt{2}}\exp[-i\omega\sqrt{\rho_0}(x-d)]. \qquad (4.3.22)$$

Concluding this subsection, it is possible to affirm that, using a set-up as the one reported in fig. 2.4, it is possible to obtain, as inputs of the open cavity, two counter-propagating field pumps $\hat{\phi}^{(\rightarrow)}_\omega(x)$ and $\hat{\phi}_{\omega}^{(\leftarrow)}(x)$ mismatched with a phase-difference $\Delta\varphi$, respectively represented by the operators (4.3.21)-(4.3.22) and satisfying the incoming waves conditions (4.3.14) and (4.3.18).

## 3. 2. Non canonical QNM's commutation rules in an open cavity: link of QNM's operators inside the cavity with NM's operators of the free space.

Let purpose to develop a formalism whereby second quantization of the scalar field $\phi(x,t)$ can be implemented in terms of the discrete QNMs [see eq. (4.2.4)], such that the coefficients $a_n(t)$ and $a_n^*(t)$ of the QNM's superposition can be interpreted respectively as generalized annihilation and creation operators of the QNMs. In analogy with the canonical quantum expansion of scalar fields in the free space, where NM's operators are multiplied by NM's functions [2], second quantization on the field expansion (4.2.4) proceeds first by promoting the scalar field $\phi(x,t)$ to a field operator denoted as $\hat{\phi}(x,t)$, so that the superposition coefficient $a_n(t)$ and $a_n^*(t)$ can be promoted as well as QNM's operators, $\hat{a}_n(t)$ and $\hat{a}_n^\dagger(t)$, multiplying QNM's functions. The e.m. field $\hat{\phi}(x,t)$ can be regarded as an operator only in the open cavity (see ref. [9]), and the QNM's operators $\hat{a}_n(t)$ and $\hat{a}_n^\dagger(t)$ obey the operator form of the motion equation (4.2.5).

The field operator $\hat{\phi}_\omega(x)$, which in Fourier's domain satisfies the property $\hat{\phi}_\omega^\dagger(x) = \hat{\phi}_{-\omega}(x)$, is expressed as a superposition of the QNM's operators $\hat{a}_n(\omega)$ in "(frequency) Heisenberg's representation", i.e.

$$\hat{\phi}_\omega(x) = \sum_{n=-\infty}^{\infty} \hat{a}_n(\omega) F_n(x) = \sum_{n=0}^{\infty} [\hat{a}_n(\omega) F_n(x) + \hat{a}_n^\dagger(\omega) F_n^*(x)] \quad , \quad \forall x \, | \, 0 \le x \le d , \qquad (4.3.23)$$

and the QNM's operators have to satisfy the property $a_n^\dagger(\omega) = a_{-n}(-\omega)$, which is similar to the expression valid for NM's operators. If the equal-time canonical quantization rule (4.3.8) is applied, the QNM's commutation relation can be derived in absence of an external pumping or in presence of two counter-propagating incoming waves [9].

Consider an open cavity in absence of an external pumping. If the equal-time canonical quantization rule (4.3.8) is applied, the commutation relation for the QNM's operators $\hat{a}_n$ and $\hat{a}_n^\dagger$ is derived (see ref. [8])

$$\left[\hat{a}_n, \hat{a}_n^\dagger\right] = \frac{\hbar}{2} \frac{I_n}{1/d} \frac{\text{Re}[\omega_n]}{|\omega_n|^2}, \qquad (4.3.24)$$

where are introduced the overlapping integrals [8]:

$$I_n = \frac{1}{d}\int_0^d \rho(x)|F_n(x)|^2 dx = \frac{\sqrt{\rho_0/d}}{2|\operatorname{Im}\omega_n|}\left[|F_n(0)|^2 + |F_n(d)|^2\right]. \qquad (4.3.25)$$

Eqs. (4.3.24) and (4.3.25) confirm that the operators $\hat{a}_n$ and $\hat{a}_n^\dagger$ can be interpreted as generalized annihilation and creation operators of the QNMs. In fact, in the conservative limit [2], the leaky cavity is characterized by narrow resonances $|Im\omega_n|<<|Re\omega_n|$; since the damping is related to the escape of energy from the two surfaces of the cavity, in eq. (4.3.25) the imaginary part $Im[\omega_n]\to 0$ vanishes with the same velocity of $\sqrt{\rho_0}/2(|F_n(0)|^2 + |F_n(d)|^2) \to 0$ and each overlapping integral (4.3.25) tends to $I_n \approx 1/d$. As a consequence, the QNM's commutator becomes $\left[\hat{a}_n, \hat{a}_n^\dagger\right] \cong \hbar/2\omega_n$, similar to the one concerning the annihilation and creation operators, $\hat{b}_n$ and $\hat{b}_n^\dagger$, of a closed cavity, if $\hat{a}_n = \sqrt{2(\omega_n/\hbar)}\hat{b}_n$ and $\hat{a}_n^\dagger = \sqrt{2(\omega_n/\hbar)}\hat{b}_n^\dagger$ are assumed.

In absence of an external pumping, the equations of motion (4.2.5) for the QNM's operators, in quantum domain, read as

$$\begin{aligned}\hat{a}_n(t) &= \hat{a}_n \exp(-i\omega_n t)\\ \hat{a}_n^\dagger(t) &= \hat{a}_n^\dagger \exp(i\omega_n^* t)\end{aligned}, \qquad (4.3.26)$$

which in Fourier's domain become:

$$\begin{aligned}\hat{a}_n(\omega) &= \hat{a}_n \sqrt{2\pi}\delta(\omega - \omega_n)\\ \hat{a}_n^\dagger(\omega) &= \hat{a}_n^\dagger \sqrt{2\pi}\delta(\omega + \omega_n^*)\end{aligned}. \qquad (4.3.27)$$

So, the QNM's operators are not observables, because satisfy the symmetry conditions $\hat{a}_n^\dagger(\omega) = \hat{a}_{-n}(-\omega) \ne \hat{a}_n(-\omega)$, and the QNM's commutation relation in "(frequency) Heisenberg's representation" becomes:

$$\left[\hat{a}_n(\omega), \hat{a}_n^\dagger(\omega')\right] = 2\pi\left[\hat{a}_n, \hat{a}_n^\dagger\right]\delta(\omega - \omega_n)\delta(\omega + \omega_n^*). \qquad (4.3.28)$$

Consider a leaky cavity $C = [0, d]$ excited by two counter-propagating field pumps $\hat{\phi}_\omega^{(\to)}(x)$ and $\hat{\phi}_\omega^{(\leftarrow)}(x)$, mismatched with a phase-difference $\Delta\varphi$ (see fig 4.2.), respectively represented by the operators (4.3.21)-(4.3.22) and satisfying the incoming waves conditions (4.3.14) and (4.3.18). The quantum principle of correspondence [2] can be applied, in frequency domain, to the two fictitious electrical currents (4.2.9) on the surfaces of the open cavity, i.e.

$$\begin{aligned}\hat{J}_0(\omega) &= 2\rho_0(-i\omega)\hat{\phi}_\omega^{(\to)}(0)\\ \hat{J}_d(\omega) &= 2\rho_0(-i\omega)\hat{\phi}_\omega^{(\leftarrow)}(d)\end{aligned}, \qquad (4.3.29)$$

so, reminding eqs. (4.3.21) and (4.3.22), the operators corresponding to the two currents on the surfaces of the cavity result

$$\hat{J}_0(\omega) = K_\omega[\hat{b}(\omega) - \hat{b}_{noise}(\omega)\exp(i\Delta\varphi)]$$
$$\hat{J}_d(\omega) = K_\omega[\hat{b}(\omega)\exp(-i\Delta\varphi) + \hat{b}_{noise}(\omega)]$$
, (4.3.30)

where is introduced the scalar constant $K_\omega = \sqrt{\hbar/\varepsilon_0 n_0^2}\sqrt{\rho_0}(\omega\sqrt{\rho_0})^3$. Promoting in operator form eq. (4.2.8), the QNM's operators are obtained in (frequency) "Heisenberg's representation"

$$\hat{a}_n(\omega) = A_n(\omega)\hat{J}_0(\omega) + B_n(\omega)\hat{J}_d(\omega), \quad (4.3.31)$$

where $\alpha_n(\omega)$ and $b_n(\omega)$ are two complex functions in frequency:

$$A_n(\omega) = \frac{F_n(0)}{2\sqrt{\rho_0}\omega_n(\omega_n - \omega)}, \quad B_n(\omega) = \frac{F_n(d)}{2\sqrt{\rho_0}\omega_n(\omega_n - \omega)}. \quad (4.3.32)$$

The link (4.3.31) between the internal QNM's operator $\hat{a}_n(\omega)$ and the external canonical NM's operators $\hat{b}(\omega)$ and $\hat{b}_{noise}(\omega)$ turns out to be useful when one is interested to evaluate expectation values for the e.m. field inside an open cavity, since, by using eqs. (4.3.31) and (4.3.30), one can perform all the average operations over the NMs of the free space.

If the leaky cavity $C = [0, d]$ presents a refractive index $n(x)$ satisfying the symmetry properties $n(d/2 - x) = n(x + d/2)$, on the surfaces $x = 0$ and $x = d$ of the open cavity the values of the QNM's functions $F_n(x)$ are such that $F_n(d) = (-1)^n F_n(0)$. In this hypothesis, the two complex functions (4.3.32) satisfy the symmetric property:

$$B_n(\omega) = (-1)^n A_n(\omega). \quad (4.3.33)$$

Inserting eq. (4.3.30) in eq. (4.3.31) and reminding eq. (4.3.33), it results [9]:

$$\hat{a}_n(\omega) = K_\omega A_n(\omega)\{\hat{b}(\omega)[1 + (-1)^n \exp(-i\Delta\varphi)] - \hat{b}_{noise}(\omega)[1 - (-1)^n \exp(-i\Delta\varphi)]\}. \quad (4.3.34)$$

Eq. (4.3.34) establishes the link between the QNMs inside a symmetric cavity and the NMs of the free space. In the particular case $\Delta\varphi = 2\pi$ (or multiple), eq. (4.3.34) becomes:

$$\hat{a}_n(\omega) = \begin{cases} 2K_\omega A_n(\omega)\hat{b}(\omega) & \text{for } n = 0, 2, \ldots \\ -2K_\omega A_n(\omega)\hat{b}_{noise}(\omega) & \text{for } n = 1, 3, \ldots \end{cases}. \quad (4.3.35)$$

As from eq. (4.3.35), the even QNM's operators are influenced only by the input field $\hat{\phi}_\omega^{(in)}(x)$ of the beam splitter *BS*, whereas in the odd QNMs there is information about the noise field $\hat{\phi}_\omega^{(noise)}(x)$. Applying eq. (4.3.34), the commutation rule involving the QNM's operator $\hat{a}_n(\omega)$ and $\hat{a}_n^\dagger(\omega)$ of a symmetric cavity can be calculated, obtaining:

$$[\hat{a}_n(\omega), \hat{a}_n^\dagger(\omega')] = 2K_\omega^2 |A_n(\omega)|^2 (-1)^n \cos(\Delta\varphi)\delta(\omega - \omega'). \quad (4.3.36)$$

If the leaky cavity is pumped by two counter-propagating incoming waves, the QNM's commutation relation is not canonical and depends on the geometry of the open cavity and the phase difference of the two incoming waves. Modification of the commutation relations with respect to the canonical ones, gives rise to a modification of the emission properties of the atom embedded in the cavity, as clearly described in the work of Ueda *et al.* [16]; therefore, an influence is expected on the emission processes due to the geometry of the cavity and the phase difference of the two incoming waves.

### 3. 3. Auto-correlation function of e.m. field inside an open cavity as QNM's superposition depending on the incoming waves.

At a time $t>0$, the quantum state of an e.m. field is defined by specifying a ket $|\psi(t)\rangle$ belonging a state space [2]. The time evolution of the state vector $|\psi(t)\rangle$ is governed by the Schroedinger equation

$$i\hbar \frac{d}{dt}|\psi(t)\rangle = \hat{H}|\psi(t)\rangle, \qquad (4.3.37)$$

where, for the quantum principle of correspondence, the classical total Hamiltonian (4.3.7) is promoted to a quantum operator $\hat{H}$ associated with the total energy of the e.m. field. Since the QNM's operators are linked to the NM's ones [see eq. (4.3.34)], in order to calculate the auto-correlation function of an e.m. field inside an open cavity, it is convenient to calculate the expectation values on the NMs of the free space, which are well known from canonical quantum electrodynamics.

In the free space, two counter-propagating field pumps with a phase-difference $\Delta\varphi$ (see fig. 4.2.), respectively represented by the operators (4.3.21)-(4.3.22) and satisfying the incoming waves conditions (4.3.14) and (4.3.18), are tuned at frequency $\omega$ and prepared in a quantum state

$$|\psi_0\rangle = |\beta(\omega), 0_\omega\rangle, \qquad (4.3.38)$$

if the beam splitter *BS* is excited by an input field $\hat{\phi}_\omega^{(in)}(x)$ lying on a coherent state $|\beta(\omega)\rangle$ of amplitude $\beta(\omega)$ and the noise field $\hat{\phi}_\omega^{(noise)}(x)$ lies on the vacuum fluctuations $|0_\omega\rangle$. The expectation values for the NM's annihilation and creation operators $\hat{b}(\omega)$ and $\hat{b}^\dagger(\omega)$ must be calculated on the initial state (4.3.38). As a consequence, the expectation values of the photon-number and Hamiltonian operators are respectively

$$\langle \hat{N}(\omega) \rangle = \langle \psi_0 | \hat{b}^\dagger(\omega) \hat{b}(\omega) | \psi_0 \rangle \qquad (4.3.39)$$

and

$$\langle \hat{H}(\omega) \rangle = \hbar \omega \langle \hat{N}(\omega) \rangle. \qquad (4.3.40)$$

Adopting the (frequency) Heisenberg's representation, the quantum variances of the two current operators (4.3.30) on the initial state (4.3.37) are operatively defined as

$$\langle \hat{J}_0^\dagger(\omega)\hat{J}_0(\omega)\rangle = \langle\psi_0|\hat{J}_0^\dagger(\omega)\hat{J}_0(\omega)|\psi_0\rangle \quad , \quad \langle \hat{J}_d^\dagger(\omega)\hat{J}_d(\omega)\rangle = \langle\psi_0|\hat{J}_d^\dagger(\omega)\hat{J}_d(\omega)|\psi_0\rangle, \quad (4.3.41)$$

which can be calculated as [9]:

$$\langle \hat{J}_0^\dagger(\omega)\hat{J}_0(\omega)\rangle = \langle \hat{J}_d^\dagger(\omega)\hat{J}_d(\omega)\rangle = \frac{\hbar}{\varepsilon_0 n_0^2}\rho_0(\omega\sqrt{\rho_0})^3\langle\hat{N}(\omega)\rangle. \quad (4.3.42)$$

The quantum cross-correlations of these two current operators, calculated on the initial state, show the properties [9]:

$$\langle \hat{J}_d^\dagger(\omega)\hat{J}_0(\omega)\rangle = \langle \hat{J}_0^\dagger(\omega)\hat{J}_d(\omega)\rangle^* = \langle \hat{J}_0^\dagger(\omega)\hat{J}_0(\omega)\rangle \exp(i\Delta\varphi). \quad (4.3.43)$$

In order to model vacuum fluctuations of the free space, the ground state $|0\rangle = |0_\omega, 0_\omega\rangle$, corresponding to $\beta(\omega) = 0$, is not filled of photons, i.e.

$$\langle\hat{N}(\omega)\rangle_{\beta=0} = 0, \quad (4.3.44)$$

but its zero of energy must be redefined [2] respect to eq. (4.3.40):

$$\langle\hat{H}(\omega)\rangle_{\beta=0} \triangleq \hbar\omega/2. \quad (4.3.45)$$

As a consequence, the two current operators (4.3.30) can be still characterized by the quantum cross-correlations (4.3.43)-(4.3.42), i.e.

$$\langle \hat{J}_d^\dagger(\omega)\hat{J}_0(\omega)\rangle = \langle \hat{J}_0^\dagger(\omega)\hat{J}_d(\omega)\rangle^* = \frac{\hbar}{\varepsilon_0 n_0^2}\rho_0(\omega\sqrt{\rho_0})^3\langle\hat{N}(\omega)\rangle \exp(i\Delta\varphi), \quad (4.3.46)$$

since, in hypothesis of vacuum fluctuations, i.e. $\langle\hat{N}(\omega)\rangle_{\beta=0} = 0$, the two currents can be retained uncorrelated, i.e. $\langle \hat{J}_d^\dagger(\omega)\hat{J}_0(\omega)\rangle_{\beta=0} = \langle \hat{J}_0^\dagger(\omega)\hat{J}_d(\omega)\rangle_{\beta=0} = 0$. Then, the quantum variances of the two current operators must be calculated as

$$\langle \hat{J}_0^\dagger(\omega)\hat{J}_0(\omega)\rangle = \langle \hat{J}_d^\dagger(\omega)\hat{J}_d(\omega)\rangle = \frac{\sqrt{\rho_0}}{\varepsilon_0 n_0^2}(\rho_0\omega)^2\left(\hbar\omega\langle\hat{N}(\omega)\rangle + \langle\hat{H}(\omega)\rangle_{\beta=0}\right), \quad (4.3.47)$$

since, in hypothesis of vacuum fluctuations, i.e. $\langle\hat{N}(\omega)\rangle_{\beta=0} = 0$, the variances of the two currents must be retained non-null, i.e. $\langle \hat{J}_0^\dagger(\omega)\hat{J}_0(\omega)\rangle_{\beta=0} = \langle \hat{J}_d^\dagger(\omega)\hat{J}_d(\omega)\rangle_{\beta=0} \propto \langle\hat{H}(\omega)\rangle_{\beta=0}$.

The open cavity is excited by two counter-propagating field pumps, respectively represented by the operators (4.3.21)-(4.3.22), in the initial state (4.3.38). So, the expectation values of the QNM's operators can be directly calculated on the initial state (4.3.38). Applying the method of ref. [2], the auto-correlation function of the e.m. field inside an open cavity can be calculated as a QNM's superposition depending on the incoming waves [9]

$$\Phi(x, x', \omega) = \langle \psi_0 | \hat{\phi}_\omega^\dagger(x) \hat{\phi}_\omega(x') | \psi_0 \rangle = \langle \hat{\phi}_{-\omega}(x) \hat{\phi}_\omega(x') \rangle =$$

$$= \langle \hat{J}_0^\dagger(\omega) \hat{J}_0(\omega) \rangle \sum_{n=-\infty}^{\infty} \sum_{n'=-\infty}^{\infty} \frac{F_n(0) F_{n'}(0) F_n(x) F_{n'}(x')}{4\rho_0 \omega_n \omega_{n'} (\omega_n - \omega)(\omega_{n'} + \omega)} +$$

$$+ \langle \hat{J}_d^\dagger(\omega) \hat{J}_d(\omega) \rangle \sum_{n=-\infty}^{\infty} \sum_{n'=-\infty}^{\infty} \frac{F_n(d) F_{n'}(d) F_n(x) F_{n'}(x')}{4\rho_0 \omega_n \omega_{n'} (\omega_n - \omega)(\omega_{n'} + \omega)} +, \qquad (4.3.48)$$

$$+ \langle \hat{J}_d^\dagger(\omega) \hat{J}_0(\omega) \rangle \sum_{n=-\infty}^{\infty} \sum_{n'=-\infty}^{\infty} \frac{F_n(0) F_{n'}(d) F_n(x) F_{n'}(x')}{4\rho_0 \omega_n \omega_{n'} (\omega_n - \omega)(\omega_{n'} + \omega)} +$$

$$+ \langle \hat{J}_0^\dagger(\omega) \hat{J}_d(\omega) \rangle \sum_{n=-\infty}^{\infty} \sum_{n'=-\infty}^{\infty} \frac{F_n(d) F_{n'}(0) F_n(x) F_{n'}(x')}{4\rho_0 \omega_n \omega_{n'} (\omega_n - \omega)(\omega_{n'} + \omega)}$$

where, in previous calculations, the property $\hat{J}_{0/d}^\dagger(\omega) = \hat{J}_{0/d}(-\omega)$ is used for the current operators.

## 4. Discussion and conclusions.

This chapter has examined the properties of a one-dimensional (1D) size-limited optical cavity, subjected to two quantized counter propagating field pumps entering through the two limiting surfaces of the optical device. The interest is strategically confined to the behaviour of the leaky cavity in its own and then the environmental field pumps are regarded as assigned external sources acting upon the optical structure under scrutiny. From a mathematical point of view, this target has been addressed exploiting the Quasi Normal Modes's (QNM's) approach in a second quantization scheme, here appropriately extended with the aim of covering the treatment of an open cavity coherently pumped by noisy field pumps. It is shown that the presence of such pumps may be globally taken into account by introducing a new physically transparent description of the boundary conditions fulfilled by the e.m. field inside the cavity. This chapter succeeds in incorporating the external counter propagating field pumps tracing back their effect to that of two assigned electrical currents existing only on the two limiting surfaces of the leaky cavity. This method enables to establish, for the first time, a direct link between the internal QNM's operators and the external Normal Mode's (NM's) ones at a generic frequency. The advantage of such kind of relations stems from the possibility of tracing back the calculation of the expectation value of any field operator inside a open cavity to that involving only field operators of free space. In particular, this route has been followed in order to get the expression of the auto-correlation function of the e.m. field inside the cavity. The main result of this chapter, beside the introduction of a peculiar language directly stemming from the fact of considering such a optical device as a forced open quantum system, is the understanding of the changes provoked by the two field pumps on some optical properties of the device. Such studies pave the way to other interesting investigations like, for example, the modified density of states inside the optical structure, as well as the consequent modification of the spontaneous and stimulated emission processes of an atom placed inside the structure. In order to explore the usefulness of these ideas and mathematical approach in a physical context of theoretical and applicative interest, the optical device can be specialized as a 1D Photonic Band Gap (PBG) of finite length.

# References.

[1] A. Lahiri, P. B. Pal, *A first book of Quantum Field Theory* (CRC Press, Calcutta, 2000); C. Cohen-Tannoudji, J. Dupont-Roc, G. Grynberg, *Photons and Atoms – Introduction to Quantum Electrodynamics* (John Wiley, New York, 1997).

[2] J. J. Sakurai, *Advanced Quantum Mechanics* (Addison-Wesley, New York, 1995); S. Weinberg, *The Quantum Theory of Fields* (Cambridge University Press, New York, 1996).

[3] L. Mandel, E. Wolf, *Optical Coherence and Quantum Optics* (Cambridge University Press, Cambridge, 1995).

[4] R. Feynman, *Quantum Electrodynamics* (Benjamin, New York, 1962).

[5] K.J. Blow, Rodney Loudon, Simon J.D. Phoenix, T.J. Shepherd, Phys. Rev. A **42**, 4102 (1990).

[6] T. Gruner and D.-G. Welsch, Phys. Rev. A **53**, 1818 (1996); Ho Trung Dung, L. Knöll, and D. G. Welsch, Phys. Rev. A **57**, 3931 (1998); S. Scheel, L. Knöll, and D.G. Welsch, Phys. Rev. A **58**, 700 (1998).

[7] A. Tip, Phys. Rev. A **57**, 4818 (1998).

[8] A. Tip, L. Knöll, S. Scheel, and D. G. Welsch, Phys. Rev. A **63**, 034806 (2001).

[9] K. C. Ho, P. T. Leung, Alec Maassen van den Brink and K. Young, Phys. Rev. E **58**, 2965 (1998).

[10] P. T. Leung, S. Y. Liu, and K. Young, Phys. Rev. A **49**, 3057 (1994); P. T. Leung, S. S. Tong, and K. Young, J. Phys. A **30**, 2139 (1997); P. T. Leung, S. S. Tong, and K. Young, J. Phys. A **30**, 2153 (1997); E. S. C. Ching, P. T. Leung, A. Maassen van der Brink, W. M. Suen, S. S. Tong, and K. Young, Rev. Mod. Phys. **70**, 1545 (1998).

[11] W. H. Louisell and L. R. Walker, Phys. Rev. **137**, B204 (1965); I. R. Senitszky, *ibid.* **119**, 670 (1960); **124**, 642 (1961); **131**, 2827 (1963); M. Lax, *ibid.* **145**, 110 (1966).

[12] H. Dekker, Phys. Lett. A **104**, 72 (1984); **105**, 395 (1984); **105**, 401 (1984); Phys. Rev. A **31**, 1607 (1985).

[13] S. Severini, A. Settimi, C. Sibilia, M. Bertolotti, A. Napoli, A. Messina, Phys. Rev. E **70**, 056614 (2004).

[14] S. Severini, A. Settimi, C. Sibilia, M. Bertolotti, M. Centini, A. Napoli, N. Messina, *Quantum counter-propagation in open cavities via Quasi Normal Modes approach*, Laser Physics, **16**, 911-920 (2006).

[15] J. D. Jackson, *Classical Electrodynamics* (John Wiley, New York, 1975).

[16] M. Ueda, I. Nobuyuki, Phys. Rev. A **50**, 89 (1994).

# Figures and captions

Figure 4.1.

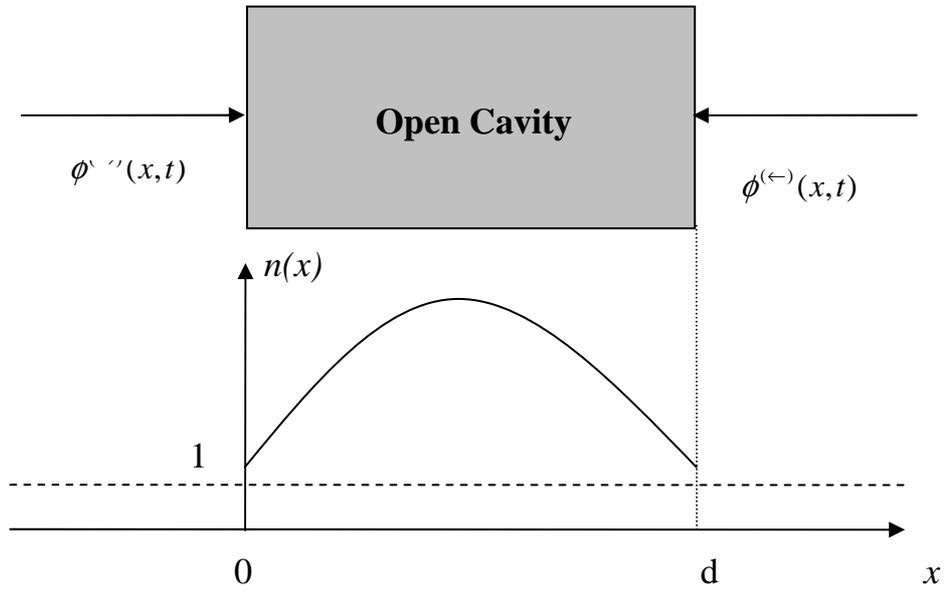

Figure 4.2.

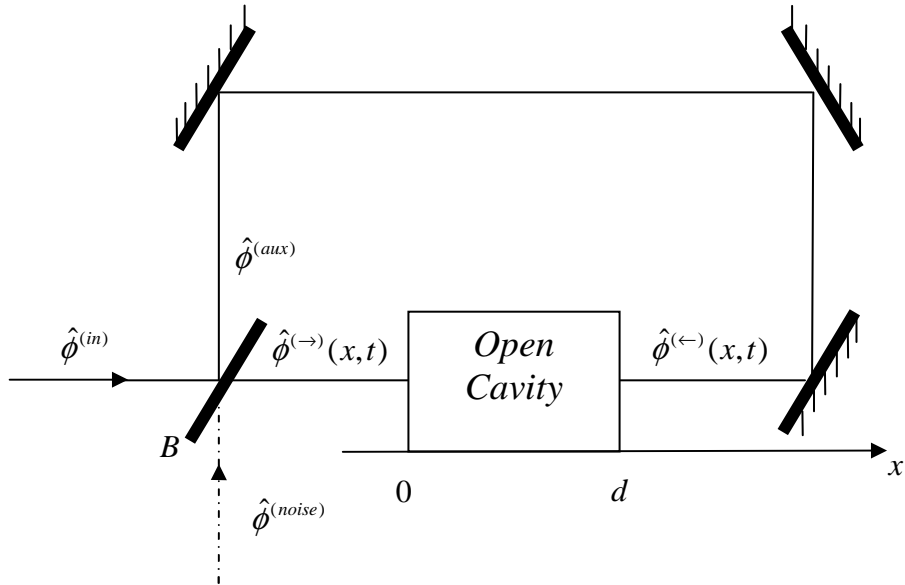

Figure 4.1. Generic counter-propagation concept in classical view: two incoming (coherent) pumps enter from both sides of the open cavity and interfere. The total field inside the cavity has a double contribution: the interference between these two coherent pumps and the interference due to the non homogeneity of the cavity.

Figure 4.2: Simple schematic representation for an operatorial realization of the quantum counter-propagation apparatus. This setup, from a classical point of view, i.e. when no scalar field $\phi^{(noise)}$ comes from the noise channel of the Beam Splitter (50/50 BS), provides, on the two surfaces of an open cavity, two identical scalar fields with the same amplitude and travelling in opposite directions with a suitable phase difference $\Delta\varphi$. The consideration of the same scheme in the quantum stand point gives rise the so called quantum counter propagation.

# Chapter 5

# Density of probability for QNMs:
# link with boundary conditions of open cavities.

## 1. Introduction.

Several attempts have been made to generalize the notion of local density of states (LDOS) and integral density of states (DOS) to the case of open cavities, i.e. structures of finite size where e.m. energy can flow in and out of the volume bounded by the surface $S$ of the cavity [1]. Quite surprisingly, the concept of density of states for a finite-size structure still lacks a simple, concise definition.

If one has a closed cavity such that the e.m. field vanishes at the edges or a cavity where periodic boundary conditions can be applied (such as a multilayer stack of infinite length), filled with a non-absorbing medium, one can expect that the e.m. energy will be conserved inside the cavity, and the problem will be Hermitian. In this case, the LDOS stands for the number of eigen-modes per unit volume and unit frequency at a point $\vec{r}$ inside the cavity. If the e.m. field can be specified by a single field component (TE or TM polarization), then the scalar Green's function can be expanded in terms of the eigen-modes of the cavity and the LDOS can be calculated through the imaginary part of the scalar Green's function: $\sigma(\vec{r},\omega) \propto -\text{Im}[G(\vec{r},\vec{r},\omega)]$ [2]. The DOS is then defined as the average LDOS inside the volume $V$ of the cavity.

So one can ask what happens if the cavity is open, such that the e.m. energy can flow in and out of the volume bounded by the surface $S$ of the cavity. This question has no easy answers fundamentally because the e.m. problem is no longer Hermitian and the cavity does not admit eigen-modes in the usual sense of the word.

Structures in which scattering or diffracting elements are arranged in such a way that their mutual distances are comparable with the wavelength of the incident wave are often referred to as Photonic Crystals (PCs), or Photonic Band Gap structures (PBGs) [3]-[5].

Although the number of experimental and theoretical reports on 1D-PBG structures is already quite large, the issue of the DOS, regarding what one means by it, and its true and otherwise implied connections to other physical or measurable quantities, such as emitted energy to name just one, is still far from being considered closed.

Besides the Green's function approach [1][2], there are at least other two different approaches to calculate the DOS which are currently used in literature. The first consists of calculating the DOS as the spatially averaged e.m. energy density stored inside the 1D-PBG [9]. The second was first proposed in ref. [10], where the DOS was defined as $\sigma_\varphi(\omega) = (1/d)(d\varphi_t/d\omega)$, being $\varphi_t(\omega)$ the phase of the transmission function $t(\omega) = |t(\omega)|\exp[i\varphi_t(\omega)]$.

D'Aguanno *et al.* have presented in ref. [8] an unified treatment of DOS and tunnelling times in finite 1D-PBGs. D'Aguanno has exploited connections and differences between the various approaches used to calculate the DOS, which include the Green function, the Wigner phase time, and the e.m. energy density, and concludes that the Green function is always the correct path to the true DOS. Ref. [8] also finds that, for an arbitrary structure, the DOS can always be found as the ratio between the power emitted by a source located inside the cavity and the power emitted by the same source in free space, regardless of absorption or dispersion.

Ho *et al.* in ref. [2] already made an essential first step towards the application of QNMs [1] to quantum electrodynamics phenomema in one side open and homogeneous cavities. In the approach of the Normal Modes (NMs) [11], the unit weight of the resonances in the LDOS emerges simply as a numerical result, and is difficult to understand theoretically; in ref. [2] the same result (in 1D) is justified analytically, and moreover, one can in principle (a) estimate the corrections due to other resonances in the LDOS (note that there is no background apart from the QNM's contributions) and (b) discuss the local DOS than the integral DOS.

In this chapter, an operative definition is proposed for the DOS of QNMs in a cavity opened from both ends [9][10], with an inhomogeneous distribution of refractive index [7]-[8]. An appendix proves the correspondence between the discrete modes of a one-dimensional cavity [2] and the continuous modes in the absence of a cavity, which approach is formulated in ref. [5]; the physical link of the DOS with the autocorrelation function of the e.m. field in absence of a cavity is examined closely. This appendix is viewed as a key passage in order to pave a reasonable way of generalization: in fact, this kind of link is retained to hold again between the DOS for QNMs and the autocorrelation function of the e.m. field inside an open cavity.

## 2. "Density" of "states".

In this section, the density of probability for the Quasi Normal Modes (QNMs) [2] [1] is operatively defined inside an open cavity which is excited by an external pumping; this operative definition, here reported, has been presented in refs. [9][10].

The quantum state of an e.m. field, excited inside an open cavity, is generally the super-position of infinite eigen-states, which are the QNMs. In this sense, a "density" of "states" (DOS) can be introduced as the density of probability that the e.m. field, excited inside the cavity, is just in one QNM, oscillating into a range of frequency. As a first step, a local density of states $\sigma^{(loc)}(x,\omega)$ can be introduced. The local DOS of QNMs defines the probability $d^2p(x,\omega)$ that an e.m. field, inside an infinitesimal cavity $(x,x+dx)$, is just in one QNM, oscillating into an infinitesimal range of frequency $(\omega,\omega+d\omega)$, i.e.

$$d^2 p(x,\omega) = \sigma^{(loc)}(x,\omega) dx d\omega. \qquad (5.2.1)$$

The (integral) density of states $\sigma(\omega)$ remains to define. The (integral) DOS for QNMs is linked to the probability $dp(\omega)$ that an e.m. field, inside a cavity of length $d$, is just in one QNM, oscillating into an infinitesimal range of frequency $(\omega,\omega+d\omega)$, i.e.

$$dp(\omega) = d \cdot \sigma(\omega) d\omega. \qquad (5.2.2)$$

If the geometry of the cavity is represented by the interval $C=[0,d]$, the (integral) DOS for QNMs is calculated as the spatial average of the local DOS for QNMs inside the cavity, i.e.

$$\sigma(\omega) = \frac{1}{d} \int_0^d \sigma^{(loc)}(x,\omega) dx. \qquad (5.2.3)$$

The universe can be viewed as a cavity of infinite length $d\to\infty$. The e.m. field inside the universe is generally a continuum of Normal Modes (NMs) [2][5]. The density of probability that the e.m. field excited inside the universe is just in the NM tuned at frequency $\omega$, i.e. $g_\omega(x) = 1/\sqrt{2\pi} \exp(i\omega\sqrt{\rho_0} x)$, is obtained as [2]

$$\sigma_{ref}^{(loc)}(x,\omega) = \sqrt{\rho_0} |g_\omega(x)|^2 = \sigma_{ref}(\omega) = \sqrt{\rho_0}/2\pi, \qquad (5.2.4)$$

where $\rho_0 = (n_0/c)^2$, being $n_0$ the refractive index of universe and $c$ the light velocity in vacuum.

The (local) DOS of NMs is the ratio between the auto-correlation function of the e.m. field in the universe and the expectation value of the e.m energy [see Appendix C]. By analogy, let propose an operative definition of the local QNM's DOS $\sigma^{(loc)}(x,\omega)$ inside an open cavity of relative index $n(x)$, which is pumped by two counter-propagating field pumps [eqs. (4.3.21) and (4.3.22)]. The local QNM's DOS $\sigma^{(loc)}(x,\omega)$ is viewed as a $n^2(x)$-modulated transfer function, if the input is the e.m energy $\langle \hat{H}(\omega) \rangle$ of the two field pumps and the output is the auto-correlation function $\Phi(x,x,\omega)$ of the e.m. field inside the cavity, i.e.

$$\sigma^{(loc)}(x,\omega) \triangleq K_\sigma \frac{\varepsilon_0 n^2(x)}{\pi} \frac{\Phi(x,x,\omega)}{\langle \hat{H}(\omega) \rangle}, \qquad (5.2.5)$$

being $K_\sigma$ is a suitable constant of normalization. Eq. (5.2.5) states that the QNM's DOS for an open cavity depends from the boundary conditions of the cavity: in fact, as from chapter 4, the autocorrelation function [eq. (4.3.48)] of the e.m. field inside the leaky cavity depends from the expectation values [eqs. (4.3.46) and (4.3.47)] of assigned current operators existing only on the two surfaces of the cavity, which, for the equivalence theorem of the e.m. source [15], are equivalent to the two actual field pumps incoming from the universe.

If the open cavity is a symmetric quarter-wave (QW) one dimensional (1D) Photonic Band Gap (PBG) structure [3] with reference frequency $\omega_{ref}$, number of periods $N$ and length $d$, then, over the $[0, 2\omega_{ref})$ range, $2N+1$ QNMs in units of $d$ can be excited [6]; so, the constant $K_\sigma$ is obtained by the normalization condition:

$$\int_0^{2\omega_{ref}} \sigma(\omega) d\omega = \frac{2N+1}{d}. \qquad (5.2.6)$$

## 2. 1. Frequencies of QNMs as resonances of DOS.

In this subsection, the physical meaning of QNM's frequencies is clarified as resonances of the DOS for an open cavity (see ref. [7]).

Let suppose that the two counter-propagating field pumps describe vacuum fluctuations in the universe: they lie on the ground state of the e.m field [as from eq. (4.3.38), specifying $\beta(\omega)=0$], so the expectation value of the photon-number is null [eq. (4.3.44)], while the e.m. energy $\langle \hat{H}(\omega) \rangle_{\beta=0}$ can not be brought to naught [eq. (4.3.45)]. As a consequence, the fictitious currents $\hat{J}_0(\omega)$ and $\hat{J}_d(\omega)$ on the two surfaces of the open cavity are characterized by some cross-correlations functions which are null [by inserting eq. (4.3.44) in eq. (4.3.46)]

$$\langle \hat{J}_d^\dagger(\omega) \hat{J}_0(\omega) \rangle = \langle \hat{J}_0^\dagger(\omega) \hat{J}_d(\omega) \rangle^* = 0, \qquad (5.2.7)$$

while their variance values [by inserting eq. (4.3.44) in eq. (4.3.47)] read as:

$$\langle \hat{J}_0^\dagger(\omega) \hat{J}_0(\omega) \rangle = \langle \hat{J}_d^\dagger(\omega) \hat{J}_d(\omega) \rangle = \frac{\sqrt{\rho_0}}{\varepsilon_0 n_0^2} (\rho_0 \omega)^2 \langle \hat{H}(\omega) \rangle_{\beta=0}. \qquad (5.2.8)$$

The local density of probability (LDOP) $\sigma_n^{(loc)}(x,\omega)$ for the $n^{th}$ QNM is the density of probability that the e.m. field, around a point $x$ of the open cavity, is excited just on the $n^{th}$ QNM, oscillating at frequency $\omega \approx Re\omega_n$. The local LDOP for the $n^{th}$ QNM, specified for vacuum fluctuations as boundary conditions, can be obtained inserting in the operative definition of the DOS (5.2.5) the auto-correlation function [see eqs. (4.3.44)-(4.3.46)] when the e.m. field consists of one QNM ($n'=-n$ [2]) and then specifying vacuum fluctuations by eqs. (5.2.7) and (5.2.8), i.e.

$$\sigma_n^{(loc)}(x,\omega) = \frac{1}{I_n}\sigma_n(\omega)\rho(x)|F_n(x)|^2, \qquad (5.2.9)$$

where is introduced the (integral) DOP $\sigma_n(\omega)$ for the $n^{th}$ QNM. In eq. (5.2.9), $\rho(x)=[n(x)/c]^2$, $F_n(x) = f_n(x)\sqrt{2\omega_n/\langle f_n|f_n\rangle}$ denotes the normalized QNM's function, being $\langle f_n|f_n\rangle$ the QNM's norm. The DOP for the $n^{th}$ QNM is the density of probability that the e.m. field, inside the cavity of length $d$, is excited just on the $n^{th}$ QNM, oscillating at frequency $\omega \approx Re\omega_n$. The DOP for the $n^{th}$ QNM, specified for vacuum fluctuations, can be calculated from eq. (5.2.3) as the spatial average of the local DOP (5.2.9) for the $n^{th}$ QNM, i.e.

$$\sigma_n(\omega) = K_\sigma \frac{d}{2\pi} \frac{I_n^2|Im\,\omega_n|}{(\omega-Re\,\omega_n)^2 + Im^2\,\omega_n}, \qquad (5.2.10)$$

where are introduced the overlapping integrals $I_n$ [see eq. (4.3.25)]. So, the DOP for the $n^{th}$ QNM (5.2.10), specified to vacuum fluctuations, is a Lorentian function, with real and imaginary parts of the $n^{th}$ QNM's frequency as parameters; the overlapping integrals $I_n$ is linked to the statistical weight in the DOS of the $n^{th}$ QNM.

In the conservative limit, the open cavity is characterized by narrow resonances $|Im(\omega_n)|<<|Re(\omega_n)|$. In this limit, the Lorentian functions (5.2.10) are so sharp that, when they are superposed, almost no aliasing occurs: i.e. the peak of the $n^{th}$ QNM's Lorentian function can not see the tails of the $(n-1)^{th}$ and $(n+1)^{th}$ QNM's Lorentian functions, and the $n^{th}$ resonance of the DOS is not shifted from the $n^{th}$ QNM's frequency. As a consequence, almost without any aliasing, the total DOS, i.e. the density of probability that the e.m. field inside the cavity is just in one of the QNMs, can be calculated as the superposition of all the DOPs (5.2.10), reminding that the e.m. field generally consists of all the QNMs. In the conservative limit, each overlapping integral can be approximated to $I_n \approx 1/d$ [2] so the total DOS, due to vacuum fluctuations, can be calculated as [7]:

$$\sigma(\omega) \cong \sum_{n=-\infty}^{\infty}\sigma_n(\omega) \cong \frac{K_\sigma}{2\pi d}\sum_{n=-\infty}^{\infty}\frac{|Im\,\omega_n|}{(\omega-Re\,\omega_n)^2 + Im^2\,\omega_n}. \qquad (5.2.11)$$

The complex frequency $\omega_n$ of the $n^{th}$ QNM well describes the $n^{th}$ peak of the DOS due to vacuum fluctuations, in the sense that: (a) the real part $Re\,\omega_n$ of the QNM's frequency corresponds to the resonance of the $n^{th}$ peak of the DOS; and (b) the imaginary part $|Im\,\omega_n|$ of the QNM's frequency is linked to the FWHM of the $n^{th}$ peak of the DOS. Moreover, the QNM's approach confirms the results of ref. [10]; this approach proves, for a QW 1D-PBG structure, that the peaks of the DOS (QNM's frequencies) are shifted from the transmission peaks (actual resonances) (see chapter 3).

## 2. 2. DOS inside an open cavity:
### link with the phase-difference of two external counter-propagating laser-beams.

Let suppose that the two field pumps describe two counter-propagating laser-beams in the universe (phase-difference $\Delta\varphi$), so they lie on a coherent state of the e.m. field [as from eq. (4.3.38) specifying that $\beta(\omega)\neq 0$]. The expectation value of the photon number is $\langle \hat{N}(\omega) \rangle$ and the one $\langle \hat{H}(\omega) \rangle$ for the e.m. energy is largest than $\langle \hat{H}(\omega) \rangle_{\beta=0}$, due to the vacuum fluctuations, i.e.

$$\langle \hat{H}(\omega) \rangle = \hbar\omega \langle \hat{N}(\omega) \rangle \gg \langle \hat{H}(\omega) \rangle_{\beta=0} = \hbar\omega/2 \,. \tag{5.2.12}$$

As a consequence, the fictitious currents $\hat{J}_0(\omega)$ and $\hat{J}_d(\omega)$ on the two surfaces of the open cavity are characterized by the variance values [inserting eq. (5.2.12) in eq. (4.3.47)]

$$\langle \hat{J}_0^\dagger(\omega)\hat{J}_0(\omega) \rangle = \langle \hat{J}_d^\dagger(\omega)\hat{J}_d(\omega) \rangle = \frac{\sqrt{\rho_0}}{\varepsilon_0 n_0^2}(\rho_0\omega)^2 \langle \hat{H}(\omega) \rangle, \tag{5.2.13}$$

and their cross-correlations functions are [here eq. (4.3.46) is suitably reported]:

$$\langle \hat{J}_d^\dagger(\omega)\hat{J}_0(\omega) \rangle = \langle \hat{J}_0^\dagger(\omega)\hat{J}_d(\omega) \rangle^* = \langle \hat{J}_0^\dagger(\omega)\hat{J}_0(\omega) \rangle exp(i\Delta\varphi) \,. \tag{5.2.14}$$

Inside the cavity, if the condition of the QNM's completeness is applied [1]

$$\sum_{n=-\infty}^{\infty} \frac{F_n(x)F_n(x')}{\omega_n} = 0 \,, \quad \forall x, x' \in [0, d], \tag{5.2.15}$$

the auto-correlation function (4.3.48) of the e.m. field reduces to

$$\Phi(x, x', \omega) = \frac{\rho_0^{3/2}}{4\varepsilon_0 n_0^2} \langle \hat{H}_\omega \rangle \left[ \sum_{n=-\infty}^{\infty} \sum_{n'=-\infty}^{\infty} \frac{F_n(0)F_{n'}(0) + F_n(d)F_{n'}(d)}{(\omega-\omega_n)(\omega+\omega_{n'})} F_n(x)F_{n'}(x') + \right. \\ \left. + \sum_{n=-\infty}^{\infty} \sum_{n'=-\infty}^{\infty} \frac{F_n(0)F_{n'}(d)\exp(i\Delta\varphi) + F_n(d)F_{n'}(0)\exp(-i\Delta\varphi)}{(\omega-\omega_n)(\omega+\omega_{n'})} F_n(x)F_{n'}(x') \right], \tag{5.2.16}$$

and the local DOS (5.2.5) can be obtained as

$$\sigma^{(loc)}(x,\omega) \triangleq K_\sigma \frac{\varepsilon_0 n^2(x)}{\pi} \frac{\Phi(x,x,\omega)}{\langle \hat{H}(\omega) \rangle} =$$

$$= K_\sigma \frac{\sqrt{\rho_0}}{4\pi} \rho(x) \left[ \sum_{n=-\infty}^{\infty} \sum_{n'=-\infty}^{\infty} \frac{F_n(0)F_{n'}(0) + F_n(d)F_{n'}(d)}{(\omega-\omega_n)(\omega+\omega_{n'})} F_n(x) F_{n'}(x) + \right.$$

$$\left. + \sum_{n=-\infty}^{\infty} \sum_{n'=-\infty}^{\infty} \frac{F_n(0)F_{n'}(d)\exp(i\Delta\varphi) + F_n(d)F_{n'}(0)\exp(-i\Delta\varphi)}{(\omega-\omega_n)(\omega+\omega_{n'})} F_n(x) F_{n'}(x) \right] \quad (5.2.17)$$

When the e.m. field consists of one QNM ($n'=-n$), eq. (5.2.9) expresses the local density of probability (LDOP) $\sigma_n^{(loc)}(x,\omega)$ in terms of the (integral) DOP for the $n^{th}$ QNM, i.e.

$$\sigma_n(\omega) = \frac{1}{d} \int_0^d \sigma_n^{(loc)}(x,\omega) dx =$$

$$= K_\sigma \frac{\sqrt{\rho_0}}{4\pi} \frac{I_n}{(\omega-\text{Re}\,\omega_n)^2 + \text{Im}^2\,\omega_n} \left\{ |F_n(0)|^2 + |F_n(d)|^2 + \right. \quad (5.2.18)$$

$$\left. + F_n(0)[F_n(d)]^* \exp(i\Delta\varphi) + F_n(d)[F_n(0)]^* \exp(-i\Delta\varphi) \right\}$$

where $I_n$ are the overlapping integrals.

In the conservative limit, the open cavity is characterized by narrow resonances $|Im(\omega_n)|<<|Re(\omega_n)|$, so each overlapping integral is $I_n \approx 1/d$ and the total DOS is the superposition of all the DOPs (5.2.18), i.e.

$$\sigma(\omega) \cong \sum_{n=-\infty}^{\infty} \sigma_n(\omega) \cong$$

$$\cong K_\sigma \frac{\sqrt{\rho_0}}{4\pi d} \sum_{n=-\infty}^{\infty} \frac{|F_n(0)|^2 + |F_n(d)|^2}{(\omega-\text{Re}\,\omega_n)^2 + \text{Im}^2\,\omega_n} + \quad (5.2.19)$$

$$+ K_\sigma \frac{\sqrt{\rho_0}}{4\pi d} \sum_{n=-\infty}^{\infty} \frac{F_n(0)[F_n(d)]^* \exp(i\Delta\varphi) + F_n(d)[F_n(0)]^* \exp(-i\Delta\varphi)}{(\omega-\text{Re}\,\omega_n)^2 + \text{Im}^2\,\omega_n}$$

If the cavity presents a refractive index $n(x)$ which satisfies the symmetry properties $n(d/2-x)=n(d/2+x)$, on the surfaces of the cavity the values of the QNM's functions are such that $F_n(d) = (-1)^n F_n(0)$. If $I_n \approx 1/d$, then $|F_n(0)|^2 \cong |\text{Im}\,\omega_n|/\sqrt{\rho_0}$ [as from eq. (4.3.25)]. From (5.2.19), it results the DOS for an open cavity, specified for two counter-propagating laser-beams as boundary conditions, with a phase difference $\Delta\varphi$ as initial condition (ref. [10]):

$$\sigma(\omega) = K_\sigma \left[ \tilde{\sigma}_1(\omega) + \tilde{\sigma}_2(\omega) \cos(\Delta\varphi) \right], \quad (5.2.20)$$

where

$$\tilde{\sigma}_1(\omega) = \frac{1}{2\pi d} \sum_{n=-\infty}^{\infty} \frac{|\text{Im}\,\omega_n|}{(\omega-\text{Re}\,\omega_n)^2 + \text{Im}^2\,\omega_n}, \quad (5.2.21)$$

$$\tilde{\sigma}_2(\omega) = \frac{1}{2\pi d} \sum_{n=-\infty}^{\infty} (-1)^n \frac{|\text{Im}\,\omega_n|}{(\omega-\text{Re}\,\omega_n)^2 + \text{Im}^2\,\omega_n}. \quad (5.2.22)$$

Eq. (5.2.21) can be interpreted as DOS due to only one laser-beam; it is a series of infinite QNM's Lorentian functions and coincides with the DOS (5.2.11), due to the vacuum fluctuations: in fact, the potential number of QNMs inside the open cavity is independent from the statistics of pumping because it is fixed only by the geometry of the cavity which acts as an e.m. filter of frequency; moreover the density of probability to excite one QNM (i.e. the DOS) can not be modified because vacuum fluctuations do not add other degrees of freedom in the universe. Eq. (5.2.22) can be interpreted as the interference term due to the two laser-beams and it is an alternate-sign series of the same Lorentian functions. The interference of the two laser-beams produces the control of the DOS inside the cavity. In fact, the phase-difference adds one degree of freedom in the universe; so, if the two counter-propagating laser-beams are in phase $\Delta\varphi=0$, the DOS is a series of the even QNM's Lorentian functions, while if the two laser-beams are opposite of phase $\Delta\varphi=\pi$, the DOS is a series of the odd Lorentian functions.

A symmetric QW 1D-PBG, with $N$ periods and $\omega_{ref}$ as reference frequency, presents $2N+1$ families of QNMs [6], i.e. $m=0,1,\ldots,2N$: the $m^{th}$ QNM's family consists of infinite QNM's frequencies $\omega_{m,k}$, $k\in\mathbb{Z}=\{0,\pm 1,\pm 2,\ldots\}$ with same imaginary part $Im(\omega_{m,k})=Im(\omega_{m,0})$ and aligned by a step $\Delta=2\omega_{ref}$, i.e. $Re(\omega_{m,k})=Re(\omega_{m,0})+k\Delta$. The DOS (5.2.21), due to one laser-beam, can be specified as [10]

$$\tilde{\sigma}_1(\omega) = \frac{1}{4d\Delta}\sum_{m=0}^{2N}\coth\left[i\frac{\pi}{\Delta}\left(\omega-\omega_{m,0}^*\right)\right]+\frac{i}{4d\Delta}\sum_{m=0}^{2N}\cot\left[\frac{\pi}{\Delta}\left(\omega-\omega_{m,0}\right)\right], \quad (5.2.23)$$

while the DOS-term (5.2.22), representing the interference of the two counter-propagating laser-beams with phase-difference $\Delta\varphi$, converges to [10]

$$\begin{aligned}\tilde{\sigma}_2(\omega) &= \frac{1}{2\pi d}\sum_{m=0}^{2N}\sum_{k=-\infty}^{\infty}(-1)^{m+k}\frac{|Im\,\omega_{m,0}|}{[\omega-(Re\,\omega_{m,0}+k\Delta)]^2+Im^2\,\omega_{m,0}}=\\ &=\frac{i}{4d\Delta}\sum_{m=0}^{2N}(-1)^m\csc\left[\frac{\pi}{\Delta}\left(\omega-\omega_{m,0}\right)\right]-\frac{i}{4d\Delta}\sum_{m=0}^{2N}(-1)^m\csc\left[\frac{\pi}{\Delta}\left(\omega-\omega_{m,0}^*\right)\right]\end{aligned}. \quad (5.2.24)$$

The number of QNMs over the $[0,\Delta)$ range is $2N+1$ in units of $d$. The constant $K$ in the DOS (5.2.20) can be determined by the condition of normalization (5.2.6), i.e.

$$K_\sigma = \frac{2N+1}{d}\frac{1}{S}, \quad (5.2.25)$$

where

$$S = S_1 + S_2\cos(\Delta\varphi), \quad S_1 = \int_0^\Delta \tilde{\sigma}_1(\omega)d\omega, \quad S_2 = \int_0^\Delta \tilde{\sigma}_2(\omega)d\omega. \quad (5.2.26)$$

In fig. 5.1, the "density" of "states" (DOS) (5.2.20) is plotted, in units of the reference DOS (5.2.4), as a function of the dimensionless frequency $\omega/\omega_{ref}$, for three different symmetric QW 1D-PBGs: (a) $\lambda_{ref}=1\mu m$, $N=6$, $n_h=2$, $n_l=1.5$; (b) $\lambda_{ref}=1\mu m$, $N=6$, $n_h=3$, $n_l=2$; (c) $\lambda_{ref}=1\mu m$, $N=7$, $n_h=3$, $n_l=2$.

In each figure, the DOS is plotted for an external pumping consisting of: one laser beam or vacuum fluctuations (dotted line) [see eq. (5.2.23)]; two counter-propagating laser-beams in phase $\Delta\varphi=0$ (continuous thin line) [see eq. (5.2.24)] ; two laser-beams opposite in phase $\Delta\varphi=\pi$ (continuous thick line). The DOS predicted by the QNM's approach is in good agreement with the DOS obtained by the other methods of literature, as Bendickson [10] and D'Aguanno [8].

A symmetric QW 1D-PBG, with *N* periods and $\omega_{ref}$ as reference frequency, presents 2*N*+1 QNMs in the $[0,2\omega_{ref})$ range., i.e. *k*=0,1,…2*N* excluding $\omega=2\omega_{ref}$. If the 1D-PBG structure is excited by two counter-propagating laser-beams with a phase-difference $\Delta\varphi$, the even (odd) QNMs, i.e. *k=0,2,…2N*, (i.e. *k=1,3,…2N-1*) increase in strength when the two laser-beams are in phase, i.e. $\Delta\varphi=0$ (opposite of phase, i.e. $\Delta\varphi=\pi$) and almost flag when the two laser-beams are opposite of phase, i.e. $\Delta\varphi=\pi$ (in phase, i.e. $\Delta\varphi=0$) [see figs. 5.1.a and 5.1.b]. If one period is added to the 1D-PBG, the QNMs next to the low frequency band-edge and to the high frequency band-edge exchange their physical response with respect to the phase-difference of the two laser-beams [see fig. 5.1.c]. The DOS due to two counter-propagating laser-beams is the physical key to discuss stimulated emission processes inside open cavities; depending the DOS on the phase-difference of the two laser-beams, a coherent control of the stimulated emission can be effected, in particular inside 1D-PBG structures .

## 3. Discussion and conclusions.

The results of this chapter put well into evidence how the "density" of "states" of an open cavity depends on the excitation condition of the cavity. The local density of states (LDOS) can be interpreted as the density of probability that the e.m. field, around the point *x* of the cavity, is excited on just one eigen-state, oscillating around the frequency $\omega$[20]. The integral density of states (DOS) is linked to the boundary and initial conditions of any optical cavity [5]. For a closed cavity, the e.m. field satisfies nodal conditions on the surfaces of the cavity, so the DOS can not depend on the external environment and it describes only the distribution in frequency for the Normal Modes (NMs) of the e.m. field, depending only on the geometry of the cavity [2]. Instead, for an open cavity, the e.m. field satisfies, as boundary conditions, incoming wave conditions on the surfaces of the cavity, so the DOS depends, by the boundary conditions, on the photon reservoir due to the external pumping [2]; if the cavity is excited by two counter-propagating field pumps, the DOS describes the density of probability that the two field pumps excite one Quasi Normal Mode (QNM) of the e.m. field [9]. In the hypothesis of spontaneous emission processes, the two pumps are modelled as vacuum fluctuations, lying on the ground state of the e.m. field, and the DOS is only a peculiarity of the cavity geometry [8]: in fact, the number of the QNMs inside the cavity is independent from the statistics of the pumping because it is fixed only by the geometry of the cavity which acts as an e.m. filter of frequency; moreover, the density of probability to excite the QNMs inside the cavity (i.e. the DOS) can not be modified because vacuum fluctuations do not add degrees of freedom in the universe. In the case of stimulated emission, the two pumps are modelled as two laser-beams, lying on a coherent state, so the DOS, besides depending on the geometry of the cavity, can be controlled by the initial condition regarding the phase-difference of the two laser-beams [10]: the interference of the two laser-beams produces the control of the DOS inside the cavity; in fact, the phase-difference of the two laser-beams adds one degree of freedom in the universe.

# Appendix C: Examining deeper the link of the LDOS with the auto-correlation function of the e.m. field.

Let prove the correspondence between the discrete modes of a one-dimensional cavity [2] and the continuous modes in the absence of a cavity, which approach is formulated in ref. [5]. This proof enables to correct some inaccuracies in the expressions of the continuous-field operators reported in ref. [5] [see chapter 4, section 3.1.]. The physical link of the LDOS with the autocorrelation function of the e.m. field in absence of a cavity is examined deeper, since this link is not clearly discussed in literature. The proof is viewed as a key passage in order to pave a reasonable way of generalization: in fact, this kind of link is retained to hold again between the local density of probability for Quasi Normal Modes (QNMs) and the autocorrelation function of the e.m. field inside an open cavity.

Let consider a one dimensional (1D) cavity $U=(-L/2,L/2)$ with refractive index $n_0$. The e.m. modes are defined as Normal Modes (NMs) if the normalization condition holds [2]

$$\int_{-L/2}^{L/2} f_n(x) f_{n'}^*(x) dx = \delta_{n,n'}, \qquad (5.C.1)$$

where $\delta_{n,n'}$ is the Kronecker's symbol, and they are represented as travelling plane waves when the cyclic conditions are applied

$$\begin{cases} f_n(-L/2) = f_n(L/2) \\ \partial_x f_n(x)\big|_{x=-L/2} = \partial_x f_n(x)\big|_{x=L/2} \end{cases}, \qquad (5.C.2)$$

so that different NMs of the cavity have frequencies given by different integer multiplies of the mode spacing, i.e.

$$\begin{cases} p_k = n\dfrac{\pi}{(L/2)} \quad, \quad n \in \mathbb{Z} = \{0, \pm 1, \pm 2 \ldots\} \\ \omega_n = \dfrac{p_n}{\sqrt{\rho_0}} \quad, \quad f_n(x) = \dfrac{1}{\sqrt{L}} \exp(ip_n x) \end{cases}, \qquad (5.C.3)$$

being $\rho_0 = (n_0/c)^2$ and $c$ the speed of the light in vacuum. The annihilation and creation operators $\hat{b}_n$ and $\hat{b}_n^\dagger$ are defined by the usual quantization rules [2]

$$\left[\hat{b}_n, \hat{b}_{n'}^\dagger\right] = \delta_{n,n'}. \qquad (5.C.4)$$

As a consequence, the e.m. field operator $\hat{E}(x,t)$ can be expressed in terms of the annihilation and creation operators $\hat{b}_n$ and $\hat{b}_n^\dagger$, i.e.

$$\hat{\phi}(x,t) = \frac{i}{\sqrt{2\varepsilon_0 n_0^2}} \sum_{n=1}^{\infty} \sqrt{\hbar\omega_n} [f_n(x)\hat{b}_n \exp(-i\omega_n t) - f_n^*(x)\hat{b}_n^\dagger \exp(i\omega_n t)], \tag{5.C.5}$$

being $\hbar$ the Planck's constant and $\varepsilon_0$ the dielectric constant in vacuum.

For planes waves which propagate parallel to the axis, the e.m. modes are separated by a step of wave numbers $\Delta p = \pi/(L/2)$ and a step of frequencies:

$$\Delta\omega = \frac{\Delta p}{\sqrt{\rho_0}} = \frac{\pi}{(L/2)\sqrt{\rho_0}}. \tag{5.C.6}$$

The eigen-functions can be re-expressed as

$$f_n(x) = \sqrt{\rho_0^{1/2}\Delta\omega}\, g_n(x) \quad , \quad g_n(x) = \frac{1}{\sqrt{2\pi}} \exp(i\omega_n \sqrt{\rho_0}\, x) \tag{5.C.7}$$

and the e.m. field operator (5.C.5) is

$$\hat{\phi}(x,t) = \frac{i}{\sqrt{2\varepsilon_0 n_0^2}} \sum_{k=1}^{\infty} \sqrt{\hbar\omega_n} \sqrt{\rho_0^{1/2}\Delta\omega} [g_n(x)\hat{b}_n \exp(-i\omega_n t) - g_n^*(x)\hat{b}_n^\dagger \exp(i\omega_n t)]. \tag{5.C.8}$$

The mode spectrum becomes continuous as $L\to\infty$ and $\Delta\omega\to 0$, so in this limit it is convenient to transform into continuous-mode operators according to:

$$\hat{b}_n \to \sqrt{\Delta\omega}\,\hat{b}(\omega). \tag{5.C.9}$$

The Kronecker's and Dirac's functions are correspondingly related by:

$$\delta_{n,n'} \to \Delta\omega\,\delta(\omega-\omega'). \tag{5.C.10}$$

Sums over discrete quantities are converted to integrals over continuous frequency according to:

$$\sum_{n=1}^{\infty} \to \frac{1}{\Delta\omega}\int_0^{\infty} d\omega. \tag{5.C.11}$$

In the usual continuous-mode form, the discrete modes (5.C.1) and (5.C.7) read as

$$\int_{-\infty}^{\infty} g_\omega(x) g_{\omega'}^*(x) dx = \delta(\omega-\omega') \quad , \quad g_\omega(x) = \frac{1}{\sqrt{2\pi}} \exp(i\omega\sqrt{\rho_0}\,x), \tag{5.C.12}$$

and the commutation relation (5.C.4) as:

$$\left[\hat{b}(\omega), \hat{b}^\dagger(\omega')\right] = \delta(\omega-\omega'). \tag{5.C.13}$$

The continuous-mode quantized electric field operators are obtained from their discrete-mode counterparts (5.C.8) by this procedure, with respective results reported in eqs. (4.3.9)-(4.3.11), where the field operators have been divided into their creation and annihilation operators parts. Only the parts of the field operator which correspond to propagation in the positive $x$ direction are included in eqs. (4.3.9)-(4.3.11) and the ranges of integration over $\omega$ are from $0$ to $\infty$. The range of integration in the above expressions is so strictly extended, since the physical frequencies are defined to be positive. However, the range over $\omega$ can be extended from $-\infty$ to $\infty$ without significant errors since most optical experiments use a narrow band source $B$, in the sense that [5]

$$B \ll \omega_c, \qquad (5.C.14)$$

being $\omega_c$ the central frequency of the bandwidth $B$. If the range of integration is extended however, it is useful to define the Fourier's transformed operators reported in eqs. (4.2.12)-(4.3.14).

Inside a 1D cavity, a local density of states (LDOS) can be introduced as the density of probability that the e.m. field, in the point $x$, is excited on just one eigen-state, oscillating around the frequency $\omega$ [2]. Consider the 1D cavity $C=(-L/2,L/2)$ of refractive index $n_0$ which holds the NMs (5.C.3), the LDOS can be calculated as:

$$\begin{aligned} \sigma(x,\omega) &= \sum_{n=-\infty}^{\infty} |f_n(x)|^2 \delta(\omega-\omega_n) = \\ &= \sum_{n=-\infty}^{\infty} \Delta p |g_n(x)|^2 \delta(\omega - \frac{p_n}{\sqrt{\rho_0}}) = \\ &= \sqrt{\rho_0} \sum_{n=-\infty}^{\infty} \Delta p |g_n(x)|^2 \delta(p_n - \omega\sqrt{\rho_0}) \end{aligned} \qquad (5.C.15)$$

If $L \to \infty$, the discrete modes of the cavity (5.C.3) become the continuous modes (5.C.12) of the universe, so the LDOS (5.C.15) reduces to

$$\begin{aligned} \sigma^{(loc)}(x,\omega) &= \sqrt{\rho_0} \int_{-\infty}^{\infty} dp |g_p(x)|^2 \delta(p - \omega\sqrt{\rho_0}) = \\ &= \sqrt{\rho_0} |g_\omega(x)|^2 = \sigma(\omega) = \frac{\sqrt{\rho_0}}{2\pi} \end{aligned} \qquad (5.C.16)$$

If the field operator $\hat{\phi}_\omega(x)$ [eq. (4.3.12)] lies on a quantum state $|\psi_0\rangle$, characterized by an expectation value $\langle \hat{H}(\omega) \rangle$ for the Hamiltonian operator [eqs. (4.3.40) and (4.3.39)], then the auto-correlation of the e.m. field

$$F(x,x',\omega) \triangleq \langle \psi_0 | \hat{\phi}_\omega^\dagger(x) \hat{\phi}_\omega(x') | \psi_0 \rangle = \pi \frac{\sqrt{\rho_0}}{\varepsilon_0 n_0^2} \langle \hat{H}(\omega) \rangle g_\omega^*(x) g_\omega(x') \qquad (5.C.17)$$

is so linked with the LDOS (5.C.16):

$$\sigma^{(loc)}(x,\omega) = \frac{1}{\pi} \varepsilon_0 n_0^2 \frac{F(x,x,\omega)}{\langle \hat{H}(\omega) \rangle}. \qquad (5.C.18)$$

This link has been obtained in a homogeneous cavity and an extension of eq. (5.C.18) is used, with the weight given by the value of the local index of refraction, in the case of inhomogeneous cavities like Photonic Crystals (PC), in which a refractive index assumes different constant values in aligned limited regions.

# References.

[1] A. A. Asatryan, K. Busch, R. C. McPhedran, L. C. Botten, C. Martijn de Sterke, and N. A. Nicorovici, Phys. Rev. E **63**, 046612 (2001); D. Felbacq and R. Smaali, Phys. Rev. B **67**, 085105 (2003).

[2] E. N. Economou, *Green's Functions in Quantum Physics* (Springer-Verlag, Berlin, 1983).

[3] S. John, Phys. Rev. Lett. **53**, 2169 (1984); S. John, Phys. Rev. Lett. **58**, 2486 (1987); E. Yablonovitch, Phys. Rev. Lett. **58**, 2059 (1987); E. Yablonovitch and T. J. Gmitter, Phys. Rev. Lett. **63**, 1950 (1989).

[4] J. Maddox, Nature (London) **348**, 481 (1990); E. Yablonovitch and K.M. Lenny, Nature (London) **351**, 278, (1991); J. D. Joannopoulos, P. R. Villeneuve, and S. H. Fan, Nature (London) **386**, 143 (1997).

[5] J. D. Joannopoulos, *Photonic Crystals: Molding the Flow of Light* (Princeton University Press, Princeton, New York, 1995); K. Sakoda, *Optical properties of photonic crystals* (Springer Verlag, Berlin, 2001); K. Inoue and K. Ohtaka, *Photonic Crystals: Physics, Fabrication, and Applications* (Springer-Verlag, Berlin, 2004).

[6] G. D'Aguanno, M. Centini, M. Scalora, C. Sibilia, Y. Dumeige, P. Vidakovic, J. A. Levenson, M. J. Bloemer, C. M. Bowden, J. W. Haus, and M. Bertolotti, Phys. Rev. E **64**, 016609 (2001).

[7] J. M. Bendickson, J. P Dowling, and M. Scalora, Phys. Rev. E **53**, 4107 (1996).

[8] G. D'Aguanno, N. Mattiucci, M. Scalora, M. J. Bloemer and A. M. Zheltikov, Phys. Rev. E. **70**, 016612 (2004).

[9] K. C. Ho, P. T. Leung, Alec Maassen van den Brink and K. Young, Phys. Rev. E **58**, 2965 (1998).

[10] P. T. Leung, S. Y. Liu, and K. Young, Phys. Rev. A **49**, 3057 (1994); P. T. Leung, S. S. Tong, and K. Young, J. Phys. A **30**, 2139 (1997); P. T. Leung, S. S. Tong, and K. Young, J. Phys. A **30**, 2153 (1997); E. S. C. Ching, P. T. Leung, A. Maassen van der Brink, W. M. Suen, S. S. Tong, and K. Young, Rev. Mod. Phys. **70**, 1545 (1998).

[11] S. C. Ching, H. M. Lai, and K. Young, J. Opt. Soc. Am. B **4**, 1995 (1987); **4**, 2004 (1987).

[12] S. Severini, A. Settimi, C. Sibilia, M. Bertolotti, M. Centini, A. Napoli, N. Messina, *Quantum counter-propagation in open cavities via Quasi Normal Modes approach*, Laser Physics, **16**, 911-920 (2006).

[13] A. Settimi, S. Severini, C. Sibilia, M. Bertolotti, M. Centini, A. Napoli, N. Messina, Phys. Rev. E **71,** 066606 (2005).

[14] S. Severini, A. Settimi, C. Sibilia, M. Bertolotti, A. Napoli, A. Messina, *Quasi Normal Frequencies in open cavities: an application to Photonic Crystals*, Acta Phys. Hung. B **23/3-4**,

135-142 (2005); A. Settimi, S. Severini, B. Hoenders, *Quasi Normal Modes description of transmission properties for Photonic Band Gap structures*, J. Opt. Soc. Am. B, **26**, 876-891 (2009).

[15] A. Settimi, S. Severini, N. Mattiucci, C. Sibilia, M. Centini, G. D'Aguanno, M. Bertolotti, M. Scalora, M. Bloemer, C. M. Bowden, Phys. Rev. E, **68**, 026614 (2003).

[16] S. Severini, A. Settimi, C. Sibilia, M. Bertolotti, A. Napoli, A. Messina, Phys. Rev. E, **70** 056614 (2004).

[17] J. J. Sakurai, *Advanced Quantum Mechanics* (Addison-Wesley, New York, 1995); S. Weinberg, *The Quantum Theory of Fields* (Cambridge University Press, New York, 1996).

[18] K.J. Blow, R. Loudon, S. J.D. Phoenix, T.J. Shepherd, Phys. Rev. A **42**, 4102 (1990).

[19] J. D. Jackson, *Classical Electrodynamics* (John Wiley, New York, 1975).

[20] A. A. Abrikosov, L. P. Gor'kov and I. E. Dzyaloshinski, *Methods of Quantum Field Theory in Statistical Physics* (Dover, New York, 1975); P. T. Leung, A. Maassen van den Brink and K. Young, *Frontiers in Quantum Physics*, Proceedings of the International Conference, edited by S. C. Lim, R. Abd-Shukor and K. H. Kwek (Springer-Verlag, Singapore, 1998).

[21] M. Centini, G. D'Aguanno, M. Scalora, M. J. Bloemer, C. M. Bowden, C. Sibilia, N. Mattiucci, and M. Bertolotti, Phys. Rev. E **67** 036617 (2003).

# Figures and captions.

Figure 5.1.a.

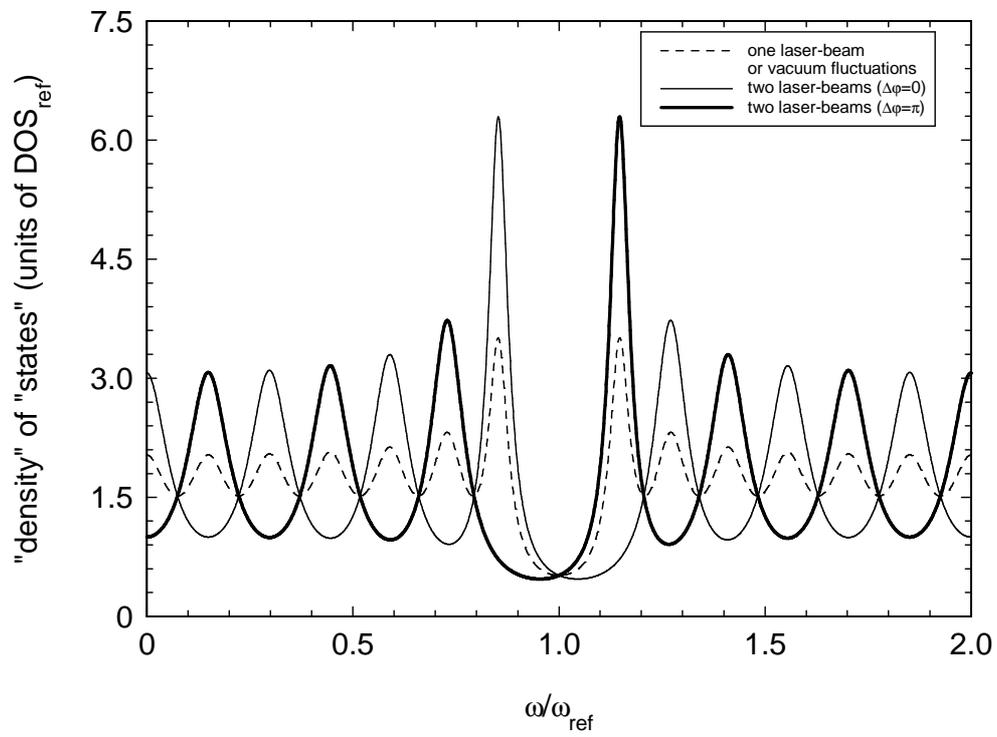

Figure 5.1.b.

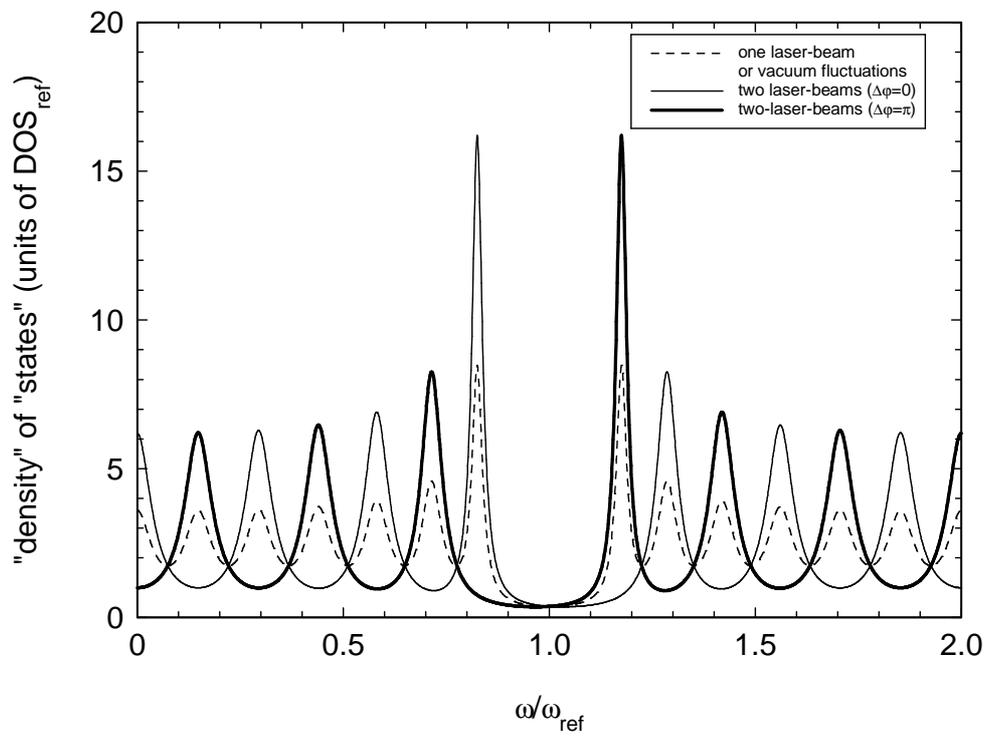

Figure 5.1.c.

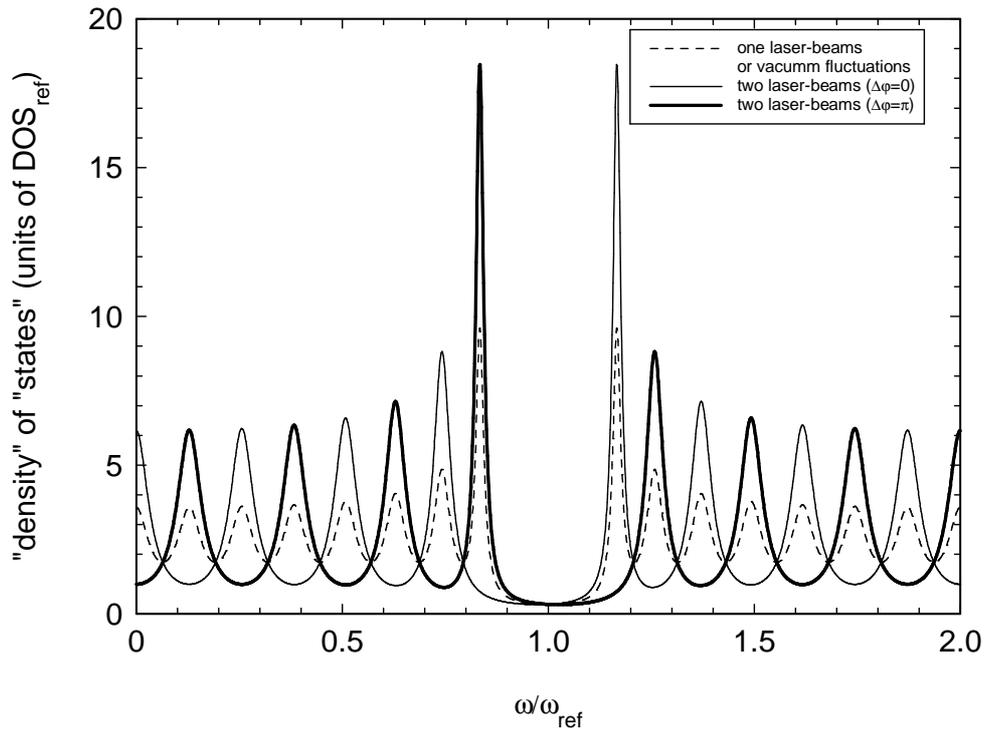

Figure 5.1. The "density" of "states" (DOS) (5.2.20) is plotted, in units of the reference DOS (5.2.4) , as a function of the dimensionless frequency $\omega/\omega_{ref}$, for three different symmetric quarter-wave (QW) one dimensional (1D) Photonic Band Gap (PBG) structures: (a) $\lambda_{ref}=1\mu m$, $N=6$, $n_h=2$, $n_l=1.5$; (b) $\lambda_{ref}=1\mu m$, $N=6$, $n_h=3$, $n_l=2$; (c) $\lambda_{ref}=1\mu m$, $N=7$, $n_h=3$, $n_l=2$. In each figure, the DOS is plotted for an external pumping consisting of: one laser beam or vacuum fluctuations (dotted line) [see eq. (5.2.23)]; two counter-propagating laser-beams in phase $\Delta\varphi=0$ (continuous thin line) [see eq. (5.2.24)] ; two laser-beams opposite in phase $\Delta\varphi=\pi$ (continuous thick line).

**Third Part:**

**QNM's approach for emission processes in open cavities.**

# Chapter 6

# Spontaneous Emission and coherent control of Stimulated Emission inside one dimensional Photonic Crystals.
# (Weak Coupling regime)

## 1. Introduction.

The behaviour of small systems (few degrees of freedom) coupled to dissipative reservoirs represents a central theme in many physical contexts. In quantum optics, and in its simplest form, the problem consists of a two-level excited atom decaying in open space to its ground state through an allowed electric dipole transition; in this condition, the emitted radiation propagates away never coming back to the atom. This is a prototype of irreversible decay of a prepared state, as well as a classical manifestation of the quantization of the e.m. field [2].
Spontaneous or stimulated emission is a fundamental process resulting from the interaction between radiation and matter. It depends not only on the properties of the excited atomic system but also on the nature of the environment to which the system is optically coupled [2][3]. It is possible to control the rate of spontaneous or stimulated emission from an excited atom by altering the photon density of states (DOS) near the resonant frequency, i.e. by modifying the accessible states into which the excited atom can radiate. If the photon DOS in the vicinity of the frequency of interest is less than that of the free space, the atomic decay will be retarded, if it is greater it will be accelerated [4],[5]; this link with the DOS has been put well into evidence in ref. [6].

An important and intriguing situation arose when it was realized that it is possible to create environments in which the spectrum of the e.m. field exhibits gaps in frequency. In other words, no radiation over some extended range of frequency can propagate in that environment; simply stated, an excited atom whose transition frequency falls in the range of that gap should at first sight never decay. The structures exhibiting such frequency gaps are referred to as photonic crystals (PC) or even photonic band gap (PBG) structures [3]-[5]; in the wake of theoretical promulgation of PBGs, concern has turned to the question of the behaviour of the atomic decay rate in such structures [10].

The photon DOS is the fundamental feature that determines the behaviour of the system 'atom-field' and characterizes the various types of environments. The form and analytical properties of the DOS dictate the type of the approximations permissible in formulating the equations governing the time evolution of the system. When the photon DOS is a smooth function of frequency over the spectral range of atomic transition, the rate of spontaneous and stimulated emission is described by the Fermi's golden rule. A substantial modification in the DOS can be effected by means of PBG structures. Petrov *et al* [11] have reported some modifications of the emission processes of dye molecules embedded in a three-dimensional solid-state PBG exhibiting a stop band in the visible range; the results are interpreted in terms of a redistribution of photon DOS in the structure [12].

Ho *et al.* in ref. [2] already made an essential first step towards the application of the Quasi Normal Mode's (QNM's) approach [1] to quantum electrodynamics phenomema in a one side open and homogeneous cavity. In ref. [2], the Feynman's propagator inside the cavity is expressed in terms of QNMs, labeled by a discrete index rather than a continuous momentum.

In ref. [8], the second quantization scheme based on the QNM's approach has been extended to double side open and inhomogeneous cavities [6], excited by vacuum fluctuations of the universe. Although a full analysis of the spontaneous emission should be performed in three dimensional geometries, ref. [8] puts into evidence that the QNM's approach, limited to 1D-PBG structures, is consistent with results already known in literature [4]-[19].

Using numerical methods, Centini et al. in ref. [5] have studied the propagation of counter-propagating pulses in finite 1D-PBG structures. Linear interference and localization effects are shown to combine to either enhancement or suppression of stimulated emission processes, depending on the initial phase difference between the input pulses.

The main object of this chapter is to give a fully quantum treatment for the dynamics of an atom inside a 1D-PBG structure, coupled with two counter-propagating laser-beams, by using the QNM's approach [10], i.e. treating the problem in the framework of open systems. The e.m. field is quantized in terms of the QNMs inside the 1D-PBG and the atom is modelled as a two levels system. In the electric dipole approximation, the atom is assumed to be weakly coupled to just one of the QNMs. As a result, the decay-time depends from the position of the dipole inside the structure, and can be controlled by the phase-difference of the two laser-beams. Such a system is relevant for a single-atom, phase-sensitive, optical memory device on atomic scale.

The chapter is organized as follows. In section 2, the emission power of an atom is calculated inside an open cavity, and a sensitivity function is defined for the atom-e.m. field coupling. In section 3, the spontaneous emission processes are described inside a symmetric quarter wave (QW) 1D-PBG structure. In section 4, the coherent control of the stimulated emission is discussed inside the same 1D-PBG. In section 5, the decay-times of the atom are compared with the dwell-times of the e.m.

radiation inside the structure [22]-[24]. Conclusions and discussion are given in section 6. In appendix, if a dipole (a two-level system) is coupled with the Normal Modes (NMs) of universe, the physical link between the emission power of the two-level system and the DOS for NMs [2] is examined closely. This appendix is viewed as a key passage in order to pave a reasonable way of generalization; if the dipole is coupled with the QNMs inside an open cavity, this kind of link is retained to hold again between the emission power of the two-level system and the DOS for QNMs.

## 2. Atomic emission power inside open cavities.

In order to discuss the emission processes of an atom inside an open cavity excited by two counter-propagating field pumps, let consider the atom as a two-level system and the e.m. field in the cavity as a superposition of Quasi Normal Modes (QNMs). The cavity $C=[0,d]$ is filled with a medium of refractive index $n(x)$. The dipole approximation [2] is assumed, so the two-level system acts as an electrical dipole of length $l$, placed in a point $x$, which oscillates orthogonally to the $x$-direction. At initial time $t=0$, the dipole is prepared in the excited state $|\psi\rangle_{t=0} = |+\rangle$, corresponding to the higher energy, and its operator of momentum $\hat{P}$ is characterized by an expectation value $\langle \hat{P} \rangle_{t=0} = \langle + | \hat{P} | + \rangle$. The dipole bandwidth is so narrow, if compared with QNM's spectrum [2], that the dipole is assumed to be (weakly) coupled with just one of the QNMs, being excited with a local density of probability $\sigma^{(loc)}(x,\omega)$ which corresponds to the local "density" of "states" (LDOS) for QNMs, [Chapter 5].

At time $t>0$, when the dipole is evolved in a superposition state $|\psi(t)\rangle$ [2], the time average for the expectation value of e.m. power $\overline{\langle \hat{W} \rangle_t}$, supplied by the dipole inside the open cavity, is inversely proportional to the relative dielectric constant $n^2(x)$ and is directly linked to the QNM's LDOS [see Appendix D]

$$\overline{\langle \hat{W} \rangle_t} = -\frac{\pi/2}{\varepsilon_0 n^2(x)} \left( \langle \hat{P} \rangle_{t=0} \big/ l \right)^2 \omega^2 \sigma^{(loc)}(x,\omega), \qquad (6.2.1)$$

where $\varepsilon_0$ is the dielectric constant in vacuum.

If the cavity extends to the whole universe, then the e.m. field is reduced to a continuum of Normal Modes (NMs); the refractive index becomes uniform, i.e. $n(x)=n_0$, as well as the DOS for NMs, i.e. $\sigma^{(loc)}_{ref}(x,\omega) = \sigma_{ref}(\omega) = \sqrt{\rho_0}/2\pi$, being $\rho_0=(n_0/c)^2$ and $c$ the speed of the light in vacuum [Appendix C]. The expectation value of the e.m. power (6.2.1), supplied by the dipole free in the universe, reduces to:

$$\overline{\langle \hat{W} \rangle_t} = -\frac{\pi/2}{\varepsilon_0 n_0^2}\left(\langle \hat{P} \rangle_{t=0}/l\right)^2 \omega^2 \sigma_{ref}(\omega) = -\frac{\sqrt{\rho_0}}{4\varepsilon_0 n_0^2}\left(\langle \hat{P} \rangle_{t=0}/l\right)^2 \omega^2. \qquad (6.2.2)$$

If the two counter-propagating field pumps are filtered at frequency $\omega \approx Re\omega_n$, just the $n^{th}$ QNM is excited and not the other QNMs, because the $n^{th}$ QNM oscillates at frequency $\omega \approx Re\omega_n$ within a narrow range $2|Im\omega_n|<<|Re\omega_n|$, so distant enough from the other QNMs [2]. The e.m. power $\overline{\langle \hat{W}_n \rangle_t}$, supplied by the dipole to the $n^{th}$ QNM,

$$\overline{\langle \hat{W}_n \rangle_t} = -\frac{\pi/2}{\varepsilon_0 n^2(x)}\left(\langle \hat{P} \rangle_{t=0}/l\right)^2 \omega^2 \sigma_n^{(loc)}(x,\omega), \qquad (6.2.3)$$

is proportional to the local density of probability (LDOP) for the $n^{th}$ QNM [here, eq. (5.2.9) is suitably reported]

$$\sigma_n^{(loc)}(x,\omega) = \frac{1}{I_n}\sigma_n(\omega)\rho(x)|F_n(x)|^2, \qquad (6.2.4)$$

which is directly linked to the (integral) density of probability (DOP) $\sigma_n(\omega)$ for the $n^{th}$ QNM. In eq. (6.2.4), $\rho(x)=[n(x)/c]^2$, $I_n$ denotes the usual overlapping integral and $F_n(x) = f_n(x)\sqrt{2\omega_n/\langle f_n | f_n \rangle}$ the normalized QNM's function, being $\langle f_n | f_n \rangle$ the QNM's norm.

In order to discuss processes of spontaneous emission, the two field pumps describe vacuum fluctuations in the universe, lying on the ground state of e.m fields [2]. The DOP for the $n^{th}$ QNM, specified for vacuum fluctuations, is expressed as [eq. (5.2.12)]:

$$\sigma_n^{(A)}(\omega) = K_n \frac{d}{2\pi} \frac{I_n^2 |Im\omega_n|}{(\omega - Re\omega_n)^2 + Im^2\omega_n}. \qquad (6.2.5)$$

The normalization constant $K_n$ can be obtained by the following condition:

$$\int_{Re\omega_n - |Im\omega_n|}^{Re\omega_n + |Im\omega_n|} \sigma_n^{(A)}(\omega) d\omega = \frac{1}{d}. \qquad (6.2.6)$$

In order to discuss processes of stimulated emission, the two field pumps describe two laser-beams in the universe, lying on a coherent state of e.m. fields [2]. When the open cavity presents a refractive index $n(x)$ which satisfies the symmetry properties $n(d/2-x)=n(d/2+x)$, the DOP for the $n^{th}$ QNM, specified for the two laser-beams, is simplified as:

$$\sigma_n^{(B)}(\omega) = \sigma_n^{(A)}(\omega)[1+(-1)^n \cos\Delta\varphi]. \qquad (6.2.7)$$

As from eq. (6.2.7), the DOP for the $n^{th}$ QNM can be controlled by the phase-difference $\Delta\varphi$ of the two external laser-beams.

## 2. 1. Sensitivity function of dipole-e.m. field coupling.

Let define the *sensitivity function* $S(x_0, \omega)$ of a *dipole-cavity coupling* as the ratio between the e.m. power (6.2.1) supplied by a dipole embedded inside an open cavity and the e.m. power (6.2.2) supplied by the same dipole free in the universe:

$$S(x_0, \omega) \triangleq \frac{\left\langle \widehat{W} \right\rangle_t}{\left\langle \widehat{W}_{ref} \right\rangle_t}. \tag{6.2.8}$$

Inserting eqs. (6.2.1) and (6.2.2) in eq. (6.2.8), it results that:

$$S(x_0, \omega) = \frac{\dfrac{\sigma^{(loc)}(x, \omega)}{\sigma_{ref}(\omega)}}{\left[\dfrac{n(x)}{n_0}\right]^2}. \tag{6.2.9}$$

If the relative dielectric constant is inhomogeneous, i.e. $n_0^2 \to n^2(x)$, the LDOS is modified, i.e. $\sigma_{ref}(\omega) \to \sigma^{(loc)}(x, \omega)$; the sensitivity function $S(x, \omega)$ can be expressed as the per-cent variation of the LDOS, $\sigma^{(loc)}(x, \omega)/\sigma_{ref}(\omega)$ in units of the per-cent variation of the relative dielectric constant $n^2(x)/n_0^2$. The meaning of the sensitivity function is ensued: eq. (6.2.9) provides a physical tool to discuss the emission processes of a dipole excited inside a cavity with respect to the case of the same dipole free in the universe.

If the *sensitivity function* $S_n(x, \omega)$ of the *dipole-$n^{th}$ QNM coupling*, i.e.

$$S_n(x, \omega) = \left[\frac{n_0}{n(x_0)}\right]^2 \frac{\sigma_n^{(loc)}(x, \omega)}{\sigma_{ref}(\omega)}, \tag{6.2.10}$$

is developed inserting the LDOS of the $n^{th}$ QNM (6.2.4), it follows that (reminding $\rho(x)=[n(x)/c]^2$ and $\rho_0=(n_0/c)^2$):

$$S_n(x, \omega) = \rho_0 \frac{|F_n(x)|^2}{I_n} \frac{\sigma_n(\omega)}{\sigma_{ref}(\omega)}. \tag{6.2.11}$$

If the dipole inside the open cavity (point $x$) is coupled to the $n^{th}$ QNM (frequency $\omega \approx Re\omega_n$), then the sensitivity function $S_n(x,\omega)$ is linked both to the normalized intensity $|F_n|^2$ of the $n^{th}$ QNM, in the point $x$, and to the DOP $\sigma_n$ for the $n^{th}$ QNM, in the frequency $\omega \approx Re\omega_n$. So, the weak coupling of a dipole to one QNM responds to the Fermi's golden rule [2].

## 3. Spontaneous emission inside a symmetric QW 1D-PBG.

In this section, the QNM's approach is applied to discuss about atomic spontaneous emission processes inside a symmetric quarter wave (QW) one dimensional (1D) Photonic Band Gap structure; this discussion, here reported, is revised respect the one reported in ref. [8] Let start with the spontaneous emission of an atom [denoted as process A], inside an open cavity at a point $x$, which is induced by vacuum fluctuations, oscillating outside the cavity at a frequency $\omega$. The conservative limit [2] is the unique case in which a single QNM can dominate, for the limit of narrow resonances $|Im\omega_n|<<|Re\omega_n|$, and vacuum fluctuations can be coupled just to one QNM, i.e. $\omega \approx Re\omega_n$. The sensitivity function (6.2.11) for the dipole-$n^{th}$ QNM coupling, in spontaneous emission, can be calculated in terms of the DOP (6.2.5) for the $n^{th}$ QNM, due to vacuum fluctuations:

$$S_n^{(A)}(x,\omega) = \rho_0 \frac{|F_n(x)|^2}{I_n} \frac{\sigma_n^{(A)}(\omega)}{\sigma_{ref}(\omega)} = $$
$$= K_n \cdot \sqrt{\rho_0} (d \cdot I_n) |F_n(x)|^2 \frac{|Im\omega_n|}{(\omega - Re\omega_n)^2 + Im^2\omega_n} \quad (6.3.1)$$

As an example, let consider a symmetric QW 1D-PBG structure with $\lambda_{ref}=1\mu m$ as reference wavelength, $N=6$ periods, consisting of two layers with refractive indices $n_h=3$ and $n_l=2$ and lengths $h=\lambda_{ref}/4n_h$ and $l=\lambda_{ref}/4n_l$. The terminal layers of the 1D-PBG are characterized by the parameters $n_h$ and $h=\lambda_{ref}/4n_h$ and the length of the structure is $d=N(h+l)+h$. In the conservative limit, a dipole can be coupled to just one of the $2N+1$ QNMs in the $[0, 2\omega_{ref})$ range, which can be enumerated, excluding $\omega=2\omega_{ref}$, as $|n\rangle = |n,0\rangle$, $n \in [0,2N]$ [6].

It is necessary to report some considerations about the QNM's functions and the QNM's DOS of a symmetric QW 1D-PBG [8], useful to discuss the spontaneous emission processes inside the structure. So, the highest e.m. field for the 1D-PBG structure with $N$ periods is found inside a central layer with low refractive index, and can be approximated to the absolute maximum of the QNM $|N+1\rangle$, with frequency close to the high-frequency band-edge. If the 1D-PBG is composed by an even (odd) number of periods, i.e. $N=2,4,...$ ($N=3,5...$), the central layer, i.e. the $(N+1)^{th}$ one, is characterized by the high (low) refractive index $n_h$ ($n_l$); so the absolute maximum of the QNM $|N+1\rangle$ is found in the adjacent layers (central layer), i.e. the $(N-1)^{th}$ and $(N+2)^{th}$ ones (just the $(N+1)^{th}$ one). In the centre $x=d/2$ of the structure, the QNM's functions present a maximum for even values of $n$ and are null for odd values of $n$ and, on the two terminal surfaces $x=0$ and $x=d$, all the QNM's functions present the same value (in modulus). The photonic band-gap is in a region $\omega/\omega_{ref}=(0.8249, 1.175)$; the DOS for QNMs is suppressed in the gap and is enhanced at the band-edge resonances.

In the figures 6.2, the spontaneous emission processes are described in terms of the sensitivity functions $S_n^{(A)}$, plotted as functions of the dimensionless space $x/d$ and the dimensionless frequency $\omega/\omega_{ref}$. In (a) is depicted the sensitivity function $S_{N-1}^{(A)}$, for the coupling dipole-$|N-1\rangle$ (where the real part of the QNM's frequency is close to the first transmission peak before the low-frequency band-edge); in (b), the sensitivity function $S_N^{(A)}$, for the coupling dipole-$|N\rangle$ (corresponding to the low-frequency band-edge, being $\omega_{I\ band-edge}/\omega_{ref} = 0.8249$); in (c) is depicted the sensitivity function $S_{N+1}^{(A)}$, for the coupling dipole-$|N+1\rangle$ (where the real part of the QNM's frequency is close to the high-frequency band-edge, being $\omega_{II\ band-edge}/\omega_{ref} = 1.175$) and, in (d), the sensitivity function $S_{N+2}^{(A)}$, for the coupling dipole-$|N+2\rangle$ (corresponding to the first transmission peak after the high-frequency band-edge). The sensitivity functions $S_N^{(A)}$ and $S_{N+1}^{(A)}$, with frequencies close to the two band edges, stand out respect to $S_{N-1}^{(A)}$ and $S_{N+2}^{(A)}$ corresponding to the two adjacent transmission peaks, because, next to the two band-edges resonances, the DOS for QNMs is enhanced and the QNM's functions present the most large intensities inside the 1D-PBG structure. Moreover, the sensitivity functions $S_n^{(A)}$ is centred around the frequency $Re\omega_n$ of the $n^{th}$ QNM and, like the QNM's function $F_n(x)$, has $n$ minima along the $x$-direction. Let note that the QNM's approach uses a realistic model for the 1D-PBG, as a finite structure with discontinuities in the refractive index, so this approach improves the results of ref. [4], obtained by the Kronig-Penney model.

In the figures 6.3, the sensitivity functions $S_N^{(A)}$ and $S_{N+1}^{(A)}$, with frequencies close to the two band-edges, and $S_{N-1}^{(A)}$, $S_{N+2}^{(A)}$, corresponding to the two adjacent transmission peaks, are plotted, as functions of the dimensionless frequency $\omega/\omega_{ref}$, when: (a) the dipole is embedded inside the $4^{th}$ period of the 1D-PBG structure, and it is exactly in the centre $x=(d/2)+(\delta/2)$ of the low-index layer, being $\delta=h+l$. The sensitivity functions $S_n^{(A)}$, $n\in[0,2N]$ are depicted, when: (b) the dipole is embedded inside the $4^{th}$ period of the 1D-PBG structure, and it is exactly in the centre $x=d/2$ of the high-index layer, being also the centre of the 1D-PBG; (c) the dipole lies on one of the two terminal surfaces $x=0$ and $x=d$ of the 1D-PBG. In all figs 6.3.a-c, the sensitivity function is strongly suppressed over the band gap $\omega/\omega_{ref}=(0.8249, 1.175)$, since the QNM's DOS is there suppressed.

When the dipole is embedded inside the $4^{th}$ period of the structure, if it is in the centre $x=(d/2)+(\delta/2)$ of the low index layer (fig. 6.3.a.): the sensitivity function $S_{N+1}^{(A)}$, corresponding to the high-frequency band-edge, is enhanced, because, in $x=(d/2)+(\delta/2)$, the eigen-function $F_{N+1}(x)$ of the QNM $|N+1\rangle$ assumes the absolute maximum [see fig. 6.2.]; moreover, the sensitivity function $S_N^{(A)}$, corresponding to the low-frequency band-edge, is almost suppressed, because, even if the QNM's DOS is large, the QNM's function $F_N(x)$ is almost null where the QNM's intensity $F_{N+1}(x)$ is maximum [8]. For the 1D-PBG with parameters above reported, i.e. $\lambda_{ref}=1\mu m$, $N=6$, $n_h=3$ and $n_l=2$, the peak of sensitivity function $S_{N+1}^{(A)}$ assumes the absolute maximum $S_{max}=18.55$ (as previously stated in some considerations); in $x=(d/2)+(\delta/2)$, the peak of the sensitivity function $S_N^{(A)}$, in units of $S_{max}$ is 0.006, the peak of the sensitivity function $S_{N-1}^{(A)}$, in units of $S_{max}$, is 0.1507 and the ratio between the peak of the sensitivity function $S_{N+2}^{(A)}$ and the peak of the sensitivity function $S_{N-1}^{(A)}$ is 0.4966.

When the dipole is embedded inside the $4^{th}$ period of the structure ($N=6$ periods), if it is in the centre of the high-index layer (i.e. $x=d/2$) [see fig. 6.1.], the sensitivity functions in the low and high frequency band-edges exchange their physical behaviour; the sensitivity function $S_N^{(A)}$, in the low-frequency band-edge (the $N^{th}$ one, an even peak) is enhanced, and the sensitivity function $S_{N+1}^{(A)}$, in the high-frequency band-edge (the $(N+1)^{th}$ one, an odd peak), is suppressed (fig. 6.3.b.). In fact, in the centre of the structure, the dipole can be coupled just to one of the QNMs $|n\rangle$ with an even $n$, because, in $x=d/2$, the QNM's function $F_n(x)$ has a maximum for even values of $n$ and is null for odd values of $n$ [see fig. 6.2.] [8]. For the 1D-PBG with parameters above reported, in $x=d/2$, the peak of $S_N^{(A)}$, in units of $S_{max}$ is 0.6905; instead, the ratio between the peak of the $S_{N-2}^{(A)}$ and the peak of $S_N^{(A)}$ is 0.3565.

When the dipole lies on the terminal surfaces $x=0$ and $x=d$ of the structure (fig. 6.3.c.): the dipole can be coupled to any of the QNMs $|n\rangle$ (in fact, any of the QNM's functions $F_n(x)$ is not null on $x=0$ and $x=d$); moreover, the peaks of the sensitivity functions $S_N^{(A)}$ and $S_{N+1}^{(A)}$, present values similar to the peaks of the sensitivity functions $S_n^{(A)}$, as it could be intuitively suggested (all the QNM's functions $F_n(x)$ have the same values on $x=0$ and $x=d$). For the 1D-PBG considered, in $x=0$ and $x=d$, the $S_{n=N,N+1}^{(A)}$ peak in units of $S_{max}$ is 0.2239; instead, the peak ratio $S_{n=N-1,N+2}^{(A)} / S_{n=N,N+1}^{(A)}$ is 1.0265.

The QNM's approach to discuss the spontaneous emission processes of a dipole inside a 1D-PBG presents the advantage to develop a general quantum treatment. This approach agrees with the theoretical results of ref. [18], which consider only a classical model. Moreover, the approach confirms the results of ref. [19], which is only a numerical investigation.

## 4. Coherent control of stimulated emission inside a symmetric QW 1D-PBG.

In the following sections, the QNM's approach is applied to discuss how a coherent control of the atomic stimulated emission processes inside a symmetric QW 1D-PBG can be effected by the phase-difference of two counter-propagating laser-beams [10]; such a system is relevant for a single-atom, phase-sensitive, optical memory device on atomic scale. Let start with the stimulated emission of an atom [denoted as process B], inside a symmetric cavity [refractive index $n(x)$ such that $n(d/2-x)=n(d/2+x)$], which is induced by two counter-propagating laser-beams (phase-difference $\Delta\varphi$). As stated in advance, the conservative limit is the unique case in which a single QNM can dominate, for the limit of narrow resonances $|Im\omega_n|<<|Re\omega_n|$, and the two laser-beams can be coupled just to one QNM, i.e. $\omega\approx Re\omega_n$. The sensitivity function (6.2.11) for the dipole-$n^{th}$ QNM coupling, in stimulated emission, can be calculated in terms of the DOP (6.2.7) for the $n^{th}$ QNM, due to the two laser-beams:

$$S_n^{(B)}(x,\omega) = \rho_0 \frac{|F_n(x)|^2}{I_n} \frac{\sigma_n^{(B)}(\omega)}{\sigma_{ref}(\omega)} =$$
$$= \rho_0 \frac{|F_n(x)|^2}{I_n} \frac{\sigma_n^{(A)}(\omega)[1+(-1)^n \cos\Delta\varphi]}{\sigma_{ref}(\omega)} = S_n^{(A)}(x,\omega)[1+(-1)^n \cos\Delta\varphi] \quad (6.4.1)$$

As from (6.4.1), the sensitivity function of the dipole-$n^{th}$ QNM coupling, for stimulated emission inside an open cavity, depends on the phase-difference of two external counter-propagating laser-beams. If the dipole is embedded in a point of the cavity in which the normalized intensity of the $n^{th}$ QNM is almost null, all the emission processes are inhibited. Otherwise, if the dipole is in a point of the cavity in which the $n^{th}$ QNM intensity is not null, it can be coupled to one of the QNMs with an even $n$, when the two laser-beams are in phase $\Delta\varphi=0$, while it can be coupled to one of the QNMs with an odd $n$, when the two laser-beams are opposite in phase $\Delta\varphi=\pi$.

As an example, let consider the symmetric QW 1D-PBG structure with parameters above reported, i.e. $\lambda_{ref}=1\mu m$, $N=6$, $n_h=3$ and $n_l=2$. As stated in advance, in the conservative limit, a dipole can be coupled to just one of the $2N+1$ QNMs in the $[0, 2\omega_{ref})$ range, which can be enumerated, excluding $\omega=2\omega_{ref}$, as $|n\rangle=|n,0\rangle$, $n\in[0,2N]$. In figure 6.4, the stimulated emission processes are described by the sensitivity functions $S_n^{(B)}$ of (6.4.1), plotted as functions of the dimensionless space $x/d$ and the dimensionless frequency $\omega/\omega_{ref}$. The sensitivity function $S_{N+1}^{(A)}$ for the coupling dipole-$|N+1\rangle$ (corresponding to the high-frequency band-edge) is depicted, when the two laser beams are nearly in phase, (a) $\Delta\varphi=\pi/4$, or nearly opposite in phase, (b) $\Delta\varphi=3\pi/4$. It is clear that the dipole, inside the 1D-PBG structure, can be actually coupled with the QNM $|N+1\rangle$, in the high-frequency band-edge (i.e. the $(N+1)^{th}$ one, an odd peak), only when the two laser-beams tend to be opposite in phase (i.e. $\Delta\varphi\to\pi$). The amplification factor of the stimulated emission respect to the spontaneous emission is identified as $1+(-1)^n \cos\Delta\varphi$, which assumes the values 0.2929 in fig 6.4.a and 1.7071 in fig 6.4.b.

## 5. Decay-times in units of dwell-times.

The spontaneous emission process of a dipole lying on a terminal surface ($x=0$) of an open cavity is characterized by a decay-time

$$\tau_n^{(A)}(0) = \frac{1}{\Delta\omega_n^{(A)}(0)}, \qquad (6.5.1)$$

being $\Delta\omega_n^{(A)}(0)$ the bandwidth of the sensitivity function $S_n^{(A)}(0,\omega)$ at half height [i.e. $S_n^{(A)}(0,\omega=\mathrm{Re}\,\omega_n)/2$]. After some algebra, reminding $|F_n(0)|^2 = |\mathrm{Im}\,\omega_n|/\sqrt{\rho_0}$, it results that:

$$\tau_n^{(A)}(0) = \frac{1}{2|\mathrm{Im}\,\omega_n|}. \qquad (6.5.2)$$

It is clear that no emission occurs [$\tau_n^{(A)}(0) \to \infty$] when the cavity is closed [$|\mathrm{Im}\,\omega_n| \to 0$].

The spontaneous emission process of a dipole in a point $x$ of the cavity is characterized by a decay-time

$$\tau_n^{(A)}(x) = \frac{1}{\Delta\omega_n^{(A)}(x)}, \qquad (6.5.3)$$

being $\Delta\omega_n^{(A)}(x)$ the bandwidth of the sensitivity function $S_n^{(A)}(x,\omega)$ at the half height of $S_n^{(A)}(0,\omega)$. After some algebra, it results that (see ref. [10]):

$$\left[\tau_n^{(A)}(x)\right]^2 \cong \frac{1}{4}\left(\frac{d \cdot I_n}{\sqrt{\rho_0}}\right)\frac{\tau_n^{(A)}(0)}{|F_n(x)|^2}. \qquad (6.5.4)$$

As from eq. (6.5.4), in case of spontaneous emission processes inside an open cavity, the decay-time of the dipole-$n^{th}$ QNM coupling depends from the position of the dipole in the cavity.

If the cavity of length $d$ is pumped by vacuum fluctuations, filtered at $\omega \approx \mathrm{Re}\,\omega_n$, it is possible to introduce the dwell-time [22][8] of vacuum fluctuations, linked to the DOP (6.2.5) for the $n^{th}$ QNM:

$$\delta t_n^{(A)}(\omega) \triangleq d \cdot \sigma_n^{(A)}(\omega) = K_n \frac{(d \cdot I_n)^2}{2\pi} \frac{|\mathrm{Im}\,\omega_n|}{(\omega-\mathrm{Re}\,\omega_n)^2 + \mathrm{Im}^2\,\omega_n} \cong \frac{K_n}{\pi}\frac{(d \cdot I_n)^2}{2|\mathrm{Im}\,\omega_n|}. \qquad (6.5.5)$$

The decay-time (6.5.2) of the dipole coupled with the $n^{th}$ QNM, when the dipole lies on a terminal surface ($x=0$), is different from the dwell-time (6.5.5) of vacuum fluctuations at $\omega \approx \mathrm{Re}\,\omega_n$; in fact, if the dipole is on one surface of a symmetric QW 1D-PBG structure with parameters $\lambda_{ref}=1\mu m$, $N=6$, $n_h=3$, $n_l=2$ and it is excited by vacuum fluctuations filtered at one band-edge, i.e. $\omega_{I\ band-edge}/\omega_{ref} \approx$

*0.8249* or $\omega_{II\ band-edge}/\omega_{ref} \approx 1.175$, then the decay-time is $\tau^{(A)}_{band-edge}(x_0 = 0) \cong 33.30$ in units of *$1/\omega_{ref}$*, and the ratio between the decay-time and the dwell-time is $\tau^{(A)}_{band-edge}(x_0 = 0)/\delta t^{(A)}_{band-edge} \cong 1.571$.

The stimulated emission process of a dipole embedded in a point $x$ of the open cavity is characterized by a decay-time

$$\tau^{(B)}_n(x) = \frac{1}{\Delta\omega^{(B)}_n(x)}, \qquad (6.5.6)$$

being $\Delta\omega^{(B)}_n(x)$ the bandwidth of the sensitivity function $S^{(B)}_n(x,\omega)$ at the half height of $S^{(A)}_n(0,\omega)$. After some algebra, it results that (ref. [10]):

$$\tau^{(B)}_n(x) \cong \frac{\tau^{(A)}_n(x)}{\sqrt{1+(-1)^n \cos\Delta\varphi}}. \qquad (6.5.7)$$

All the bandwidths $\Delta\omega^{(A)}_n(0)$, $\Delta\omega^{(A)}_n(x)$ and $\Delta\omega^{(B)}_n(x)$ are referred to the half length of $S^{(A)}_n(0,\omega)$, so the decay-times (6.5.2), (6.5.4) and (6.5.7) run from the same instant. As from eq. (6.5.7), in case of stimulated emission inside an open cavity pumped by two counter-propagating laser-beams, the decay-time of the dipole-$n^{th}$ QNM coupling in a point of the cavity can be controlled by the phase-difference of the two laser-beams.

Figure 6.5 refers to an excited dipole, embedded inside a symmetric QW 1D-PBG structure with parameters $\lambda_{ref}=1\mu m$, $N=6$, $n_h=3$, $n_l=2$, pumped by two counter-propagating field pumps filtered at low-frequency band-edge (fig. 6.5.a.) or at high-frequency band-edge (fig. 6.5.b.). The decay-time $\tau$, in units of the decay-time $\tau_{ref}$ for spontaneous emission when the dipole lies on one terminal surface ($x=0$) of the structure [eq.(6.5.2)], is plotted as a function of the dimensionless position $x/d$ of the dipole, being $d$ the length of the 1D-PBG. Several cases are shown: the spontaneous emission [see eq.(6.5.4)], when the two field pumps describe vacuum fluctuations (———); and the stimulated emission [see eq.(6.5.7)], when the two pumps describe two input laser-beams, nearly in phase (– – –) or opposite of phase (— – —). So, in the low-frequency (high-frequency) band-edge, all the emission processes are enhanced if the dipole is inside the layers with high (low) refractive index; while, the stimulated emission can be inhibited increasing (reducing) the phase-difference of the two laser-beams if the dipole is inside the layers with low (high) refractive index. In fact, next to the low-frequency (high-frequency) band-edge, the DOS is minimum if the two laser-beams are opposite of phase (in phase) [fig.5.1.b.]. Figure 6.5.a.bis. (fig. 6.5.b.bis.) is a magnification of fig. 6.5.a. (fig. 6.5.b.) for the decay-time when the dipole is in the $3^{th}$ and $4^{th}$ periods of the 1D-PBG. If the dipole is centred into the structure, the decay-time is accelerated (tends to be highly retarded) in the low-frequency (high-frequency) band-edge. In fact, in the centre of the structure, the QNM corresponding to the low-frequency (high-frequency) band-edge is maximum (tends to zero) [as stated in some considerations after eq. (6.3.1)].

If an open cavity of length $d$ is pumped by two counter-propagating laser-beams tuned at the frequency $\omega \approx Re\omega_n$ (phase-difference $\Delta\varphi$), it is possible to introduce the dwell-time of the two laser-beams, linked to the DOS (6.2.7) for the $n^{th}$ QNM [10]:

$$\delta t_n^{(B)} \triangleq d \cdot \sigma_n^{(B)} = d \cdot \sigma_n^{(A)}[1+(-1)^n \cos\Delta\varphi] = \delta t_n^{(A)}[1+(-1)^n \cos\Delta\varphi] . \qquad (6.5.8)$$

The decay-time (6.5.7) for stimulated emission of the dipole, coupled with the $n^{th}$ QNM, and the dwell-time (6.5.8) of the two laser-beams, tuned at the $n^{th}$ transmission resonance, are dual functions; in fact, the stimulated emission is inhibited [ $\tau_n^{(B)}(x_0) \to \infty$ ] when the two laser-beams are reflected by the cavity [ $\delta t_n^{(B)} = 0$ ], and it is enhanced [ $\tau_n^{(B)}(x_0)$ is minimum] when the two laser-beams "stand" in the cavity [ $\delta t_n^{(B)}$ is maximum].

Figure. 6.6. refers to the excited dipole embedded inside the symmetric QW 1D-PBG of fig. 6.5., pumped by two counter-propagating laser-beams filtered at low-frequency band-edge (fig. 6.6.a.) or at the high-frequency band-edge (fig. 6.6.b.). The decay-time for stimulated emission (———), in units of the decay-time for spontaneous emission [eq. (6.5.7)], is compared with the dwell-time (– – –) for the two laser-beams, in units of the dwell-time for vacuum fluctuations [eq. (6.5.8)]. The decay-time for stimulated emission and the dwell-time for the two laser-beams are compared on different scales as functions of the phase-difference between the two laser-beams. In the low-frequency (high-frequency) band-edge, the decay-time ratio is rising (slopes down) and so the dwell-time ratio slopes down (is rising) when the phase-difference of the two laser-beams is increasing from $\Delta\varphi=0$ to $\Delta\varphi=\pi$; in fact, the decay-time [figs. 6.5.a. and 6.5.b.] tends to the maximum (minimum) and the DOS [fig. 5.1.b.] tends to the minimum (maximum) when the laser-beams are going opposite of phase. Then, in the low-frequency (high-frequency) band-edge, the decay-time ratio tends to infinity and so the dwell-time ratio is null when the phase-difference of the two laser-beams is $\Delta\varphi=\pi$ ($\Delta\varphi=0$); in fact, the dipole is not coupled to the QNM corresponding to the low-frequency (high-frequency) band-edge when the two laser-beams are opposite in phase (in phase) [as stated in some considerations after eq. (6.4.1)].

# 6. Discussion and conclusions.

Non-Hermitian Hamiltonians and the ensuing complex eigen-values figure prominently in Siegman's work on dissipative cavity quantum electro-dynamics (CQED) [4], elaborating the Fox and Lee's works [5], which consider eigen-value problems for complex symmetric operators. However, refs. [4] and [5] deal with transverse modes in the semi-classical limit, only considering (a) *c*-number fields with some effective quantum noise, and (b) the limit $\lambda<<d$, being $\lambda$ the field wavelength and $d$ the length of the cavity.

The present chapter pertains to a different regime; several numerical simulations have pointed out the physical link between the sensitivity function of an atom-e.m. field coupling and the density of states (DOS) for Quasi Normal Modes (QNMs) inside an open cavity.

As main object, a fully quantum treatment is given for the dynamics of an atom inside a 1D-PBG structure, coupled with two counter-propagating laser-beams, by using the QNM's approach, i.e. treating the problem in the framework of open systems. The e.m. field has been quantized in terms of the QNMs inside the 1D-PBG and the atom has been modelled as a two levels system. In the electric dipole approximation, the atom has been assumed to be weakly coupled to just one of the QNMs. As a result, the decay-time depends from the position of the dipole inside the structure, and can be controlled by the phase-difference of the two laser-beams. Such a system is relevant for a single-atom, phase-sensitive, optical memory device on atomic scale.

# Appendix D. Examining deeper the link of atomic emission power with density of states (DOS).

Let consider, according to classical electro-dynamics [15], a dipole which is radiating an e.m. field in the universe, and then let discuss, applying second quantization rules [2], the coupling of a two-level system with normal modes (NMs) of the universe; the physical link between the emission power of the dipole and the "density" of "states" (DOS) for NMs is examined deeper: in fact, this link is not clearly discussed in literature. This proof is viewed as a key passage in order to pave a reasonable way of generalization; if the two-levels system is coupled with the Quasi Normal Modes (QNMs) inside an open cavity, this kind of link is retained to hold again between the emission power of the dipole and the DOS for QNMs. Besides, the proof enables to solve some self-contradictions in the expressions of the emission power which are reported in refs. [24] and [8].

## D. 1. Current density producing an e.m. field in the universe.

Let refer to a Cartesian frame *xyz*. The universe is modelled as an infinite one dimensional cavity, $x \in C = (-L/2, L/2)$, $L \to \infty$, which is filled with a linear, isotropic and homogeneous dielectric, dissipative for an electrical conductibility σ, with a dielectric constant ε=$ε_0ε_r$ and a magnetic permeability μ=$μ_0μ_r$≈$μ_0$. No charge distribution is present and a current distribution $J_y(x,t)$, which oscillates along the direction of the *y*-axis, produces an e.m. plane wave $E_y(x,t)$ and $H_z(x,t)$, which is a transverse electric and magnetic field respect to the *x*-axis.

Under the monochromatic condition, the current density oscillates just at one frequency

$$J_y(x,t) = \frac{1}{2}[\tilde{J}_y(x,\omega)\exp(-i\omega t) + c.c.], \qquad (6.D.1)$$

and so the e.m. field

$$E_y(x,t) = \frac{1}{2}[\tilde{E}_y(x,\omega)\exp(-i\omega t) + c.c.]$$
$$H_z(x,t) = \frac{1}{2}[\tilde{H}_z(x,\omega)\exp(-i\omega t) + c.c.] \qquad (6.D.2)$$

satisfies the equations [15]

$$\frac{\partial_x^2 \tilde{E}_y}{\partial x^2} + \rho_0 \omega^2 \tilde{E}_y(x,\omega) = -i\omega\mu_0 \tilde{J}_y(x,\omega)$$
$$\tilde{H}_z(x,\omega) = \frac{1}{i\omega\mu_0}\frac{\partial_x \tilde{E}_y}{\partial x} \qquad (6.D.3)$$

where $\rho_0=(n_0/c)^2$, being $n_0 \cong \sqrt{\varepsilon_r}$ the refractive index of the dielectric material and $c = 1/\sqrt{\mu_0\varepsilon_0}$ the speed of the light in vacuum.

The Green's function of the e.m. problem (6.D.3) is defined as the solution of the equation [15]

$$\frac{\partial_x^2 G}{\partial x^2} + \rho_0\omega^2 G(x,x',\omega) = -\delta(x-x'), \qquad (6.D.4)$$

where $\delta(x-x')$ is the Dirac's delta function and $x$, $x'$ stand for observation-point and the source-point respectively. The Green's function $G(x,x',\omega)$ is named as the kernel of the e.m. problem (6.D.3) because it allows to obtain the corresponding solution:

$$\tilde{E}_y(x,\omega) = i\omega\mu_0 \int_{-\infty}^{\infty} \tilde{J}_y(x',\omega) G(x,x',\omega) dx'. \qquad (6.D.5)$$

In universe, an energetic balance is settled between the current distribution and the e.m. plane wave. In time domain, the e.m. power supplied by the current density $J_y(x,t)$ to the e.m. field $E_y(x,t)$ is defined as [15]

$$W(t) = \int_{-\infty}^{\infty} w(x,t) dx = -\frac{1}{2}\int_{-\infty}^{\infty} J_y(x,t) E_y(x,t) dx, \qquad (6.D.6)$$

so, in the period $T=2\pi/\omega$, the time average of the e.m. power (6.D.6) is calculated as:

$$\overline{W} = \frac{1}{T}\int_{-T/2}^{T/2} W(t) dt. \qquad (6.D.7)$$

Instead, in frequency domain, a complex power supplied by the current density $\tilde{J}_y(x,\omega)$ to the e.m. field $\tilde{E}_y(x,\omega)$ can be defined [15]

$$\widetilde{W}(\omega) = \int_{-\infty}^{\infty} \tilde{w}(x,\omega) dx = -\frac{1}{2}\int_{-\infty}^{\infty} \tilde{J}_y^*(x,\omega)\tilde{E}_y(x,\omega) dx, \qquad (6.D.8)$$

such that the time-average of the e.m. power (6.D.7) can be calculated as the real part of the complex power (6.D.8):

$$\overline{W} = \mathrm{Re}[\widetilde{W}(\omega)]. \qquad (6.D.9)$$

### D. 2. Electrical dipole.

Let consider an electrical dipole of length $l$ and charge $q(t)=(1/2)\,[q_0 exp(-i\omega t)+c.c.]$. The momentum of the dipole is $p(t)=lq(t)=(1/2)\,[p_0 exp(-i\omega t)+c.c.]$, being $p_0=q_0 l$. The intensity of the electrical current which flows inside the dipole is defined as $I(t) = dq/dt = 1/2[\tilde{I}(\omega)\exp(-i\omega t)+c.c.]$, being $\tilde{I}(\omega) = -i\omega(p_0/l)$. If the dipole is placed in the $x_0$ point and oscillates along the direction of the y-axes, then, in frequency domain, the density of the current $\tilde{J}_y(x,\omega)$, such that $\int_{-\infty}^{\infty} \tilde{J}_y(x,\omega)dx = \tilde{I}(\omega)$, is operatively defined as:

$$\tilde{J}_y(x,\omega) = \tilde{I}(\omega)\delta(x-x_0) = -i\omega(p_0/l)\delta(x-x_0). \tag{6.D.10}$$

It results that, the e.m. field produced by the dipole is obtained inserting eq. (6.D.10) in eq. (6.D.5)

$$\tilde{E}_y(x,\omega) = \mu_0 (p_0/l)\omega^2 G(x,x_0,\omega), \tag{6.D.11}$$

the complex power supplied by the dipole to the universe is obtained inserting eqs. (6.D.10) and (6.D.11) in eq. (6.D.8)

$$\widetilde{W}(\omega) = -i\frac{\mu_0}{2}\left(\frac{p_0}{l}\right)^2 \omega^3 G(x_0,x_0,\omega), \tag{6.D.12}$$

and the time-average of the power supplied by the dipole is obtained inserting eq. (6.D.12) in eq. (6.D.9):

$$\overline{W} = \frac{\mu_0}{2}\left(\frac{p_0}{l}\right)^2 \omega^3 \operatorname{Im} G(x_0,x_0,\omega). \tag{6.D.13}$$

As from eq. (6.D.13), the time average of the power $\overline{W}$ supplied by the electrical dipole to the universe depends on the resonance $\omega$ of the dipole and it is linked to the square modulus of the momentum $p_0$ and then to the position of the dipole $x_0$ via the imaginary part of the universe Green's function $G(x_0, x_0, \omega)$. It is easy to verify that the physical link (6.D.13) holds also for a dipole inside an inhomogeneous cavity, if the Green's function $G(x, x', \omega)$ is referred to a cavity which is filled with a medium of refractive index $n(x)$.

### D. 3. Feynman's propagator.

To solve for the kernel equation (6.D.4), let introduce the Fourier's space-transform of the Green's function

$$G(x,x',\omega) = \frac{1}{\sqrt{2\pi}} \int_{-\infty}^{\infty} \widetilde{G}(k,\omega)\exp[ik(x-x')]dk, \tag{6.D.14}$$

and the Fourier's representation of the Dirac's delta function [14]:

$$\delta(x) = \frac{1}{2\pi} \int_{-\infty}^{\infty} \exp(ikx)dk. \tag{6.D.15}$$

If the Fourier's space-transforms (6.D.14) and (6.D.15) are applied on equation (6.D.4), the space-transformed Green's function is obtained:

$$\widetilde{G}(k,\omega) = \frac{1}{\sqrt{2\pi}}\frac{1}{k^2 - \rho_0\omega^2} \quad , \quad k^2 \neq \rho_0\omega^2. \tag{6.D.16}$$

However, there is an ambiguity; let put this expression into the Fourier's transform (6.D.14) and try to perform the integration over the variable $k$: the integrand presents the poles $k_{\pm} = \pm\omega\sqrt{\rho_0}$ and therefore the integration cannot be performed over the real axis for the poles $k_{\pm}$.

To get around this ambiguity, let adopt the Feynman's prescription [4]; the Green's function is replaced by a complex propagator: so, in the transformed space-domain

$$\tilde{\Delta}_F(k,\omega) \triangleq \frac{1}{\sqrt{2\pi}}\frac{1}{k^2 - \rho_0\omega^2 + i\eta'} = \frac{1}{\sqrt{2\pi}}\frac{1}{k^2 - (\omega\sqrt{\rho_0} - i\eta)^2} =$$

$$= \frac{1}{\sqrt{2\pi}}\frac{1}{2(\omega\sqrt{\rho_0} - i\eta)}\left[\frac{1}{k - (\omega\sqrt{\rho_0} - i\eta)} - \frac{1}{k + (\omega\sqrt{\rho_0} - i\eta)}\right], \quad k \in \mathbb{R}, \quad (6.D.17)$$

$$i\eta' = \eta^2 + i\eta 2\omega\sqrt{\rho_0}, \quad \eta, \eta' \in \mathbb{R}^+$$

and in the effective space domain

$$\Delta_F(x,x',\omega) \triangleq \frac{1}{\sqrt{2\pi}}\int_{-\infty}^{\infty}\tilde{\Delta}_F(k,\omega)\exp[ik(x-x')]dk. \quad (6.D.18)$$

If the parameter $\eta$ tends to zero, the complex propagator reduces, in transformed space-domain, to the Green's function

$$\tilde{G}(k,\omega) \triangleq \lim_{\eta\to 0}\tilde{\Delta}_F(k,\omega), \quad (6.D.19)$$

and, in effective space domain, to the Feynman's propagator which is not the simple Green's function:

$$G_F(x,x',\omega) \triangleq \lim_{\eta\to 0}\Delta_F(x,x',\omega) = \lim_{\eta\to 0}\frac{1}{\sqrt{2\pi}}\int_{-\infty}^{\infty}\tilde{\Delta}_F(k,\omega)\exp[ik(x-x')]dk \neq$$

$$\neq G(x,x',\omega) = \frac{1}{\sqrt{2\pi}}\int_{-\infty}^{\infty}\tilde{G}(k,\omega)\exp[ik(x-x')]dk = \frac{1}{\sqrt{2\pi}}\int_{-\infty}^{\infty}\lim_{\eta\to 0}\tilde{\Delta}_F(k,\omega)\exp[ik(x-x')]dk.$$

(6.D.20)

The complex analysis [14] provides the following result

$$\lim_{\eta\to 0}\int_{-\infty}^{\infty}\frac{\exp(-ikx)}{k+i\eta}dk = -2\pi i\theta(x), \quad (6.D.21)$$

being $\theta(x)$ the unit step function. To calculate the Feynman's propagator (6.D.20), the result (6.D.21) is used, after inserting eq. (6.D.17) into eq. (6.D.18):

$$G_F(x,x',\omega) = -\frac{i}{2\omega\sqrt{\rho_0}}\left\{\exp[-i\omega\sqrt{\rho_0}(x'-x)]\theta(x'-x) + \exp[-i\omega\sqrt{\rho_0}(x-x')]\theta(x-x')\right\}. \quad (6.D.22)$$

The e.m. field inside the universe consists on a continuum of NMs:

$$g_\omega(x) = \frac{1}{\sqrt{2\pi}}\exp(i\omega\sqrt{\rho_0}\,x), \quad \omega \in \mathbb{R}. \quad (6.D.23)$$

The DOS for NMs is defined as the density of probability that the e.m. field, excited inside the universe, consists of one NM, tuned at the frequency $\omega$, so the DOS for NMs is calculated as [Appendix C]

$$\sigma^{(loc)}(x,\omega) = \sqrt{\rho_0}\,|g_\omega(x)|^2 = \sqrt{\rho_0}/2\pi \ . \tag{6.D.24}$$

It results that, in frequency domain, the Feynman's propagator (6.D.22) of the universe can be expressed in terms of the NMs (6.D.23)

$$G_F(x,x',\omega) = -\frac{i\pi}{\omega\sqrt{\rho_0}}\left[g_\omega^*(x')g_\omega(x)\theta(x'-x) + g_\omega^*(x)g_\omega(x')\theta(x-x')\right], \tag{6.D.25}$$

and it can be linked to the NM's DOS (6.D.24)

$$\sigma^{(loc)}(x,\omega) = -\frac{1}{\pi}\rho_0\omega\,\mathrm{Im}\,G_F(x,x,\omega), \tag{6.D.26}$$

being $\theta(x=x') \triangleq 1/2$. As from eq. (6.D.26), the DOS is linked to the imaginary part of the Feynman's propagator.

### D. 4. Two-levels system.

Let quantize an atom as a two-levels system [2]. The atomic Hamiltonian $\hat{H}_0$ admits only two eigen-values $E_-$ and $E_+ > E_-$, corresponding to two eigen-states $|-\rangle$ and $|+\rangle$, i.e.

$$\hat{H}_0|\pm\rangle = E_\pm|\pm\rangle. \tag{6.D.27}$$

The atom is assumed to be coupled with the e.m. field in the universe, which is quantized as a continuum of NMs. The total Hamiltonian $\hat{H}$ is the sum between the unperturbed Hamiltonian $\hat{H}_0$ of the atom and the perturbation $\hat{H}_{int}$ of the atom-e.m. field coupling, i.e.

$$\hat{H} = \hat{H}_0 + \hat{H}_{int}. \tag{6.D.28}$$

Let adopt the dipole approximation [2], so the atom acts as an electrical dipole of length $l$ which is placed in the $x_0$ point and oscillates along the direction of the $y$-axes [see section D.2.]. To discuss emission processes, at initial time $t=0$, the dipole is assumed to be in the unperturbed state $|+\rangle$ corresponding to the higher energy $E_+$, i.e.

$$|\psi\rangle_{t=0} = |+\rangle, \tag{6.D.29}$$

so, the momentum of the dipole is described by an observable $\hat{P}$, which is characterized by the expectation value:

$$\langle\hat{P}\rangle_{t=0} = \langle+|\hat{P}|+\rangle. \tag{6.D.30}$$

If the dipole is weakly coupled with the e.m. field of the universe, i.e. $\hat{H}_{int} \ll \hat{H}_0$, at time $t>0$, the state of the dipole $|\psi(t)\rangle$ can be represented as a linear superposition of the unperturbed states $|-\rangle$ and $|+\rangle$, i.e.[2]

$$|\psi(t)\rangle = c_1(t)|-\rangle + c_2(t)|+\rangle. \quad (6.D.31)$$

The Ehrenfest's theorem can be applied under the dipole and weak-coupling approximations. So, according to eq. (6.D.13), the time average for the expectation value of the e.m. power in the state (6.D.31), supplied to one NM (frequency $\omega$) of the universe by the electrical dipole (length $l$, point $x_0$, initial momentum $\langle \hat{P} \rangle_{t=0}$), can be expressed in terms of the imaginary part of the Feynman's propagator, i.e.

$$\overline{\langle \hat{W} \rangle}_t = \frac{\mu_0}{2}\left(\langle \hat{P} \rangle_{t=0}/l\right)^2 \omega^3 \operatorname{Im} G_F(x_0, x_0, \omega). \quad (6.D.32)$$

As well as eq. (6.D.13) [see comments], eq. (6.D.32) holds also for a two-levels system inside a inhomogeneous cavity, if the Feynman's propagator $G_F(x, x', \omega)$ satisfies the boundary conditions of a cavity which is filled with a medium of refractive index $n(x)$.

Recalling that $\rho_0 = (n_0/c)^2 = n_0^2 \mu_0 \varepsilon_0$, the e.m. power (6.D.32) supplied by the electrical dipole to one NM of the universe can be expressed in terms of the NM's DOS (6.D.26), i.e.

$$\overline{\langle \hat{W} \rangle}_t = -\frac{\pi/2}{\varepsilon_0 n_0^2}\left(\langle \hat{P} \rangle_{t=0}/l\right)^2 \omega^2 \sigma^{(loc)}(x_0, \omega), \quad (6.D.33)$$

and, inserting eq. (6.D.24), it results that the e.m. power (6.D.33) is uniform in the space:

$$\overline{\langle \hat{W}_{ref} \rangle}_t = -\frac{\sqrt{\rho_0}}{4\varepsilon_0 n_0^2}\left(\langle \hat{P} \rangle_{t=0}/l\right)^2 \omega^2. \quad (6.D.34)$$

The physical link (6.D.34) holds also for QNMs inside an open cavity, if the local QNM's DOS satisfies the boundary conditions of a cavity filled with a medium of refractive index $n_0$ [20]. Moreover, the physical link (6.D.34) can be generalized for QNMs inside an inhomogeneous cavity of refractive index $n(x)$,

$$\overline{\langle \hat{W} \rangle}_t = -\frac{\pi/2}{\varepsilon_0 n^2(x_0)}\left(\langle \hat{P} \rangle_{t=0}/l\right)^2 \omega^2 \sigma^{(loc)}(x_0, \omega), \quad (6.D.35)$$

if the QNMs oscillate at wavelengths so short that the refractive index of the cavity can be considered as smooth [6]

$$\left|\frac{dn(x)}{dx}\right| \ll \frac{4\pi}{\lambda}. \quad (6.D.36)$$

# References.

[1] J. J. Sakurai, *Advanced Quantum Mechanics* (Addison-Wesley, New York, 1995); S. Weinberg, *The Quantum Theory of Fields* (Cambridge University Press, New York, 1996).

[2] E. M. Purcell, Phys. Rev. **69**, 681 (1946); D. Kleppner, Phys. Rev. Lett. **47**, 233 (1981).

[3] E. A. Hinds, Adv. At., Mol., Opt. Phys. **28**, 237 (1991).

[4] M. Lewenstein, J. Zakrzewski, T. M. Mossberg, and J. Mostowski, J. Phys. B **21**, L9 (1988); M. Lewenstein, J. Zakrzewski, and T. M. Mossberg, Phys. Rev. A **38**, 1075 (1988).

[5] S. Haroche, in *New Trends in Atomic Physics,* edited by G.Grynberg and R.Stora, Proceedings of the Les Houches Summer School of Theoretical Physics, XXXVIII, 1982 (North-Holland, Amsterdam, 1982), p.193.

[6] Z. Huang, C. Lei, D. G. Deppe, C. C. Lin, C. J. Pinzone, and R. D. Dupuis, Appl. Phys. Lett. **61**, 2961 (1992); Q. Deng and D. G. Deppe, Phys. Rev. A **53**, 1036 (1996).

[7] S. John, Phys. Rev. Lett. **53**, 2169 (1984); S. John, Phys. Rev. Lett. **58**, 2486 (1987); E. Yablonovitch, Phys. Rev. Lett. **58**, 2059 (1987); E. Yablonovitch and T. J. Gmitter, Phys. Rev. Lett. **63**, 1950 (1989).

[8] J. Maddox, Nature (London) **348**, 481 (1990); E. Yablonovitch and K.M. Lenny, Nature (London) **351**, 278, 1991; J. D. Joannopoulos, P. R. Villeneuve, and S. H. Fan, Nature (London) **386**, 143 (1997).

[9] J. D. Joannopoulos, *Photonic Crystals: Molding the Flow of Light* (Princeton University Press, Princeton, New York, 1995); K. Sakoda, *Optical properties of photonic crystals* (Springer Verlag, Berlin, 2001); K. Inoue and K. Ohtaka, *Photonic Crystals: Physics, Fabrication, and Applications* (Springer-Verlag, Berlin, 2004).

[10] G. Kurizki and A. Z. Genack, Phys. Rev. Lett. **61,** 2269 (1988); G. Kurizki, Phys. Rev. A **42**, 2915 (1990).

[11] E. P. Petrov, V. N. Bogomolov, I. I. Kalosha, and S. V. Gaponenko, Phys. Rev. Lett. **81**, 77 (1998).

[12] S. V. Gaponenko, Phys. Rev. B **65**, 140303 (2002).

[13] K. C. Ho, P. T. Leung, Alec Maassen van den Brink and K. Young, Phys. Rev. E **58**, 2965 (1998).

[14] P. T. Leung, S. Y. Liu, and K. Young, Phys. Rev. A **49**, 3057 (1994); P. T. Leung, S. S. Tong, and K. Young, J. Phys. A **30**, 2139 (1997); P. T. Leung, S. S. Tong, and K. Young, J. Phys. A **30**, 2153 (1997); E. S. C. Ching, P. T. Leung, A. Maassen van der Brink, W. M. Suen, S. S. Tong, and K. Young, Rev. Mod. Phys. **70**, 1545 (1998).

[15] S. Severini, A. Settimi, C. Sibilia, M. Bertolotti, A. Napoli, A. Messina, Phys. Rev. E **70**, 056614 (2004).

[16] A. Settimi, S. Severini, N. Mattiucci, C. Sibilia, M. Centini, G. D'Aguanno, M. Bertolotti, M. Scalora, M. Bloemer, C. M. Bowden, Phys. Rev. E **68**, 026614 (2003).

[17] J. P. Dowling and C. M. Bowden, Phys. Rev. A **46**, 612 (1992).

[18] J. P. Dowling, J. of Lightwave Technol. **17**, 2142 (1999).

[19] M. Scalora, J. P. Dowling, M. Tocci, M. J. Bloemer, C. M. Bowden, J. W. Haus, Appl. Phys. B: Lasers Opt. **60**, S57 (1995).

[20] M. Centini, G. D'Aguanno, M. Scalora, M. J. Bloemer, C. M. Bowden, C. Sibilia, N. Mattiucci, and M. Bertolotti, Phys. Rev. E **67**, 036617 (2003).

[21] A. Settimi, S. Severini, C. Sibilia, M. Bertolotti, M. Centini, A. Napoli, N. Messina, Phys. Rev. E **71,** 066606 (2005).

[22] E. P. Wigner, Phy. Rev. **98**, 145 (1955).

[23] G. D'Aguanno, N. Mattiucci, M. Scalora, M. J. Bloemer and A. M. Zheltikov, Phys. Rev. E **70**, 016612 (2004).

[24] G. D'Aguanno, N. Mattiucci, M. Centini, M. Scalora and M. J. Bloemer, Phys. Rev. E **69**, 057601 (2004).

[25] J. D. Jackson, *Classical Electrodynamics* (John Wiley and Sons, New York, 1975).

[26] G. F. Carrier, M. Krook, C. E. Pearson, *Functions of a complex variable – theory and technique* (McGraw-Hill Book Company, New York, 1983).

[27] R. Feynman, *Quantum Electrodynamics* (Benjamin, New York, 1962).

[28] A. A. Abrikosov, L. P. Gor'kov and I. E. Dzyaloshinski, *Methods of Quantum Field Theory in Statistical Physics* (Dover, New York, 1975); P. T. Leung, A. Maassen van den Brink and K. Young, *Frontiers in Quantum Physics*, Proceedings of the International Conference, edited by S. C. Lim, R. Abd-Shukor and K. H. Kwek (Springer-Verlag, Singapore, 1998).

[29] A. E. Siegman, Phys. Rev. A **39**, 1253 (1989); **39**, 1264 (1989).

[30] A. G. Fox and T. Li, Bell Syst. Tech. J. **40**, 453 (1961).

**Figures and captions.**

Figure 6.1.

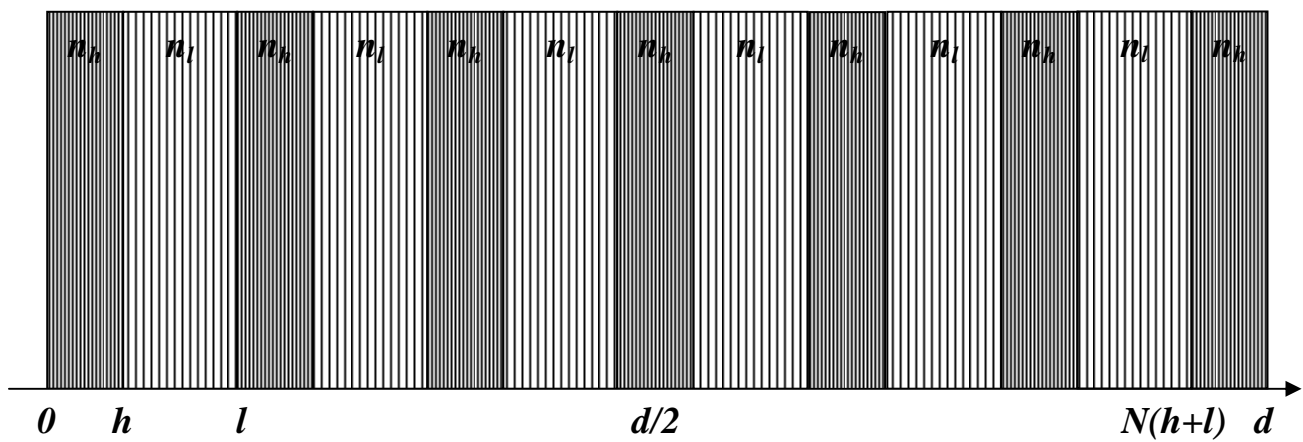

Figure 6.2.a.

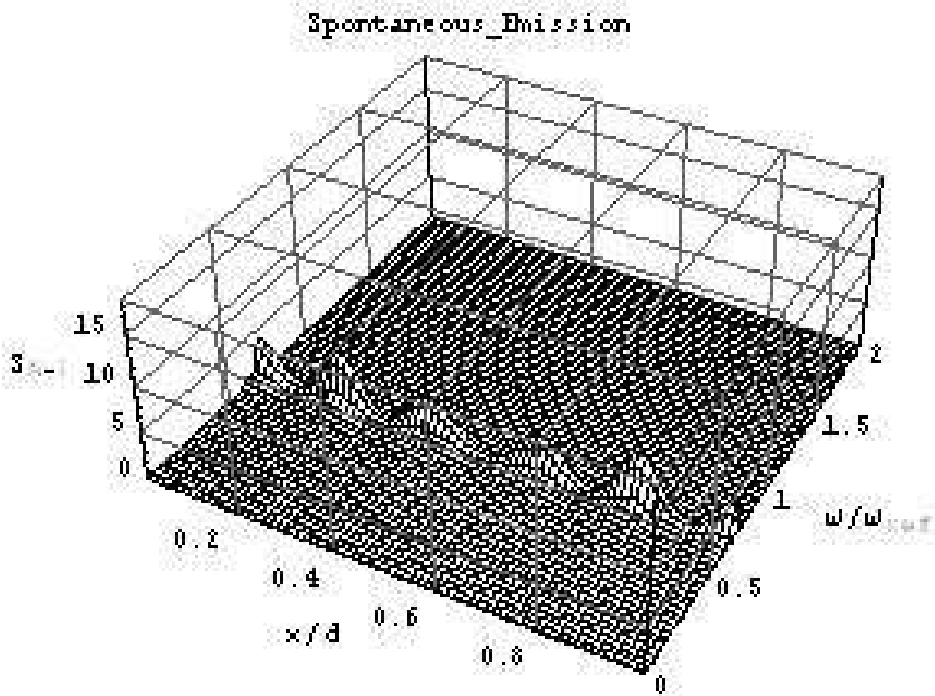

Figure 6.2.b.

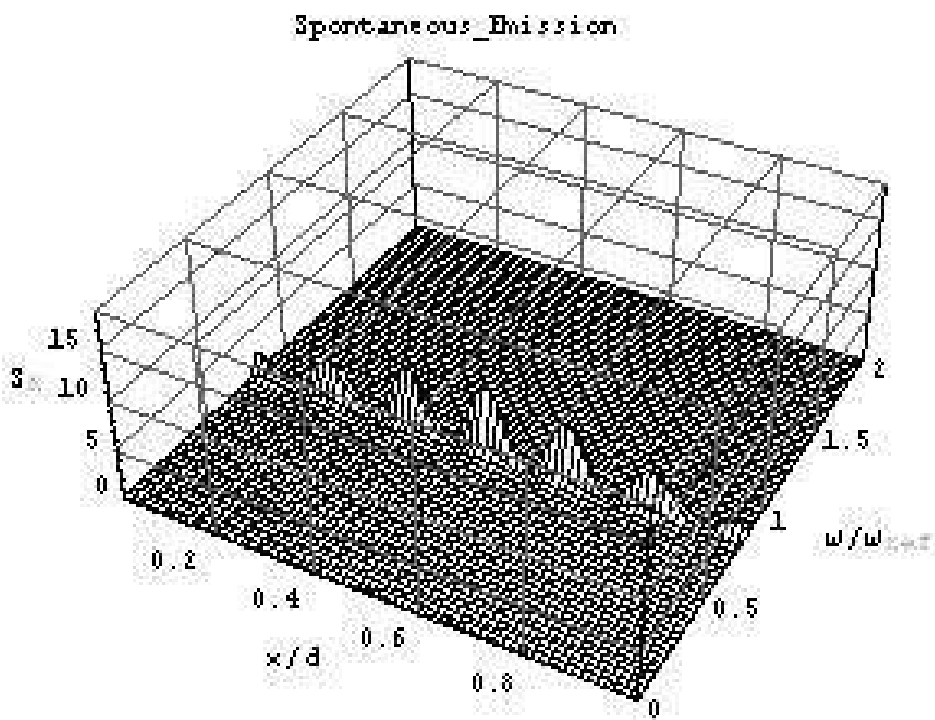

Figure 6.2.c.

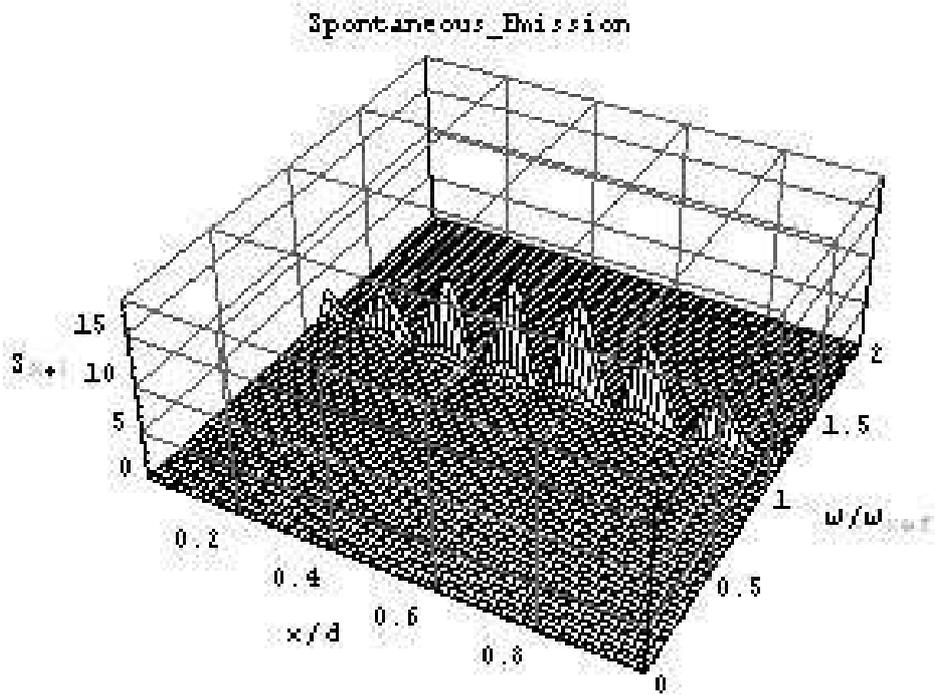

Figure 6.2.d.

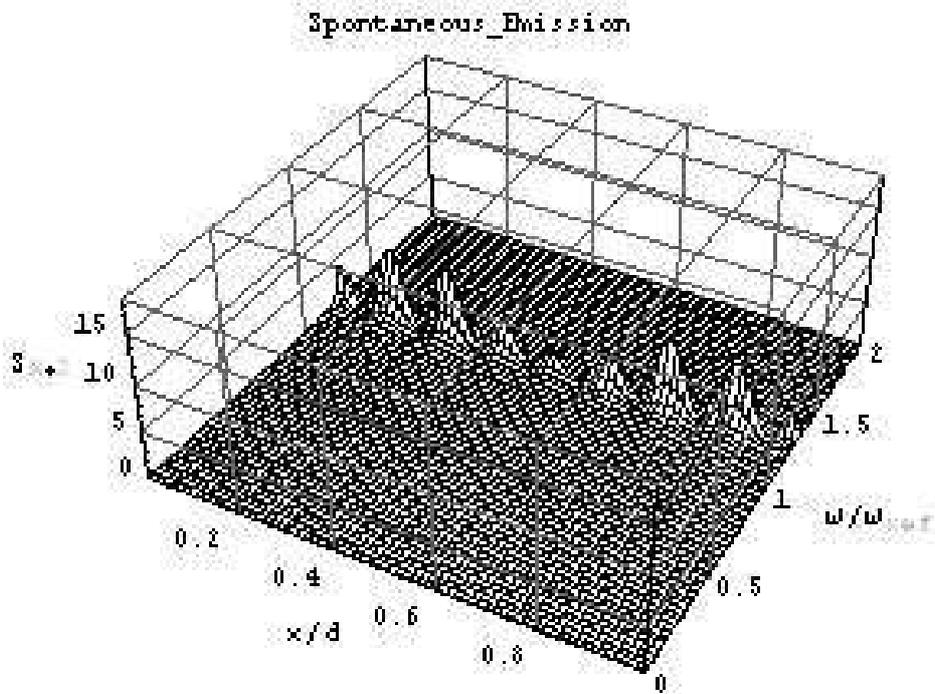

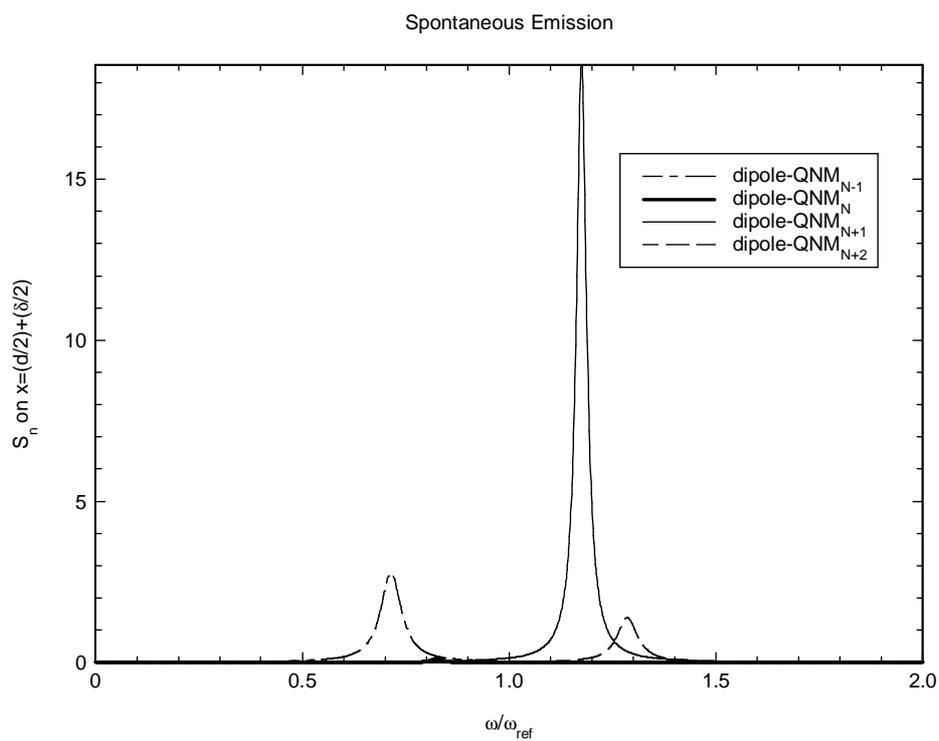

Figure 6.3.a.

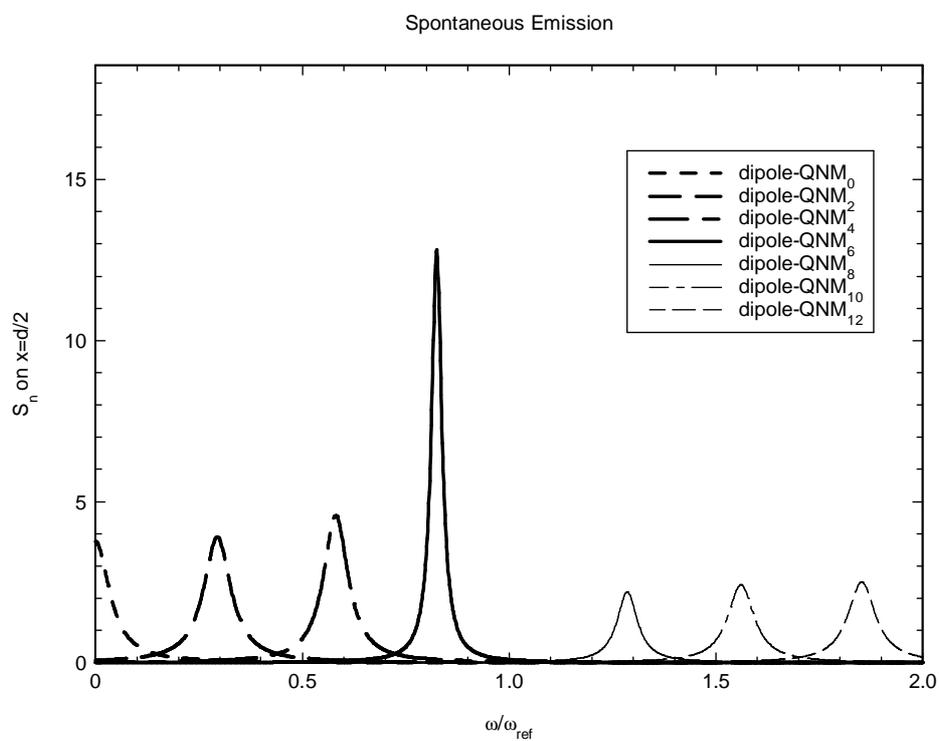

Figure 6.3.b.

Figure 6.3.c.

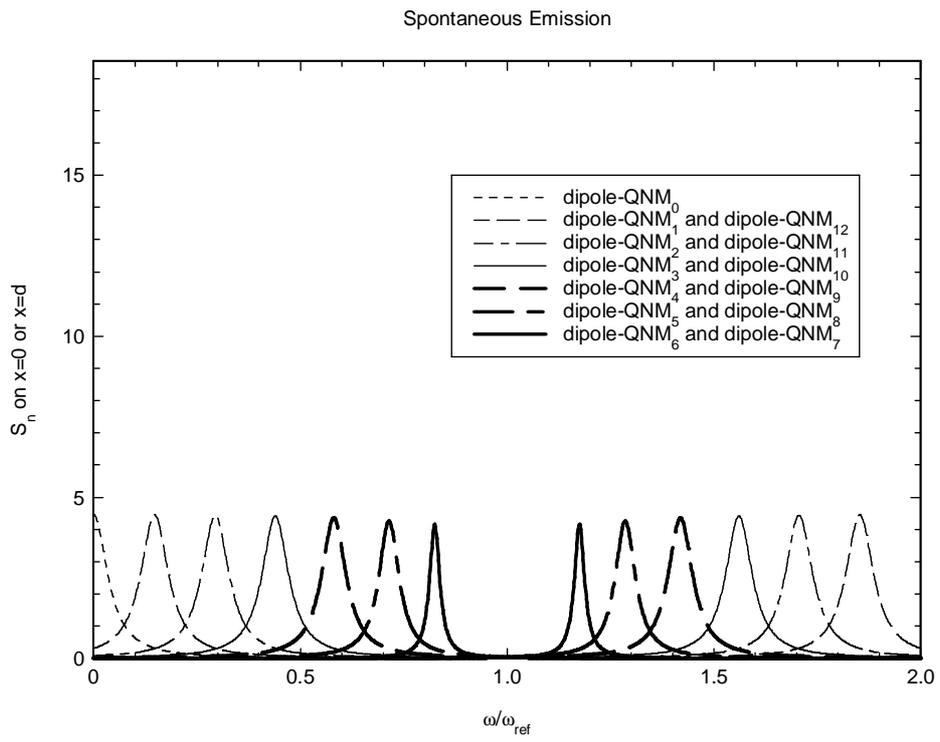

Figure 6.4.a.

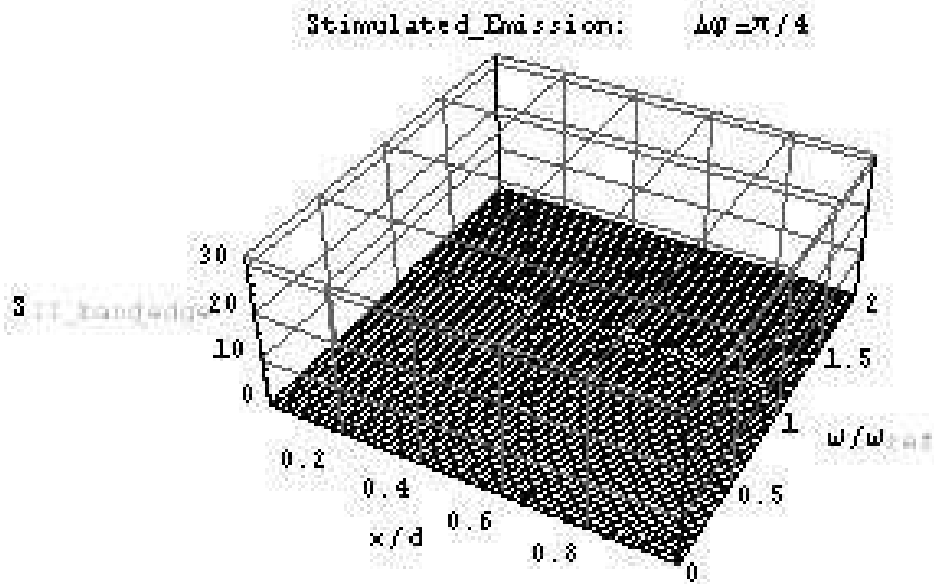

Figure 6.4.b.

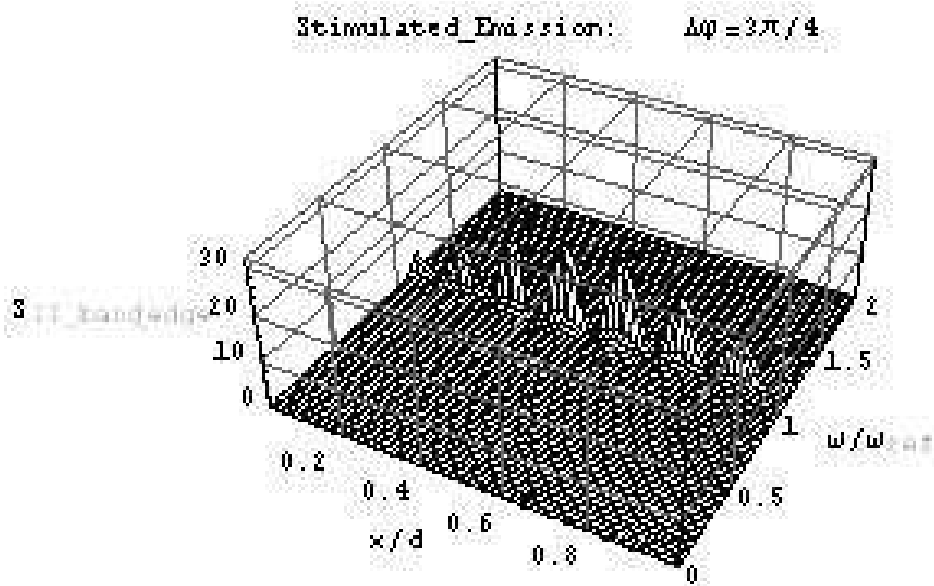

Figure 6.5.a.

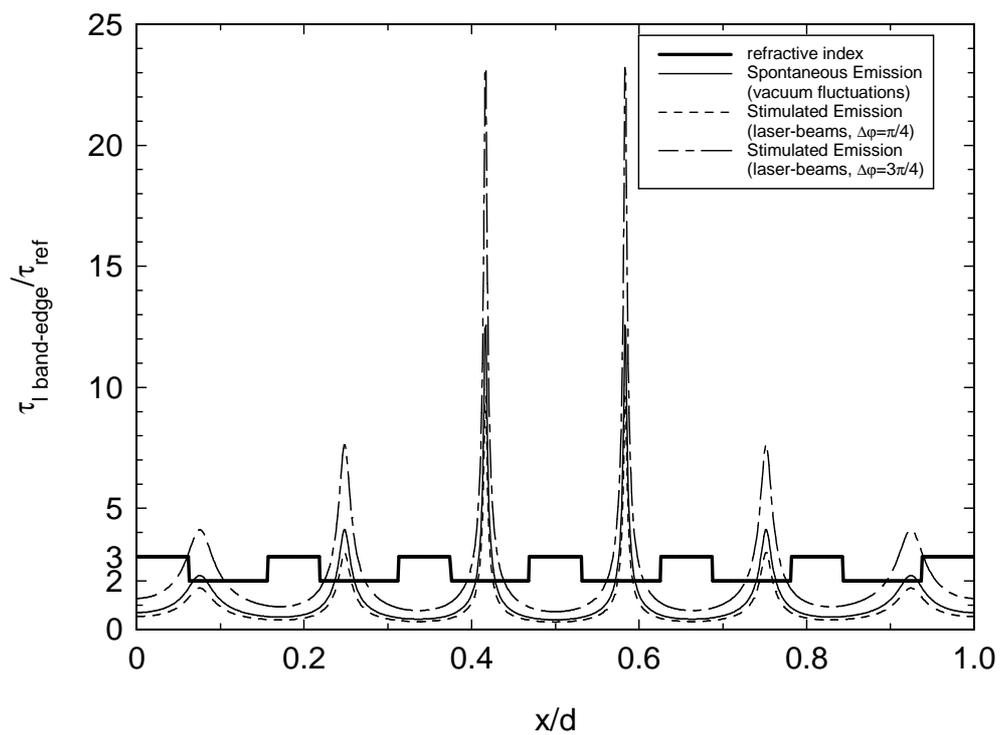

Figure 6.5.a.bis.

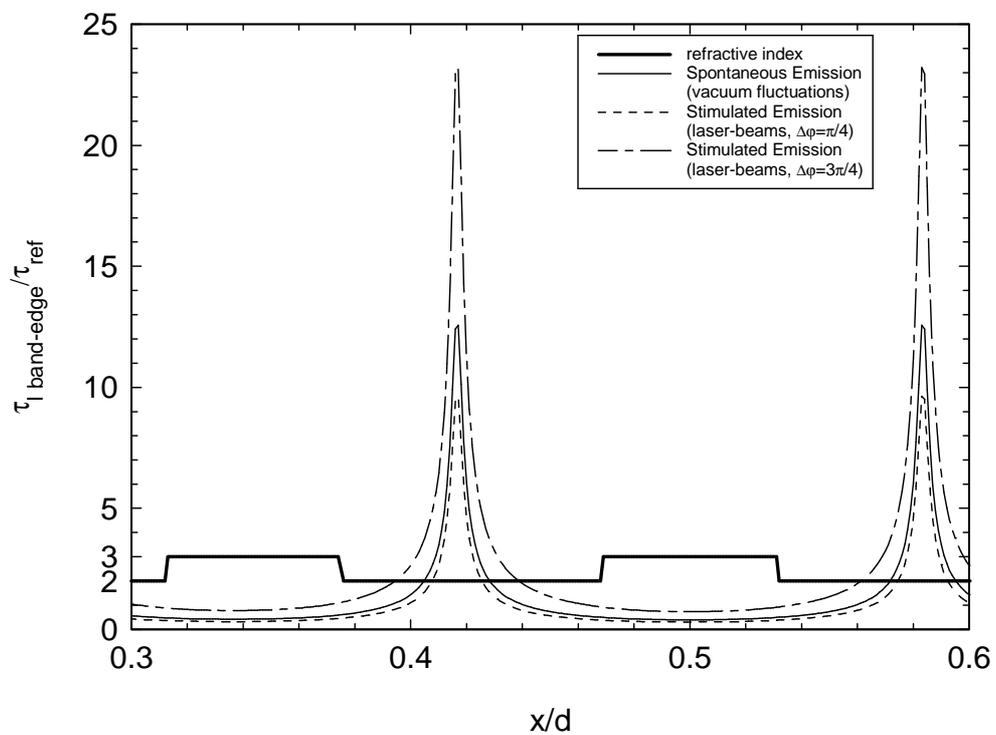

Figure 6.5.b.

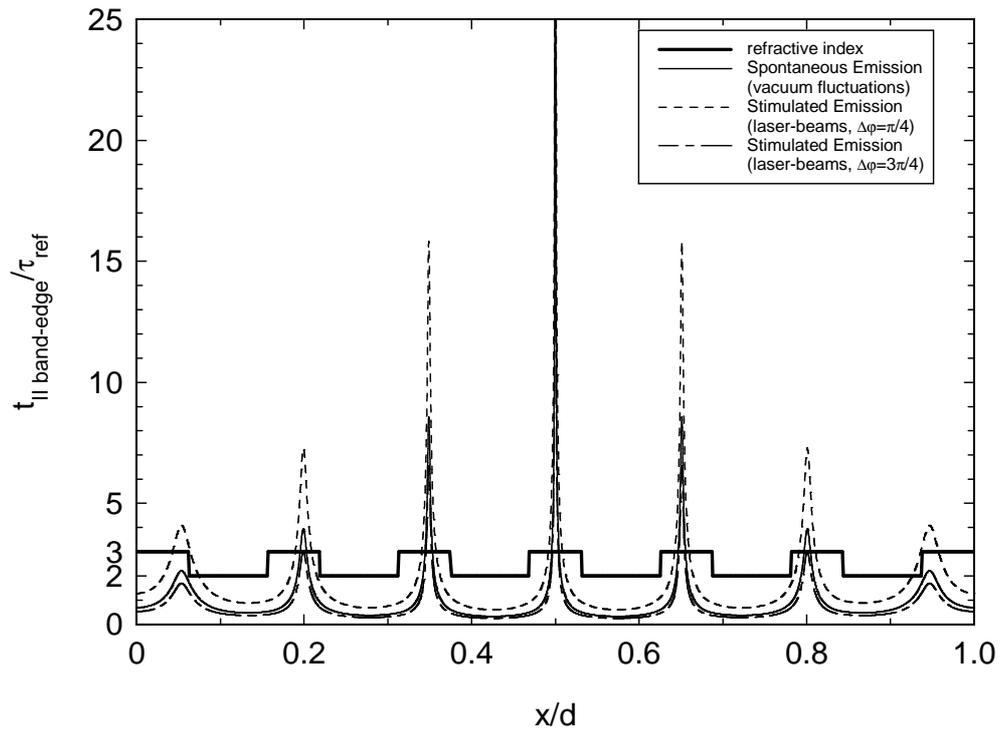

Figure 6.5.b.bis.

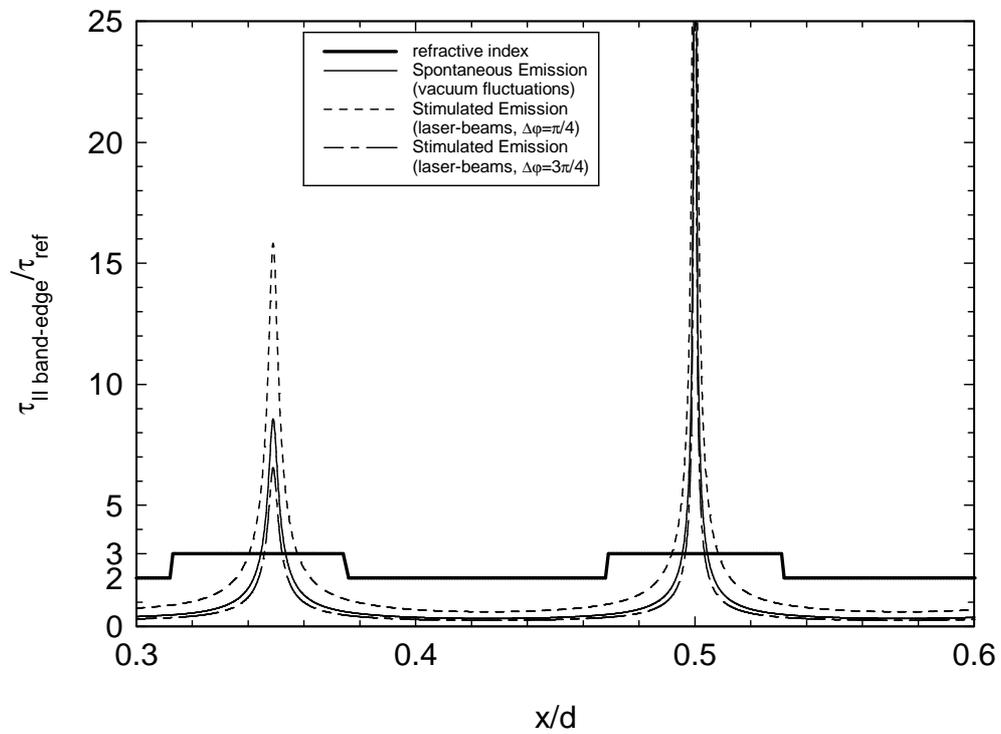

Figure 6.6.a.

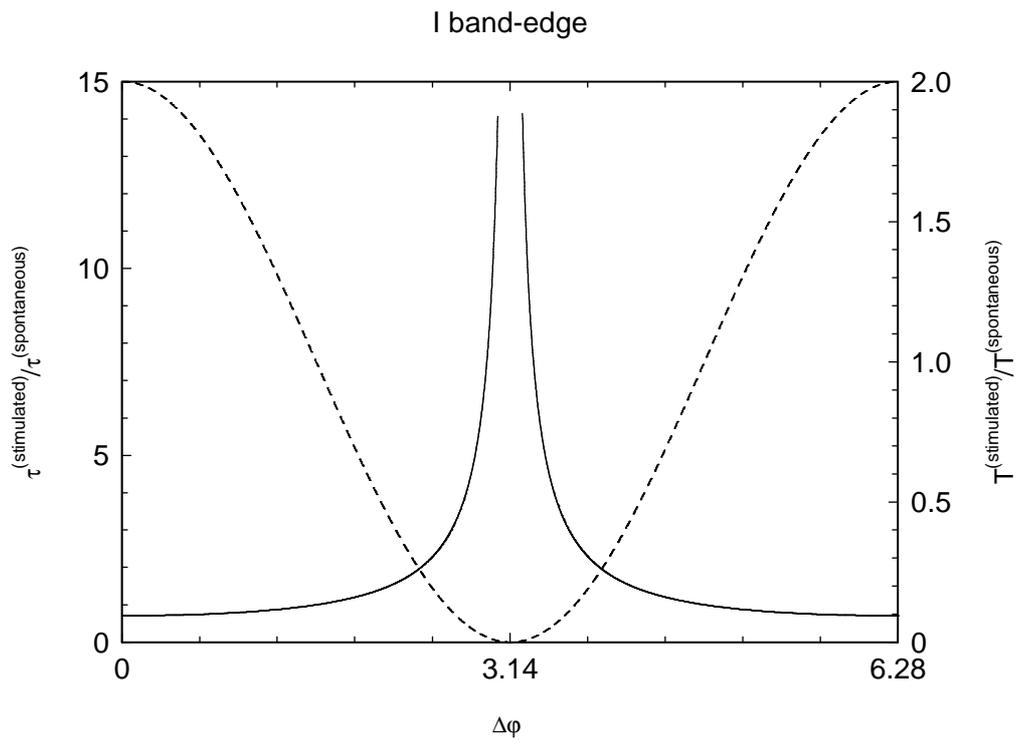

Figure 6.6.b.

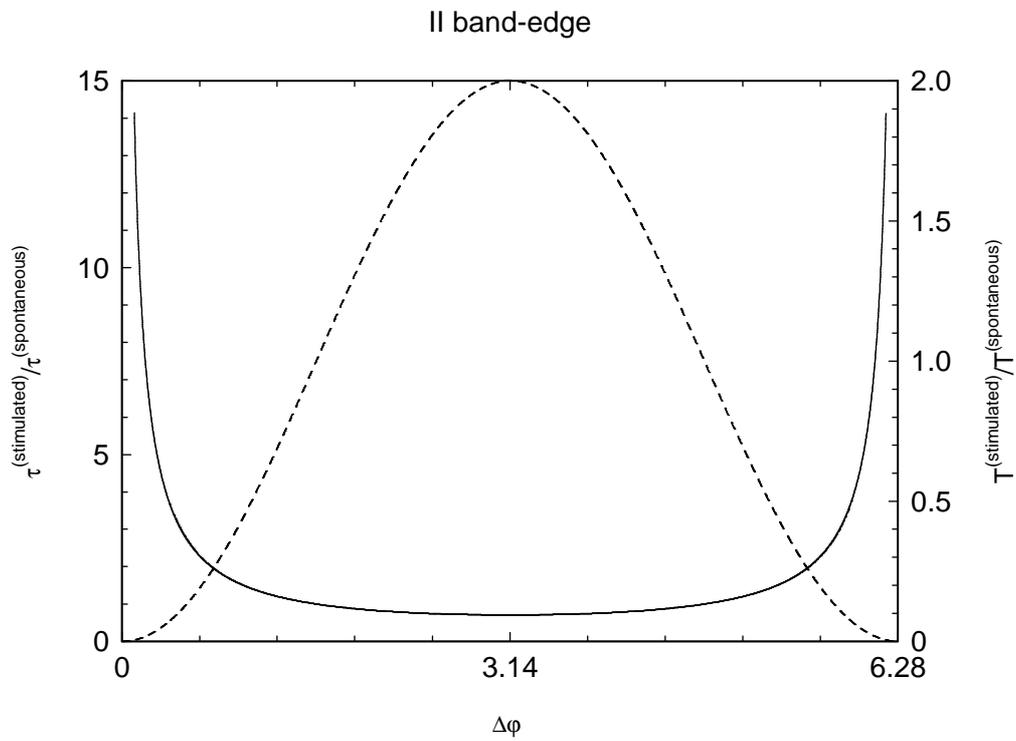

Figure 6.1. Symmetric quarter-wave (QW) one dimensional (1D) Photonic Band Gap (PBG) structure with $\lambda_{ref}=1\mu m$ as reference wavelength, $N=6$ periods, consisting of two layers with refractive indices $n_h=3$ and $n_l=2$ and lengths $h=\lambda_{ref}/4n_h$ and $l=\lambda_{ref}/4n_l$. Terminal layers of the symmetric QW 1D-PBG structure with parameters: $n_h$ and $h=\lambda_{ref}/4n_h$. Length of the 1D-PBG structure: $d=N(h+l)+h$. In the conservative limit, a dipole can be coupled to just one of the $2N+1$ Quasi Normal Modes (QNMs) in the $[0, 2\omega_{ref})$ range, which can be enumerated, excluding $\omega=2\omega_{ref}$, as $|n\rangle=|n,0\rangle$, $n\in[0,2N]$.

Figure 6.2. With reference to a dipole embedded inside the symmetric QW 1D-PBG structure of fig. 6.1, spontaneous emission processes are described in terms of the sensitivity functions $S_n^{(A)}$ [eq. (6.3.1)], plotted as functions of the dimensionless space $x/d$ and the dimensionless frequency $\omega/\omega_{ref}$. In (a) is depicted the sensitivity function $S_{N-1}^{(A)}$, for the coupling dipole-$|N-1\rangle$ (where the real part of the QNM's frequency is close to the first transmission peak before the low-frequency band-edge); in (b), the sensitivity function $S_N^{(A)}$, for the coupling dipole-$|N\rangle$ (corresponding to the low-frequency band-edge, being $\omega_{I\,band\text{-}edge}/\omega_{ref} = 0.8249$); in (c) is depicted the sensitivity function $S_{N+1}^{(A)}$, for the coupling dipole-$|N+1\rangle$ (where the real part of the QNM's frequency is close to the high-frequency band-edge, being $\omega_{II\,band\text{-}edge}/\omega_{ref} = 1.175$) and, in (d), the sensitivity function $S_{N+2}^{(A)}$, for the coupling dipole-$|N+2\rangle$ (corresponding to the first transmission peak after the high-frequency band-edge).

Figure 6.3. With reference to the dipole embedded inside the symmetric QW 1D-PBG structure of fig. 6.1, in order to describe for the best the spontaneous emission processes of fig. 6.2, the sensitivity functions $S_N^{(A)}$ and $S_{N+1}^{(A)}$, with frequencies close to the two band-edges, and $S_{N-1}^{(A)}$, $S_{N+2}^{(A)}$, corresponding to the two next transmission peaks, are plotted, as functions of the dimensionless frequency $\omega/\omega_{ref}$, when: (a) the dipole is inside the $4^{th}$ period of the 1D-PBG structure, and it is exactly in the centre $x=(d/2)+(\delta/2)$ of the low-index layer, being $\delta=h+l$. The sensitivity functions $S_n^{(A)}$, $n\in[0,2N]$ are depicted, when: (b) the dipole is inside the $4^{th}$ period of the 1D-PBG structure, and it is exactly in the centre $x=d/2$ of the high-index layer, being also the centre of the 1D-PBG; (c) the dipole lies on one of the two terminal surfaces $x=0$ and $x=d$ of the 1D-PBG.

Figure 6.4. With reference to a dipole embedded inside the symmetric QW 1D-PBG structure of fig. 6.1, stimulated emission processes can be described in terms of the sensitivity functions $S_n^{(B)}$ [eq. (6.4.1)], plotted as functions of the dimensionless space $x/d$ and of the dimensionless frequency $\omega/\omega_{ref}$. The 1D-PBG structure is excited by two counter-propagating laser-beams, mismatched by a phase-difference $\Delta\varphi$. The sensitivity function $S_{N+1}^{(A)}$ for the coupling dipole-$|N+1\rangle$ (corresponding to the high-frequency band-edge) is depicted, when the two laser beams are nearly in phase, (a) $\Delta\varphi=\pi/4$, or nearly opposite in phase, (b) $\Delta\varphi=3\pi/4$.

Figure 6.5. An excited dipole is embedded inside a symmetric QW 1D-PBG structure with parameters $\lambda_{ref}=1\mu m$, $N=6$, $n_h=3$, $n_l=2$, pumped by two counter-propagating field pumps filtered at low-frequency band-edge (fig. 6.5.a.) or at high-frequency band-edge (fig. 6.5.b.). The decay-time $\tau$, in units of the decay-time $\tau_{ref}$ for spontaneous emission when the dipole lies on one terminal surface ($x=0$) of the 1D-PBG structure [eq.(6.5.2)], is plotted as a function of the dimensionless position $x/d$ of the dipole. Several cases are shown: the spontaneous emission [see eq.(6.5.4)], when the two field pumps describe vacuum fluctuations (———); and the stimulated emission [see eq.(6.5.7)], when the two pumps describe two input laser-beams, nearly in phase (− − −) or opposite

of phase (— – —). Fig. 6.5.a.bis. (fig. 6.5.b.bis.) is a magnification of fig. 6.5.a. (fig. 6.5.b.) for the decay-time in the $3^{th}$ and $4^{th}$ periods of the 1D-PBG.

Figure 6.6. The excited dipole, embedded inside the symmetric QW 1D-PBG of fig. 6.1, is pumped by two counter-propagating laser-beams filtered at low-frequency band-edge (fig. 6.6.a.) or at the high-frequency band-edge (fig. 6.6.b.). In order to describe for the best the stimulated emission processes of fig. 6.5, the decay-time for stimulated emission (———), in units of the decay-time for spontaneous emission [eq. (6.5.7)], is compared with the dwell-time (– – –) for the two laser-beams, in units of the dwell-time for vacuum fluctuations [eq. (6.5.8)]. The decay-time for stimulated emission and the dwell-time for the two laser-beams are compared on different scales as functions of the phase-difference between the two laser-beams.

# Chapter 7

# Coherent Control of Stimulated Emission inside 1D-PBG structures. (Strong Coupling regime)

## 1. Introduction.

It has been demonstrated theoretically [2] and experimentally [2] that emission processes, which had been believed to be an inherent nature of an atom, can be modified in a cavity whose size is comparable with the wavelength of the emitted light.

Abrupt changes in the density of states (DOS) and photon localization effects may drastically modify the emission dynamics [3]. This modification takes the form of long time memory effects and non-Markovian behaviour in the atom-reservoir interaction. Strong modification in the DOS can be effected by means of photonic crystals [4][5]. It has been suggested that this would be accompanied by the inhibition of single-photon emission, classical light localization, a photon-atom bound state, fractionalized single-atom inversion, and anomalously large vacuum Rabi splitting [6].

Lai et al [7] have rigorously investigated the spontaneous electromagnetic decay of a two-level atom coupled to a narrow cavity resonance in terms of the (Hermitian) modes of the "universe" [1] rather then the (dissipative) Quasi Normal Modes (QNM) of the open cavity [1][2]. Special attention is paid to the strong-coupling regime (atomic line-width $\Gamma$ nearly equal to cavity resonance width $\gamma$), in which there are significant corrections to the golden rule. In particular, spontaneous decay is most rapid for intermediate values of the quality factor $Q$ of the cavity resonance.

In the present chapter, the emission processes in a 1D open cavity are analysed on the basis of the theory presented in ref. [7]. In particular, the stimulated emission is discussed, in strong coupling regime, for an atom, embedded inside a 1D-PBG structure, which is pumped by two counter-propagating laser beams. As in ref. [7], quantum electro-dynamics [2] is applied to model the atom-e.m. field coupling, by considering the atom a two level system, the e.m. field a superposition of normal modes, the coupling in dipole approximation, and the equations of motion in Wigner-Weisskopf's and rotating wave approximations. With respect to ref. [7], the QNM'a approach [7]- [10] is adopted for an open cavity, interpreting the local density of states (LDOS) as the local density of probability to excite one QNM of the cavity; so this LDOS depends on the phase difference of the two laser beams.

As main results of this chapter, the strong coupling regime is defined as the one associated to high values of the LDOS; according to the results of the literature, the emission probability of the atom decays with an oscillatory behaviour, so that the atomic emission spectrum is splitted into two peaks (Rabi splitting); the novelty with respect to previous discussions available in literature is that the phase difference of the two laser beams allow to produce a coherent control of both the oscillations for the atomic emission probability and the Rabi's splitting for the emission spectrum.

The chapter is organized as follows. In section 2, the coupling of an atom to an e.m. field is modelled via quantum electrodynamics equations; it is retained useful to recall the example of an atom in the free space. In section 3, the atom is embedded inside a leaky cavity and the emission processes of the atom are modelled. In section 4, the emission probability of the atom is discussed in strong coupling regime. In section 5, the emission spectrum of the atom is discussed, also in terms of its poles. In section 6, some criteria are proposed to design an active delay line. In section 7, the conclusions are reported.

## 2. Coupling of an atom to an e.m. field.

Let consider, inside a one dimensional (1D) universe $U = \{x | x \in [-L/2, L/2] \, , \, L \to \infty\}$ of refractive index $n_0$, an atom, coupled to an e.m. field in the point $x_0 \in U$, and an open cavity $C = \{x | x \in (0, d) \, , \, d < L\}$, with an inhomogeneous refractive index $n(x)$.

The atom is quantized as a two level system, oscillating at resonance $\Omega$ [1], and the e.m. field is in terms of the 1D universe modes,

$$\begin{cases} g_\lambda(x) = \dfrac{1}{\sqrt{L}} \exp\left(i\omega_\lambda \sqrt{\rho_0} x\right) \\ \omega_\lambda = \lambda \dfrac{\pi}{(L/2)\sqrt{\rho_0}} \quad , \quad \lambda \in \mathbb{Z} \end{cases}, \qquad (7.2.1)$$

where $\rho_0 = (n_0/c)^2$, being $c$ the speed of the light in vacuum. The atom is modelled by a not permanent dipole operator $\mu$ [1] along the polarization direction of the e.m. field, and the coupling of the atom to the e.m. field by the electric dipole approximation [2].

At the initial time ($t=0$), the atom is prepared in the excited state $|+\rangle$ and the e.m. field in the vacuum state $|\{0\}\rangle = \prod_{\lambda=-\infty}^{\infty}|0_\lambda\rangle$. With the above initial conditions, the system evolution may be described using the basis states listed below, with their corresponding eigen values [7]:

$$\begin{aligned}|+,\{0\}\rangle = |+\rangle|\{0\}\rangle &, \quad \varepsilon_+ \\ |-,1_\lambda\rangle = |-\rangle|1_\lambda\rangle &, \quad \varepsilon_- + \hbar\omega_\lambda, \quad \lambda \in \mathbb{Z}\end{aligned} \quad (7.2.2)$$

where $|+,\{0\}\rangle$ denotes the state for which the atom is in the upper state, and no photon exists in any e.m. mode; and $|-,1_\lambda\rangle$ denotes the state for which the atom is in the lower state, one photon exists in the $\lambda^{th}$ e.m. mode, while no photon exists in all the other e.m. modes.

## 2. 1. Quantum electrodynamics equations.

After a time $t>0$, the system of the atom coupled to the e.m. field lies on a quantum state $|\psi(t)\rangle$ which can be calculated on the basis states (7.2.2), applying the quantum principle of superposition [1], i.e.

$$|\psi(t)\rangle = c_+(t)|+,\{0\}\rangle + \sum_{\lambda=-\infty}^{\infty} c_{-,\lambda}(t)|-,1_\lambda\rangle, \quad (7.2.3)$$

where are introduced the probability coefficients $c_+(t)$ and $c_{-,\lambda}(t)$ such that $c_+(0)=1$ and $c_{-,\lambda}(0)=0$. The rotating wave approximation [2] is assumed. The time evolution equations for the probability coefficients $c_+(t)$ and $c_{-,\lambda}(t)$ [7],

$$\begin{aligned}\frac{dc_+}{dt} &= \frac{\langle +|\mu|-\rangle}{\hbar} \sum_{\lambda=1}^{\infty} \sqrt{\frac{\hbar\omega_\lambda}{2\varepsilon_0 n_0^2}} g_\lambda(x_0) c_{-,\lambda}(t) \exp\left[-i(\omega_\lambda - \Omega)t\right] \\ \frac{dc_{-,\lambda}}{dt} &= -\frac{\langle -|\mu|+\rangle}{\hbar} \sqrt{\frac{\hbar\omega_\lambda}{2\varepsilon_0 n_0^2}} g_\lambda^*(x_0) c_+(t) \exp\left[i(\omega_\lambda - \Omega)t\right]\end{aligned}, \quad (7.2.4)$$

can be solved, giving the time evolution equation for the probability coefficient $c_+(t)$,

$$\frac{dc_+}{dt} = -\frac{M}{2\varepsilon_0 n_0^2 \hbar^2} \frac{1}{L} \sum_{\lambda=1}^{\infty} \hbar\omega_\lambda \int_0^t c_+(\tau) \exp\left[i(\omega_\lambda - \Omega)(\tau - t)\right] d\tau. \quad (7.2.5)$$

where $\varepsilon_0$ is the dielectric constant in vacuum, $\hbar = h/2\pi$, being $h$ the Planck's constant, and $M = |\langle +|\mu|-\rangle|^2$.

A correspondence can be established between the discrete modes of a 1D cavity with length $L$ and the continuous modes of a boundless universe with infinite length $L\to\infty$. The mode spectrum becomes continuous as $L\to\infty$, because of $\Delta\omega_\lambda = \pi/(L/2)\sqrt{\rho_0} \approx d\omega \to 0$, and, in this limit, sums over discrete quantities are converted to integrals over continuous frequency, i.e.

$$\frac{1}{L}\sum_{\lambda=1}^{\infty} \Rightarrow \int_0^{\infty} d\omega \sigma^{(loc)}(x_0,\omega), \qquad (7.2.6)$$

where $\sigma^{(loc)}(x,\omega)$ is the local density of states (LDOS) which can be interpreted as the density of probability that the e.m. field, in the point $x$, is excited on just one eigen-state, oscillating around the frequency $\omega$[20]. The range of integration over $\omega$ in the expression (7.2.6) strictly extends only from $0$ to $\infty$, since the physical frequencies are defined to be positive. However, the range can be extended from $-\infty$ to $\infty$ without significant errors, since most optical experiments use a narrow band source $B$ [5], i.e. $B<<\omega_c$ being $\omega_c$ is the central frequency of the bandwidth $B$. So, the time evolution equation (7.2.5) becomes

$$\frac{dc_+}{dt} = \int_0^t d\tau \cdot S^{(kernel)}(x_0,t-\tau)c_+(\tau), \qquad (7.2.7)$$

being $S^{(kernel)}(x, t)$ the kernel function, defined as:

$$S^{(kernel)}(x,t) = -\frac{M}{2\varepsilon_0 n_0^2 \hbar^2} \int_{-\infty}^{\infty} d\omega \hbar \omega \sigma^{(loc)}(x,\omega)\exp\left[-i(\omega-\Omega)t\right]. \qquad (7.2.8)$$

As from (7.2.8), the kernel function depends very strongly on the LDOS $\sigma^{(loc)}(x,\omega)$ which can be re-interpreted as the photon density of states in the reservoir; in essence, eq. (7.2.8) is a measure of the photon reservoir's memory of its previous state, on a time scale of evolution of the atomic system, so $S^{(kernel)}(x, t)$ can be interpreted as memory kernel of the reservoir.

**2. 2. Atom in the free space.**

If the atom is localized in a point $x_0$ which is outside the open cavity, i.e. $x_0<0$ and $x_0>d$, then the LDOS $\sigma^{(loc)}(x,\omega)$ is referred to the free space [2], [10] [see eq. (5.2.4)]. In the free space, the emission probability of the atom decays exponentially,

$$|c_+(t)|^2 = \exp(-\Gamma_0 t) \quad , \quad t \geq 0, \qquad (7.2.9)$$

where $\Gamma_0$ is the atomic decay rate:

$$\Gamma_0 = \frac{M}{\hbar}\frac{\sqrt{\rho_0}}{\varepsilon_0 n_0^2}\Omega \left/ \left[1+\frac{1}{4}\left(\frac{M}{\hbar}\frac{\sqrt{\rho_0}}{\varepsilon_0 n_0^2}\right)^2\right]\right.. \qquad (7.2.10)$$

Since the free space is an infinitely broad photon reservoir (flat spectrum), its response should be instantaneous and the memory effect, associated with emission dynamics, is infinitesimally short compared to all times of interest. Such interactions are said to be Markovian [2]. In other words, in the free space, the population of the excite state eventually decay to the ground level, independent of the strength of a driving field. This is a general result valid for almost any broadband smoothly varying LDOS.

Introducing the atomic parameter

$$R = \frac{\Gamma_0}{\Omega} \cong \frac{M}{\hbar} \frac{\sqrt{\rho_0}}{\varepsilon_0 n_0^2}, \qquad (7.2.11)$$

which will be interpreted as degree of coupling, the kernel function (7.2.8) can be re-expressed as:

$$S^{(kernel)}(x,t) = -\frac{R}{2\sqrt{\rho_0}} \int_{-\infty}^{\infty} d\omega \cdot \omega \sigma^{(loc)}(x,\omega) \exp[-i(\omega-\Omega)t]. \qquad (7.2.12)$$

## 3. Atom inside an open cavity.

Let suppose $0<x_0<d$, so the atom is embedded in a point $x_0$ of the open and inhomogeneous cavity with refractive index $n(x)$. If the atom, with resonance $\Omega$, is coupled to the $n^{th}$ QNM, oscillating at frequency $Re\omega_n$, then the atom - $n^{th}$ QNM coupling is characterized by the frequency detuning:

$$\Delta_n = \frac{Re\omega_n - \Omega}{R}. \qquad (7.3.1)$$

The local density of probability $\sigma_n^{(loc)}(x,\omega)$ for the $n^{th}$ QNM is linked to the integral one $\sigma_n(\omega)$ by eq. (6.2.4) and the emission processes of the atom are characterized by a kernel function $S^{(kernel)}(x,t)$ which can be developed as [eq. (7.2.12)]:

$$S_n^{(kernel)}(x,t) = -\frac{R}{2\sqrt{\rho_0}} \frac{1}{I_n} \rho(x) |F_n(x)|^2 \int_{-\infty}^{\infty} d\omega \cdot \omega \sigma_n(\omega) \exp[-i(\omega-\Omega)t], \forall x | 0<x<d. (7.3.2)$$

where $\rho(x)=[n(x)/c]^2$, $I_n$ denote the usual overlapping integrals, and $F_n(x) = f_n(x)\sqrt{2\omega_n/\langle f_n | f_n \rangle}$ are the normalized QNM's functions, being $\langle f_n | f_n \rangle$ and $\omega_n$ respectively the QNM's norm and frequencies.

### 3. 1. Spontaneous emission: DOS corresponding to vacuum fluctuations.

If the open cavity is pumped by vacuum fluctuations, which are filtered at the atomic resonance $\Omega$, next to $n^{th}$ QNM's frequency of the cavity ($\Omega \approx Re\omega_n$), then the integral density of probability $\sigma_n^{(A)}(\omega)$ for the $n^{th}$ QNM is expressed by eqs. (6.2.5)-(6.2.6) and the spontaneous emission of the atom is characterized by a time evolution equation [eq.(7.2.7)] where the kernel function $S_n^{(kernel)}(x,t)$ can be specified as [eq. (7.3.2)]

$$S_{n,A}^{(kernel)}(x,t) = -\frac{R}{2\sqrt{\rho_0}} \frac{1}{I_n} \rho(x)|F_n(x)|^2 K_n \frac{d}{2\pi} I_n^2 \exp(i\Omega t) \cdot$$
$$\cdot i\sqrt{2\pi} \left[ \frac{1}{\sqrt{2\pi}} \int_{-\infty}^{\infty} (-i\omega) \frac{|\text{Im}\,\omega_n|}{(\omega - \text{Re}\,\omega_n)^2 + \text{Im}^2\,\omega_n} \exp(-i\omega t) d\omega \right]. \qquad (7.3.3)$$

where $K_n$ is the constant of normalization for $\sigma_n^{(A)}(\omega)$, given by eq. (6.2.6). It is easy to transform the following signal $x(t)$ in the Fourier domain [14],

$$x(t) = \frac{1}{2} \frac{d}{dt}\left[ e^{-(|\text{Im}\,\omega_n|+i\text{Re}\,\omega_n)|t|} \right]$$
$$\Rightarrow X(\omega) = \frac{1}{\sqrt{2\pi}} \int_{-\infty}^{\infty} x(t)\exp(i\omega t)dt = -\frac{i\omega}{\sqrt{2\pi}} \frac{|\text{Im}\,\omega_n|}{\text{Im}^2\,\omega_n + (\omega - \text{Re}\,\omega_n)^2}. \qquad (7.3.4)$$

Inserting eq.(7.3.4) in eq. (7.3.3), it results the kernel function of the spontaneous emission:

$$S_{n,A}^{(kernel)}(x,t) = -\frac{1}{4} \frac{R}{\sqrt{\rho_0}} K_n (d \cdot I_n) \omega_n \rho(x)|F_n(x)|^2 \exp\left[-i(\omega_n - \Omega)t\right] \quad, \quad \forall t \geq 0. \qquad (7.3.5)$$

First applying the "theorem of derivation under the integral sign" [14] to eq. (7.2.7),

$$\frac{d^2 c_+}{dt^2} = S_{n,A}^{(kernel)}(x_0, t=0) c_+(t) + \int_0^t d\tau \frac{\partial S_{n,A}^{(kernel)}}{\partial t} c_+(\tau), \qquad (7.3.6)$$

and then deriving in time eq. (7.3.5), sampled in the $x_0$ point,

$$\frac{\partial S_{n,A}^{(kernel)}}{\partial t} = -i(\omega_n - \Omega) S_{n,A}^{(kernel)}(x_0, t) \quad, \quad t \geq 0, \qquad (7.3.7)$$

it results, after some algebra, a second order differential equation in time for the probability of spontaneous emission:

$$\frac{d^2 c_+}{dt^2} + i(\omega_n - \Omega)\frac{dc_+}{dt} - S_{n,A}^{(kernel)}(x_0, t=0) c_+(t) = 0$$
$$c_+(0) = 1 \quad, \quad \left.\frac{dc_+}{dt}\right|_{t=0} \propto c_{-,n}(0) = 0 \qquad (7.3.8)$$

### 3. 2. Stimulated emission:
**DOS depending on the phase difference of two counter-propagating laser-beams.**

If the open cavity is pumped by two counter propagating laser beams with a phase difference $\Delta\varphi$, which are tuned at the atomic resonance $\Omega$, next to the $n^{th}$ QNM's frequency of the cavity ($\Omega \approx Re\omega_n$), then the density of probability $\sigma_n^{(B)}(\omega)$ for the $n^{th}$ QNM, corresponding to the two laser beams, is linked to the one $\sigma_n^{(A)}(\omega)$, due to vacuum fluctuations, by eq (6.2.7), and the stimulated emission of the atom is characterized by a kernel function $S_n^{(kernel)}(x,t)$, which can be specified as [eq. (7.3.3)]

$$S_{n,B}^{(kernel)}(x,t) = S_{n,A}^{(kernel)}(x,t)\left[1+(-1)^n \cos\Delta\varphi\right]. \qquad (7.3.9)$$

In terms of the frequency detuning for the atom - $n^{th}$ QNM coupling (7.3.1), the quantity $(\omega_n-\Omega)$ can be re-expressed as $(\omega_n-\Omega)=(Re\omega_n-\Omega)+iIm\omega_n=R\Delta_n+iIm\omega_n$ and the second order differential equation in time for the probability of emission processes as

$$\frac{d^2 c_+}{dt^2} + i(R\Delta_n + i\operatorname{Im}\omega_n)\frac{dc_+}{dt} - S_n^{(kernel)}(x_0,t=0)c_+(t) = 0. \qquad (7.3.10)$$

## 4. Emission probability of the atom.

The algebraic equation associated to the Cauchy problem (7.3.10), with initial conditions as in eq. (7.3.8), i.e.

$$p^2 + (R\Delta_n + i\operatorname{Im}\omega_n)p + S_n^{(kernel)}(x_0,t=0) = 0, \qquad (7.4.1)$$

is solved by two roots,

$$p_{1,2} = \frac{(R\Delta_n + i\operatorname{Im}\omega_n)}{2}\left[-1 \pm \sqrt{1 - \frac{4S_n^{(kernel)}(x_0,t=0)}{(R\Delta_n + i\operatorname{Im}\omega_n)^2}}\right], \qquad (7.4.2)$$

which allow to express the particular integral of the differential equation (7.3.10):

$$c_+(t) = \frac{p_2}{p_2 - p_1}\exp(ip_1 t) - \frac{p_1}{p_2 - p_1}\exp(ip_2 t). \qquad (7.4.3)$$

The atom – $n^{th}$ QNM coupling is established in strong coupling regime when the particular integral (7.4.3) presents an oscillatory behaviour, and then the two roots (7.4.2) of the associated algebraic equation (7.4.1) are complex conjugated [7].

### 4. 1. Strong coupling regime.

Let consider the spontaneous emission in order to discuss the atom – $n^{th}$ QNM coupling in strong regime. Since the following condition is valid,

$$4\left|\frac{S_{n,A}^{(kernel)}(x_0,t=0)}{(R\Delta_n + i\operatorname{Im}\omega_n)^2}\right| < \frac{R}{d \cdot I_n}\frac{\sigma_n^{(A)}(x_0,\Omega)}{\sigma_{free\text{-}space}} \cong R\frac{\sigma_n^{(A)}(x_0,\Omega)}{\sigma_{free\text{-}space}}, \qquad (7.4.4)$$

the atom – $n^{th}$ QNM coupling is in strong regime if [7]:

$$4\left|\frac{S_{n,A}^{(kernel)}(x_0,t=0)}{(R\Delta_n+i\,\mathrm{Im}\,\omega_n)^2}\right|>1 \quad \Rightarrow \quad \frac{\sigma_n^{(A)}(x_0,\Omega)}{\sigma_{free\text{-}space}}>\frac{1}{R}\ . \tag{7.4.5}$$

As from eq. (7.4.5), the strong coupling regime is established when the LDOS $\sigma_n^{(A)}(x_0,\Omega)$ inside the open cavity and sampled at the atomic resonance, in units of the LDOS $\sigma_{free\text{-}space}$ referred to the free space, exceeds the inverse of the atomic parameter $R$ [eq. (7.2.11)]; the parameter $R$ can be interpreted as degree of coupling: in fact, the more $R$ is large, the more eq. (7.4.5) is satisfied.

In strong coupling regime (7.4.5), the two roots (7.4.2) become complex conjugated,

$$p_{1,2}\cong -\frac{R\Delta_n+i\,\mathrm{Im}\,\omega_n}{2}\pm i\sqrt{S_{n,A}^{(kernel)}(x_0,t=0)}\,, \tag{7.4.6}$$

and the particular integral (7.4.3) presents an oscillatory behaviour:

$$c_+(t)\cong \exp\left(-i\frac{R\Delta_n+i\,\mathrm{Im}\,\omega_n}{2}t\right)\cdot$$
$$\cdot\left\{\cosh\left[\sqrt{S_{n,A}^{(kernel)}(x_0,t=0)}\,t\right]+i\frac{R\Delta_n+i\,\mathrm{Im}\,\omega_n}{2\sqrt{S_{n,A}^{(kernel)}(x_0,t=0)}}\sinh\left[\sqrt{S_{n,A}^{(kernel)}(x_0,t=0)}\,t\right]\right\} \tag{7.4.7}$$

In fact [eq. (7.3.5)]: $S_{n,A}^{(kernel)}(x_0,t=0)=-s_{n,A}^{(kernel)}(x_0,t=0)$ so $\cosh(i\cdot)=\cos(\cdot)$ and $\sinh(i\cdot)=i\sin(\cdot)$. The oscillatory behaviour may be interpreted as the emission and re-absorption of one photon; the net decay rate is then determined by the rate of photon leakage, i.e. $|Im\omega_n|/2$.

Now, let consider the stimulated emission; again the atom – $n^{th}$ QNM coupling can be discussed in strong regime. If the two counter propagating laser beams have a phase difference $\Delta\varphi$, the atom – $n^{th}$ QNM coupling is characterized by the kernel function (7.3.9). In hypothesis of strong coupling, expressed by a condition similar to eq.(7.4.5), the particular integral (7.4.3) presents an oscillatory behaviour,

$$c_+(t)\cong \exp\left(-i\frac{R\Delta_n+i\,\mathrm{Im}\,\omega_n}{2}t\right)\cdot$$
$$\cdot\left\{\cosh\left[\sqrt{S_{n,B}^{(kernel)}(x_0,t=0)}\,t\right]+i\frac{R\Delta_n+i\,\mathrm{Im}\,\omega_n}{2\sqrt{S_{n,B}^{(kernel)}(x_0,t=0)}}\sinh\left[\sqrt{S_{n,B}^{(kernel)}(x_0,t=0)}\,t\right]\right\}, \tag{7.4.8}$$

where $S_{n,B}^{(kernel)}(x_0,t=0)$ is linked to the phase difference $\Delta\varphi$ by eq. (7.3.9); the number of oscillations for the atomic emission probability depends on the position of the atom inside the cavity and can be controlled by the phase-difference of the two laser-beams.

The following condition,

$$1+(-1)^n\cos\Delta\varphi=0\,, \tag{7.4.9}$$

is verified whether the atom is coupled to an odd QNM, i.e. $n=1,3,...$ and the two laser beams are in phase, i.e. $\Delta\varphi=0$, or the atom is coupled to an even QNM, i.e. $n=0,2,...$ and the two laser beams are opposite in phase, i.e. $\Delta\varphi=\pi$. If eq. (7.4.9) is satisfied, the emission probability is over-damped inside the whole cavity; in spite of the strong coupling, no oscillation occurs [eq. (7.4.8)]:

$$S_{n,B}^{(kernel)}(x_0, t=0) = 0 \quad \Rightarrow \quad |b_+(t)|^2 = \exp\left(\frac{|Im\,\omega_n|^2}{4}t\right)\left[\left(1+\frac{|Im\,\omega_n|}{2}t\right)^2 + \left(\frac{R\Delta_n}{2}t\right)^2\right]. \quad (7.4.10)$$

## 5. Emission spectrum of the atom.

At the initial time ($t=0$), the atom, located in the point $x_0$, is at the upper state and no photons exists in any normal mode, i.e. $c_+(x_0,t=0)=1$; when the atomic decay has occurred ($t=\infty$), the coefficient of probability $c_{-,\lambda}(x_0,t)$ to find the atom in the lower state, one photon existing in the $\lambda^{th}$ e.m. mode and no photon in all the other modes can be derived from eq. (7.2.4):

$$c_{-,\lambda}(x_0, t=\infty) = -\frac{\langle -|\mu|+\rangle}{\hbar}\sqrt{\frac{\hbar\omega_\lambda}{2\varepsilon_0 n_0^2}}g_\lambda^*(x_0)\int_0^\infty c_+(x_0,t)\exp[i(\omega_\lambda - \Omega)t]dt. \quad (7.5.1)$$

In terms of the Laplace transform for the probability coefficient $c_+(x_0,t)$,

$$C_+(x_0, s) = \frac{1}{\sqrt{2\pi}}\int_0^\infty c_+(x_0,t)\exp(-st)dt, \quad (7.5.2)$$

eq. (7.5.1) is so re-expressed:

$$c_{-,\lambda}(x_0, t=\infty) = -\frac{\langle -|\mu|+\rangle}{\hbar}\sqrt{\frac{\hbar\omega_\lambda}{2\varepsilon_0 n_0^2}}g_\lambda^*(x_0)\sqrt{2\pi}C_+[x_0, s=i(\omega_\lambda - \Omega)]. \quad (7.5.3)$$

If the decay has occurred ($t=\infty$), the atomic emission spectrum can be defined as the density of probability that the atom in the point $x_0$ has emitted at the frequency $\omega$[7],

$$W(x_0, \omega) = \sum_{\lambda=1}^\infty |c_{-,\lambda}(x_0, t=\infty)|^2 \delta(\omega-\omega_\lambda), \quad (7.5.4)$$

where $\delta(t)$ is the delta distribution of Dirac. Inserting eq. (7.5.3) into eq. (7.5.4), it results:

$$W(x_0, \omega) = \frac{M}{2\varepsilon_0 n_0^2\hbar^2}\frac{1}{L}\sum_{\lambda=1}^\infty \hbar\omega_\lambda 2\pi|C_+[x_0, i(\omega_\lambda - \Omega)]|^2 \delta(\omega-\omega_\lambda). \quad (7.5.5)$$

If, according to eq. (7.2.6), sums over discrete quantities are converted to integrals over continuous frequency, applying the proprieties of the Dirac's distribution, the emission spectrum (7.5.5) can be reduced as

$$W(x_0, \omega) = K'\cdot\frac{R}{2\sqrt{\rho_0}}\omega\sigma^{(loc)}(x_0, \omega)2\pi|C_+[x_0, i(\omega-\Omega)]|^2, \quad (7.5.6)$$

being $K'$ a suitable constant of normalization and $\sigma^{(loc)}(x,\omega)$ the local density of states (DOS). The atomic parameter $R$ is defined in eq. (7.2.11).

The frequency range can be extended from $-\infty$ to $\infty$ without significant errors, since most optical experiments use a narrow band source [5]; the normalization constant $K'$ can be derived applying on the emission spectrum (7.5.6) the condition as density of probability, so expressed:

$$\int W(x_0, \omega) d\omega = 1. \qquad (7.5.7)$$

Let suppose $0 < x_0 < d$ and $n(x_0) > n_0$, so that the atom is embedded inside the open cavity with inhomogeneous refractive index $\rho(x) = [n(x)/c]^2$. The atom, with resonance $\Omega$, is assumed to be coupled with the $n^{th}$ QNM, oscillating at the frequency $Re\omega_n$; the atom - $n^{th}$ QNM coupling is characterized by the frequency detuning $\Delta_n$ of eq. (7.3.1). The normalization condition (7.5.7), for the atomic emission spectrum $W_n(x_0, \omega)$, reduces to:

$$\int_{\Omega - |Im\omega_n|}^{\Omega + |Im\omega_n|} W_n(x_0, \omega) d\omega = 1. \qquad (7.5.8)$$

As from eq. (6.2.4), the local density of probability $\sigma_n^{(loc)}(x, \omega)$ for the $n^{th}$ QNM is proportional to the (integral) one $\sigma_n(\omega)$. Inserting eq. (6.2.4) in eq. (7.5.6), the atomic emission spectrum [eq. (7.5.6)],

$$W_n(x_0, \omega) = K'_n \cdot \frac{R}{2\sqrt{\rho_0}} \omega \sigma_n^{(loc)}(x_0, \omega) 2\pi \left| C_+^{(n)}\left[x_0, i(\omega - \Omega)\right] \right|^2, \qquad (7.5.9)$$

is specified as:

$$W_n(x_0, \omega) = K'_n \cdot \frac{R}{2\sqrt{\rho_0}} \frac{1}{I_n} \rho(x_0) |F_n(x_0)|^2 \omega \sigma_n(\omega) 2\pi \left| C_+^{(n)}\left[x_0, i(\omega - \Omega)\right] \right|^2, \qquad (7.5.10)$$

where $K'_n$ is a normalization constant satisfying eq. (7.5.8).

The emission processes of the atom are characterized by a kernel function $S_n^{(kernel)}(x, t)$ which, in case of spontaneous emission, can be specified as in eq. (7.3.5), and, for stimulated emission, as in eq. (7.3.9). Inserting eq. (7.3.5) in eq. (7.5.10), the emission spectrum (7.5.10) reads as:

$$\frac{W_n(x_0, \omega)}{K'_n} = -\frac{2}{d \cdot I_n^2} \frac{S_{n,A}^{(kernel)}(x_0, t=0)}{\omega_n} \omega \frac{\sigma_n(\omega)}{K_n} 2\pi \left| C_+^{(n)}\left[x_0, i(\omega - \Omega)\right] \right|^2. \qquad (7.5.11)$$

As from eq. (7.5.11), the emission spectrum $W_n(x_0, \omega)$ depends on the density of probability $\sigma_n(\omega)$, as well as on the initial value of the kernel function $S_{n,A}^{(kernel)}(x_0, t=0)$.

Laplace transforming the Cauchy problem (7.3.10) with initial conditions as in eq. (7.3.8), it results [14]:

$$C_+^{(n)}(x_0, i\xi) = -\frac{i}{\sqrt{2\pi}} \frac{\xi + R\Delta_n + i\,Im\,\omega_n}{\xi^2 + (R\Delta_n + i\,Im\,\omega_n)\xi + S_n^{(kernel)}(x_0, t=0)}, \qquad (7.5.12)$$

where $\xi$ is the shifted frequency $(\omega - \Omega)$, being $\Omega$ the atomic resonance.

## 5. 1. Poles of the emission spectrum.

The emission spectrum of the atom can be described in terms of the poles, $p_1$ and $p_2$, which solve eq. (7.4.1) and are expressed in eq. (7.4.2). In strong coupling regime [eq. (7.4.5)], the atomic emission spectrum $W_n(x_0,\xi)$, function of the shifted frequency $\xi =(\omega-\Omega)$, is characterized by two peaks, approximately centred in the resonances $Re\ p_1$ and $Re\ p_2$ and with bandwidths linked to $2|Im\ p_1|$ and $2|Im\ p_2|$; so, a Rabi's splitting occurs, being the two peaks separated by:

$$\Delta\xi = \operatorname{Re} p_1 - \operatorname{Re} p_2. \quad (7.5.13)$$

Let consider the stimulated emission. Two counter propagating laser beams have a phase difference $\Delta\varphi$, so the emission spectrum $W_n(x_0,\xi)$ is described by a kernel function $S_{n,B}^{(kernel)}(x_0, t=0)$ linked to $\Delta\varphi$ [see eq. (7.3.9)]; as result, the Rabi's splitting, besides depending on the position of the atom inside the cavity, can be controlled by the phase-difference of the two laser-beams.

If the operative condition defined by eq (7.4.9) is almost verified, i.e. $S_{n,B}^{(kernel)}(x_0, t=0) \approx 0$, the spectrum $W_n(x_0,\omega)$, function of the pure frequency $\omega$, is reduced to two almost superimposed pulses: a Lorentian function centred in the $n^{th}$ QNM's frequency $Re\omega_n$, with a bandwidth $2|Im\ \omega_n|$, superimposed to a Dirac's distribution in the atomic resonance $\Omega\approx Re\omega_n$, i.e. [see eqs. (6.2.5) and (6.2.7)]

$$W_n(x_0,\omega) \approx K_n^{(1)}\sigma_n^{(II)}(\omega) + K_n^{(2)}\delta(\omega-\Omega) \rightarrow \delta(\omega-\Omega), \quad (7.5.14)$$

where $K_n^{(1)}$ and $K_n^{(2)}$ are normalization constants satisfying condition (7.5.8). In fact, the two poles, $\omega_1$ and $\omega_2$, are so simplified [eqs. (7.3.1) and (7.4.1)-(7.4.2)]:

$$\begin{aligned}\omega_1 &\approx \operatorname{Re}\omega_n + i\operatorname{Im}\omega_n \\ \omega_2 &\approx \Omega\end{aligned}. \quad (7.5.15)$$

## 6. Criteria to design an active delay line.

Let specify a Photonic Crystal (PC) as a symmetric Quarter-Wave (QW) one dimensional (1D) Photonic Band Gap (PBG) cavity; the Quasi Normal Mode's (QNM's) approach has been applied to this kind of structure in refs. [7] [6] and subsequent papers. Consider a symmetric QW 1D-PBG cavity with parameters $\lambda_{ref}=1\mu m$, $N=5$, $n_h=2$, $n_l=1.5$ (fig. 7.1.); a "toy 1D-PBG" cavity is chosen because, it is retained sufficient to discuss some criteria for a design of an active delay line. Locate an atom in the centre of the 1D-PBG, i.e. $x_0=d/2$ (fig. 7.1.). As discussed in ref. [6], for a symmetric QW 1D-PBG cavity with $\omega_{ref}$ as reference wavelength and $N$ periods, the [0, $2\omega_{ref}$) range includes $2N+1$ QNMs which, excluding $\omega=2\omega_{ref}$ are identified as $|n\rangle$, $n\in[0,2N]$. If the atom is located in the centre $x_0$ of the 1D-PBG cavity, it can be coupled to just one of the QNMs with an even $n$: in fact, in the centre of the 1D-PBG, the QNM's intensity $|F_n|^2$ has a maximum for even values of $n$ and is almost null for odd value of $n$.

The active cavity, consisting of the 1D-PBG cavity plus the atom, is characterized by a global transmission spectrum $G(x_0,\omega)$ which is the product between the transmission spectrum of the 1D-PBG $|t(\omega)|^2$ and the emission spectrum $W(x_0,\omega)$ of the atom [in units of *sec*], i.e.

$$G(x_0,\omega) = W(x_0,\omega)|t(\omega)|^2 \text{ [in units of } sec\text{]}. \tag{7.6.1}$$

A "density of coupling" (DOC) for the active cavity $\sigma_C(x_0,\omega)$ can be defined as the density of probability that the atom, embedded in the point $x_0$, is coupled to just one QNM, oscillating around the frequency $\omega$; the DOC $\sigma_C(x_0,\omega)$ [in units of $sec^2/m$] is the product between the DOS $\sigma(\omega)$ [in units of $sec/m$] and the atomic emission spectrum $W(x_0,\omega)$. Moreover, an "acceleration of coupling" $v_C(x_0,\omega)$ inside the active cavity can be introduced as:

$$a_C(x_0,\omega) = \frac{1}{\sigma_C(x_0,\omega)} = \frac{1}{W(x_0,\omega)\sigma(\omega)} = \frac{v(\omega)}{W(x_0,\omega)} \text{ [in units of } m/sec^2\text{]}. \tag{7.6.2}$$

In order to design the active cavity as an ideal delay line, the input pulse must be retarded with an high amplification but without any distortion: in a narrow pass band, the global transmission (7.6.1) must assume an high value and the acceleration of coupling (7.6.2) an almost constant value.

As reported above, the atom embedded in the centre of the symmetric QW 1D-PBG cavity with $N=5$ periods (fig. 7.1.) can be coupled only to one QNM, oscillating next to an even transmission peak $n=0,2,...,2N$; let suppose that the atom is coupled to the $(N+1)^{th}$ QNM, next to the high frequency band edge, so the quality factor of the 1D-PBG cavity is

$$Q_{N+1} = \frac{\Omega}{|\text{Im}\,\omega_{N+1}|}, \tag{7.6.3}$$

being $\Omega$ the resonance frequency of the atom. Assume to be in strong coupling regime; the directories for the active delay line can be satisfied by a suitable value of the coupling degree,

$$R = \frac{\Gamma_0}{\Omega}, \tag{7.6.4}$$

being $\Gamma_0$ the decay rate of the atom in vacuum, and by a suitable value of the frequency detuning for the atom - $(N+1)^{th}$ QNM coupling,

$$\Delta_{N+1} = \frac{\text{Re}\,\omega_{N+1} - \Omega}{R}. \tag{7.6.5}$$

In case of spontaneous emission, if a perfect tuning is assumed, i.e. $\Delta_{N+1}=0$, the atom oscillates at the frequency of the $(N+1)^{th}$ QNM, i.e. $\Omega=Re\omega_{N+1}$; there exists (figs. 7.2.a. and 7.2.b.) a suitable value of the coupling degree $R$, i.e. $R^*=0.002506$, such that the two poles [eqs. (7.4.2) and (7.3.5)] of the emission spectrum of the atom are distinct for $R>R^*$ but coincident for $0<R<R^*$. In other words, for $R>R^*$, a Rabi's splitting (fig. 7.3.a.) occurs in the atomic emission spectrum [eqs. (7.5.11)-(7.5.12) and (7.3.5)], and so an oscillatory behaviour (fig. 7.4.a.) is present in the atomic emission probability [eq. (7.4.7) and (7.3.5)]; however, for $0<R<R^*$, the emission spectrum consists of two superimposed peaks, and so the emission probability is over-damped. In order to find a Rabi's splitting in strong coupling regime, consistent with experiments $(\Gamma_0 \sim |Im\omega_{N+1}|)$ [20], let employ the degree of coupling:

$$R = R_{N+1} = \frac{1}{Q_{N+1}}. \tag{7.6.6}$$

The two poles of the spontaneous emission spectrum, shifted of the atomic resonance $\Omega$, are $\xi_1=0.06383+i0.01770$ and $\xi_2=-0.06383+i0.01995$ in units of $\omega_{ref}$ (figs. 7.2.a. and 7.2.b.). They describe, in resonance and band width, the two peaks of the emission spectrum, whose maxima are $W_1=21.87$ and $W_2=15.66$ in units of $\omega_{ref}$ (fig. 7.3.a.). Assuming the emission probability almost decayed after the second oscillation, the decay time is $\tau=94.3$ in units of $1/\omega_{ref}$ (fig. 7.4.a.). So far, the active cavity has been designed just as a not ideal optical amplifier; an input pulse is amplified but distorted: as plotted in figs. 7.5.a. and 7.5.b, in the case of spontaneous emission, there exists a narrow pass band, i.e. $\xi = \omega - \Omega \approx (-0.06, 0.06)$ (in units of $\omega_{ref}$), where the global transmission spectrum assumes relative high values, i.e. $G_{C,N+1} \in (G_{min}, G_{max}) = (2.881, 14.43)$ in units of $1/\omega_{ref}$, but the acceleration of coupling is subject to a modulation around the value $v_{C,N+1}=0.03445$ in units of $\omega_{ref}/v_{ref}$.

Let consider now the case of stimulated emission; the atom inside the symmetric QW 1D-PBG cavity is excited by two counter propagating laser beams, so that the active delay line can be realized adding, as new degree of freedom, the phase difference $\Delta\varphi$ of the two laser beams. In case of stimulated emission, if a perfect tuning is again assumed, i.e. $\Delta_{N+1}=0$, the atom oscillates still at the frequency of the $(N+1)^{th}$ QNM, i.e. $\Omega=Re\omega_{N+1}$. There exists (figs. 7.2.c. and 7.2.d.) a suitable range of the phase difference $\Delta\varphi$, i.e. $(\Delta\varphi_1, \Delta\varphi_2) = (2.747, 3.524)$ in units of rads, such that the two poles [eqs. (7.4.2) and (7.3.9)] of the emission spectrum of the atom are distinct for $\Delta\varphi<\Delta\varphi_1$ and $\Delta\varphi>\Delta\varphi_2$ but coincident for $\Delta\varphi_1 <\Delta\varphi<\Delta\varphi_2$. In other words, for $\Delta\varphi<\Delta\varphi_1$ and $\Delta\varphi>\Delta\varphi_2$, a Rabi's splitting (fig. 7.3.a.) occurs in the atomic emission spectrum [eqs. (7.5.11)-(7.5.12) and (7.3.9)], and so an oscillatory behaviour (fig. 7.4.a.) is present in the atomic emission probability [eqs. (7.4.8) and (7.3.9)]; however, for $\Delta\varphi_1 <\Delta\varphi<\Delta\varphi_2$, the emission spectrum consists of two superimposed peaks, and so the emission probability is over-damped. As a consequence, the Rabi's splitting and the oscillations of the decay time can be controlled by the phase difference of the two laser beams. In order to obtain an ideal delay line by stimulated emission, the two laser beams must exceed in quadrature, i.e. $\Delta\varphi>\pi/2$; with respect to spontaneous emission, the emission spectrum is characterized by a narrower Rabi's splitting and the emission probability by a longer decay time. So, the active cavity, consisting of the 1D-PBG plus the atom, acts as an delay line, since the delay time of the active cavity is linked with the decay time of the atom (ref. [21]). Moreover, as discussed above, the phase difference must not exceed $\Delta\varphi_l=2.747$ (in units of rads), otherwise the Rabi's splitting tends to zero; by increasing the phase difference with respect to $\Delta\varphi \approx \pi/2$, in the time domain, the decay time becomes more long still and, in the frequency domain, the global transmission (7.6.1) is characterized by an high gain but the acceleration of coupling (7.6.2) by a much too narrow pass band. So, if $\Delta\varphi \rightarrow \Delta\varphi_l$, the active cavity acts as a delay line which is active but not ideal. Let conclude that the 1D-PBG cavity should be pumped by two laser beams exceeding in quadrature of a tilt angle as:

$$\Delta\varphi = \frac{\pi}{2} + \frac{\pi}{10}. \qquad (7.6.7)$$

The two poles of the stimulated emission spectrum, shifted of the resonance $\Omega$, are $\xi_1=0.05205+i0.01787$ and $\xi_2=-0.05205+i0.01978$ (in units of $\omega_{ref}$) (figs. 7.2.c.and 7.2.d.); with respect to spontaneous emission, the two poles are nearer by $\Delta\xi=0.02356$. They describe, in resonance and band width, the two peaks of the stimulated emission spectrum, whose maxima are $W_1=14.36$ and $W_2=10.93$ (in units of $\omega_{ref}$) (fig. 7.3.a.); with respect to spontaneous emission, the two maxima are lower by $\Delta W_1=7.51$ and $\Delta W_2=4.73$. Assuming the emission probability almost decayed after the second oscillation, the decay time of stimulated emission is $\tau=113.5$ (in units of 1/$\omega_{ref}$) (fig. 7.4.a.); with respect to spontaneous emission, this time is longer by $\Delta\tau=19.2$. So, the phase difference of the two laser beams allows to control the decay time of the atom and then the delay time of the active cavity [21]. Now, a delay line has been designed, but it is not quite ideal. An input pulse is retarded, amplified but a little distorted; as plotted in figs. 7.5.a. and 7.5.b, with respect to spontaneous emission, there exists a narrower pass band, i.e. $\xi = \omega - \Omega \approx (-0.04, 0.04)$, where the global transmission spectrum assumes similar values, i.e. $G_{C,N+1} \in (G_{min}, G_{max}) = (3.270, 9.24)$ (in units of $\omega_{ref}$), and, above all, the acceleration of coupling is subject to a slight modulation, now around the value $v_{C,N+1}=0.05310$ (in units of $\omega_{ref}/v_{ref}$).

Let finally consider stimulated emission in presence of some detuning: the atom inside the symmetric QW 1D-PBG cavity is still coupled to the $(N+1)^{th}$ QNM, but does not oscillate any more at the $(N+1)^{th}$ QNM's frequency. The design of the active delay line can be improved varying, as last degree of freedom, the frequency detuning for the atom - $(N+1)^{th}$ QNM coupling (7.6.5). In order to improve the active delay line, let propose to apply the maximum detuning. The atomic resonance $\Omega$ is dropped inside the photonic band gap next to the $(N+1)^{th}$ QNM's frequency $Re\omega_{N+1}$; the atom is still coupled only to the $(N+1)^{th}$ QNM if the atomic resonance is at the most

$$\Omega = \text{Re}\,\omega_{N+1} - |\text{Im}\,\omega_{N+1}|, \qquad (7.6.8)$$

so that the detuning is maximum:

$$\Delta_{N+1} = \frac{\text{Re}\,\omega_{N+1} - \Omega}{R_{N+1}} = \frac{|\text{Im}\,\omega_{N+1}|}{R_{N+1}}. \qquad (7.6.9)$$

The two poles of the stimulated emission spectrum in detuning case, shifted of the resonance $\Omega$, have the real parts *Re($\xi_1$)= 0.03738* and *Re($\xi_2$)= –0.07380* and the imaginary parts *Im($\xi_1$)= 0.01176* and *Im($\xi_2$)= 0.02588* (both in units of $\omega_{ref}$) (see figs. 7.2.c.and 7.2.d.); with respect to the perfect tuning, the real parts are lower by *ΔRe($\xi_1$)=0.01467* and *ΔRe($\xi_2$)=0.02175*, while one imaginary part is lower by *ΔIm($\xi_1$)=0.00611* and the other is higher of *ΔIm($\xi_2$)=0.0061*. They describe, in resonance and band width, the two peaks of the stimulated emission spectrum in the detuning case, whose maxima are *$W_1$=38.83* and *$W_2$=0.2974* (in units of $\omega_{ref}$) (fig. 7.3.b.); with respect to the perfect tuning, the first peak is higher by *Δ$W_1$=24.47* and the second peak is lower by *Δ$W_2$=10.63*. Assuming the atomic emission probability almost decayed after the second oscillation, the decay time of stimulated emission (linked the delay time of the active cavity) in the detuning case is *τ=111.2* (in units of 1/ $\omega_{ref}$) (fig. 7.4.b.); with respect to the perfect tuning, the emission probability (and so the input pulse) is slightly warped and retarded by *Δτ=2.3*.

At the end, an almost ideal delay line has been designed. An input pulse is retarded, amplified and almost not distorted; as plotted in figs. 7.5.c. and 7.5.d, with respect to stimulated emission in detuning case, there exists a still narrower pass band, i.e. *$\xi = \omega - \Omega \approx$ (0.02, 0.06)*, where the global transmission spectrum assumes higher values, i.e. $G_{C,N+1} \in (G_{\min}, G_{\max}) = (6.005, 36.77)$ (in units of $1/\omega_{ref}$); above all, the acceleration of coupling is not subject to any modulation, but is almost constant, i.e. $v_{C,N+1} \cong 0.007182$ (in units of $\omega_{ref}/v_{ref}$): as from fig. 5.d., the modulation of the coupling acceleration is shifted over the not used frequency range *$\xi \approx$ (-0.07,-0.04)*.

# 7. Conclusions.

In the present chapter, the stimulated emission has been discussed, in strong coupling regime, for an atom, embedded inside a one dimensional (1D) Photonic Band Gap (PBG) structure, which is pumped by two counter-propagating laser beams. Quantum electro-dynamics has been applied to model the atom-e.m. field coupling, by considering the atom a two level system, the e.m. field a superposition of normal modes, the coupling in dipole approximation, and the equations of motion in Wigner-Weisskopf's and rotating wave approximations. Besides, the Quasi Normal Mode's (QNM's) approach has been adopted for an open cavity, interpreting the local density of states (LDOS) as the local density of probability to excite one QNM of the cavity; so, this LDOS depends on the phase difference of the two laser beams. As main results of this chapter, the strong coupling regime has been defined as the one associated to high values of the LDOS; according to the results of the literature, the emission probability of the atom decays with oscillatory behaviour, so that the atomic emission spectrum is splitted into two peaks (Rabi's splitting); the novelty with respect to previous discussions available in literature is that the phase difference of the two laser beams allows to produce a coherent control of both the oscillations for the atomic emission probability and the Rabi's splitting for the emission spectrum.

# References.

[1] E. M. Purcell, Phys. Rev. **69**, 681 (1946); D. Kleppner, Phys. Rev. Lett. **47**, 233 (1981).

[2] K. H. Drexhage, *Progress in Optics*, edited by E. Wolf (North-Holland, New York, 1974), Vol. 12; P. Goy, J. M. Raimond, M. Gross, and S. Haroche, Phys. Rev. Lett. **50**, 1903 (1983); R. G. Hulet, E. S. Hilfer, and D. Kleppner, Phys. Rev. Lett. **55**, 2137 (1985); W. Jhe, A. Anderson, E. A. Hinds, D. Meschede, L. Moi, and S. Haroche, Phys. Rev. Lett. **58**, 666 (1987); D. J. Heinzen, J. J. Childs, J. E. Thomas, and M. S. Feld, Phys. Rev. Lett. **58**, 1320 (1987); F. De Martini, G. Innocenti, G. R. Jacobovitz, and P. Mataloni, Phys. Rev. Lett. **59**, 2955 (1987).

[3] S. John, Phys. Rev. Lett. **53**, 2169 (1984); S. John, Phys. Rev. Lett. **58**, 2486 (1987); E. Yablonovitch, Phys. Rev. Lett. **58**, 2059 (1987); E. Yablonovitch and T. J. Gmitter, Phys. Rev. Lett. **63**, 1950 (1989).

[4] J. Maddox, Nature (London) **348**, 481 (1990); E. Yablonovitch and K.M. Lenny, Nature (London) **351**, 278, 1991; J. D. Joannopoulos, P. R. Villeneuve, and S. H. Fan, Nature (London) **386**, 143 (1997).

[5] J. D. Joannopoulos, *Photonic Crystals: Molding the Flow of Light* (Princeton University Press, Princeton, New York, 1995); K. Sakoda, *Optical properties of photonic crystals* (Springer Verlag, Berlin, 2001); K. Inoue and K. Ohtaka, *Photonic Crystals: Physics, Fabrication, and Applications* (Springer-Verlag, Berlin, 2004).

[6] S. John and J. Wang, Phys. Rev. Lett. **64**, 2418 (1990); Phys. Rev. B **43**, 12772 (1991); S. John, *Confined Electrons and Photons*, edited by E. Burstein and C. Weisbuch (Plenum, New York, 1995); R. F. Nabiev, P. Yeh, and J. J. Sanchez-Mondragon, Phys. Rev. A **47**, 3380 (1993).

[7] H. M. Lai, P. T. Leung, and K. Young, Phys. Rev. A **37**, 1597 (1988).

[8] A. Lahiri, P. B. Pal, *A first book of Quantum Field Theory* (CRC Press, Calcutta, 2000); C. Cohen-Tannoudji, J. Dupont-Roc, G. Grynberg *Photons and Atoms – Introduction to Quantum Electrodynamics*, (John Wiley, New York, 1997).

[9] P. T. Leung, S. Y. Liu, and K. Young, Phys. Rev. A **49**, 3057 (1994); P. T. Leung, S. S. Tong, and K. Young, J. Phys. A **30**, 2139 (1997); P. T. Leung, S. S. Tong, and K. Young, J. Phys. A **30**, 2153 (1997); E. S. C. Ching, P. T. Leung, A. Maassen van der Brink, W. M. Suen, S. S. Tong, and K. Young, Rev. Mod. Phys. **70**, 1545 (1998).

[10] K. C. Ho, P. T. Leung, Alec Maassen van den Brink and K. Young, Phys. Rev. E **58**, 2965 (1998).

[11] J. J. Sakurai, *Advanced Quantum Mechanics* (Addison-Wesley, New York, 1995); S. Weinberg, *The Quantum Theory of Fields* (Cambridge University Press, New York, 1996).

[12] S. Severini, A. Settimi, C. Sibilia, M. Bertolotti, A. Napoli, A. Messina, *Quasi Normal Frequencies in open cavities: an application to photonic crystals*, Acta Phys. Hung. B **23/3-4**, 135-142 (2005); A. Settimi, S. Severini, B. Hoenders, *Quasi Normal Modes description of transmission properties for Photonic Band Gap structures*, J. Opt. Soc. Am. B, **26**, 876-891 (2009).

[13] A. Settimi, S. Severini, N. Mattiucci, C. Sibilia, M. Centini, G. D'Aguanno, M. Bertolotti, M. Scalora, M. Bloemer, C. M. Bowden, Phys. Rev. E, **68**, 026614 (2003).

[14] S. Severini, A. Settimi, C. Sibilia, M. Bertolotti, A. Napoli, A. Messina, *Quantum counter-propagation in open cavities via Quasi Normal Modes approach*, Laser Physics, **16**, 911-920 (2006).

[15] S. Severini, A. Settimi, C. Sibilia, M. Bertolotti, A. Napoli, A. Messina, Phys. Rev. E **70**, 056614 (2004).

[16] A. Settimi, S. Severini, C. Sibilia, M. Bertolotti, M. Centini, A. Napoli, N. Messina, Phys. Rev. E **71,** 066606 (2005).

[17] A. A. Abrikosov, L. P. Gor'kov and I. E. Dzyaloshinski, *Methods of Quantum Field Theory in Statistical Physics* (Dover, New York, 1975); P. T. Leung, A. Maassen van den Brink and K. Young, *Frontiers in Quantum Physics*, Proceedings of the International Conference, edited by S. C. Lim, R. Abd-Shukor and K. H. Kwek (Springer-Verlag, Singapore, 1998).

[18] K. J. Blow, R. Loudon, S. J. D. Phoenix, and T. J. Shepherd, Phys. Rev A **42**, 4102 (1990).

[19] G. F. Carrier, M. Krook, C. E. Pearson, *Functions of a complex variable – theory and technique* (McGraw-Hill Book Company, New York, 1983).

[20] S. Hughes and H. Kamada, Phys. Rev. B **70**, 195313 (2004).

[21] L. Allen and J. H. Eberly, *Optical Resonance and Two-Level Atoms* (Dover, New York, 1987); R. Bonifacio and L. A. Lugiato, Phys. Rev. A **11**, 1507 (1975); K. Sakoda and J. W. Haus, Phys. Rev. A **68**, 053809 (2003).

**Figures and Captions.**

Figure 7.1.

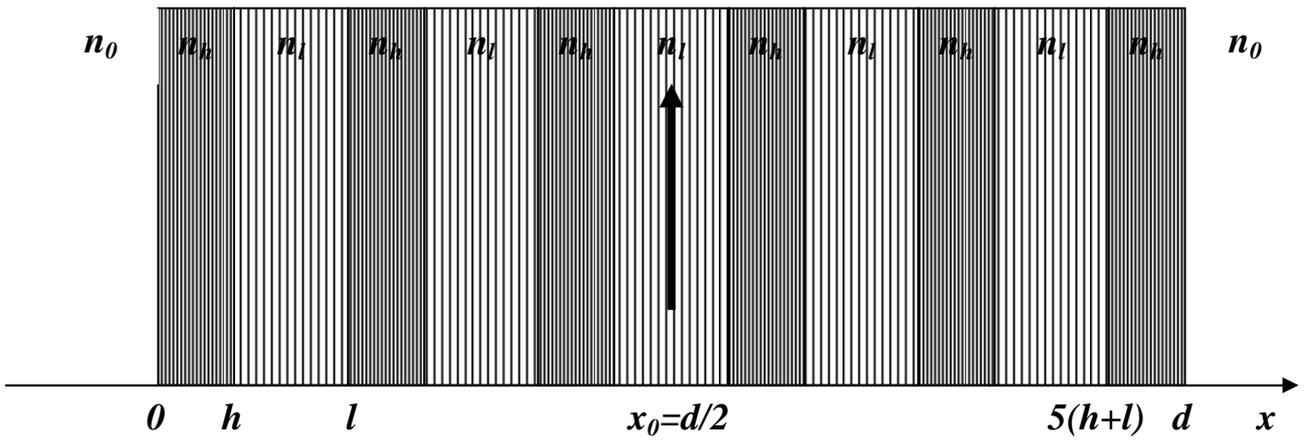

Figure 7.2.a.

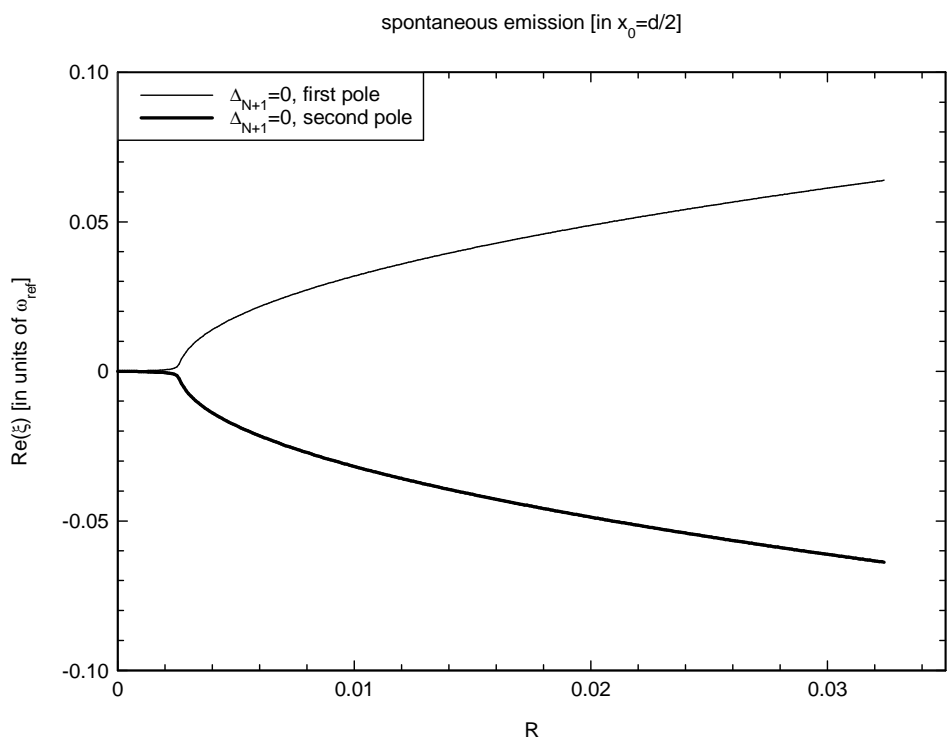

Figure 7.2.b.

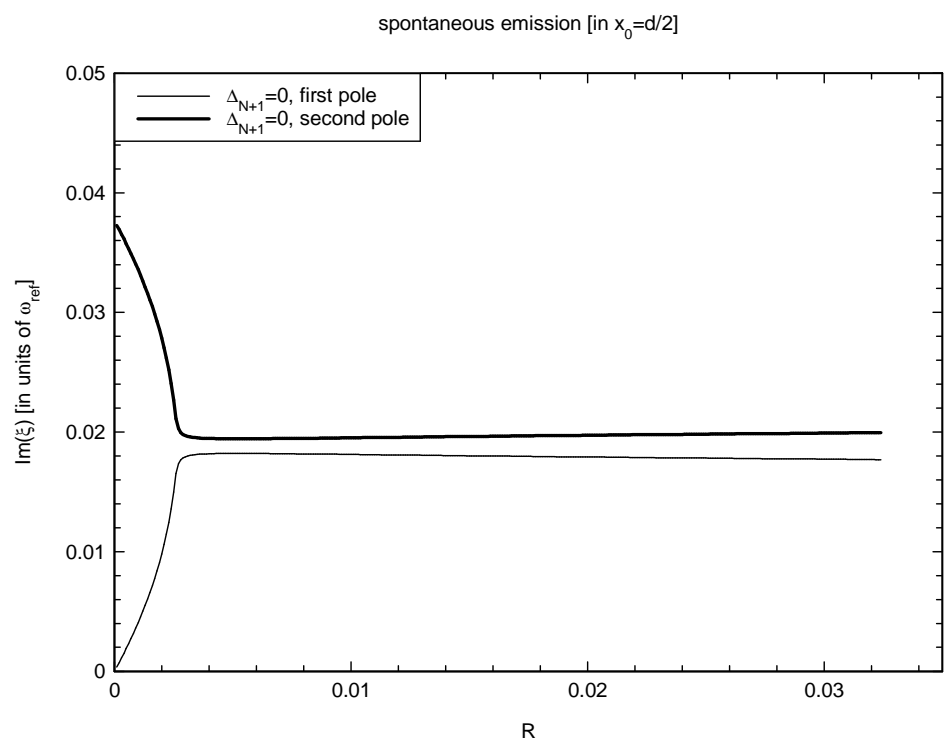

Figure 7.2.c.

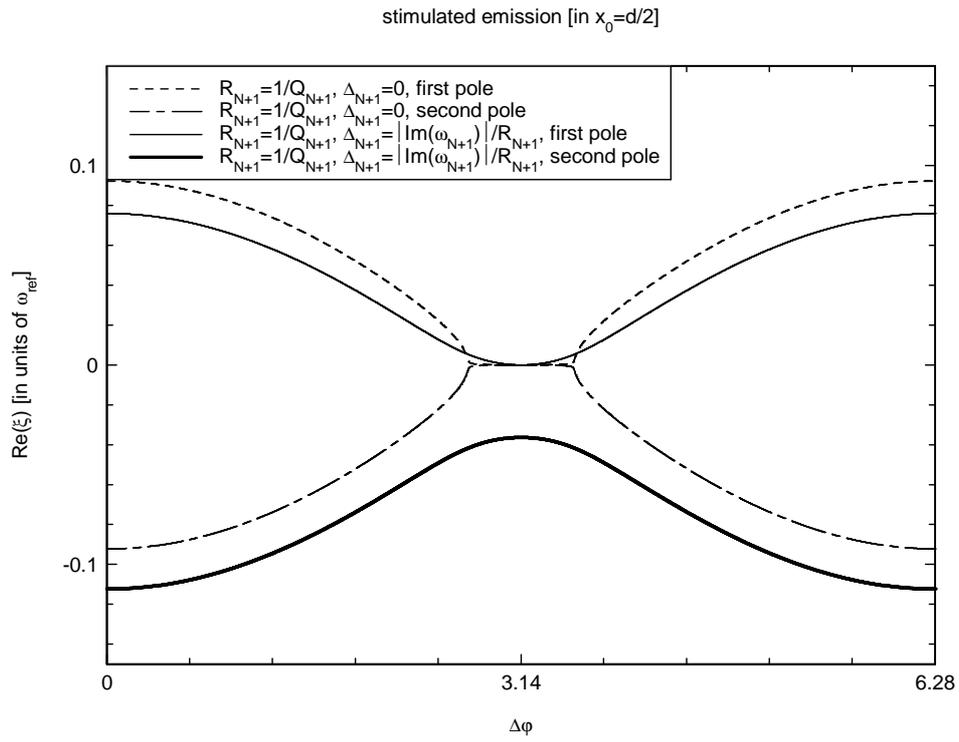

Figure 7.2.d.

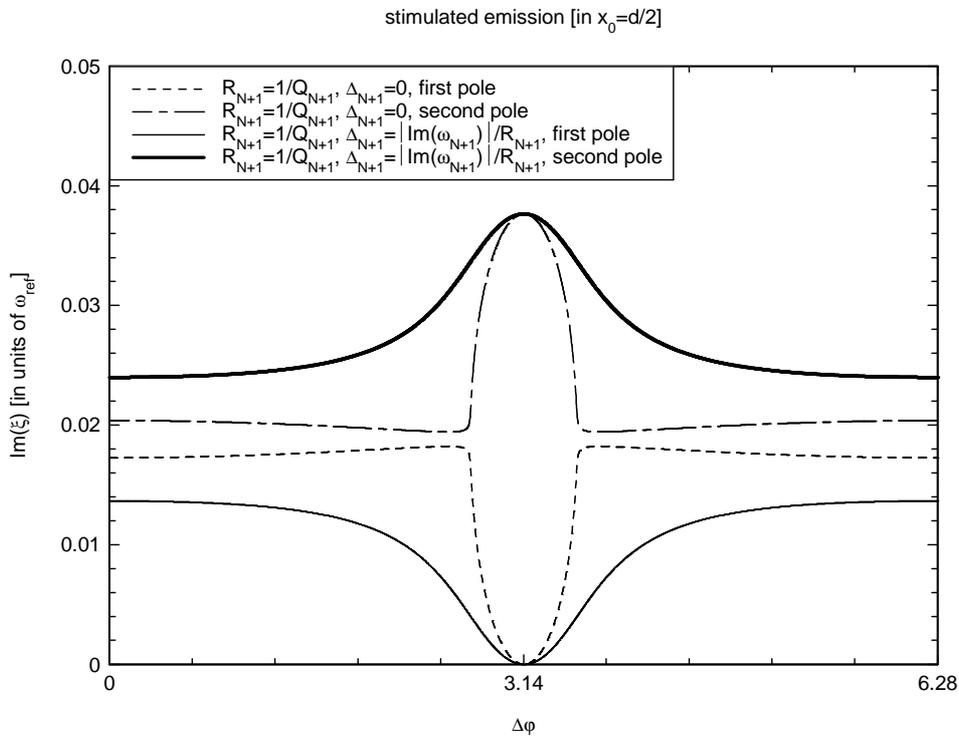

Figure 7.3.a.

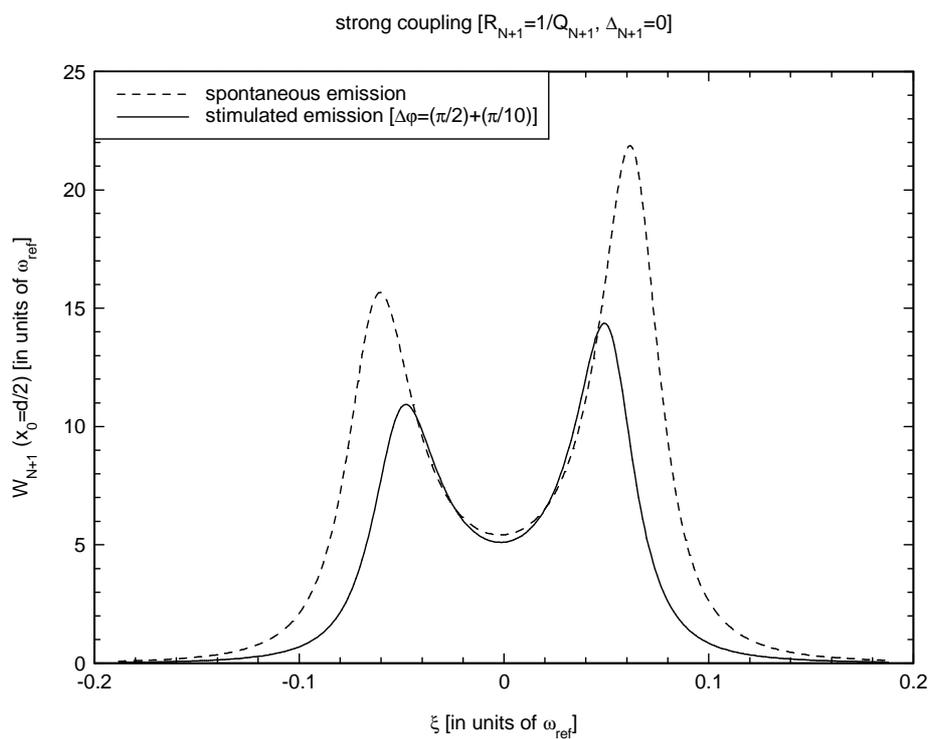

Figure 7.3.b.

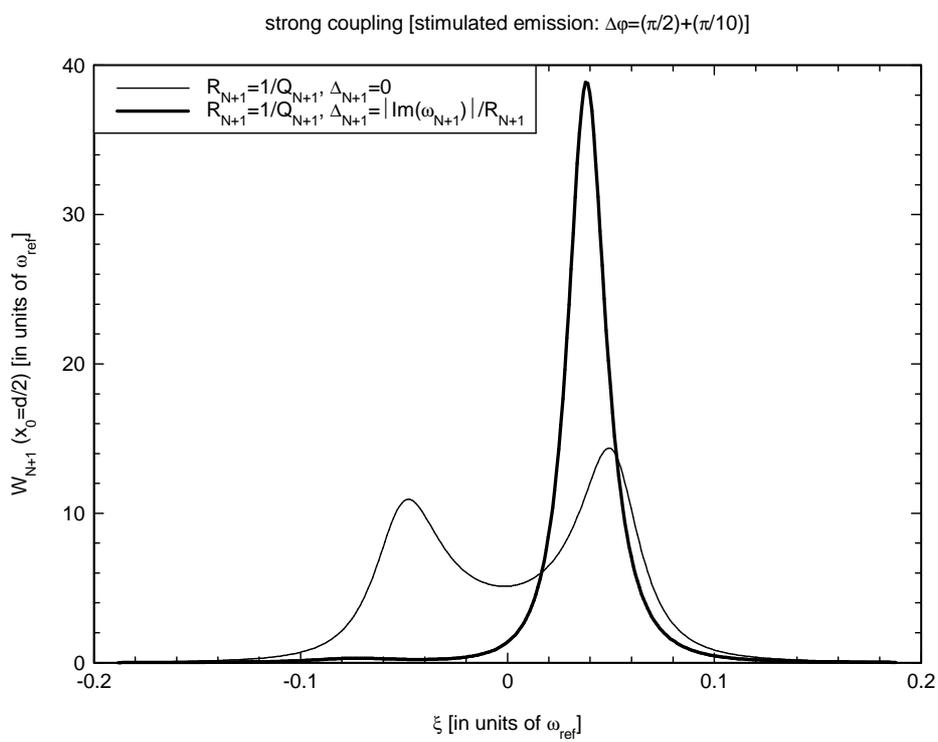

Figure 7.4.a.

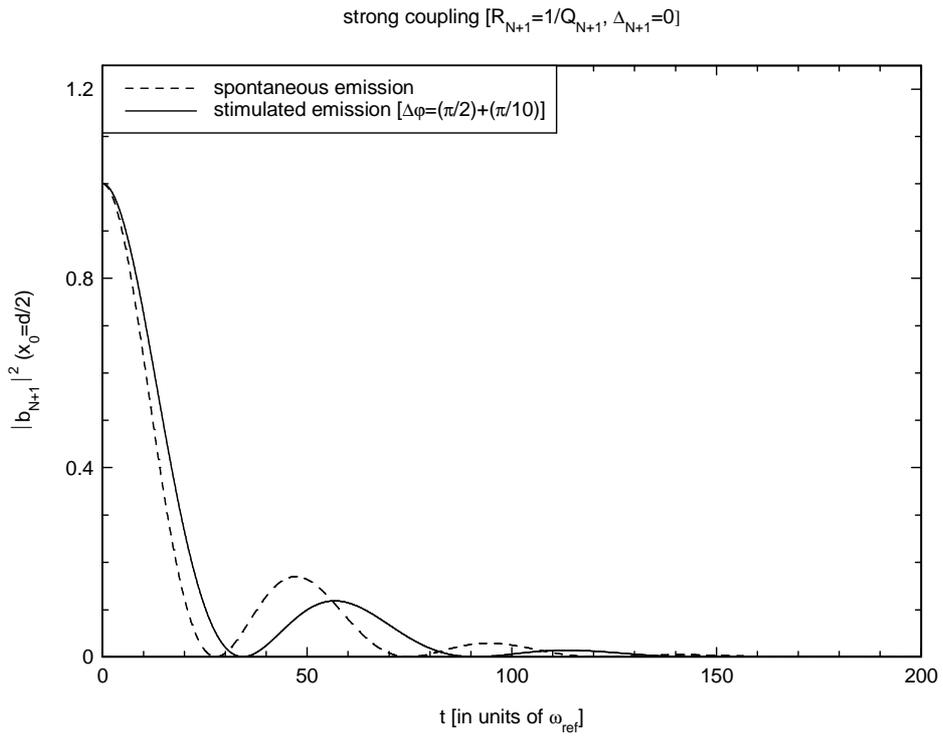

Figure 7.4.b.

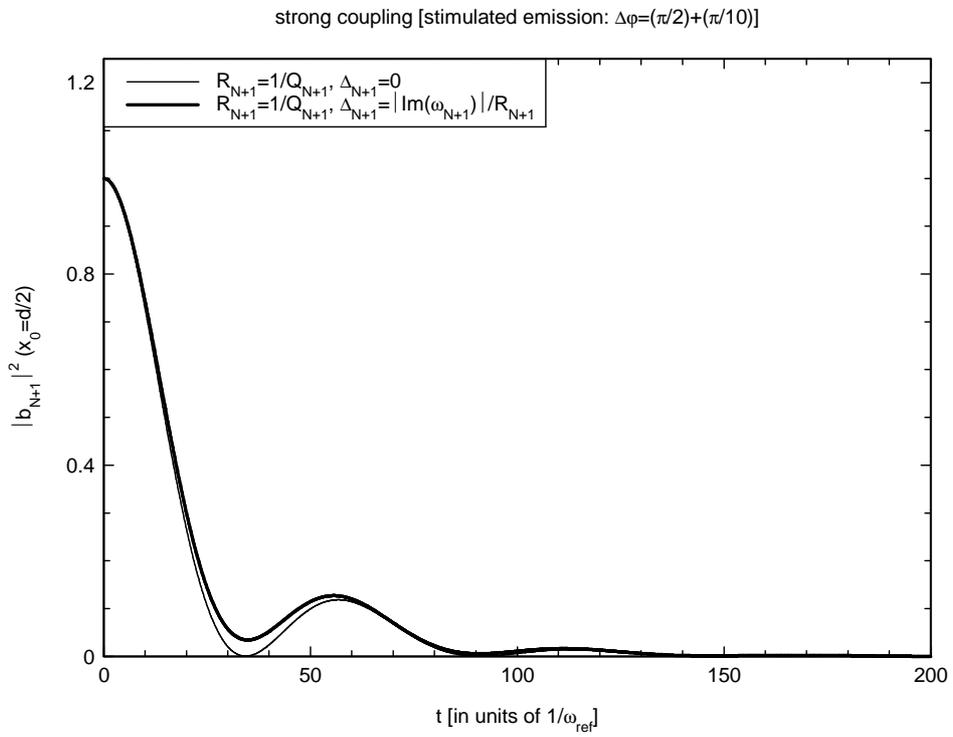

Figure 7.5.a.

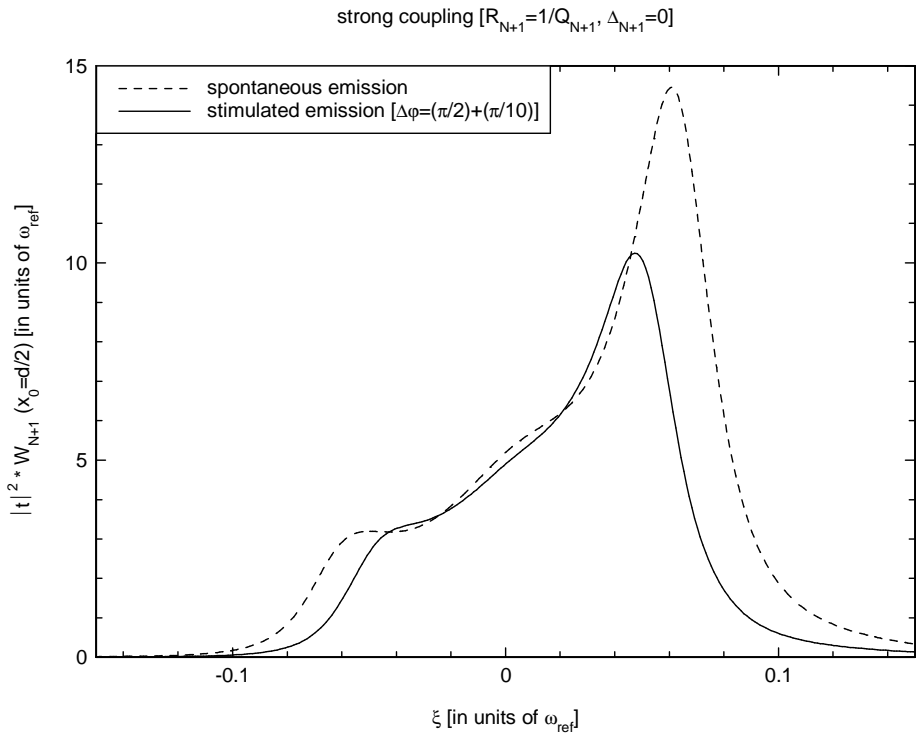

Figure 7.5.b.

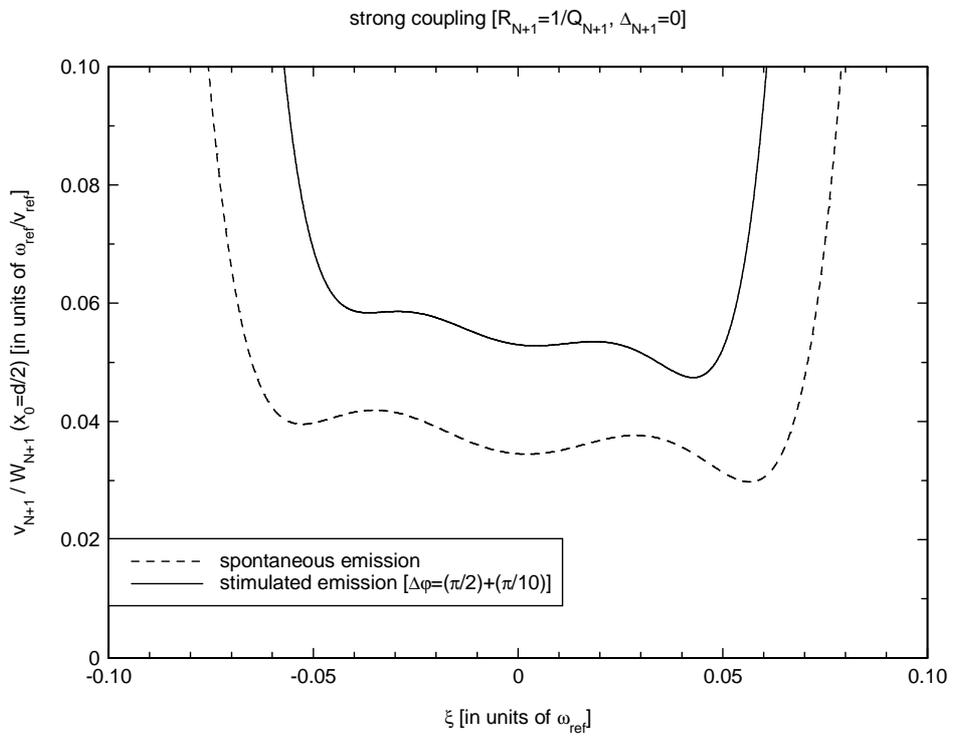

Figure 7.5.c.

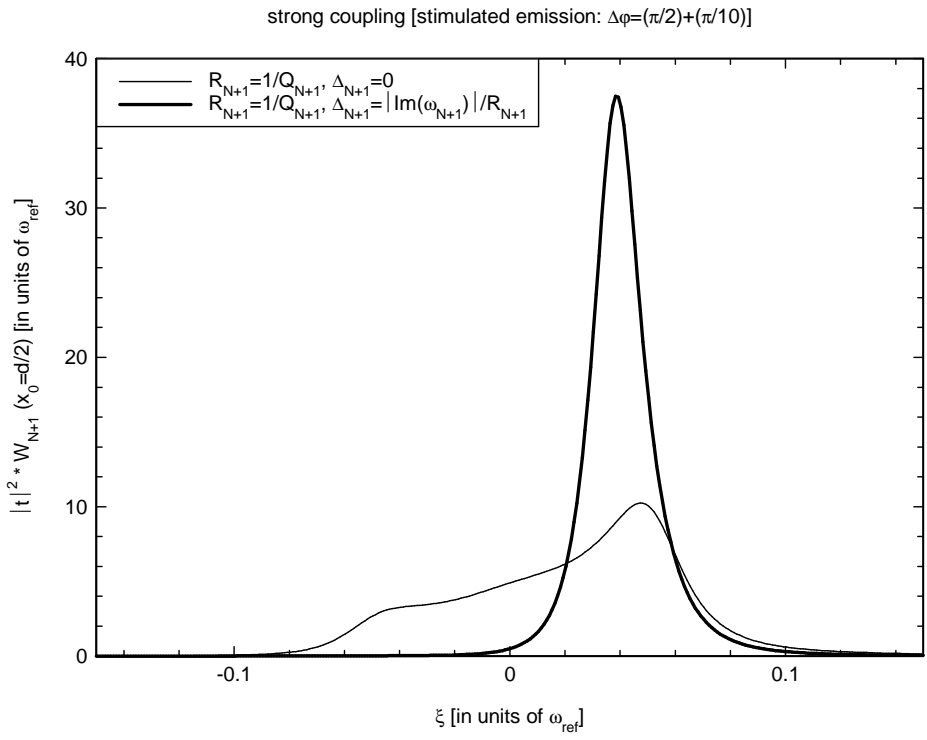

Figure 7.5.d.

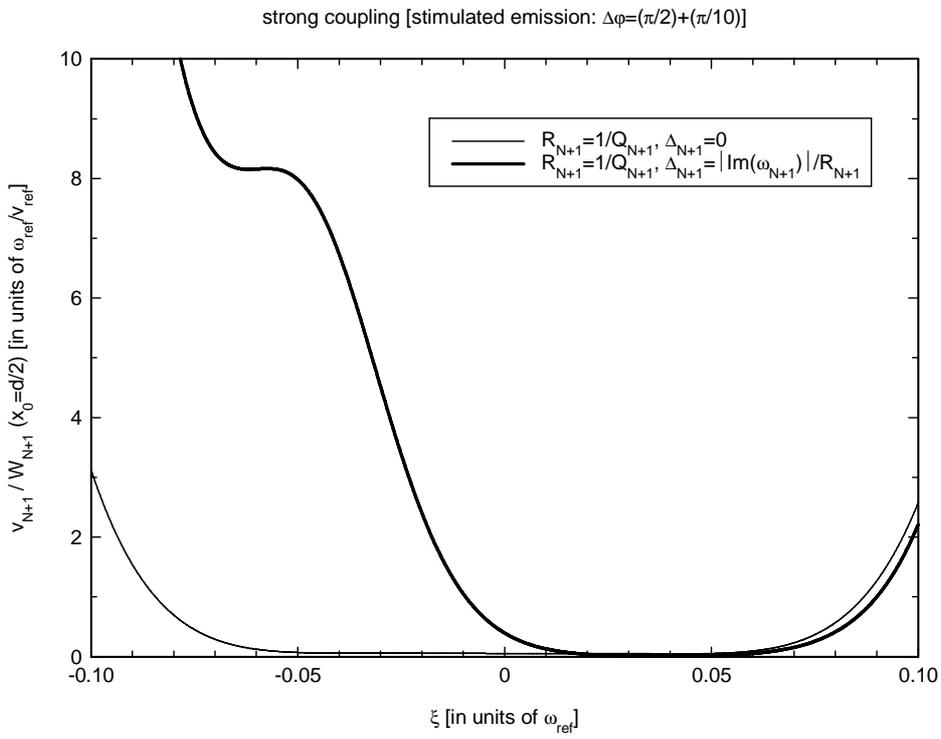

Figure. 7.1. Symmetric Quarter-Wave (QW) one dimensional (1D) Photonic Band Gap (PBG) structure with $\lambda_{ref}=1\mu m$ as reference wavelength, $N=5$ periods, consisting of two layers with refractive indices $n_h=2$ and $n_l=1.5$ and lengths $h=\lambda_{ref}/4n_h$ and $l=\lambda_{ref}/4n_l$. Terminal layers of the symmetric QW 1D-PBG structure with parameters: $n_h$ and $h=\lambda_{ref}/4n_h$. Length of the 1D-PBG structure: $d=N(h+l)+h$. One atom is embedded in the centre of the 1D-PBG, i.e. $x_0=d/2$.

Figure. 7.2. If the atom embedded inside the 1D-PBG structure of fig. 7.1. oscillates at the $(N+1)^{th}$ Quasi Normal Mode (QNM), next to the high-frequency band edge [i.e. in perfect tuning $\Delta_{N+1}=0$, see eq. (7.6.5)], the spontaneous emission in strong coupling regime can be characterized by the two poles of the emission spectrum of the atom [eqs. (7.4.2) and (7.3.5)], which poles are shifted of the atomic resonance $\Omega$; the real (a) and imaginary (b) parts, in units of the 1D-PBG reference frequency $\omega_{ref}$, are plotted as functions of the coupling degree $R=\Gamma_0/\Omega$, which is the ratio between the atomic decay-rate in vacuum $\Gamma_0$ and the resonance $\Omega$ [eq. (7.6.4)].

If two counter-propagating laser beams are tuned at the resonance $\Omega$ and the atom is coupled to the $(N+1)^{th}$ QNM, [i.e. $Q_{N+1}=\Omega //Im[\omega_{N+1}]/$, see eq. (7.6.3)], the stimulated emission in strong coupling [for $R_{N+1}=1/Q_{N+1}$, eq. (7.6.6)] can be characterized by the two poles of the atomic emission spectrum [eqs. (7.4.2) and (7.3.9)]; whether the atom oscillating at the $(N+1)^{th}$ QNM frequency [i.e. in perfect tuning $\Delta_{N+1}=0$] or at a frequency in the band gap next to the high frequency band edge [i.e. in detuning case $\Delta_{N+1}=/Im[\omega_{N+1}]//R_{N+1}$, eq. (7.6.9)], the real (c) and imaginary (d) parts, in units of the 1D-PBG reference frequency $\omega_{ref}$, are plotted as functions of the phase difference $\Delta\varphi$ between the two laser beams.

Figure. 7.3. Referring to the atom embedded inside the 1D-PBG structure of fig. 7.1, the emission spectrum of the atom, in units of the 1D-PBG reference frequency $\omega_{ref}$, is plotted as a function of the dimensionless shifted frequency $\xi=(\omega-\Omega)/\omega_{ref}$. The atom is coupled to the $(N+1)^{th}$ QNM frequency and the emission processes occur in strong coupling regime [for $R_{N+1}=1/Q_{N+1}$]. In tuning hypothesis (a), the atomic emission spectrum for spontaneous processes [eqs. (7.5.11)-(7.5.12)] is compared with the stimulated emission spectrum [eqs. (7.5.11)-(7.5.12)] when the 1D-PBG is pumped by two laser beams with a suitable phase difference: $\Delta\varphi=(\pi/2)+(\pi/10)$ [eq. (7.6.7)]. In the case of stimulated processes (b) the atomic emission spectrum in perfect tuning is compared with the emission spectrum in detuning case.

Figure. 7.4. Referring to the atom embedded inside the 1D-PBG structure of fig. 7.1, the emission probability of the atom is plotted as a function of the normalized time $\omega_{ref}t$. Referring to the operative conditions of fig. 7.3: in tuning hypothesis (a), the atomic emission probability for spontaneous processes [eqs.(7.4.7)] is compared with the stimulated emission probability [eqs. (7.4.8)] when the 1D-PBG is pumped by two laser beams with a suitable phase difference: $\Delta\varphi=(\pi/2)+(\pi/10)$; in the case of stimulated processes (b), the atomic emission probability in perfect tuning is compared with the emission probability in detuning case.

Figure. 7.5. In order to design a delay line using the active cavity composed of the 1D-PBG structure plus the atom (fig. 7.1.), the delay line can be characterized by a global transmission [eq.(7.6.1)] and by a "coupling acceleration" [eq. (7.6.2)] of the e.m. field; the global transmission, in units of $\omega_{ref}$, and the coupling acceleration, in units of $\omega_{ref}/v_{ref}$, being $v_{ref}$ the group velocity of the e.m. field in vacuum, are plotted as functions of the dimensionless shifted frequency $\xi=(\omega-\Omega)/\omega_{ref}$. Referring to the operative conditions of fig. 7.3: in perfect tuning, as from fig. 7.5.a. (fig. 7.5.b.), the global transmission (the coupling acceleration) of the active delay line for spontaneous emission is compared with the global transmission (the coupling acceleration) for stimulated emission when the 1D-PBG is pumped by two laser beams with a suitable phase difference: $\Delta\varphi=(\pi/2)+(\pi/10)$; in the case of stimulated emission, as from fig. 7.5.c. (fig. 7.5.d.), the global transmission (the coupling acceleration) of the active delay line in perfect tuning is compared with the global transmission (the coupling acceleration) in detuning case.

# DISCUSSION AND CONCLUSIONS

The e.m. field inside an open cavity can be obtained by suitable methods as the transfer matrix [1] or the ray method [2]. The representation of the e.m. field inside an open cavity can be given also as a superposition of Quasi Normal Modes (QNMs) which describe the coupling between the cavity and the environment. The importance of the QNM's approach lies in the fact that it is possible to recover the orthogonal representation of the e.m. field, as it is necessary to consider quantum and non linear processes. Several methods of description for open cavities are discussed in literature, but the initial hypothesis is different: for example, the QNM's approach is quite different from the pseudo-mode's one introduced by Dalton ed al. in ref. [3]. In fact, the pseudo-modes are obtained by a Fano's transformation of the NMs, they are defined as pseudo-modes because they depend on the external pumping, and they use an ordinary metric; the QNMs are actual modes because they are defined in absence of an external pumping, and they use a specific definition of complex metric.

The present Ph-D thesis has examined the properties of a one dimensional (1D) size-limited optical cavity, subjected to two quantized counter propagating field pumps entering through the two limiting surfaces of the optical device. The interest is strategically confined to the behaviour of the leaky cavity in its own and then the environmental field pumps are regarded as assigned external sources acting upon the optical structure under scrutiny. From a mathematical point of view, this target has been addressed exploiting the QNM's approach in a second quantization scheme, here appropriately extended with the aim of covering the treatment of an open cavity coherently pumped by noisy field pumps. It is shown that the presence of such pumps may be globally taken into account by introducing a new physically transparent description of the boundary conditions fulfilled by the e.m. field inside the cavity. The external counter propagating field pumps are incorporated, tracing back their effect to that of two assigned electrical currents, existing only on the two limiting surfaces of the leaky cavity. This method enables to establish, for the first time, a direct link between the internal QNM's operators and the external NM's ones at a generic frequency. The advantage of such kind of relations stems from the possibility of tracing back the calculation of the expectation value of any field operator inside an open cavity to that involving only field operators of the free space. In particular, this route has been followed in order to get the expression of the auto-correlation function of the e.m. field inside the cavity. The main result of this Ph-D thesis, beside the introduction of a peculiar language directly stemming from the fact of considering such a optical device as a forced open quantum system, is to understand the changes provoked by the two field pumps on some optical properties of the device. Such studies lead to other interesting investigations like, for example, the modified density of states (DOS) inside the optical structure, as well as the consequent modification of the spontaneous and stimulated emission processes of an atom placed inside the structure. In order to explore the usefulness of these ideas and mathematical approach in a physical context of theoretical and applicative interest, the optical device is specialized as a 1D Photonic Crystal (PC) of finite length.

Non-Hermitian Hamiltonians and the ensuing complex eigen-values figure also in Siegman's works on dissipative cavity quantum electro-dynamics (CQED) [4], elaborating Fox and Lee's works [5], which consider eigen-value problems for complex symmetric operators. However, refs. [4] and [5] deal with transverse modes in the semi-classical limit, only considering (a) *c*-number fields with some effective quantum noise, and (b) the limit $\lambda<<d$, being $\lambda$ the field wavelength and $d$ the length of the cavity.

The present Ph-D thesis pertains to a different regime. A fully quantum treatment is given for the dynamics of an atom inside a 1D-PBG, coupled with two counter-propagating laser-beams, by using the QNM's approach, i.e. treating the problem in the framework of open systems. As main result, the decay-time depends from the position of the atom inside the cavity, and can be controlled by the phase-difference of the two laser-beams. Such a system is relevant for a single-atom, phase-sensitive, optical memory device on atomic scale.

The strong coupling regime has been defined as the one associated to high values of the local DOS. According to the results of the literature, the emission probability of the atom decays with oscillatory behaviour, so that the atomic emission spectrum is splitted into two peaks (Rabi's splitting); the novelty with respect to previous discussions available in literature is that the phase difference of the two laser beams allows to produce a coherent control of both the oscillations for the atomic emission probability and the Rabi's splitting for the emission spectrum. Some criteria have been proposed to design the active cavity consisting of the 1D-PBG plus the atom as an active delay line: suitable phase differences between the two laser beams allow to obtain a high transmission in a narrow pass band for a delayed pulse.

This Ph-D thesis paves the way to extend the classic and quantum Quasi Normal Mode's theory also in nonlinear optical regime, for two, three dimensional Photonic Crystals and more general optical devices. In fact, the QNM's theory is retained the natural approach to study open cavities, useful to confirm the results of the literature for PBGs in classic ambit and in nonlinear regime; besides, even if the QNM's theory may involve boring calculations, it is retained one of the more effective methods of physical analysis to exam deeper the phenomena of nonlinear cavity quantum electrodynamics, and then to propose new applications in CQED.

[1] M. Born and E. Wolf, *Principles of Optics* (Macmillan, New York, 1964).

[2] A.W. Crook, J. Opt. Soc. Am. A **38**, 954 (1948).

[3] B. J. Dalton, Stephen M. Barnett, and B. M. Garraway, Phys. Rev. A. **64**, 053813 (2001).

[4] A. E. Siegman, Phys. Rev. A **39**, 1253 (1989); **39**, 1264 (1989).

[5] A. G. Fox and T. Li, Bell Syst. Tech. J. **40**, 453 (1961).